\documentclass{aa}
\usepackage{graphicx}
\usepackage{caption}
\usepackage{txfonts}
\usepackage{epstopdf}
\usepackage{color}
\usepackage{multirow}
\usepackage{lscape}
\usepackage{amsmath}
\usepackage{amssymb}
\usepackage{upgreek}
\usepackage{hyperref}
\begin{document}

\title{NELIOTA: New results and updated statistics after 6.5 years of lunar impact flashes monitoring}
     \author{A. Liakos\inst{1}
             \and
              A. Z. Bonanos\inst{1}
             \and
              E. M. Xilouris\inst{1}
             \and
              D. Koschny\inst{2}
             \and \\
             I. Bellas-Velidis\inst{1}
             \and
             P. Boumis\inst{1}
             \and
             A. Maroussis\inst{1}
            \and
            R. Moissl\inst{3}
            }
   \institute{Institute for Astronomy, Astrophysics, Space Applications and Remote Sensing, National Observatory of Athens,\\
              Metaxa \& Vas. Pavlou St., GR-15236, Penteli, Athens, Greece \\
              \email{alliakos@noa.gr}
             \and
              Lunar and Planetary Exploration, Technical University of Munich, 85521 Ottobrunn, Germany
              \and
              Near Earth Object Coordination Centre, European Space Research Institute (ESA/ESRIN), Largo Galileo Galilei 1, 00044 Frascati (Roma), Italy
              }

           \date{Received XX January 2024; accepted XX January 2024}


\abstract
{We present results of the NELIOTA campaign for lunar impact flashes observed with the 1.2~m Kryoneri telescope. From August 2019 to August 2023, we report 113 validated and 70 suspected flashes. For the validated flashes, we calculate the physical parameters (masses, radii) of the corresponding projectiles, the temperatures developed during the impacts, and the expected crater sizes. For the multiframe flashes we present light curves and thermal evolution plots. Using the whole sample of NELIOTA that encompasses 192 validated flashes in total from 2017, the statistics of the physical parameters of the meteoroids, the peak temperatures of the impacts and the expected crater sizes has been updated. Using this large sample, empirical relations correlating the luminous energies per photometric band were derived and used to roughly estimate the parameters of 92 suspected flashes of the NELIOTA archive. For a typical value of the luminous efficiency, we found that the majority ($>75\%$) of the impacting meteoroids have masses between 1-200~g, radii between 0.5-3~cm and produced craters up to 3.5~m. 85\% of the peak temperatures of the impacts range between 2000 and 4500~K. Statistics regarding the magnitude decline and the cooling rates of the multiframe flashes are also presented. The recalculation of the appearance frequency of meteoroids, lying within the aforementioned ranges of physical parameters, on the Moon yields that the total lunar surface is bombarded with 7.4 sporadic meteoroids per hour and up to 12.6 meteoroids per hour when the Earth-Moon system passes through a strong meteoroid stream. By extrapolating these rates on Earth, the respective rates for various distances from its surface are calculated and used to estimate the probability of an impact of a meteoroid with a hypothetical infrastructure on the Moon, or with a satellite orbiting Earth for various impact surfaces and duration times of the missions.}



\maketitle
%

\section{Introduction}
\label{sec:intro}

The increasing number of artificial satellites and space missions, especially during the last decade and the following few years, increased the interest for the study of meteoroids, since the probability of an impact with a space vehicle is not negligible anymore (e.g., the micrometeoroid impact on JWST six months after its launch). In addition, during the last decade, space agencies focussed more and more on the construction of bases on Moon (e.g., the NASA-Artemis mission) and, later, on other planets (e.g., Mars). Due to the lack of atmosphere in the aforementioned `targets', meteoroids may hit directly the established infrastructure and cause serious damage. Therefore, the space vehicle designers and the civil engineers should be aware of the appearance frequency of these objects as well as of their affect on infrastructure (i.e., due to the impact or the increased temperature) in order to calculate the probability of being hit and establish appropriate mitigation measures.

Lunar impact flashes (hereafter LIFs) have been studied for more than 25 years and they are caused by the larger meteoroids of cm--to--dm sizes. These object sizes can be considered as the most dangerous and they are those that produce observable LIFs from the Earth. On the contrary, the majority of the meteors observed on Earth are produced by smaller size meteoroids and only those that produce bolide or fireball events can be compared in size with those that produce the observed LIFs. Fireballs and bolides are rarely observed from a given site (i.e., the sky coverage is approximately $6\times10^4$~km$^2$) and large networks are needed for their detection. Meteors follow different physics (i.e., evaporation due to the atmospheric friction) than that of LIFs and can provide us with important information, such as velocities, origin, meteorite remnants. The Moon, however, is used as an extremely large impact surface (i.e., the near side of the Moon is $1.9\times10^7$~km$^2$) for studying the impacts of larger size meteoroids and asteroids. Observations of LIFs are very important as they can provide the size frequency distribution of meteoroids and also directly affect space industry.

In the absence of an atmosphere on the Moon, meteoroids directly impact its surface. Their kinetic energy is converted to: a)~heating the impacted material, b)~kinetic energy of the ejecta, and c)~crater excavation. The heating of the material produces light emission (i.e., luminous energy $LE$), which generates the so-called impact flash. According to the results of \citet{LIA20}, the peak temperatures of the flashes vary between 1500-6000~K, while their thermal evolution in time, in general, shows rapid decrease \citep[see also][]{BOU12}. The majority of the flashes last less than 66~ms and they mostly emit in the near-infrared passbands \citep{LIA20}.

One of the major open questions in the study of LIFs concerns the luminous efficiency $\eta$ of the LIFs that is the fraction of the luminous over the kinetic energy. Using data from LIFs observed during meteoroid streams (i.e., known velocities of the projectiles from the respective meteor showers on Earth), many research teams \citep{BEL00a, BEL00b, ORT06, MOS11, SWI11, BOU12, SUG14, MAD15a, MAD17, MAD18, MAD19a} have resulted in values between $5\times10^{-4}<\eta<5\times10^{-3}$. The NELIOTA team, using dualband observations, determined that $\eta$ is wavelength dependent and found correlations between $\eta$ per band ratio and the peak temperature of the flash \citep{LIA20}.

The `Near-Earth objects Lunar Impacts and Optical TrAnsients' (NELIOTA\footnote{\url{https://neliota.astro.noa.gr/}}) project began in 2015 at the National Observatory of Athens (NOA) and was funded by the European Space Agency (ESA). The observations began on February 2017 and officially ended in mid-August 2023. The NELIOTA team has published so far three papers in peer-reviewed journals, namely, \citet[][hereafter Paper~I]{BON18}, \citet[][hereafter Paper~II]{XIL18}, \citet[][hereafter Paper~III]{LIA20}, while other research works were also based on the NELIOTA data \citep{AVD19, MUN20, AVD21}.

In this paper, we present the detailed results for the LIFs observed from NELIOTA after July 2019 following the same methods described in Paper~III. Section~\ref{sec:OBS} briefly describes the observational strategy, the followed validation procedure, and the observations efficiency statistics of the programme. Section~\ref{sec:CAT} hosts the catalog of the observed LIFs during the aforementioned time-period that includes their photometric measurements and coordinates. The physical characteristics of the meteoroids produced these LIFs, the peak temperatures of the impacts and the expected crater sizes are given in Section~\ref{sec:PHY}. Additionally, in this section, using all the available data from the inception of NELIOTA, statistics for the physical properties of these objects and the impacts are also presented. Updated correlations between the properties of meteoroids and LIFs are given in Section~\ref{sec:COR}. In Section~\ref{sec:FREQ}, using the whole sample of NELIOTA, we derive the detection rate of LIFs and calculate the appearance frequency of the meteoroids on Moon and, by extrapolation, in the vicinity of the Earth. Moreover, we calculate the meteoroid impact probability with a potential lunar infrastructure and with an orbiting satellite. Section~\ref{sec:DIS} contains a discussion about the observations, results, and the future studies of LIFs, based on our experience, while Section~\ref{sec:SUM} contains a summary of the current results and concluding remarks.

\section{Observations and efficiency statistics of the campaign}
\label{sec:OBS}

The long-term monitoring of the Moon for LIFs utilized the 1.2~m Kryoneri telescope\footnote{\url{https://kryoneri.astro.noa.gr/en/}} (prime focus optical design, f/2.8) and a twin fast-frame camera system installed on a dichroic beam splitter (see Paper~II for details). We observe the nightside of the Moon in the $R$ and $I$~passbands simultaneously (with a synchronization accuracy of 6~ms) with a frame rate of 30~fps. The exposure time of the cameras is 23~ms followed by a readout time of 10~ms. Our set-up has an angular Field-of-View (FoV) of $16'\times14.4'$ and covers a mean lunar surface of $3\times10^6$~km$^2$. The Moon is observed during approximately the illumination phases 0.1$\pm0.01$ and 0.45$\pm0.01$ depending on the Earth-Moon distance. The observations are divided into 15~min lunar chunks separated by standard stars observations chunks. The telescope, cameras and the rest subsystems are controlled by a custom developed software (NELIOTA-OBS), which is also used for the observation planning and the data storing.

The data are reduced with a second custom developed software (NELIOTA-DET), which detects the potential events in the difference images. The validation is based on visual inspection by an expert user and follows strictly the NELIOTA validation flowchart \citep[Paper~III/Fig.~3 and][]{LIA19}. Briefly, detections in both cameras that do not appear as  moving in successive frames are characterized as validated LIFs. Multi- or single-frame detections (with similar photometric profiles, i.e., images, with the standard stars) in $I$~passband only are characterized as suspected LIFs of class 1 (SC1) and 2 (SC2), respectively.

We assume that false detections coming from satellite glints shorter than our exposure time are so rare that they are negligible. For this, we have checked all validated and suspected flashes against two publicly available databases, one from spacetrack.org\footnote{\url{https://www.space-track.org}} and the other one by Bill Gray\footnote{\url{https://www.github.com/BillGray/Tles/}}. We checked all event times for possible satellites within 0.25~degrees to the Moon center. For the validated events, this is the case in 14\% of the cases. We assume that about 90\% of all satellites are in orbits low enough that they should have been recognised by their movement, even during a short light reflection on one of their surfaces. This leads to a very first rough estimate for a false detection percentage of $0.14~*~(1~-~0.9)~<~1.4\%$. Only a small percentage of these will show a glint that can be mistaken for an impact flash, further reducing the probability of this to happen. For the rest of the paper, we thus neglect any satellite contamination. A more detailed estimate of any false detections is beyond the scope of this paper.

Aperture photometry is used for the photon flux calculation per band for the validated and suspected LIFs based on the standard star observations. Using highly detailed maps of the Moon (taken from the Virtual Moon Atlas Software\footnote{\url{https://ap-i.net/avl/en/start}}) and the localization tool of the NELIOTA-DET software, we estimate the selenographic coordinates of the LIFs with an error of $0.5\degr$. The data of the validated and suspected LIFs are, subsequently, uploaded into the project's website\footnote{\url{https://neliota.astro.noa.gr/DataAccess}} within 24~hours of observations, by employing another custom developed software (NELIOTA-ARC), and made publicly available.

\begin{figure}[t]
\centering
\begin{tabular}{c}
\includegraphics[width=8.65cm]{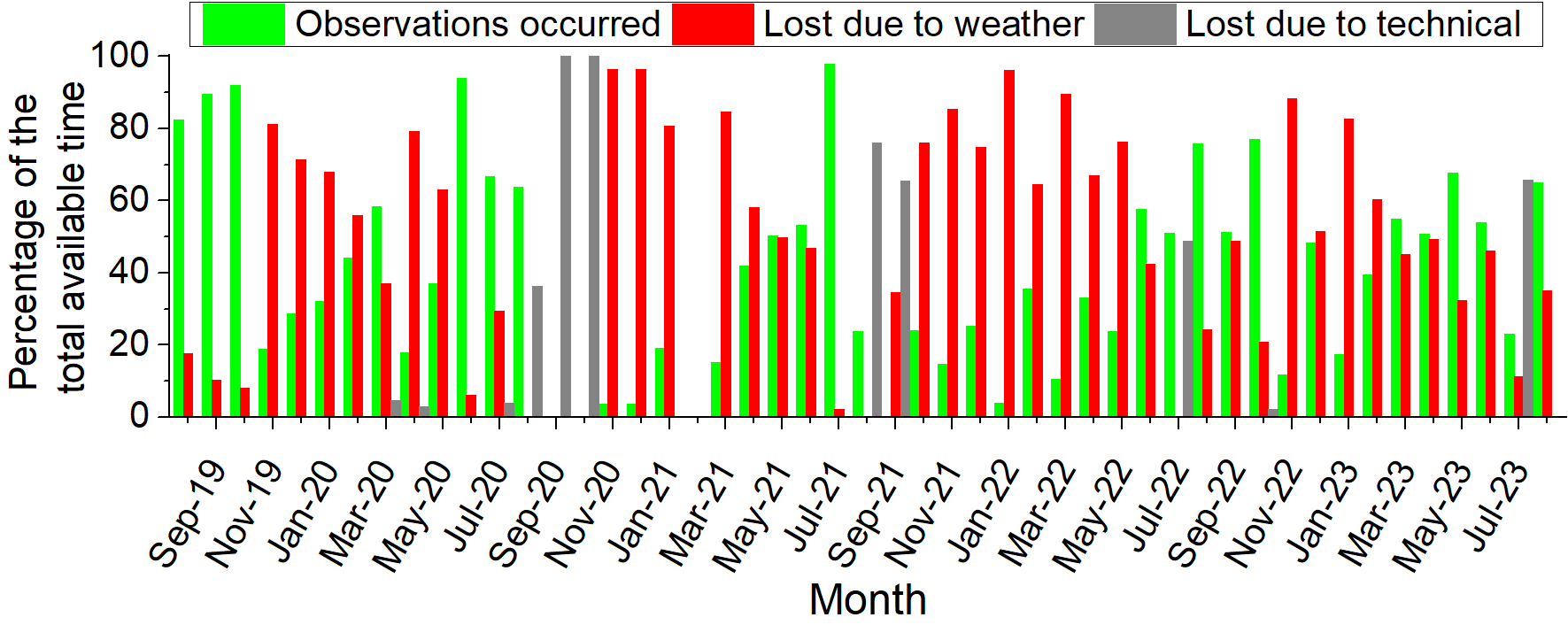}\\
\includegraphics[width=8.65cm]{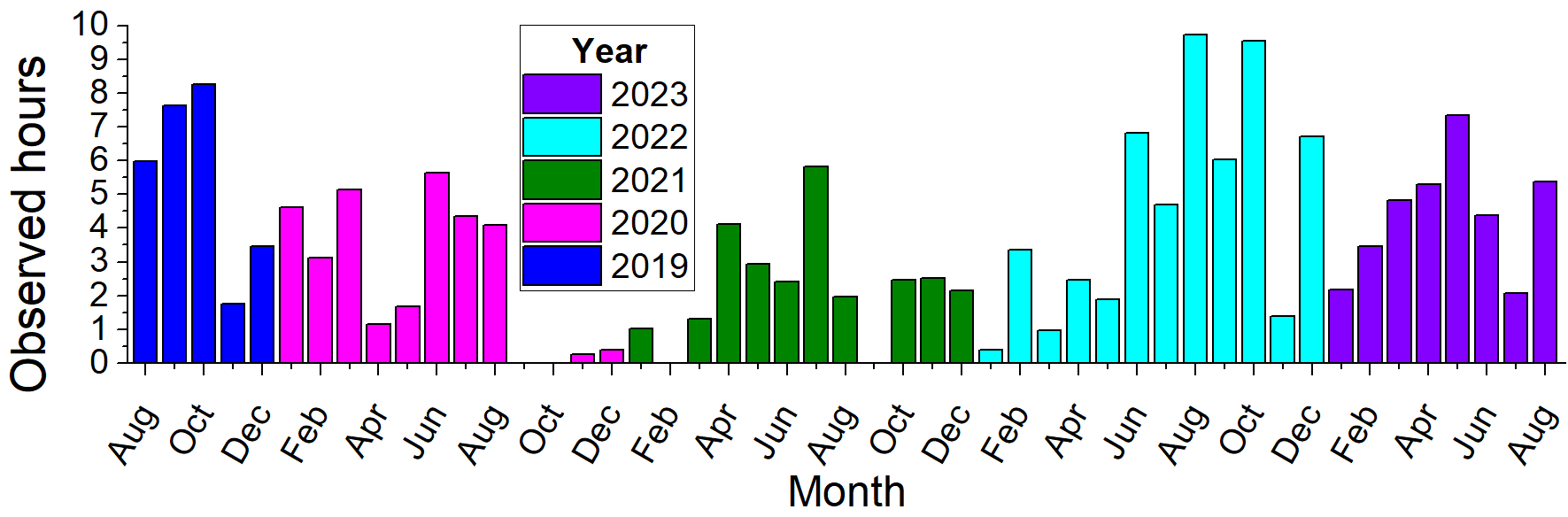}\\
\includegraphics[width=8.65cm]{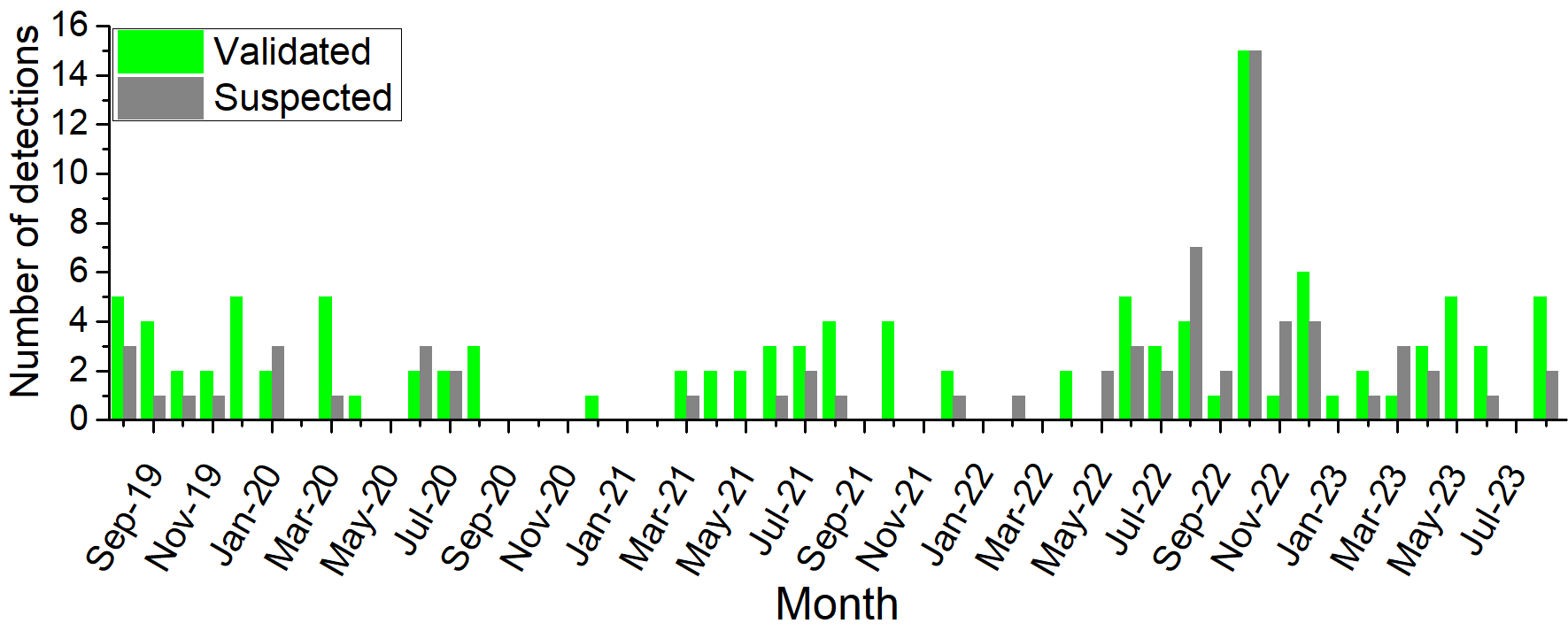}\\
\end{tabular}
\caption{Observational statistics of the NELIOTA campaign between August 2019 and mid-August 2023. The upper plot shows the percentages of: a) the observed time, b) the time lost due to bad weather conditions, and c) the time lost due to technical reasons for each month of the campaign. The middle and lower plots show the distributions of the monthly observed hours (exc. readout time and standard stars observations) and the LIF detections, respectively.}
\label{fig:obs-stats}
\vspace{0.3cm}
\includegraphics[width=4.5cm]{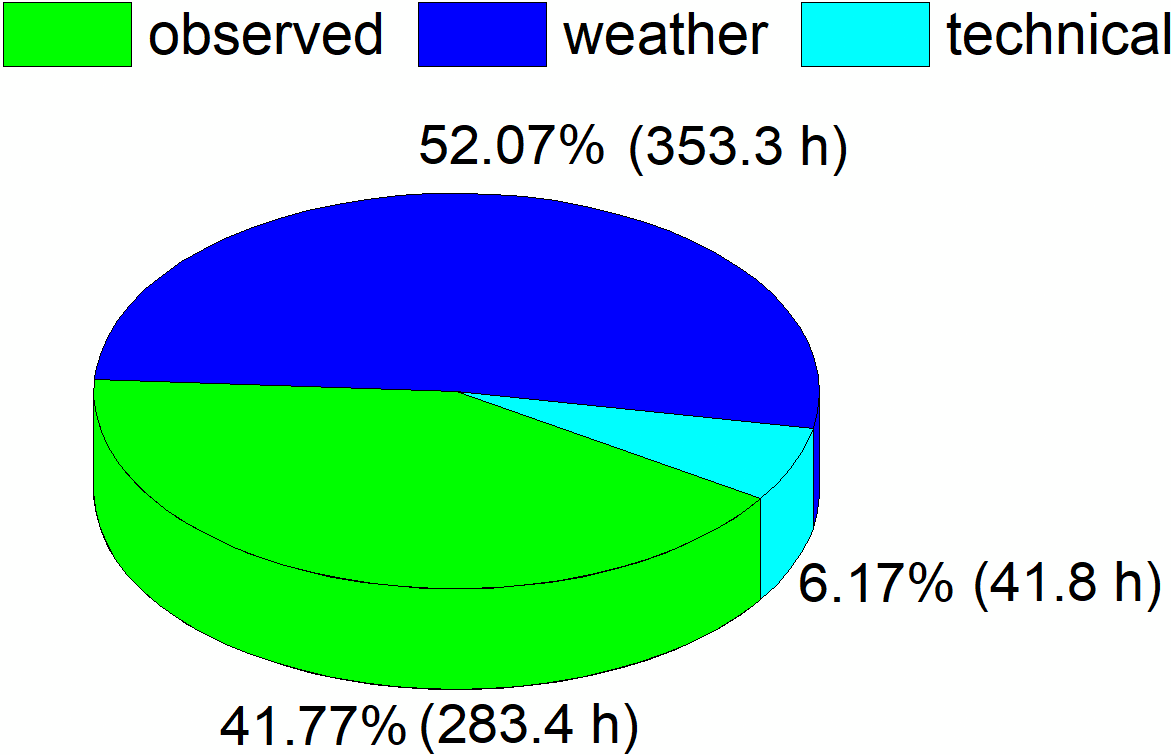}\\
\caption{Observations efficiency of the NELIOTA campaign between February 2017 and mid August 2023.}
\label{fig:pie}
\end{figure}

The observational statistics regarding the observed hours, number of detections and efficiency of the campaign for the period August 2019 - mid August 2023 are plotted in Fig.~\ref{fig:obs-stats}. The total available time for this time period was approximately 679~h. It should be noticed that approximately 35-40\% of this available time is a-priori lost due to the readout time of the cameras (30\%) and the standard stars observations (5-10\%), cf. Paper~III. Therefore, the true available time on Moon was actually 429.56~h. From this true available time, the Moon was observed for a total of 172.9~h (40.3\%) (see Fig.~\ref{fig:obs-stats}-upper and middle panels), while 217.2~h (50.6\%) and 39.5~h (9.1\%) were lost due to bad weather conditions and technical issues, respectively (see Fig.~\ref{fig:obs-stats}-upper panel). Within this time period, 113 validated and 70 suspected flashes were detected (see Fig.~\ref{fig:obs-stats}-lower panel). The respective statistics for the period before August 2019 can be found in Paper~III.

The total observing efficiency of NELIOTA is presented in a pie chart (Fig.~\ref{fig:pie}) that shows the absolute values and the percentages of the observing time, the time lost due to bad weather conditions, and the time lost due to technical reasons since 2017. It should be noticed again, that the values concern the true time durations, meaning that readout time and time spent for standard stars are excluded. Moreover, we clarify that this pie chart concerns only the time windows of the nights (i.e., 20~min up to 5~h) dedicated to lunar observations and is not the global statistics of observations of the Kryoneri Observatory. In total, we have observed the Moon for 283.4~h (287 nights spent) which corresponds to 41.8\% of the total true available time (678.5~h in 466 nights). The 52.1\% of the total true available time was lost due to bad weather conditions and 6.2\% due to technical issues.

\section{Updated catalog and statistics of LIFs}
\label{sec:CAT}

The observational results for all the detected flashes between August 2019 - mid August 2023 are given in Table~\ref{tab:list}. The LIFs are presented in chronological order (increasing number) and for each one we list: a)~the date and the UT timing of the detection, b)~the characterization for its validity (for details see Paper~III), c)~the maximum duration ($t_{\rm max}$ in ms), d)~the peak magnitudes in $R$ and $I$~passbands, and e)~the selenographic coordinates, namely latitude ($Lat.$) and longitude ($Long.$). The light curves of the multiframe LIFs are illustrated in Fig.~\ref{fig:LCs1}.

It should be noted that in these tables there are a few cases of LIFs that, although detected in both passbands (\#115, \#119, \#149, \#178, \#263), they are characterized as SC2. This is due to their very low $R-I$ index (less than 0.15), which leads to peak temperatures greater than 8500~K. As found in Paper~III and will be shown later in Section~\ref{sec:PHY}, the typical evolved temperatures of LIFs are much lower than this value. An alternative explanation is that they are simply GEO satellites/debris crossing the FoV. Therefore, we consider these events as less likely suspected LIFs. Event \#137 presented a very strange photometric behaviour. It was detected in five frames in both bands with a tiny magnitude variation from frame to frame and the $R-I$ indices varied between 0.28 and $-0.26$. Moreover, a slight displacement between successive frames was detected but it was too small to be discarded directly as a moving object. We estimate that this event is more likely a GEO satellite/debris. Event \#203 had a very elongated photometric profile and although it was relatively bright in the $I$~passband (8.84~mag), it was detected only in one set of frames. Its $R-I$ index is 0.38~mag, which corresponds to a relatively high temperature of approximately 5800~K. Therefore, again, we cannot be certain for its nature and we simply prefer to consider it as a suspected LIF.

\begin{figure}[t]
\centering
\begin{tabular}{c}
\includegraphics[width=8.65cm]{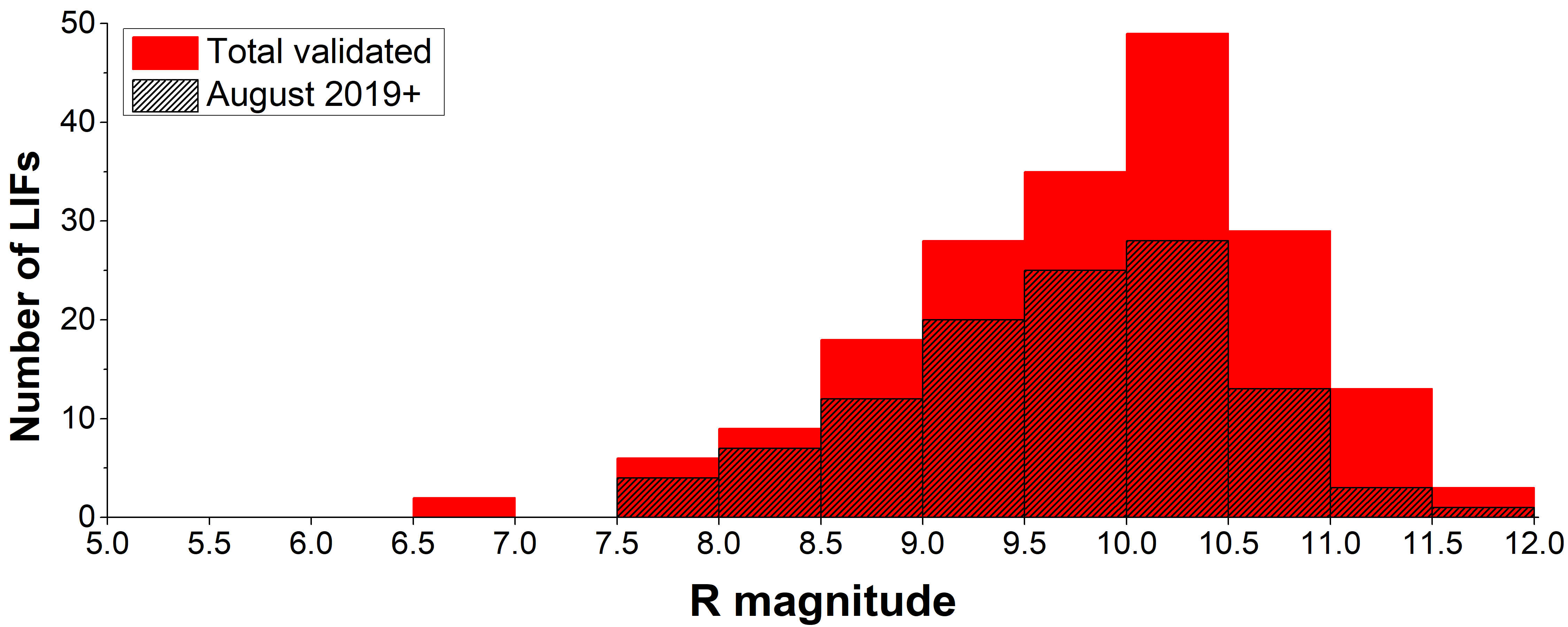}\\
\includegraphics[width=8.65cm]{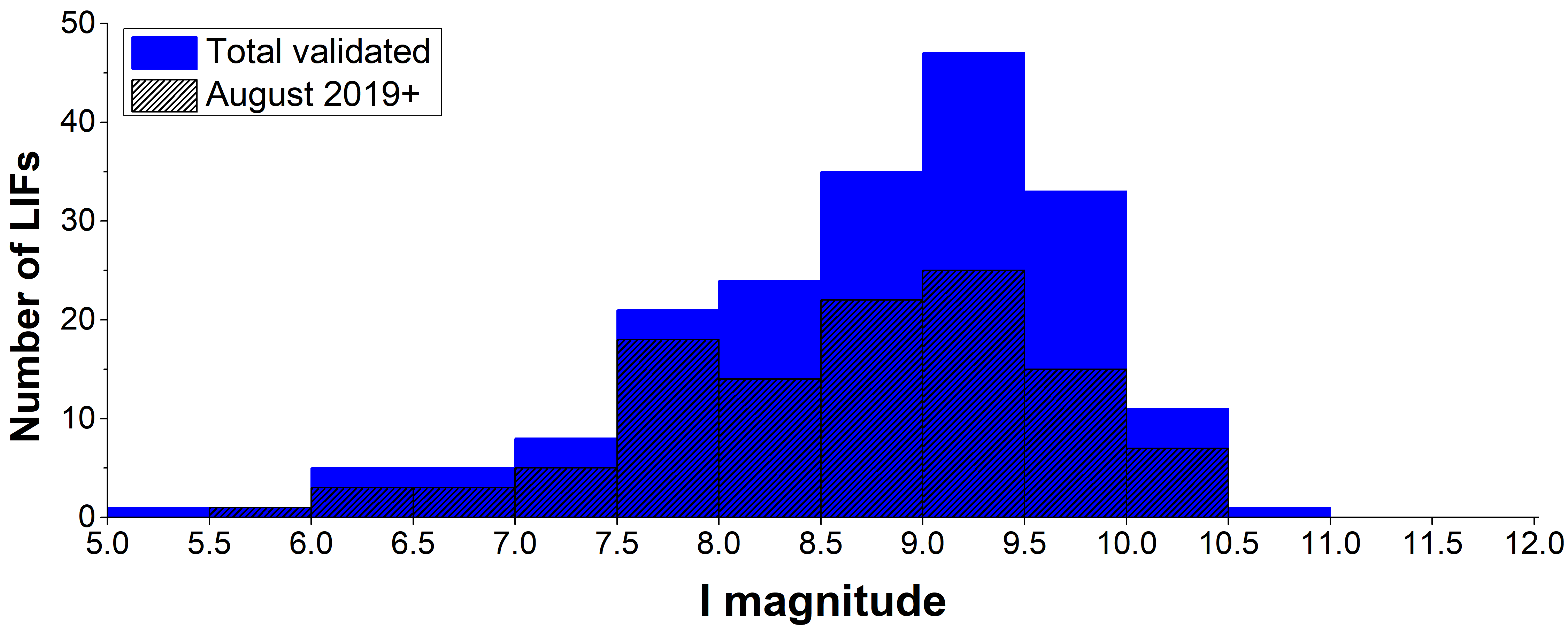}\\
\includegraphics[width=8.65cm]{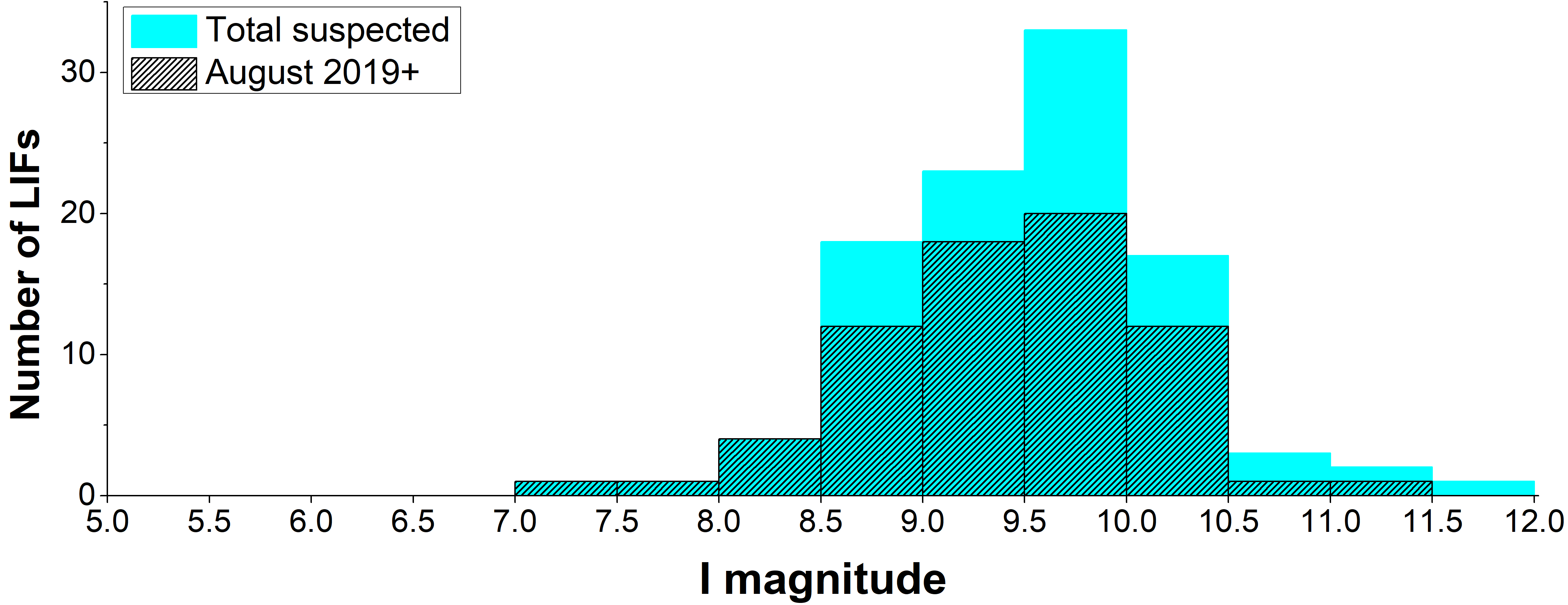}\\
\end{tabular}
\caption{Magnitude distributions of LIFs detected by NELIOTA since 2017 (colored patterns) and after July 2019 (black patterns). Upper and middle panels show the $R$ and $I$~passband magnitude distributions for the validated flashes. The lower panel is the $I$-mag distribution for the suspected flashes.}
\label{fig:obs_mags}
\end{figure}

The magnitude distributions of the validated and the suspected LIFs for the given time-period are shown in Fig.~\ref{fig:obs_mags}. For reasons of completeness, this figure illustrates also the respective magnitude distributions since the start of NELIOTA operations in 2017. It should be noticed, that there are not any suspected flashes detected only in the $R$~passband. Arguments for this can be found in Paper~III/Section~3.1. As can be seen from the $R$~passband distribution, the vast majority (73.4\%) of the LIFs detected by NELIOTA range between 9-11~mag. The LIFs between 10-11~mag are 40.6\% of the whole sample, while those between 11-12~mag are 8.3\%. Therefore, our completeness limit is between 9-9.5~mag and 9.5-10~mag in the $I$~passband for the validated and suspected LIFs, respectively, and between 10-10.5~mag in the $R$~passband. It should be noted that the latter percentages show that approximately the half of the observed LIFs cannot be detected by small-size telescopes (i.e., with diameters of the order of a few tens of centimeters). These fainter LIFs are numerous and, as can be seen in the following section, their majority were produced by low-mass meteoroids, whose kinetic energies cannot be considered negligible in case of an impact with a space vehicle or a lunar base. As shown in the works of \citet{BRO10} and \citet{SUG14}, the lower the kinetic energies of the meteoroids, the higher their collision probability. The latter, assuming similar travel velocities of the sporadic meteoroids, can be also used for the mass population determination, that is the lower the masses of the meteoroids, the higher their populations. The latter is more or less what is expected from the evolution of the minor bodies. Obviously, the reason for not detecting more LIFs between 11-12~mag is the observation limit of our setup. The $I$~passband distribution of the validated LIFs (Fig.~\ref{fig:obs_mags}-middle panel) shows clearly that the same LIFs in the $R$~passband are statistically one magnitude brighter in $I$. This confirms that the LIFs are relatively cool events, as concluded in Papers~I and III. It should be noticed that in the $I$~passband magnitude distribution of the validated LIFs (Fig.~\ref{fig:obs_mags}-middle panel), there is only a small number of detections between 10-11~mag. This is because, as mentioned before, the respective light emission in $R$~passband is very close to our observational limit, hence they cannot be validated. On the other hand, the magnitude distribution of the suspected LIFs (Fig.~\ref{fig:obs_mags}-lower panel) shows that 19.6\% range between 10-11~mag and 3.2\% between 11-12~mag. The reason that we did not detect more suspected LIFs fainter than 10~mag is again our observational limit superimposed by the difficulty to distinguish faint suspected LIFs from low-energy cosmic rays or faint slow-moving satellites/debris.

\begin{figure}[t]
\centering
\includegraphics[width=9cm]{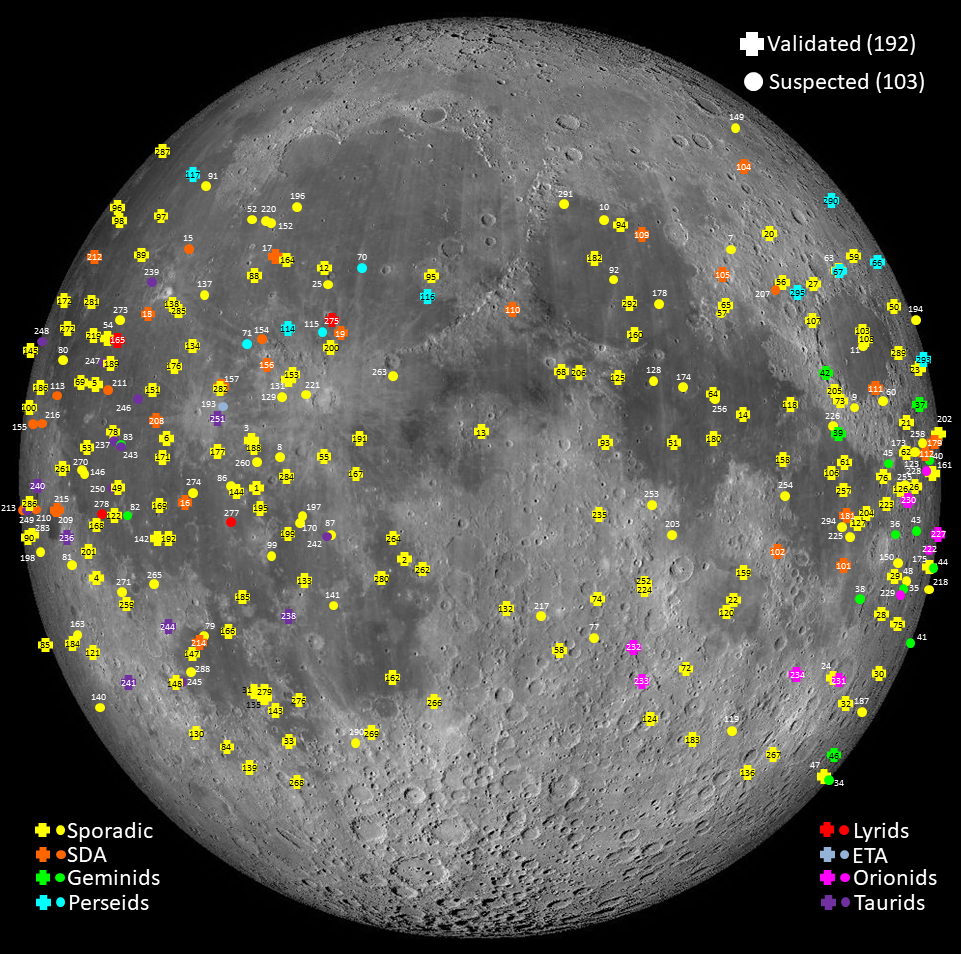}
\caption{Locations of LIFs detected by the NELIOTA campaign since February 2017. Detections after July 2019 have numbers greater than 113. Crosses and filled circles denote the validated and the suspected flashes, respectively. Different colors denote association of the projectiles with active meteoroid streams (see text for details and Paper~III).}
\label{fig:localization}
\end{figure}

The locations of the detected LIFs on Moon are shown in Fig.~\ref{fig:localization}. This figure shows all detections of NELIOTA, while those that correspond to the time-period after July 2019 have an ID number greater than 113. LIFs with colors other than yellow denote that the impacting meteoroids were associated with active meteoroid streams during the time of the observations. For the latter, the method of \citet{ORT15} and \citet{MAD15b, MAD15a} was implemented (cf. Paper~III). In this figure, LIFs that appear exactly on the limb or slightly out of it have longitude values greater than this background lunar map includes. Their locations were included in our FoV due to the lunar libration at the timing of the observations.

\begin{figure}
\centering
\includegraphics[width=8.5cm]{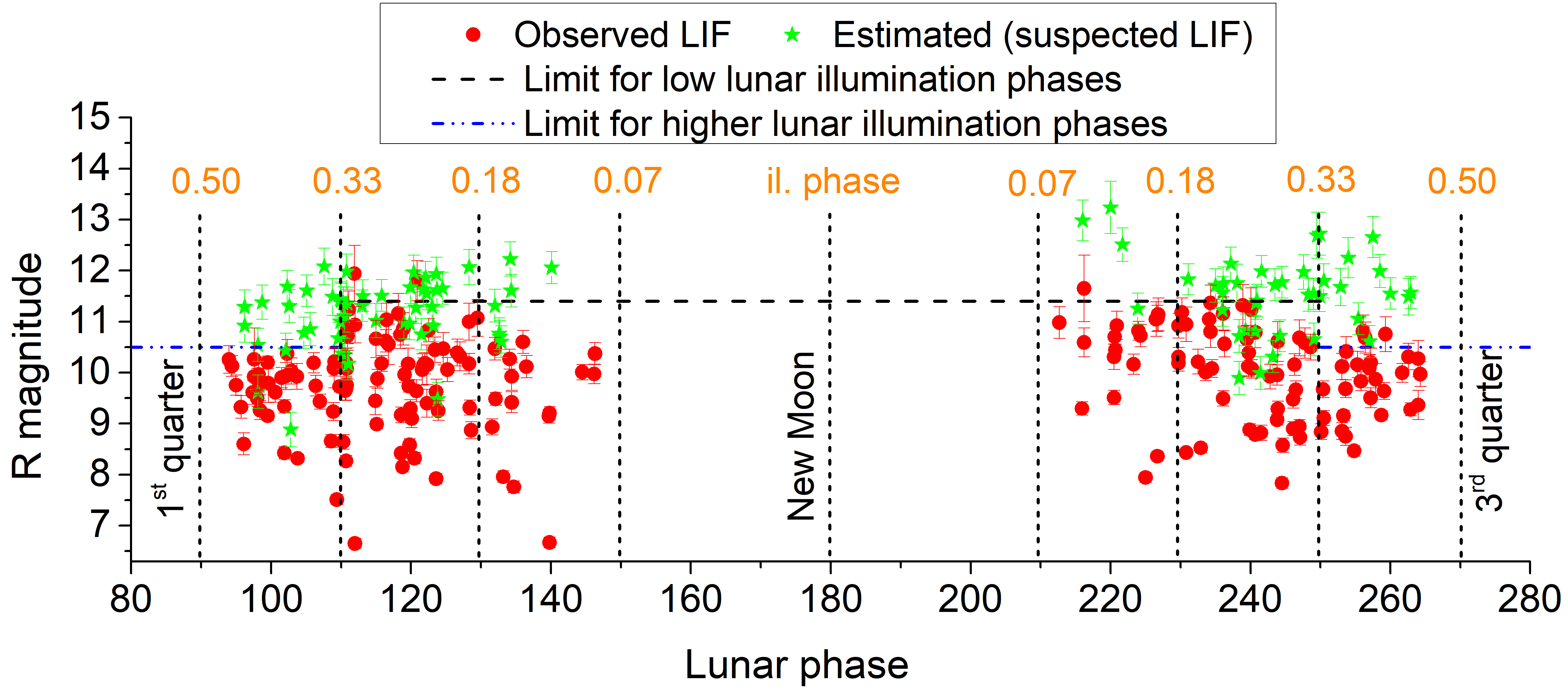}\\
\caption{$R$~passband magnitudes of validated (red points) and suspected (green stars) LIFs, observed since February 2017, against the lunar phase. Horizontal dashed lines indicate the empirically set detection limits of NELIOTA in $R$ filter according to the lunar phase (see text for details). Vertical short-dashed lines denote the range of the lunar phases during which NELIOTA was operating. For convenience, the respective illumination phases are also indicated (orange).}
\label{fig:Stat_R_phase}
\vspace{0.1cm}
\includegraphics[width=8.5cm]{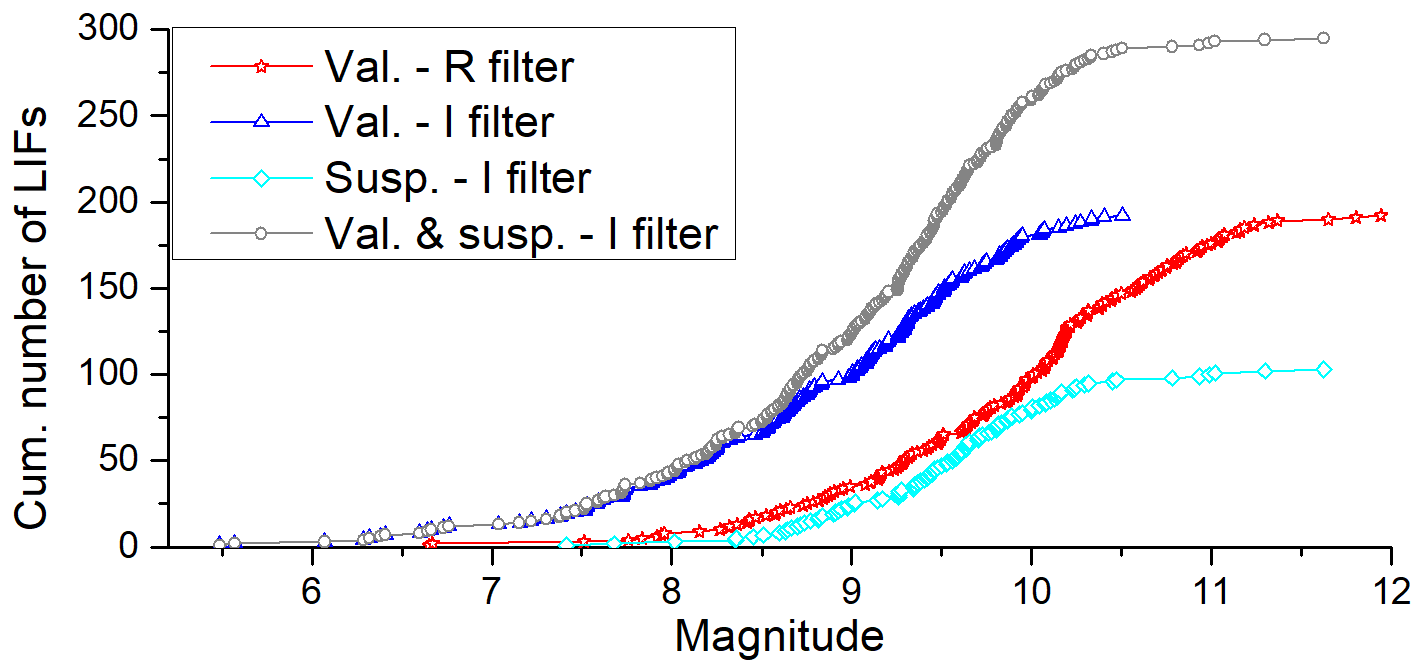}\\
\caption{Cumulative number of LIFs detected by the whole NELIOTA campaign (283.4~h of lunar monitoring) according to their peak magnitude values. Different colors denote different observed passbands and validation characterizations.}
\label{fig:Stat_Cum}
\vspace{0.1cm}
\includegraphics[width=8.5cm]{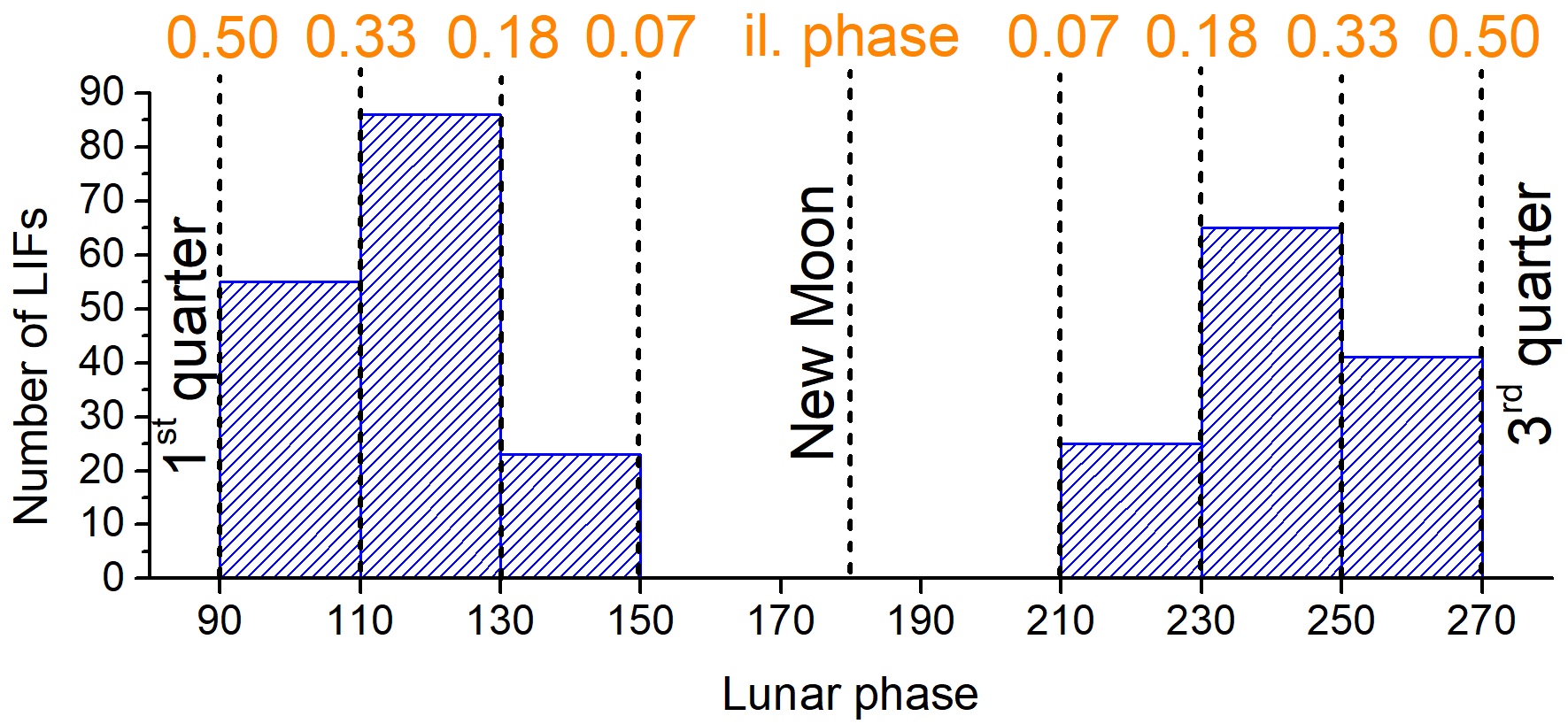}\\
\caption{Distribution of validated and suspected LIFs ($I$~passband), observed since February 2017, as a function the lunar and illumination phases.}
\label{fig:Stat_Hist}
\vspace{0.1cm}
\includegraphics[width=8.5cm]{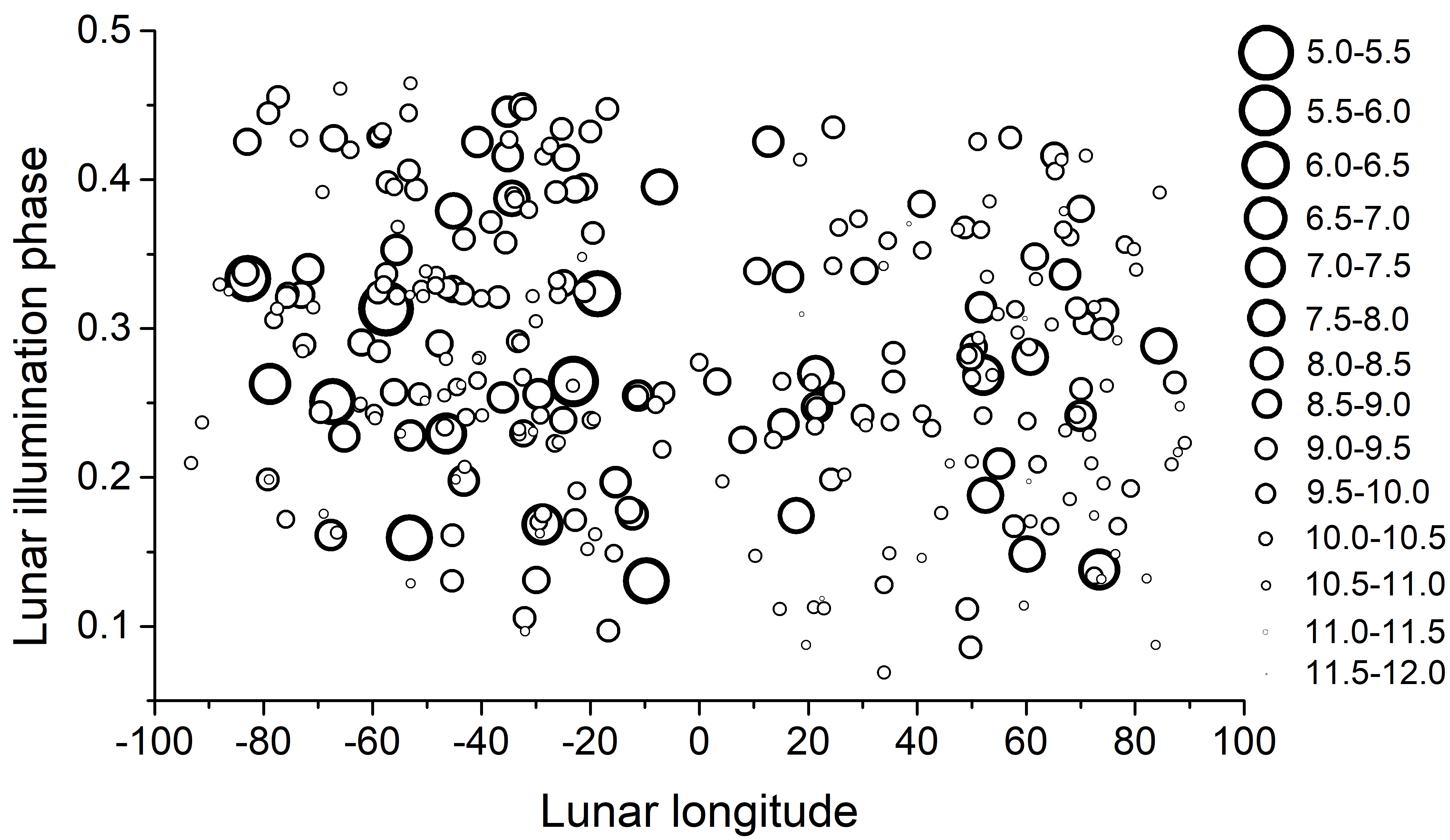}\\
\caption{Distribution of validated and suspected LIFs ($I$~passband) against lunar longitude and illumination phase. The bigger the symbol the brighter the LIF.}
\label{fig:Stat_Bubble}
\end{figure}

The detection limits of NELIOTA are not easy to be accurately determined. Although in Paper~II we showed that using this equipment we can reach a magnitude of 12 in $R$~passband, this limit varies because it depends on a) the lunar illumination phase, b) the local seeing conditions, c) the airmass of the Moon, d) the distance of the flash from the lunar terminator, that is, the dayside, and e) the Earth-Moon distance. Therefore, since the determination of our true detection limit is quite complex, we chose to present some graphs in order to obtain a qualitative view. The $R$ magnitude detections of the validated LIFs as well as the estimated magnitudes of the suspected LIFs (see Section~\ref{sec:CALC}) against the lunar phase are illustrated in Fig.~\ref{fig:Stat_R_phase}. Based on the validated LIFs, we clearly see that during the lunar phases in the ranges $110\degr$-$150\degr$ and $210\degr$-$250\degr$ that correspond to illumination phases 0.07-0.33, the upper limit of the detections is $\sim11.4$~mag. However, there are three validated LIFs above this limit that probably concern extremely good observing conditions. Beyond these lunar phase values (i.e., $>250\degr$ and $<110\degr$), which correspond to illumination phases greater than 0.33, the detection limit drops approximately 1~mag. Similarly to the previous range, there are also two validated LIFs above this threshold. Regarding the suspected (estimated $R$ magnitudes) LIFs, their majority lie well above our detection limits regardless the lunar phase. However, there are many expected detections below our limits. For these suspected flashes we argue that the aforementioned parameters (b, c, and d) prevented their detection or their $R$ magnitudes were actually fainter than estimated.

The turning points of the accumulation distributions of LIFs in Fig.~\ref{fig:Stat_Cum} can be used to roughly determine the detection limit of NELIOTA. For the $R$~passband, the turning point is at $m_{\rm R}\sim11.2$~mag, and after that there is a slight increasing part that reaches up to $\sim12$~mag. Therefore, we can estimate that, indeed, NELIOTA can marginally detect LIFs up to that magnitude, but, in general, can easily detect LIFs up to 11.2~mag in $R$~passband. Regarding the limiting $I$ magnitude, we use the LIFs of the suspected and validated LIFs because, as mentioned before, the suspected LIFs were probably too cool to be detected in $R$~passband as well. Hence, their inclusion provides more complete information regarding the limiting $I$ magnitude. The grey data points distribution in Fig.~\ref{fig:Stat_Cum} has a turning point at $\sim10.2$~mag and, similarly to $R$~passband, this is followed by a slight increasing part that ends to $\sim11.6$~mag. Therefore, we can plausibly conclude that our setup can typically detect LIFs up to $m_{\rm I}\sim10.2$~mag, and in very good observing conditions (e.g., low illumination phase, low airmass) can reach up to $\sim11.6$~mag.

The distribution of the detected LIFs in the $I$ filter against the lunar and illumination phases is illustrated in Fig.~\ref{fig:Stat_Hist}. The highest signal-to-noise ratio for a given magnitude of a LIF will be obtained close to new Moon, that is for illumination phases less than 0.18 (3-4~days past or before new Moon). Even though the background brightness due to Earthshine is the highest close to new Moon, the straylight from the illuminated part will be the lowest, hence the contrast of a LIF can be very high, even for a faint one. However, the duration of the observations during these phase parts is very short, typically less than 1~h. It should be noted that the Earthshine produces approximately 7-10 orders of magnitude less luminous energy flux than a typical LIF with a peak temperature between 2000-5000~K \citep{BOU12}, so it can be considered negligible. On the other hand, for illumination phases greater than 0.33 (5-7~days past or before new Moon) the observation duration is the longest possible (order of a few hours). However, the straylight of the dayside of the Moon increases the lunar background that much, so the LIFs are very hard to detect (see also Fig.~\ref{fig:Stat_R_phase}). In addition, when the Moon is close to the apogee (i.e., increased photons flux per pixel), the telescope has to point further away from the lunar terminator in order to avoid extremely high background values. Because of the latter, less lunar surface is covered. Therefore, according to Figs.~\ref{fig:Stat_R_phase} and \ref{fig:Stat_Hist}, the best balance between the observations duration and lunar phases is achieved between approximately 0.20-0.35 illumination phases.

The distribution of the magnitudes of the suspected and validated LIFs in the $I$~passband against the lunar illumination phase and their longitude is plotted in Fig.~\ref{fig:Stat_Bubble}. This plot was made in order to examine any dependence of the LIF detections on their distance to the lunar terminator. It should be noticed that this distribution is biased due to the fact that our setup covers only a portion of the full Moon disk. The majority of the detections were made between $\pm20\degr$ and $\pm80\degr$ longitude values. This can be explained by the lunar area coverage of our setup and the distance of the detections from the lunar terminator. Longitudes between $-20\degr$ and $20\degr$ are always closer to the lunar terminator, thus, the lunar background is always higher, hence the detection is more difficult. Moreover, regions close to the lunar center are included only when the Moon is close to the apogee, implying that more lunar surface is included in the FoV. On the other hand, when Moon is close to the apogee, more light is gathered in the pixels, hence, again, increased lunar background values are met. Detections in longitudes greater than $+80\degr$ and less than $-80\degr$ depend on the lunar libration at the time of the observations. We also see that LIFs brighter than 9~mag can be detected under all circumstances. More or less the same stands for LIFs between 9-10~mag with a slight difference that they seem to be favored by illumination phases between 0.2-0.35 (due to longer observations duration) and are mostly located in regions with longitudes greater than $20\degr$ and less than $-20\degr$. The majority of the LIFs fainter than 10~mag can be detected mostly during lunar illumination phases lower than 0.35 and at longitudes greater than $30\degr$.

It should be noticed that there are slightly more LIFs detections in the western hemisphere of the Moon (see Figs.~\ref{fig:localization} and \ref{fig:Stat_Bubble}). Particularly, 163 out of 295 LIFs (validated and suspected) were observed in this hemisphere, which results in a percentage of 55.3\%. This is explained by the slightly longer observing time spent in this hemisphere, totaly by chance, that is 145.97~h out of 283.4~h (51.5\%). Therefore, we justify that the detections of LIFs are independent of the observed hemisphere, hence, it can be concluded that the Moon is bombarded homogeneously.

\section{Physical parameters}
\label{sec:PHY}

\subsection{Calculation methods}
\label{sec:CALC}

We proceed to compute the physical properties of the impacting projectiles (i.e., masses, radii) that produced the observed LIFs after July 2019, the temperatures evolved during their impacts, and the sizes of the expected craters. The total luminous energies of the LIFs were calculated as described in Paper~III (Section~6.2), then their temperatures according to the photon fluxes from the $R$ and $I$~passbands (i.e., their magnitudes; see Papers~I and III).

The peak magnitudes and peak temperatures of the detected LIFs are given in Tables~\ref{tab:list}-\ref{tab:ResultsSusp}. It should to be noticed that these temperature values are not always derived by the respective peak magnitudes given in the same table. There are indeed cases (LIF \#191, \#209) whose peak magnitudes resulted in a cooler temperature value than that derived from other sets of $R$ and $I$ frames. The magnitudes and the temperatures of the multiframe flashes are given in Table~\ref{tab:multi}, their light curves in Fig.~\ref{fig:LCs1}, and their temperature curves in Fig.~\ref{fig:TCs1}.

For the calculation of the kinetic energy ($KE_{\rm p}$) of the meteoroids we assumed various values for the luminous efficiency $\eta$ within the range $5\times10^{-4}<\eta<5\times10^{-3}$ \citep{BEL00a, BEL00b, ORT06, MOS11, SWI11, BOU12, SUG14, MAD15a, MAD17, MAD18, MAD19a}. Subsequently, for the cases associated with meteoroid streams, the mass values were derived from the $KE_{\rm p}$ assuming velocities ($V_{\rm p}$) according to the streams of origin. For sporadic meteoroids, an impact velocity of 17~km~s$^{-1}$ was assumed \citep{SUG14}. For the radii calculation we, again, assumed densities ($\rho_{\rm p}$) according to the parent bodies of the stream meteoroids and a value of 1.8~g~cm$^{-3}$ for the sporadics. These assumed values were taken by \citet{BAD09} and for convenience of the reader are also listed in the appendix (Table~\ref{tab:streams}). The expected crater sizes were calculated using the scaling law of \citet{GAU74} \citep[cf.][]{MEL89} with the assumption of an impacting angle of 45$\degr$.

The results for the validated flashes are given in Table~\ref{tab:ResultsReal}, which are divided into two main parts and organized as follows: The first part (first five columns) contains the ID number of the LIF, its possible association with a meteoroid stream (see Section~\ref{sec:OBS} and Paper~III for details), the calculated luminous energy per observed passband ($LE_{\rm R}$ and $LE_{\rm I}$) and the peak temperature of the impact. The second part of this table lists the physical parameters of the impacting meteoroids, which are their kinetic energy ($KE_{\rm p}$), mass ($m_{\rm p}$), radius ($r_{\rm p}$), and the expected crater size ($d_{\rm c}$) for three different values of the luminous efficiency $\eta$. The complete formalism for the aforementioned methods is given in Paper~III/Sections~6.2.1 and 6.4. The respective results for the first 79 validated LIFs can be also found in that work.

\begin{figure}
\centering
\includegraphics[width=8.8cm]{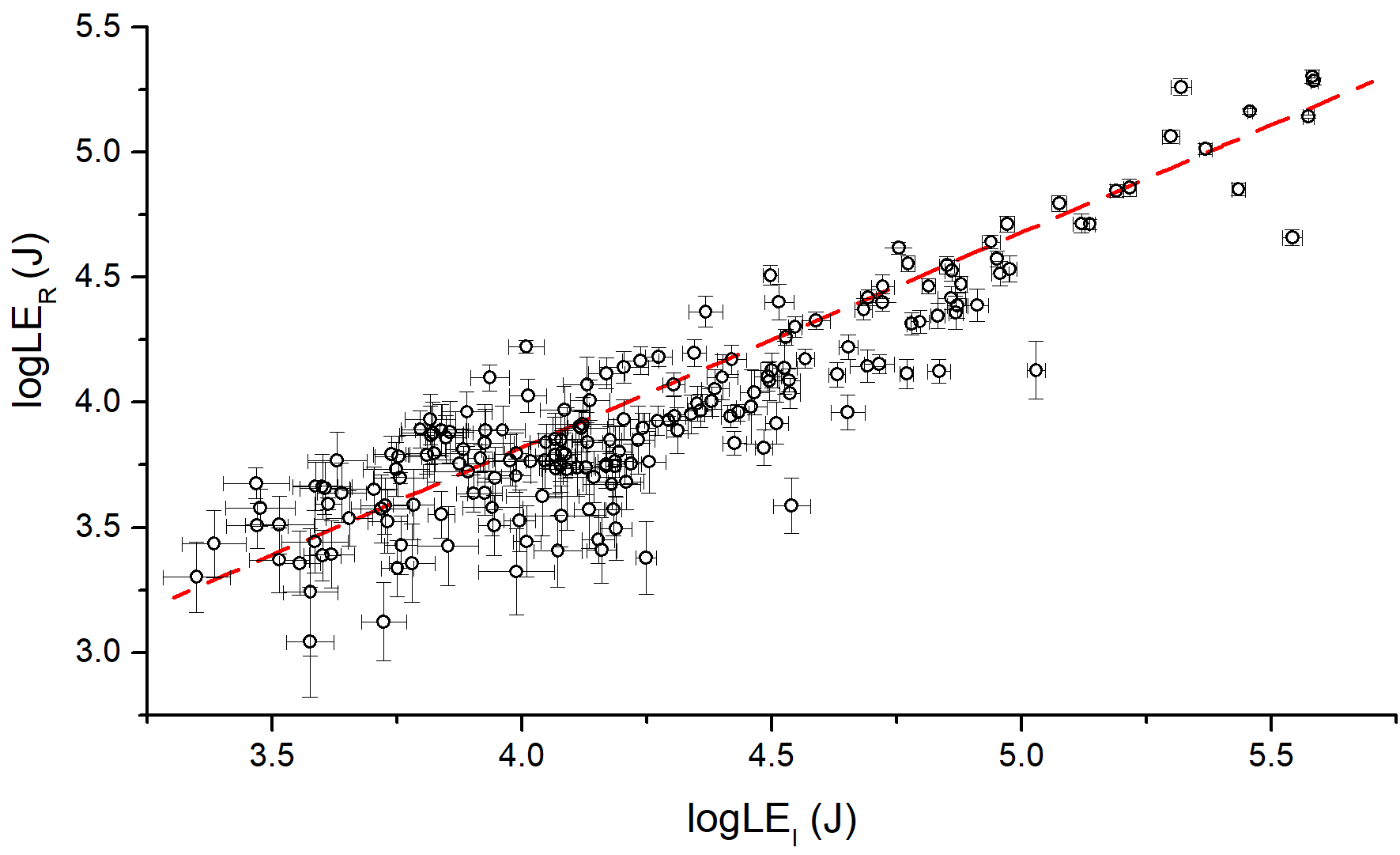}\\
\caption{Correlation between emitted energies in $R$ and $I$~passbands of the validated LIFs detected by NELIOTA since 2017.}
\label{fig:EREICOR}
\end{figure}

For a rough estimation of the physical parameters of the suspected flashes we based on the updated luminous energies per band correlation of the validated LIFs using the whole sample of LIFs detected by NELIOTA (cf. Paper~III). The data points are plotted in Fig.~\ref{fig:EREICOR} and using linear fitting we derive the following relation:

\begin{equation}
\log LE_{\rm R} = 0.38(9) + 0.86(2)\log LE_{\rm I}~{\rm with}~r=0.96,
\label{eq:LELR}
\end{equation}
where $r$ is the (Pearson's) correlation coefficient. Comparing Eq.~\ref{eq:LELR} with that from Paper~III, we notice that using larger sample, the two relations are in agreement but we constrained more (i.e., smaller error values) both the intercept and the slope, while $r$ is increased by $\sim6.7\%$. Using the above relation and the observed $LE_{\rm I}$ of the suspected LIFs, we are able to estimate the emitted $LE_{\rm R,~est}$. Subsequently, we calculate the photon flux in the $R$~passband and convert it to apparent magnitude $m_{\rm R,~est}$ (for method see Paper~III/Secs~6.1 and 6.2.2). Therefore, based on the same assumptions for $\eta$, $V_{\rm p}$, and $\rho_{\rm p}$, as in the cases of the validated LIFs, we are able to roughly estimate the temperature ($T_{\rm est}$) developed during the impact, the kinetic energy ($KE_{\rm p,~est}$), the mass ($m_{\rm p,~est}$), and the radius ($r_{\rm p,~est}$) of the impacting meteoroids. Finally, we also estimate the size of the expected formed craters ($d_{\rm c,~est}$). The aforementioned values are listed in Table~\ref{tab:ResultsSusp} for $\eta=1.5\times10^{-3}$. Despite the fact that the parameters of the first 33 suspected LIFs (Table~\ref{tab:ResultsSusp}) had been also presented in Paper~III, they are re-calculated and presented herein for reasons of consistency due to the update of the luminous energies relation (eq.~\ref{eq:LELR}). The SC2 LIFs \#52, \#70, \#71, \#81, \#115, \#119, \#137, \#149, \#178, \#203, \#263 were excluded from this table since their nature is rather ambiguous (see Paper~III for the first four ones and Section~\ref{sec:OBS} for the rest).

\subsection{Statistics of the physical properties of the meteoroids and impacts}
\label{sec:STAT}

\begin{figure}
\centering
\includegraphics[width=8.8cm]{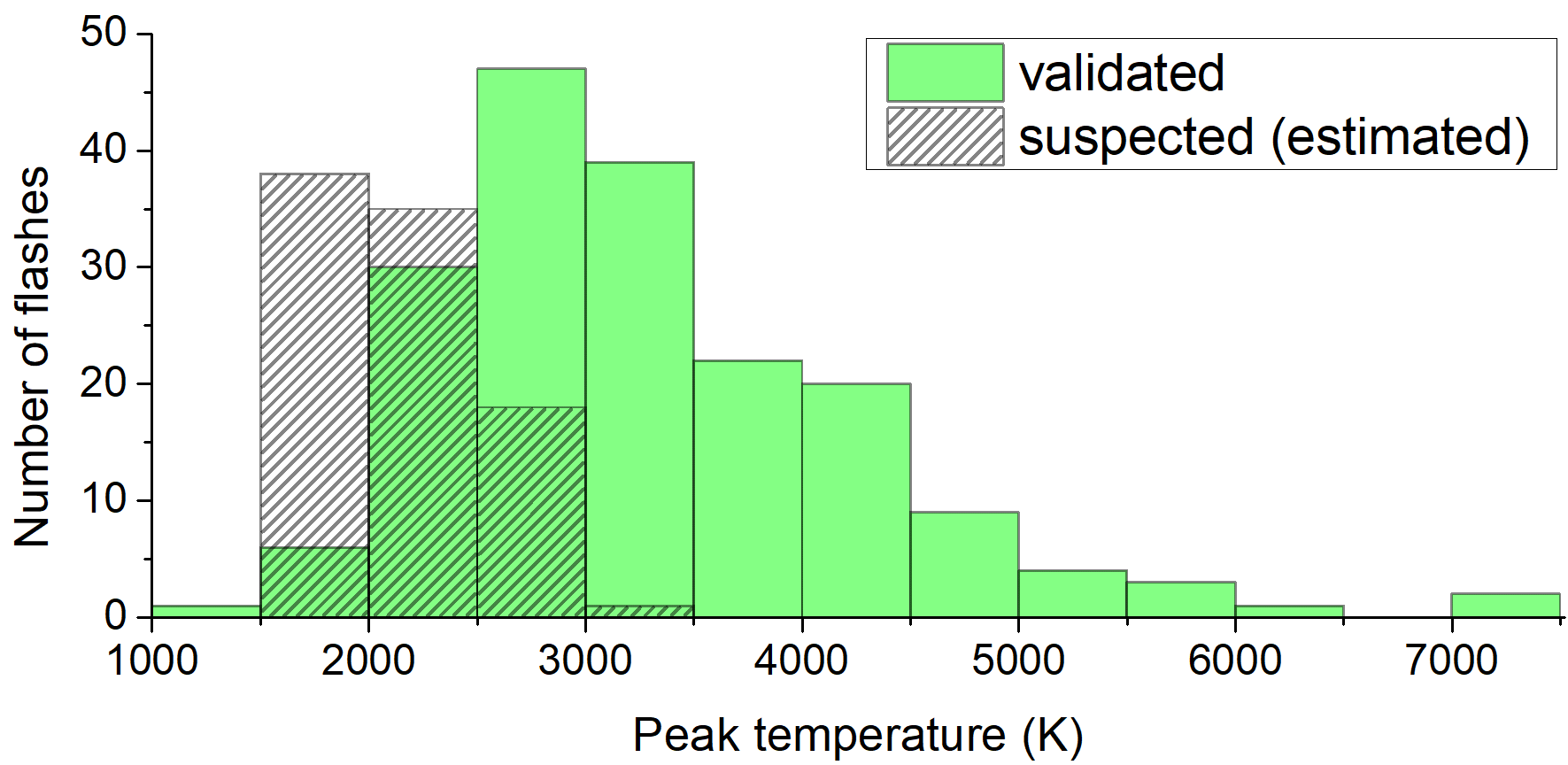}\\
\caption{Peak temperature distributions of validated and suspected LIFs detected by NELIOTA since 2017.}
\label{fig:DisT}
\vspace{0.3cm}
\includegraphics[width=8.8cm]{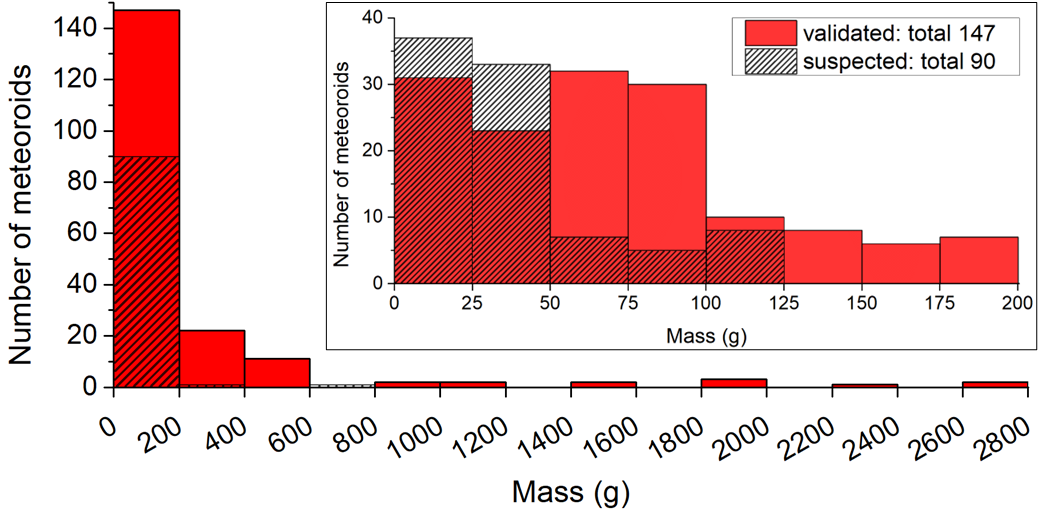}\\
\caption{Mass distributions of meteoroids producing the observed LIFs for different ranges (samples) and for $\eta=1.5\times10^{-3}$.}
\label{fig:DisM}
\vspace{0.3cm}
\includegraphics[width=8.8cm]{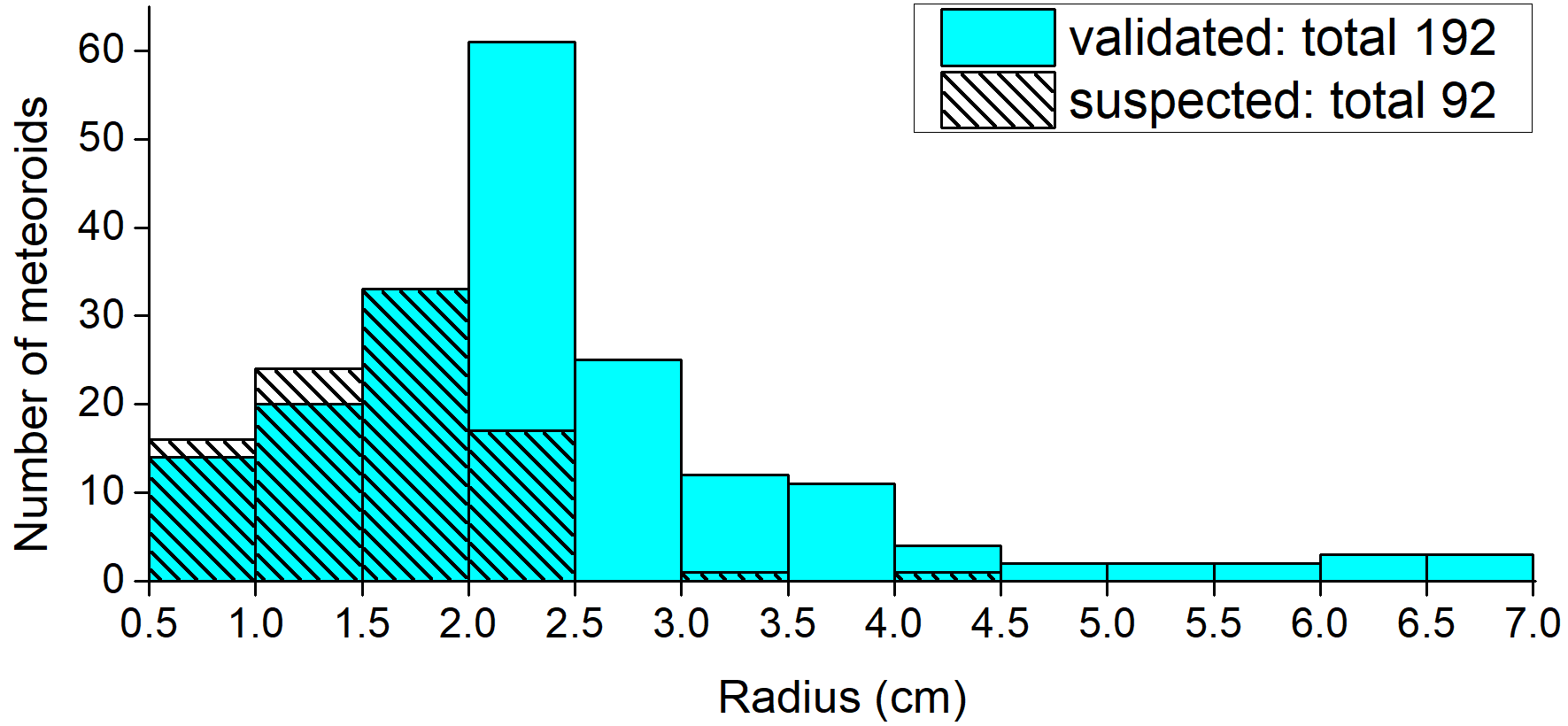}\\
\caption{Radius distribution of meteoroids producing the observed LIFs for $\eta=1.5\times10^{-3}$.}
\label{fig:DisR}
\centering
\vspace{0.3cm}
\includegraphics[width=8.8cm]{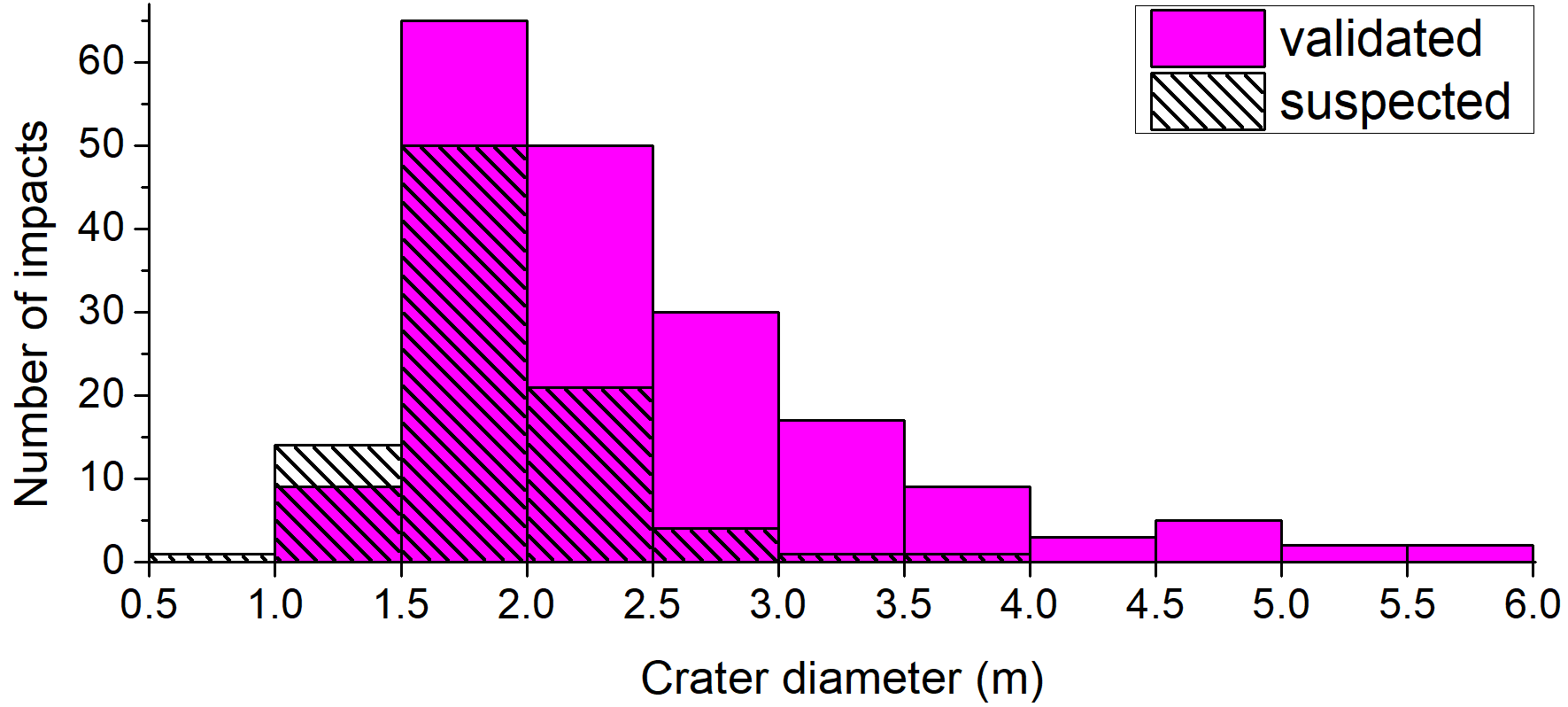}\\
\caption{Distribution of expected crater sizes of the observed LIFs.}
\label{fig:DisC}
\end{figure}

The distribution of the peak temperatures of the validated LIFs and the respective (estimated) one of the suspected LIFs are shown in Fig.~\ref{fig:DisT}. A significant portion (45.8\%) of the validated LIFs evolved peak temperatures between 2500-3500~K. It should to be noticed that 84.9\% of the whole sample of the validated LIFs range between 2000-4500~K. Only 3.6\% of this sample resulted in temperatures lower than 2000~K, and only 11.5\% higher than 4500~K. The drop off of low temperature values is due to our observing setup. Low temperature values result, in general, from LIFs with faint $R$ magnitudes, that is $R-I>1.91$~mag. So, given that the majority of the suspected flashes range between 9-10~mag in the $I$~passband (see Fig.~\ref{fig:obs_mags}), it is very possible that the LIFs were not detected in the $R$~passband due to observations limitations (see also Fig.~\ref{fig:Stat_R_phase}). The estimated peak temperatures of the suspected LIFs (see Section~\ref{sec:CALC} for details) in the same figure show a distribution with a peak between 1500-2000~K. It should be noted that these temperatures can be considered as upper limits, since we estimated the expected magnitudes of these flashes in the $R$~passband using eq.~\ref{eq:LELR} (see Sect.~\ref{sec:CALC} for details). The $LE_{\rm I}$ of the suspected flashes (i.e., those that were not detected in the $R$~passband) are within the range shown in Fig.~\ref{fig:EREICOR}. The true $LE_{\rm R}$ and $R$ magnitude values should be lower and fainter than the estimated ones, respectively, implying higher $R-I$ indices, hence lower temperatures for the suspected LIFs. We note that the peak of the flash (i.e., brightest magnitude and highest temperature phase) might have occurred during the readout of the camera and might have been missed in the observations. Nevertheless, the current values of the peak temperatures of the validated LIFs can be considered as lower limits and a detailed discussion for this is made in Section~\ref{sec:DIS}.

The meteoroid mass distribution for $\eta=1.5\times10^{-3}$ and for various mass ranges is illustrated in Fig.~\ref{fig:DisM}. The choice of this $\eta$ value is based on the fact that it stands almost between the two extreme values ($5\times10^{-4}<\eta<5\times10^{-3}$) and is widely used as reference. However, in Table~\ref{tab:ResultsReal}, the masses, radii, and crater sizes have also been calculated for both these extreme $\eta$ values. For convenience, we notify the reader that the extreme $\eta$ values rescale the masses (Fig.~\ref{fig:DisM}) by a factor or a fraction of $\sim3$. This figure shows that the derived meteoroid masses range from $\sim2$~g up to $\sim2.7$~kg, but it can be clearly seen that the vast majority weighs less than 200~g. The distribution for the latter range is plotted in the internal panel of the same figure. A fraction of 60.4\% of the meteoroids that produced validated LIFs weigh up to 100~g and they are distributed almost homogeneously into bins of 25~g. The 76.6\% of the whole sample have masses less than 200~g, and the 88\% less than 400~g. Moreover, as expected from their magnitudes (mass-magnitude correlation, cf. Paper~III and Section~\ref{sec:COR}), the masses of the meteoroids of the suspected LIFs have systematically lower values, typically less than 50~g.

Similarly to the mass distribution plots, the radius distributions of the meteoroids produced the observed LIFs are plotted in Fig.~\ref{fig:DisR} for $\eta=1.5\times10^{-3}$. Note that the plotted radius values in this figure have a multiplication factor of $\sim1.49$ with respect to those that can be derived using $\eta=5\times10^{-3}$ and a division factor of $\sim1.44$ for those with $\eta=5\times10^{-4}$. This figure shows that the radii of the projectiles range between 0.7-6.95~cm. The vast majority (69.8\%) of the validated meteoroids and almost all of the suspected have radii less than 3~cm. Therefore, and in order to have an analogy with the real life item sizes, the majority of the validated meteoroids have sizes roughly between a sphere with a diameter similar to the coin of 2\texteuro~(2.575~cm) and a tennis ball (6.54~cm). Only 1.5\% have sizes less than 2.8~cm in diameter (i.e., spheres with sizes similar to the smaller coins of 1\texteuro). The 11.9\% of the whole sample have sizes between tennis and baseball balls (diameters 6-8~cm), and only 4.1\% may reach the size of a shotput ball (13~cm).

The distribution of the expected crater sizes due to the observed meteoroid impacts (for an impacting angle of $45\degr$) is illustrated in Fig.~\ref{fig:DisC}. For the validated LIFs, the size range is estimated to be between 1-6~m. However, there is a clear peak between 1.5-2~m for both validated and suspected LIFs.

\begin{figure}
\centering
\includegraphics[width=8.8cm]{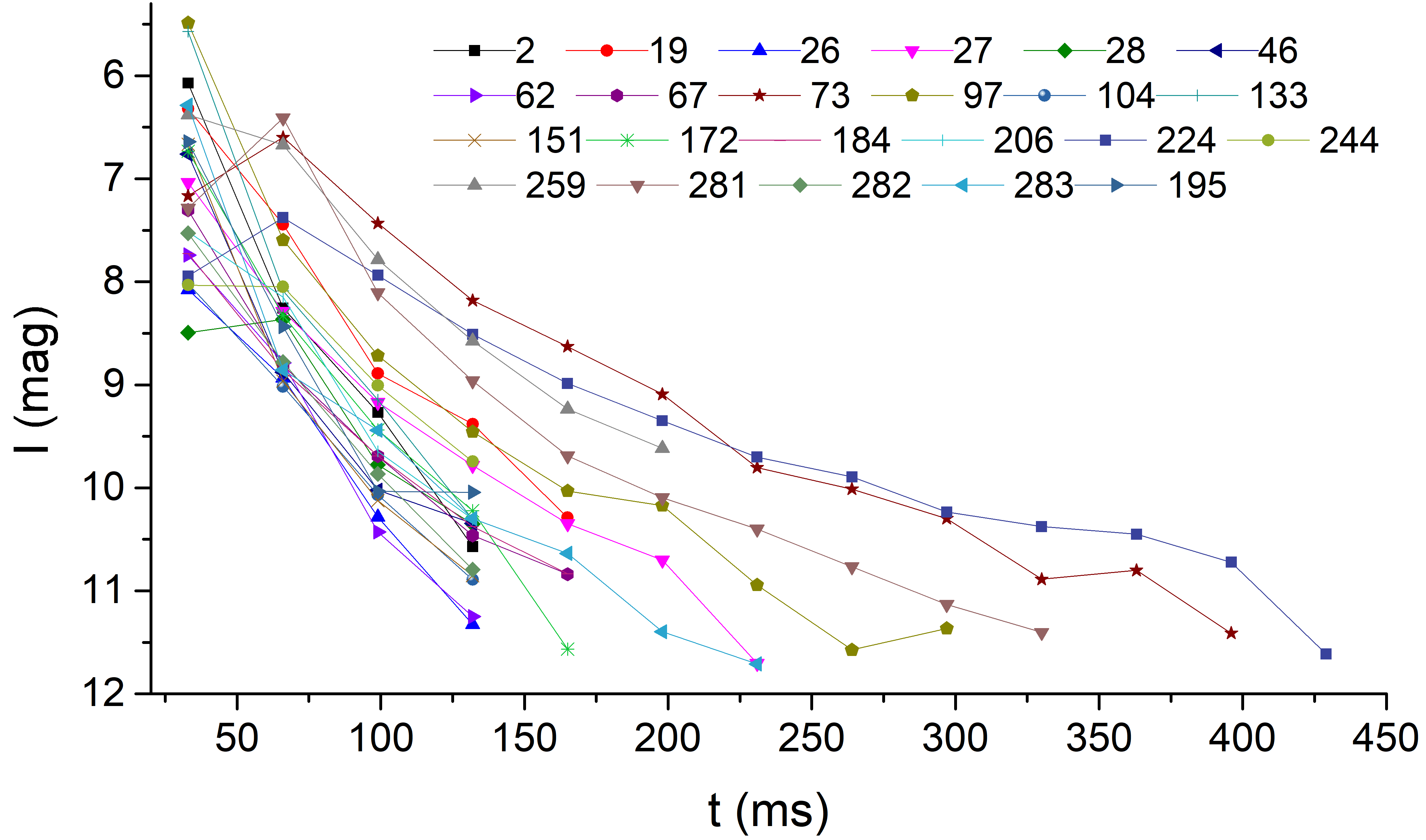}\\
\caption{Light curves of multiframe LIFs lasting more than four frames in the $I$-passband.}
\label{fig:MFLIFS}
\vspace{0.3cm}
\includegraphics[width=8.8cm]{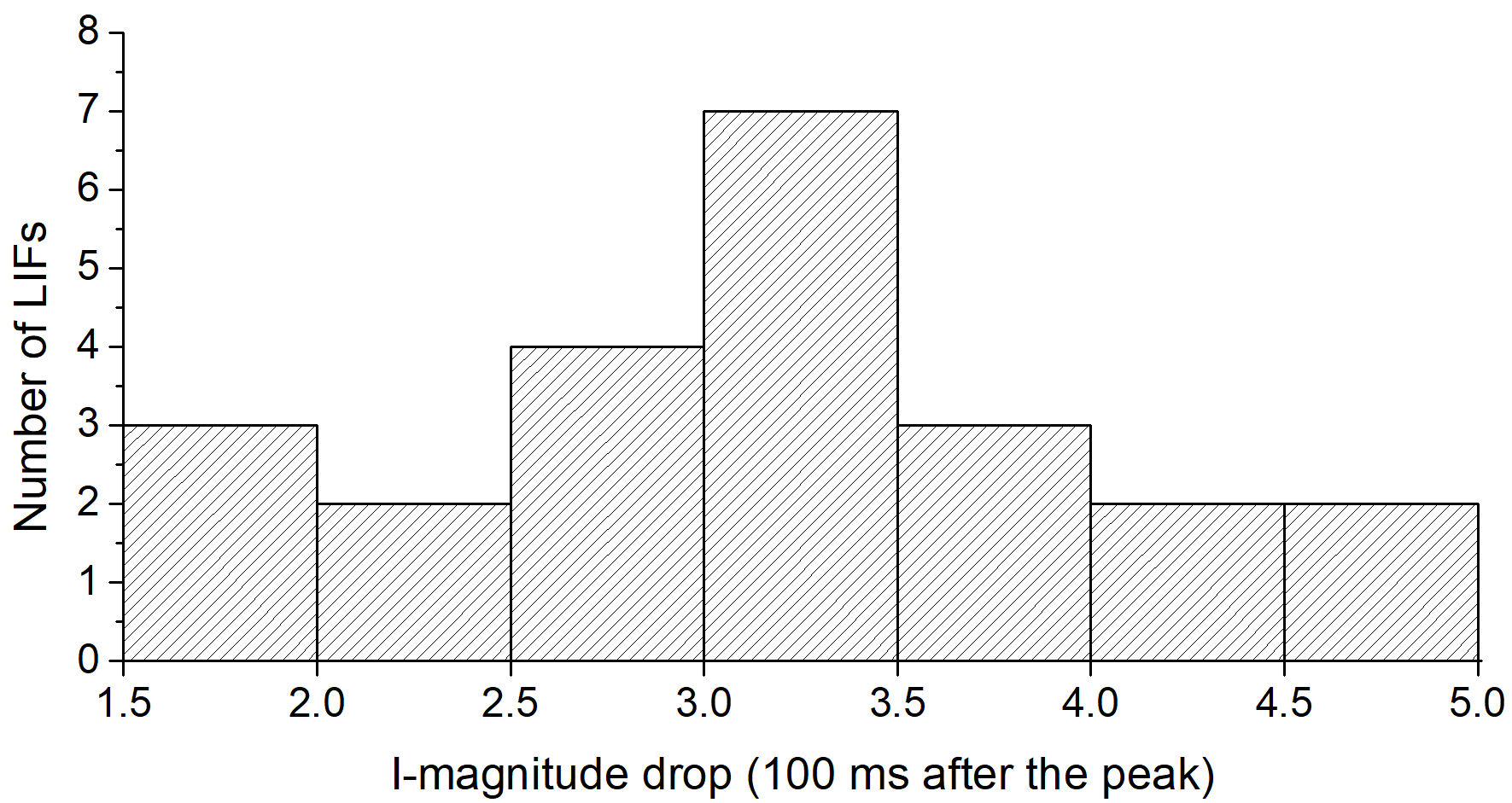}\\
\caption{Histogram of magnitude drop in the $I$-passband 100~ms after the peak magnitude. The sample consists of the LIFs shown in Fig.~\ref{fig:MFLIFS}.}
\label{fig:HISTMFLIFS}
\end{figure}

Based on our relatively large sample of multiframe LIFs, we attempted to make a comparison of their light curve profiles in order to check whether any systematic behaviour occurs. We included only the LIFs for which at least four frames in the $I$-passband were recorded and they are illustrated in Fig.~\ref{fig:MFLIFS}. The selection of the sample is based on the frame multiplicity of the LIFs in order to minimize the error of not recording all the emitted light of the first and last frames due to the readout time of the camera. The individual light curves (with the error bars) for LIFs with ID number less than 114 are given in Paper~III, while the rest are plotted in Fig.~\ref{fig:LCs1}. These profiles all present a similar exponential decrease, which we attempt to quantify by measuring the brightness decrease 100~ms after the observed peak brightness. The histogram of the decline rate distribution (Fig.~\ref{fig:HISTMFLIFS}) appears to be a normal distribution with a peak value around $\sim3$~mag. Additional correlations between their masses and temperatures and their decline rates were attempted but the results were negative.

\begin{figure}
\centering
\includegraphics[width=8.8cm]{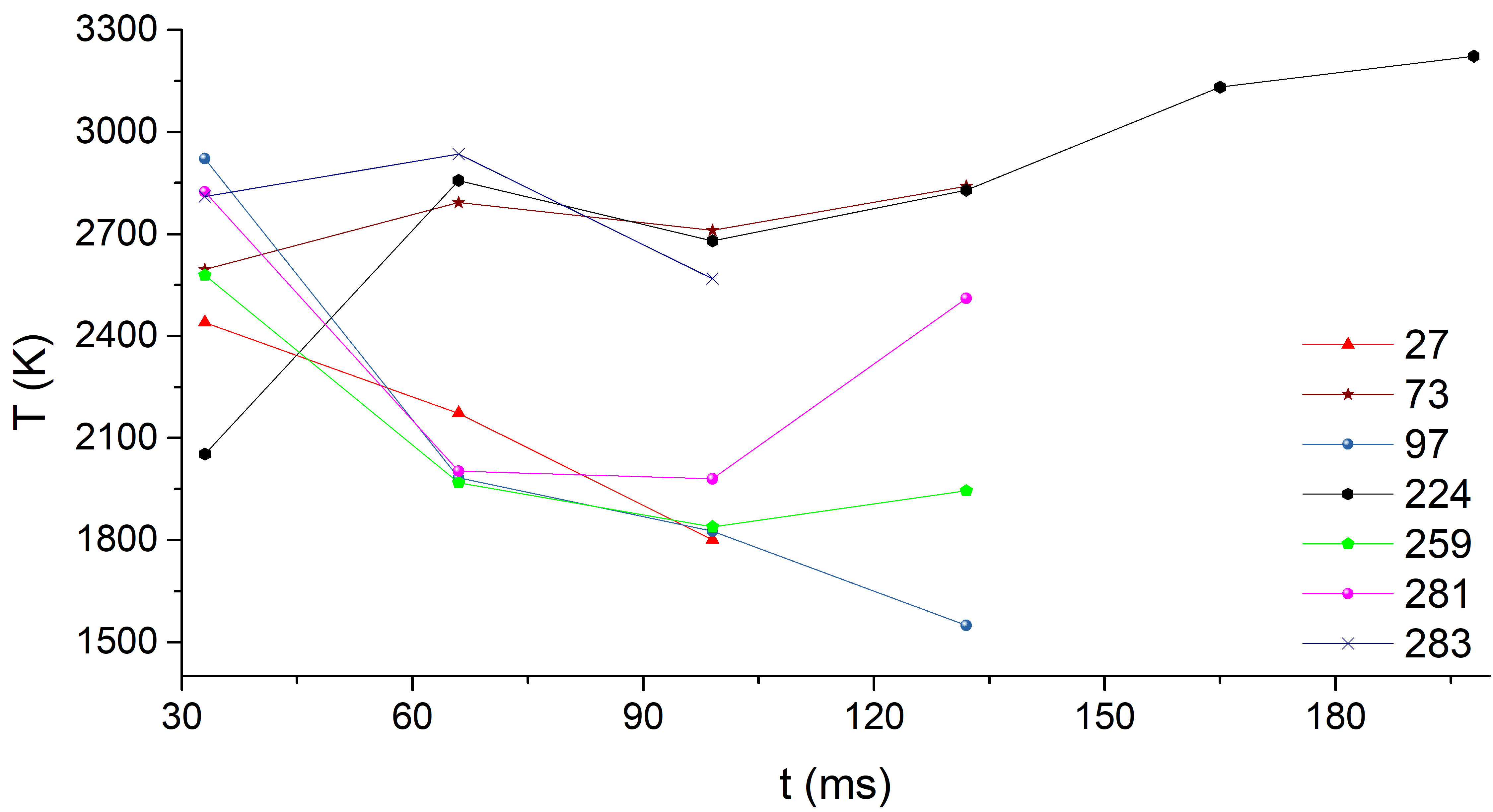}\\
\caption{Temperature curves of multiframe LIFs lasting at least three frames in both passbands.}
\label{fig:MFLIFST}
\vspace{0.3cm}
\includegraphics[width=8.8cm]{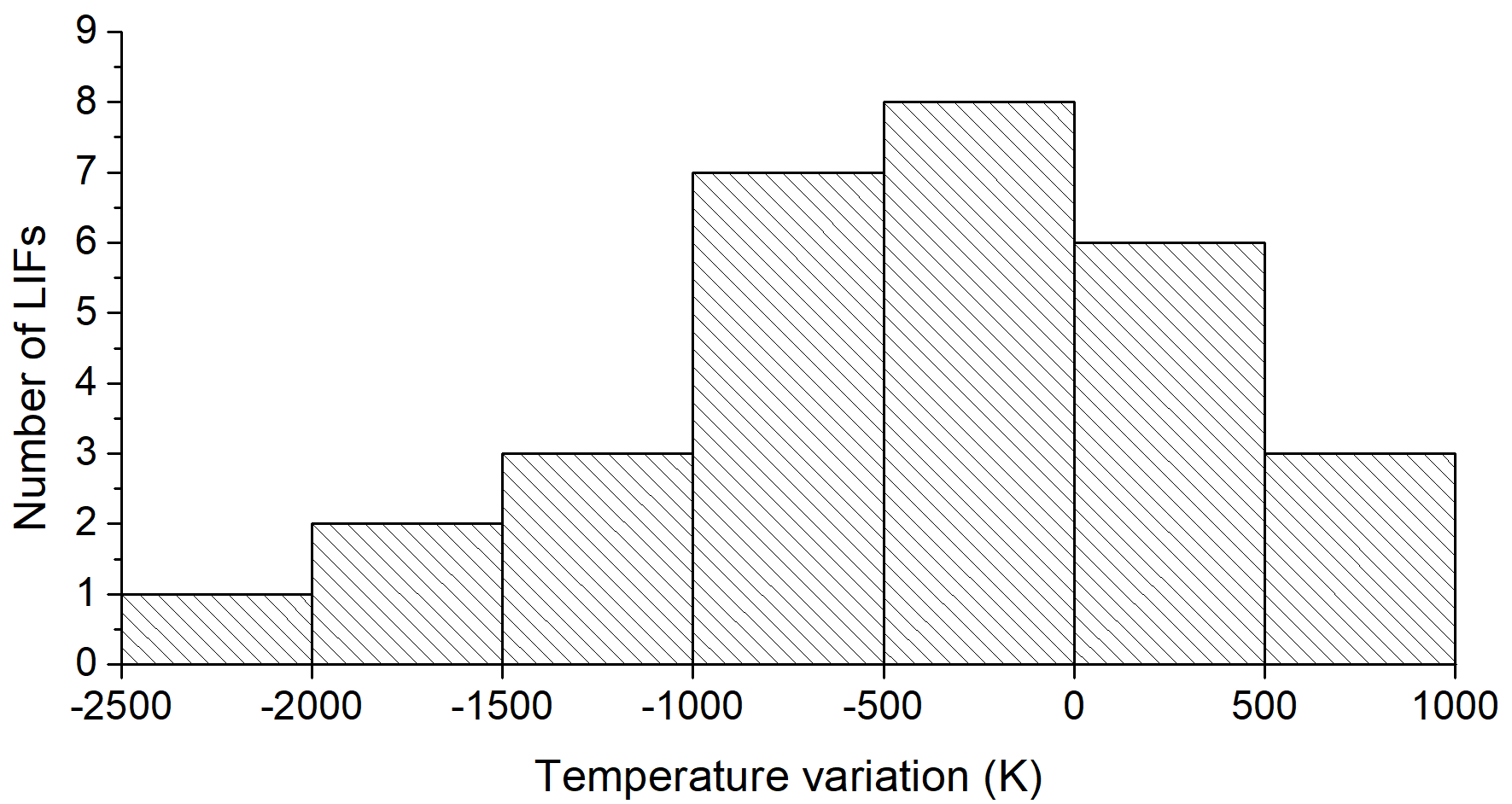}\\
\caption{Histogram of temperature variation of 30 multiframe LIFs in both passbands. The variation concerns the first 33~ms after the peak temperature.}
\label{fig:HISTMFLIFST}
\end{figure}

Regarding the temperature evolution of the LIFs, we notice that most of the LIFs presented in this work show a temperature decrease (Fig.~\ref{fig:TCs1}), but the cooling rate is not the same. Moreover, there are six cases (LIF \#172, \#191, \#209, \#224, \#283, \#287), whose temperature increased between the first and the second set of frames indicating that the recording of the impact began before the occurrence of the maximum temperature. Finally, there are also a few cases (LIF \#224, \#259, \#281), in which after an initial decrease of the temperature, a constant temperature time period occurred. In addition, a temperature increase was observed before they completely vanish in the $R$~passband. Similar to the previous plot, we tried to find any trends in the temperature evolution behaviour of the LIFs. For this, only the multiframe LIFs in both passbands were used (i.e., 30~LIFs in total). In order to avoid a dense diagram, we plotted in Fig.~\ref{fig:MFLIFST} only those that lasted more than two frames in both passbands. Individual temperature evolution curves for LIFs with ID number greater than 113 are illustrated in Fig.~\ref{fig:TCs1} and the rest are shown in Paper~III. The histogram of all the aforementioned multiframe LIFs regarding their temperature variation between the peak temperature and the next 33~ms is given in Fig.~\ref{fig:HISTMFLIFST}. This plot shows that the majority of these LIFs have relatively slow cooling rates (i.e., between 0 and $-1000$~K). On the contrary, there are many that after their peak temperature not only did not show any cooling trend, but, on the contrary, they presented a slight increase. However, it should be noted that the temperature errors are quite large and that our sample is relatively small to provide concrete conclusions. For the cases of LIFs with constant or increasing temperature, we assume that this thermal behaviour is somehow connected to the material, for example the cooler plume initially covering the hotter plume.

\section{Correlations between physical parameters}
\label{sec:COR}

\begin{figure}
\centering
\includegraphics[width=8.8cm]{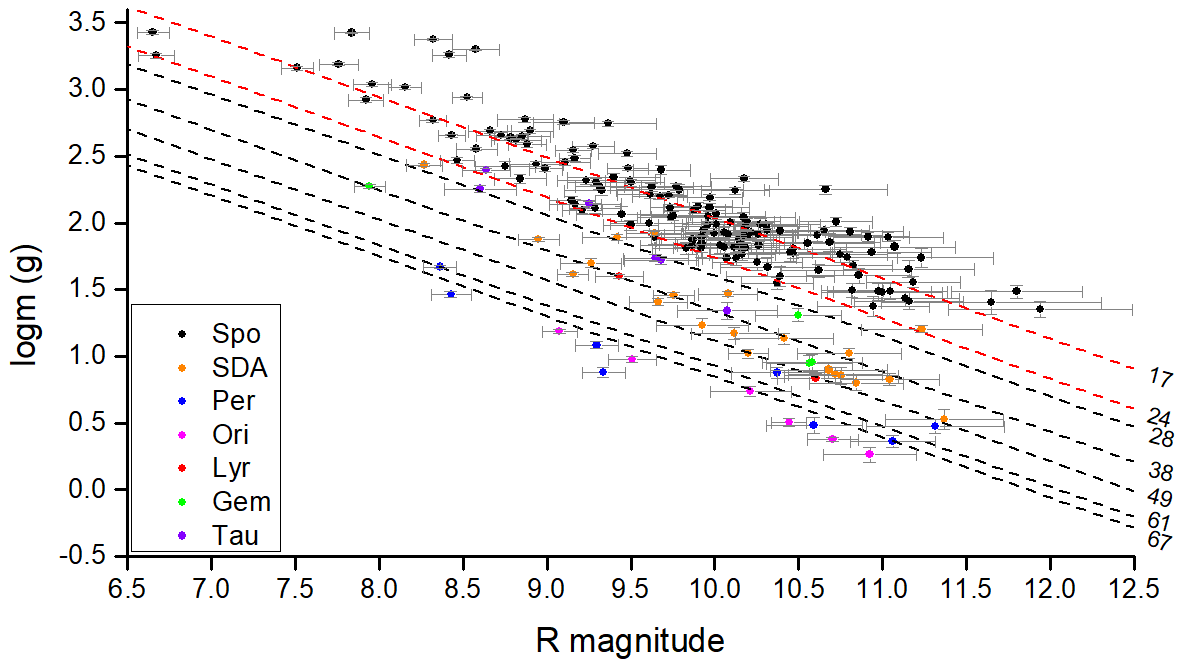}\\
\includegraphics[width=8.8cm]{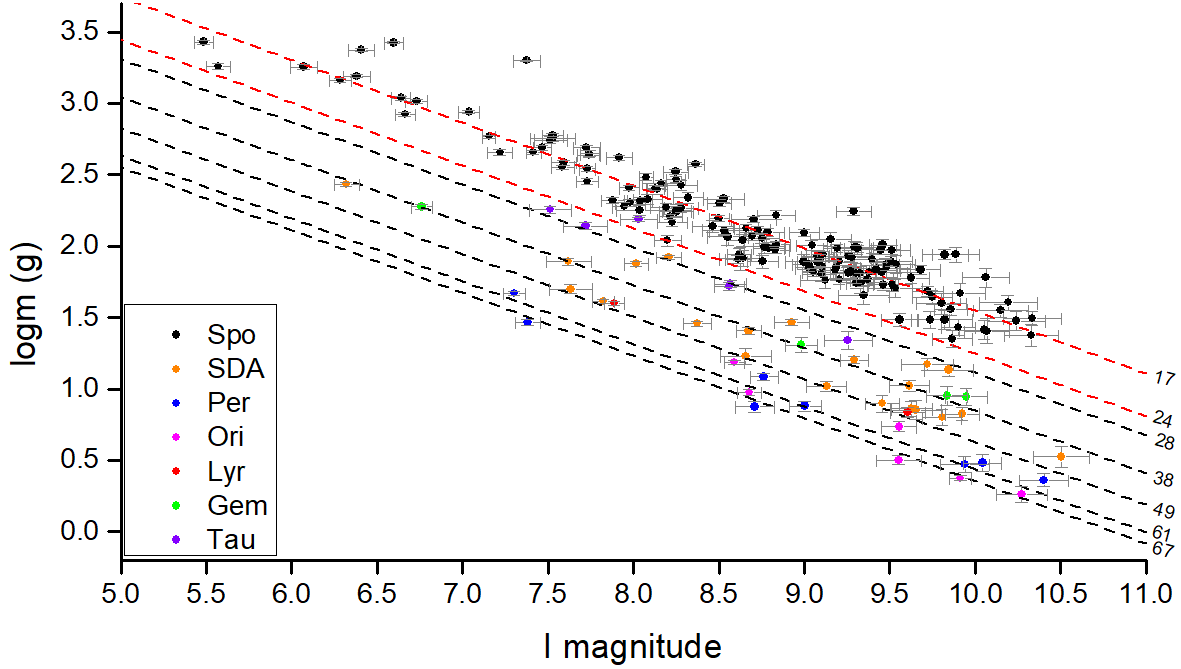}\\
\caption{Correlations between peak magnitudes in $R$ (upper panel) and $I$ (lower panel) passbands and meteoroid masses for $\eta=1.5\times10^{-3}$. For the plotting of the sporadic meteoroids an impact velocity of 17~km~s$^{-1}$ was assumed. The masses of the stream meteoroids are based on the velocities given in Table~\ref{tab:alpha}. The assumed velocities are indicated next to each fitting curve.}
\label{fig:massmag}
\vspace{0.3cm}
\includegraphics[width=8.8cm]{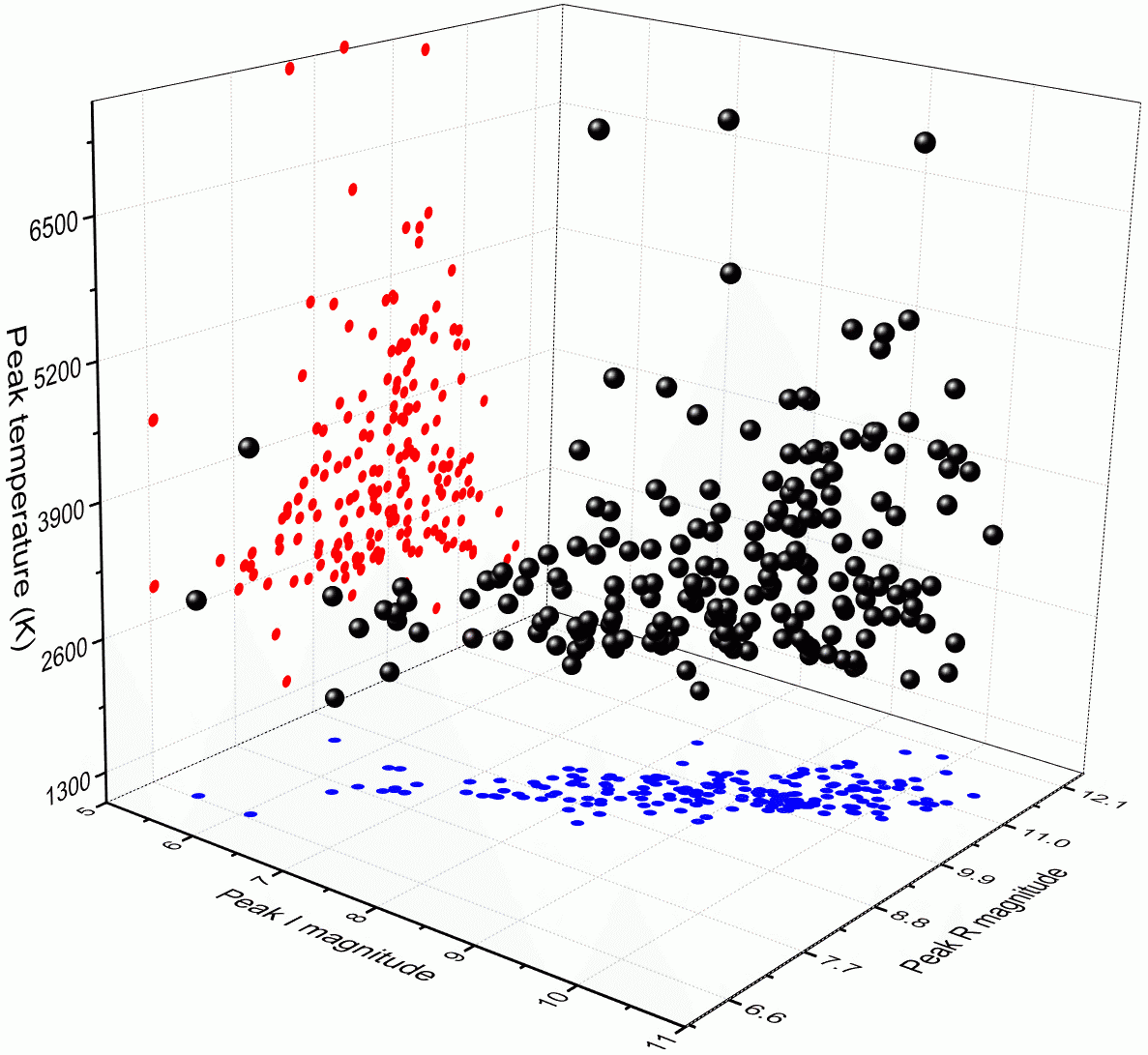}\\
\caption{Three-dimensional correlation of peak magnitudes in $R$ and $I$~passbands with the peak temperatures of the LIFs.}
\label{fig:MagT}
\end{figure}

The mass-magnitude correlation of LIFs is quite useful since it provides a direct sense of the meteoroid mass that produced the observed LIF. However, the mass calculation is not straightforward since the lost energy during the readout time of the camera and the $LE$ of all the observed passbands should be taken into account (cf. Paper~III). Using our sample, we are able to derive empirical relations between masses ($m_{\rm p}$) of the projectiles and peak magnitudes of the LIFs ($m_{\rm R}$, $m_{\rm I}$). These relations can be used from other observers to estimate the masses of the meteoroids producing LIFs as long as they use the $R$ or $I$ filters. The meteoroid masses against their respective peak magnitudes in the observed passbands are plotted in Fig.~\ref{fig:massmag}. The data points in this figure form parallel layers based on their velocities. Therefore, in order to calculate the correlation between the aforementioned quantities, we used only the data of the sporadic LIFs (black points in Fig.~\ref{fig:massmag}). Using $\eta=1.5\times10^{-3}$ and various values for the impact velocities, different masses are derived for the sporadic meteoroids, hence, different linear fittings were made. The fitting curves are plotted in the same figure and correspond to $V_{\rm p}=17$ and 24~km~s$^{-1}$ (i.e., extreme values of the sporadic meteoroids) as well as to the velocities of the stream meteoroids (see Table~\ref{tab:streams}). The slope values of the linear fittings for the data of each passband were the same regardless the assumed velocity. On the contrary, the intercept value ($A$) was changing. Therefore, we derive the following relations:
\begin{equation}
\log m_{\rm p} = A_{\rm R} - 0.45(2)m_{\rm R},~{\rm with}~r=0.91,
\label{eq:mass-magR}
\end{equation}
for the $R$~passband and
\begin{equation}
\log m_{\rm p} = A_{\rm I} - 0.44(1)m_{\rm I},~{\rm with}~r=0.95,
\label{eq:mass-magI}
\end{equation}
for the $I$~passband. The intercept values ($A$) for each assumed velocity and for each passband are listed in Table~\ref{tab:alpha}.

\begin{table}
\centering
\caption{Intercept values of Eqs.~\ref{eq:mass-magR}-\ref{eq:mass-magI} according to the impact velocity of the meteoroids based on their origin.}
\label{tab:alpha}
\begin{tabular}{cccc}
\hline	
\hline							
	Origin &	$V_{\rm p}$	&	$A_{\rm R}$	&	$A_{\rm I}$ 	\\
	&	(km~s$^{-1}$)	&	 (mag)	&	 (mag)	\\
\hline							
SPO	&	17	&	6.6(2)	&	5.9(1)	\\
SPO	&	24	&	6.3(2)	&	5.6(1)	\\
Tau	&	28	&	6.1(2)	&	5.5(1)	\\
Gem-SDA	&	38	&	5.9(2)	&	5.2(1)	\\
Lyr	&	49	&	5.6(2)	&	5.0(1)	\\
Per	&	61	&	5.5(2)	&	4.8(1)	\\
Ori	&	67	&	5.4(2)	&	4.7(1)	\\
\hline							
\end{tabular}
\end{table}

Similarly to Paper~III, we attempt to find a possible correlation between the physical parameters of the meteoroids and the peak temperatures of the LIFs using the whole sample of the validated flashes. First, we begin with a correlation between the observed peak magnitudes and the peak temperatures of the observed LIFs. This attempt is illustrated in Fig.~\ref{fig:MagT} that shows the 3D correlation of the peak $R$ and $I$ magnitudes with the temperature. It should to be noticed that the horizontal plane (i.e., $I$--$R$~mag--blue points) of this figure corresponds more or less to the plot of Fig.~\ref{fig:EREICOR}, since the $LE$ of each band is connected to the respective peak magnitude (cf. Paper~III). The projection of the sample on the $R$ mag vs $T$ plane (red points) is also shown. We notice that for brighter $R$ magnitudes (i.e., $<8$~mag) the LIFs tend to have lower temperatures. For fainter LIFs, we notice that there is a slightly increasing trend of the temperature. However, we argue that this trend is apparent and biased from our detection limits. Low-temperature LIFs (i.e., $T<3000$~K) are easier to be detected when they are bright, since their emission in $R$~passband will be always under our detection limit. On the contrary, similar temperature LIFs but fainter than approximately 10~mag in the $R$ filter, although they can be potentially detected in $I$~passband, they cannot be validated. We may also notice that the aforementioned trend begins after $R\sim8$~mag. This is due to the fact that higher temperatures LIFs emit more in $R$ filter, thus, are easier to detect.

\begin{figure}
\includegraphics[width=8.8cm]{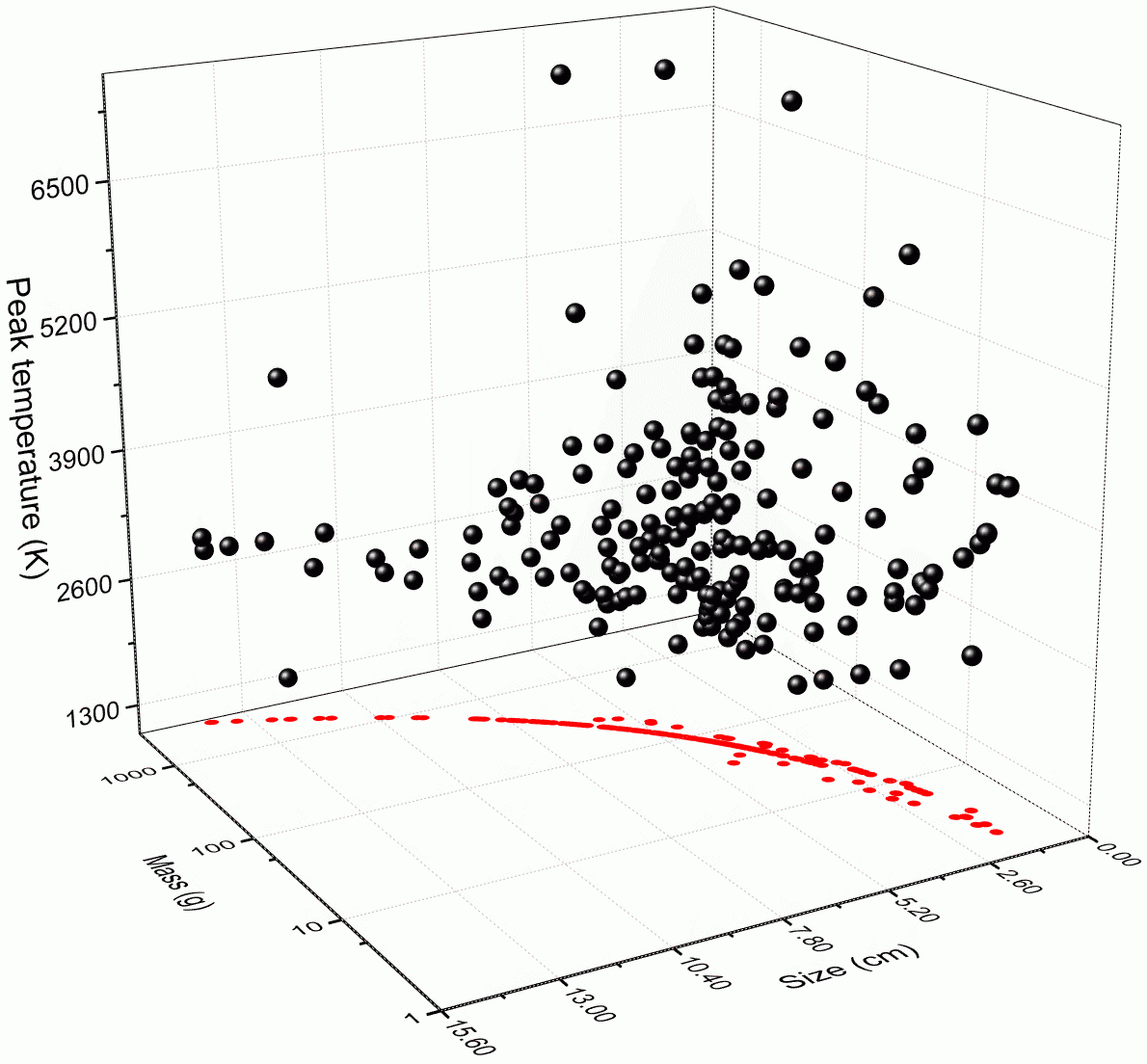}\\
\caption{Three-dimensional correlation of mass and size of the meteoroids with the peak temperature evolved during the impacts.}
\label{fig:3DMRT}
\end{figure}

One of the conclusions of Paper~III concerned the noncorrelation between the masses of the meteoroids and the impact temperatures regardless the impact site. Again, using an almost triple sample of validated LIFs, we update this diagram, but, again, no obvious correlation has been found (Fig.~\ref{fig:MT}). Moreover, we searched for a correlation between the meteoroid sizes and peak temperatures, since their contact surface may play a role. In agreement with the mass-temperature plot, this correlation shows that the peak temperature is more or less independent of the impact site no matter the size of the projectile (Fig.~\ref{fig:RT}). However, we notice a peak slightly before $T=3000$~K and a steep decrease of the sizes with the increasing temperature, something that is not that clear in the previous plot. For the aforementioned plots, we assigned colors to the data points according to visual inspection of the impact area. Dark areas on Moon corresponds mostly to the maria regions, while grey or bright to more rocky/mountainous regions. However, the 3D plot illustrated in Fig.~\ref{fig:3DMRT} combines the above two diagrams and shows that the evolved temperature is independent of or even anticorrelated to the mass and size of the projectiles. A short discussion on this result is given in Section~\ref{sec:DIS}.

\section{Meteoroid appearance frequency, impact probability and hazard estimation}
\label{sec:FREQ}

In this section we focus on the LIF detection rates of NELIOTA that lead to the calculation of the meteoroid appearance frequency on the Moon and around Earth. In Table~\ref{tab:rates} we list the total observed hours on the Moon and the number of the respective LIF detections during normal periods of time (\textit{Spo}) and during the passages of the Earth-Moon system through meteoroid streams (\textit{Str}). Moreover, for each case we list both the number of validated LIFs and the number of the validated and suspected LIFs. The latter values can be considered as the upper limits. In order to be more direct, the respective detection rates (\textit{Det.~Rate}) for each case have been calculated and listed in the same table. These rates strongly depend on our FoV, hence the covered lunar area by our setup (see Section~\ref{sec:OBS}), and are given only for a direct sense of what we can achieve with such equipment. In order for these rates to be comparable with others of similar campaigns \citep{SUG14, REM15}, they are normalized to the covered lunar surface (i.e., per squared kilometer). As mentioned in Paper~III, the covered lunar area is not constant, since it depends on the Moon-Earth distance (i.e., variation of the pixel scale) and the lunar illumination phase (i.e., phases greater than 0.35 require repositioning of the telescope in order to avoid brighter areas). Therefore, the normalised detection rates (\textit{Norm.~Rate}) in Table~\ref{tab:rates} have been calculated using an approximate lunar covered area of $3.11\times10^{6}$~km$^2$ by our setup that corresponds to an Earth-Moon distance of $3.84\times10^{5}$~km. The slight discrepancy between the current detection rates and those from Paper~III (Table~2) are clearly due to the current larger sample used.

\begin{table}
\centering
\caption{Flash detection rates of NELIOTA according to the origin of the meteoroids.}
\label{tab:rates}
\scalebox{0.94}{
\begin{tabular}{c ccc ccc}
\hline													
\hline													
	&	\multicolumn{ 3}{c}{Validated}					&	\multicolumn{ 3}{c}{Validated and suspected}					\\
\hline													
	&	Spo	&	Str	&	Sum	&	Spo	&	Str	&	Sum	\\
\hline													
Obs. Hours	&	237.8	&	45.6	&	283.4	&	237.8	&	45.6	&	283.4	\\
Detections	&	145	&	47	&	192	&	208	&	87	&	295	\\
Det. Rate$^a$	&	0.6	&	1.0	&	0.7	&	0.9	&	1.9	&	1.0	\\
Norm. Rate$^b$	&	1.96	&	3.32	&	2.18	&	2.81	&	6.14	&	3.35	\\
\hline																																						
\end{tabular}}
\tablefoot{$^a$In units of LIF~h$^{-1}$, $^b$in units of $10^{-7}$ LIF~h$^{-1}$~km$^{-2}$, Spo = sporadic, Str = stream.}
\end{table}

\begin{figure}
\centering
\includegraphics[width=8.8cm]{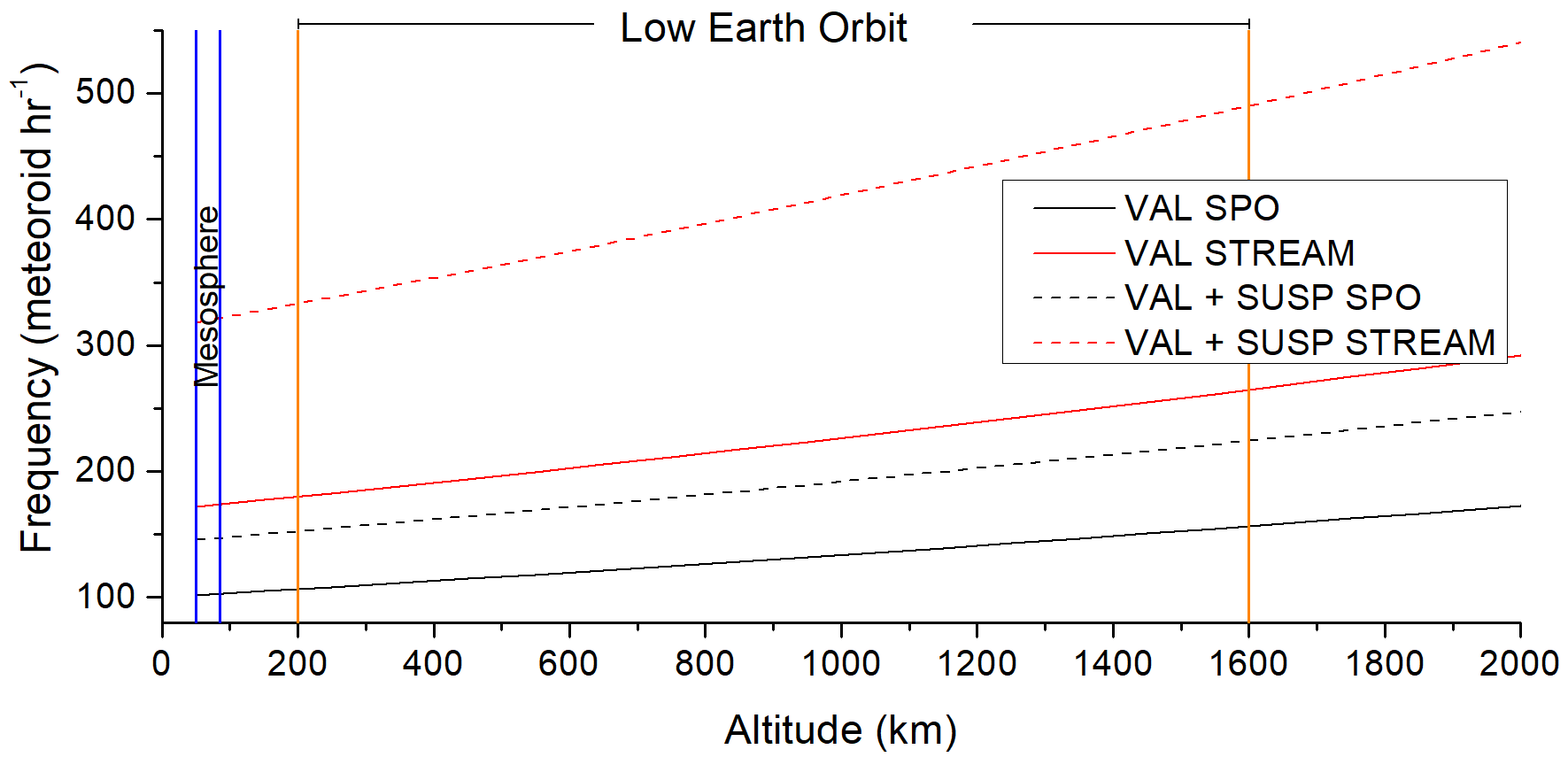}\\
\includegraphics[width=8.8cm]{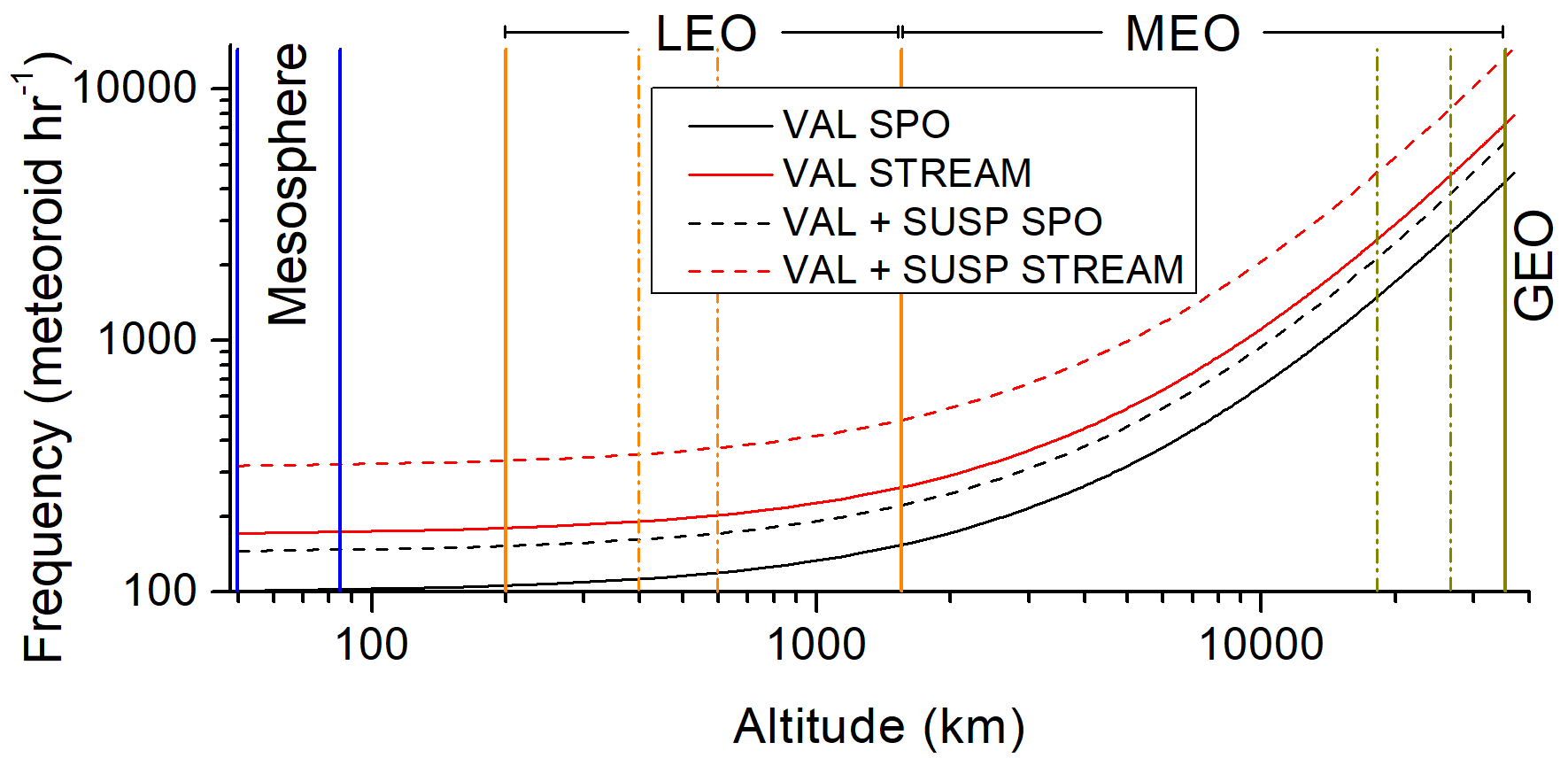}\\
\caption{Appearance frequency distributions of meteoroids around Earth up to an altitude of 2000~km (upper panel) and up to 36000~km (lower panel). Black and red lines denote the sporadic and the stream meteoroids, respectively. Solid lines correspond to the frequencies based on only the validated flashes, while dashed lines those that are based on the validated and suspected flashes. Vertical solid lines indicate the boundaries of the mesosphere, LEO, MEO, and GEO zones. Vertical dashed lines in the lower panel indicate satellite crowed orbit zones.}
\label{fig:freqs}
\end{figure}

As concluded in Section~\ref{sec:OBS}, we can plausibly assume that the Moon, hence Earth, is bombarded homogeneously by meteoroids. Using this assumption and the sporadic and the stream normalised LIF detection rates of Table~\ref{tab:rates}, we calculate and list in Table~\ref{tab:freqs} the meteoroid appearance frequencies on the lunar surface and in the vicinity of the Earth. In this table, we present the calculations for both the validated and the validated and the suspected LIFs. Regarding the frequency appearance close to Earth, the calculations concern some typical distances from Earth's surface, such as the upper zone of the mesosphere (80~km), where meteors occur, the Low-Earth Orbit zone (LEO) at an altitude of 500~km, the Medium-Earth Orbit zone (MEO) at an altitude of 20000~km, and the Geosynchronous/Geostationary Earth-Orbit zone at 36000~km. The aforementioned meteoroid appearance frequencies are plotted in Fig.~\ref{fig:freqs}, which show how many meteoroids per hour enter a sphere with radius that has the size of the sum of the radius of the Earth and the distance from its surface (altitude).

\begin{table}
\centering
\caption{Appearance frequencies of meteoroids on the Moon and around Earth in units of meteoroid~h$^{-1}$.}
\label{tab:freqs}
\scalebox{0.97}{
\begin{tabular}{c c| c| cccc}
\hline													
\hline													
	&		&	Moon	&	\multicolumn{4}{c}{Earth}							\\
\hline													
	&		&	Sur	&	MS	&	LEO	&	MEO	&	GEO	\\
Altitude (km)	&		&	0	&	80	&	500	&	20000	&	36000	\\
\hline													
\multirow{ 2}{*}{Validated}	&	Spo	&	7.4	&	102	&	116	&	1712	&	4420	\\
	&	Str	&	12.6	&	173	&	197	&	2898	&	7481	\\
\hline													
Validated and	&	Spo	&	10.7	&	147	&	167	&	2456	&	6341	\\
suspected	&	Str	&	23.3	&	321	&	364	&	5364	&	13848	\\
\hline																																									
\end{tabular}}
\tablefoot{Spo = sporadic, Str = stream, Sur = surface, MS = Mesosphere, LEO/MEO/GEO = Low/Medium/Geosynchronous Earth Orbit.}
\end{table}

\begin{figure}
\centering
\includegraphics[width=8.8cm]{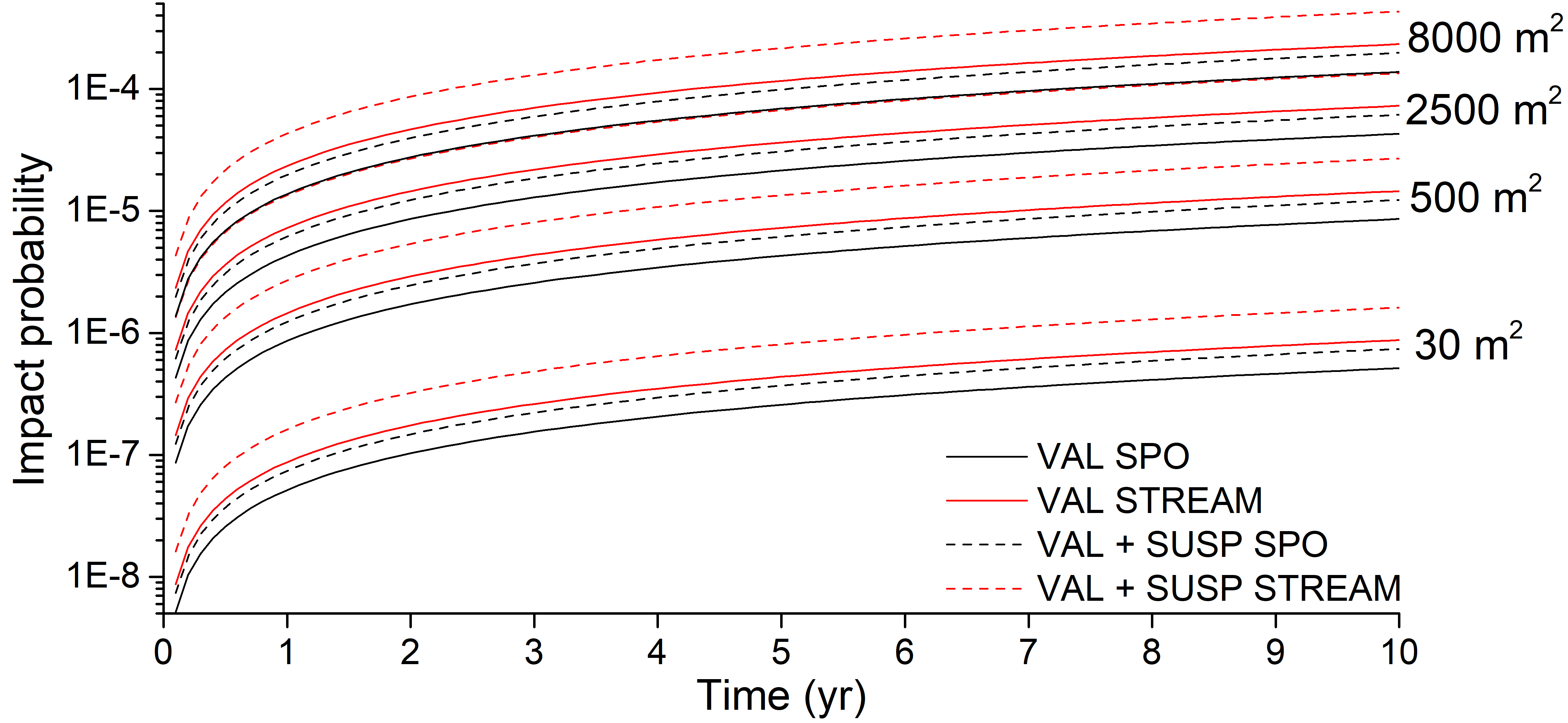}\\
\caption{Impact probability of a meteoroid with a potential infrastructure on the lunar surface or with a satellite in orbit around Earth for various impact surface areas (30~m$^2$--small size building or satellite, 500~m$^2$, 2500~m$^2$--medium size building or satellite, 10000~m$^2$--large size building or ISS size satellite). Colors and line styles denote the same as in Fig.~\ref{fig:freqs}.}
\label{fig:impacts}
\vspace{0.5cm}
\includegraphics[width=8.8cm]{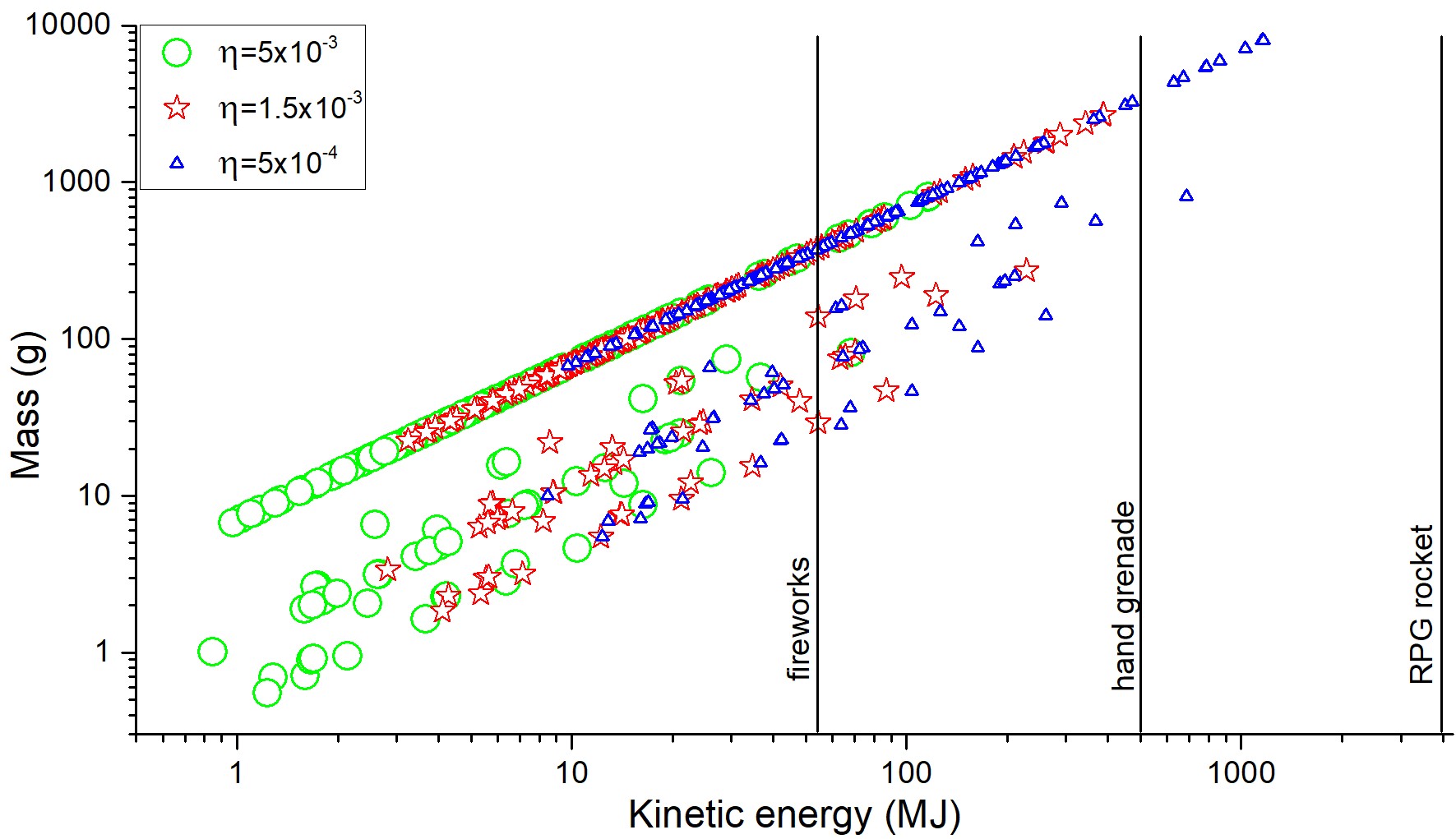}\\
\caption{Kinetic energies against masses for the meteoroids producing LIFs observed by NELIOTA for various $\eta$ values. Representative energies are also indicated for estimating the damage that a meteoroid impact may cause. We note that the muzzle energies of typical weapons (e.g., pistols, rifles) or stronger military weapons (e.g., machine guns) are less than 3-5~KJ, i.e., two order of magnitudes lower than the minimum energy that the plot covers.}
\label{fig:masKE}
\end{figure}

Based on these appearance frequency rates, we attempt to calculate the probability of a meteoroid impact with a hypothetical infrastructure on the Moon or an orbiting satellite for various dimensions of the impacting target (effective surface) and the duration of the mission. Obviously, the meteoroid size (effective surface) is negligible in comparison with that of the impacting area, hence, it is not taken into account. Therefore, the impact probability can be calculated using the following formula:
\begin{equation}
P_{\rm impact}=\frac{S_{\rm impact}}{S_{\rm total}}~f_{\rm app}~t,
\label{eq:prob}
\end{equation}
where $S_{\rm impact}$ is the effective impacting surface, $S_{\rm total}$ the total surface of the sphere (e.g., Moon) with a radius equal to that used to derive the respective calculated appearance frequency of the meteoroids $f_{\rm app}$ (Table~\ref{tab:freqs}), and $t$ the time of the mission.

Regarding the impact on lunar potential infrastructures (e.g., buildings), we calculated the probabilities for different surface areas ranging from a small building (30~m$^2$) up to a large-size building with dimensions of a (european) football field (8000~m$^2$). For a meteoroid collision with a satellite, we took into account the size of the satellites with fully deployed solar panels, therefore, the calculations were made for the following effective surface areas: 30~m$^2$ (e.g., Starlink satellite), 500~m$^2$, 2500~m$^2$, and 8000~m$^2$ (e.g., International Space Station). Obviously, the calculated probabilities depend only on the effective surface areas and the respective appearance frequency and not on the distance (orbit). The collision probabilities against time are drawn in Fig.~\ref{fig:impacts} for sporadic and stream meteoroids based on the detection of the validated and the validated and suspected LIFs.

The kinetic energies against the masses of the meteoroids producing the observed LIFs are illustrated in Fig.~\ref{fig:masKE}. Although these quantities are correlated via a power law, it is useful to plot them in order to get a rough idea of the range of the meteoroid carrying energies, hence, the potential damage on a satellite or on a lunar base that can be caused by a meteoroid impact. The velocities are based on the same aforementioned assumptions regarding the origin of the meteoroids, while the data points are calculated for various $\eta$ values. Similarly to Fig.~\ref{fig:massmag}, we notice layers of data points corresponding to different origins of meteoroids (i.e., to different velocities). We notice that the $KE$ of the projectiles can range between 0.8~MJ and 1.05~GJ. In this plot, we also indicate some representative energies of manmade explosions. Therefore, we notice that the majority of the meteoroids may cause serious damage to both satellites and potential infrastructure on the Moon.

\section{Discussion}
\label{sec:DIS}

NELIOTA was monitoring the Moon for LIFs continuously for 6.5~yr using the largest telescope in the world (1.2~m Kryoneri telescope) dedicated to this kind of studies. This project was the longer systematic campaign for LIFs worldwide and produced an extremely large amount of data (more than 215~TB of lunar images) that were publicly available within 24~h after each observing night. Due to the large mirror size, it extended our knowledge for LIFs up to two magnitudes in $R$ band, in comparison with other campaigns, while due to the dualband observations produced coherent statistics about the evolved temperatures during the impacts. The NASA Meteoroid Environment Office mentions 443 candidate impact flash events between 2005-2020 in their official webpage\footnote{\url{https://www.nasa.gov/centers/marshall/news/lunar/lunar_impacts.html}}, but no details are given. Therefore, the detection efficiency of NELIOTA can be compared only with teams that presented their results in a published paper, such as \citet{SUG14} from NASA Meteoroid Environment Office and \citet{REM15}. The former team published a detection rate of $1.03\times10^{-7}$~meteoroids~h$^{-1}$~km$^{-2}$ and the latter $1.09\times10^{-7}$~meteoroids~h$^{-1}$~km$^{-2}$. It should be reminded that both teams used smaller size telescopes (30-50~cm). Sadly, these teams did not present any rates according to the meteoroid stream activity, so their rates are the simple division of the number of detections over the observed hours and covered lunar area. In order to be directly comparable with these teams, we presented in Table~\ref{tab:rates} the respective rates (Sum). We claim that in the conservative scenario that is based only on the validated flashes, NELIOTA has a detection rate of $2.18\times10^{-7}$~meteoroids~h$^{-1}$~km$^{-2}$, while if the suspected LIFs are taken into account too, this rate yields a value of $3.35\times10^{-7}$~meteoroids~h$^{-1}$~km$^{-2}$. These detection rates are at least two times higher than those of the aforementioned teams, which proves undoubtedly that NELIOTA has played a significant role in understanding the near-Earth meteoroid environment.


The temperature calculation is based on the assumption that LIFs emit as black bodies, which is also supported by spectroscopic observations \citep{YAN21, YAN22}. Therefore, although this is currently the only way for estimating the temperatures, future laboratory experiments provide a way to validate these estimates. The temperature calculation of LIFs is not straightforward and, except for the black body emission assumption, it depends also on the time resolution of the observing setup. As can be seen in Fig.~\ref{fig:LCs1}, all multiframe LIFs in both bands vanish faster in $R$ than in the $I$~passband. Obviously, this proves that they mostly emit in $I$, hence, our general impression that their vast majority are relatively cool is correct. The 39.6\% of the validated flashes (76 out of 192) were detected in only one frame in $R$ and in two frames in $I$ band. That means that the emission in $R$ lasted less than 33~ms (i.e., sum of the the exposure and the readout time of one frame). Therefore, if we had used half of the frame rate (i.e., 15~fps) and the total exposure was 56~ms, we would have measured more photons in the $I$~passband but the same in $R$. Thus, this would result in a lower peak temperature (i.e., higher $R-I$ index) that leads to an underestimation of the peak temperature. Therefore, thinking inversely, the higher the time resolution (higher frame rates) the more accurate the peak temperature estimation.

The time resolution also plays a significant role in the underestimation of the $LE$, and hence the $E_{\rm kin}$, $m_{\rm p}$, $r_{\rm p}$, etc. In the case of single frame LIFs, we assume that all the luminous energy was emitted within the exposure time of the camera. However, it is likely that in some cases there was unrecorded LIF emission during the readout of the cameras, which leads to an underestimation of the meteoroid's mass. The calculation of the total energy flux of the multiframe LIFs, hence the corresponding meteoroid parameters, by using the energy correction method as described in Paper~III, can be considered as more accurate, since only a small fraction of the emitted energy could be neglected (i.e., the LIF began to emit during the readout time before the first frame).

Constraining of the luminous efficiency $\eta$ has proven to be a very difficult and tricky process, thus we preferred to perform our study based on its uncertainties. The $\eta$ constraining can be achieved using only LIFs originating from stream meteoroid impacts \citep[e.g.,][]{MOS11}, because their velocities are well determined from meteor showers. However, although this might sound easy, practically it is not. The first difficulty concerns the distinguishing between the stream and the sporadic meteoroids as shown by \citet{ORT15} and \citet{MAD15b, MAD15a}. This distinction can be made only during a few days around the maximum activity of the stream and probably only for the strong streams (i.e., those with high zenithal hourly rates) such as Perseids, Gemininds, Lyrids, Taurids, Orionids, SDA. Except the narrow time window for observations for each stream, the subradiant point of the stream on Moon should be on its nightside. More specifically, the subradiant point should be ideally on the near-nightside of the Moon. If not, then it should be on its far-nightside or on its near-dayside in order for at least a small portion of the impact cone to be on the near-nightside. Moreover, according to the currently known observing setups of LIF observers, the illumination phase of the Moon should be less than 60\% (for NELIOTA up to 45\%) for the near-nightside to be observable (glare affect) and, in addition, its elevation should be appropriate for pointing the telescope. In all the aforementioned conditions, we should also add the weather constrains for a given observing site. However, another, and more important, difficulty concerns the meteoroid flux density distribution of the streams that varies from year to year for a given stream and it is not easily found either in the literature or online. Therefore, the combination of data of the same stream obtained in different years produces ambiguous results. We claim that in order to satisfy the above conditions, except the last one, the LIFs observers should increase around the globe in order to minimize the weather and elevation conditions constrains. For this, and as a first step, the NELIOTA team has developed and publicly released the Flash Detection Software (FDS\footnote{\url{https://kryoneri.astro.noa.gr/en/flash-detection-software/}}), which can be used from both amateurs and professional astronomers with a variety of equipment for the observations of LIFs. Regarding the meteoroid flux density of the streams, a unified network of ground based meteor stations that would provide the values for each stream seems to be the most appropriate direction of planning.

NELIOTA has significantly contributed to the studies of LIFs and the meteoroids. The observations ended in mid-August 2023 due to end of the funded program. The NELIOTA team plans to seek for new funding to re-initiate the operations. Observations of LIFs will continue using smaller size telescopes by other groups and hopefully amateur observers will join the effort using the FDS software. However, we argue that there is urgent need of a global LIF database establishment in order to host the data and distinguish the real LIF events from other false positives. We also highly recommend to establish a link between meteor, fireball, bolide and LIF observers for observations during meteor showers. The latter would certainly benefit the studies for the constraining of the luminous efficiency of the LIFs. We conclude that, although NELIOTA has gathered a sufficient sample of LIFs, a larger sample is needed, especially during the maximum activity of the meteoroid streams. Another promising direction for the future is observing in $J$-band during daytime \citep{SHE23}. Multiband observations with more than two filters are certainly desirable since they will provide better estimation of the emitted luminous energy. More sensitive (e.g., quantum efficiency >80-90\%), higher speed (e.g., 100-200~fps) cameras with shorter readout times (e.g., 1-2~ms) on multiband (e.g., two or more passbands) beam splitters attached to meter size telescopes will allow for a better time resolution of the LIFs, hence for a better estimation of their peak temperatures. All these steps are recommended to take place in the following 2-3 years before the planned launch of the `Lunar Meteoroid Impact Observer' (LUMIO) mission\footnote{\url{https://www.eoportal.org/satellite-missions/lumio}} \citep{TOP18, CIP18, TOP23} in 2026. LUMIO, similar to the NELIOTA setup, will be equipped with two cameras attached to a dichroic beam splitter and will observe in the visible and near infrared bands with a frame rate of 15-20~fps. It will continuously observe the Moon for approximately 15 days per month, when lunar illumination phase is less than 50\%. Even though LUMIO is anticipated to observe hundreds of LIFs, its position at the Earth-Moon L2 area cannot permanently solve the aforementioned problem with the subradiant point of a stream on the Moon. Moreover, due to the opposition of the spacecraft with respect to Earth, it will observe the Moon at opposite phases than the observers on Earth. Therefore, we argue that ground based observations from various sites should continue as they are complementary to those of LUMIO.

\section{Summary and conclusions}
\label{sec:SUM}

This work presented detailed results for 113 validated and 70 suspected lunar impact flashes observed by NELIOTA between August 2019 and mid-August 2023. From this sample, 99 were multiframe flashes and their light curves were presented. From the latter LIFs, 15 of them were multiframe flashes in both passbands, hence, their temperature curves were presented as well. Physical properties of the impacting meteoroids, temperatures of the impacts and expected crater sizes were calculated for these events. These results were combined with the previous of the NELIOTA team (Paper~III) yielding a total sample of 192 validated and 103 suspected LIFs observed during a total of 283.4~h between February 2017 - mid-August 2023. An extensive statistical analysis was presented regarding the properties of the projectiles and their impacts on Moon as well as their appearance frequencies on Moon and around Earth. These frequencies were further used to estimate the impact probabilities of a meteoroid, lying within specific physical properties ranges, with a hypothetical lunar infrastructure or with an orbiting satellite. The statistics of the energies of the projectiles were compared with known catastrophic energies in order to estimate the possible damage on a space vehicle or building.

Using the recommended values of luminous efficiency $\eta$, we derived the statistics of the masses and the radii of the impacting meteoroids produced validated LIFs. For the suspected LIFs, we made rough estimation of the meteoroid parameters based on the $LE$ per band correlation as yielded from the validated LIFs. Our statistics, for $\eta=1.5\times10^{-3}$, showed that the majority (75-80\%) of the impacting meteoroids that produced validated LIFs have masses less than 200~g and radii less than 3~cm (i.e., smaller than a tennis ball) and produced craters with sizes between 1.5-3~m. Empirical relations between the magnitudes of the LIFs in each observed band and the meteoroid masses were derived for various meteoroid velocities.

The vast majority ($\sim85\%$) of the impacts produced peak temperatures within the range 2000-4500~K confirming that they are relatively cool events. However, no clear correlation between the temperature and the mass or the size of the meteoroids was found. Moreover, for the multiframe LIFs in both passbands, we found that there is no unique behaviour after the observed maximum temperature, meaning that most of them exhibit a drop off after the peak, but many of them presented constant or a slight increase in temperature. These results indicate that melting and/or complicated thermal processes of the plume probably play a significant role to the thermal evolution and the peak temperature.

The appearance frequency of meteoroids around Earth and on Moon, based on the NELIOTA LIF detection rates, allowed for the estimation of the probability of a potential impact of a meteoroid with a hypothetical infrastructure on Moon (e.g., lunar bases) or with a satellite. It should be clarified that the calculated probabilities concern only meteoroids with the aforementioned physical properties (i.e., a few grams up to several hundreds of grams and sizes of the order of a few centimeters) and not their global population (e.g., micro- or nano- meteoroids). The impact probability is directly proportional to the potential impact surface area and the duration of the mission. Small buildings on Moon (e.g., with surface areas less than 400~m$^2$) have tiny probabilities of being hit ($2-9\times10^{-5}$) even within a 10-year time period. However, for larger buildings (surface $\sim$8000~m$^2$) or group of buildings (dependent of each other) the probability increases and reaches the value of 0.9\% in the extreme scenario (based on validated and suspected LIFs during streams). More or less, the same stands for the orbiting satellites. The larger the surface area of the panels or the spacecraft itself the higher the impact probability. For small size satellites (e.g., Starlink size or cubeSats) the impact probability is extremely low (order of $10^{-7}$) for a mission duration of five years, but reaches the 1-2\% for International Space Station sizes for longer periods of time. On the other hand, although these probabilities seem very low, the damages that can be caused may be catastrophic for the missions, since the transferred energy of the meteoroids could reach the order of 1~GJ (i.e., similar to hand grenade explosion), but typically is of the order of 10-100~MJ (i.e., energy released by fireworks). Undoubtedly, along with the estimated probability of a meteoroid impact, the latter should be taken into account by the space industry for the shielding of space vehicles of buildings.

\begin{acknowledgements}
The authors gratefully acknowledge financial support by the European Space Agency under the NELIOTA program (up to January 2021), contract No.~4000112943, and the Consolidating Activities Regarding Moon, Earth and NEOs (CARMEN) project (August 2021-July 2023), No.~4000134667, via a subcontract of NOA with 6ROADS. The observations of August 2023 were funded by Europlanet 2024 RI after a successful telescope time application by D.~V.~K., A.~L., and A.~Z.~B. Europlanet 2024 RI has received funding from the European Union’s Horizon 2020 research and innovation programme under grant agreement No~871149. The authors wish to thank the referee M.~Yanagisawa for his fruitful comments that improved the quality of the paper. This study is based on observations made with the 1.2~m Kryoneri telescope, Corinthia, Greece, which is operated by the Institute for Astronomy, Astrophysics, Space Applications and Remote Sensing of the National Observatory of Athens, Greece.
\end{acknowledgements}

%
%

\bibliographystyle{aa} 
\bibliography{references_NELIOTA_IV} 

\begin{appendix}

\section{Tables of results}
\label{sec:AppResults}

\begin{sidewaystable*}
\centering
\caption{Photometric results and locations of the detected flashes by NELIOTA after July 2019. Errors are included in parentheses alongside magnitude values and correspond to the last digit(s). The error in the determination of the location is set as 0.5$\degr$.}
\label{tab:list}
\scalebox{0.93}{
\begin{tabular}{cccc cccc | cccc cccc}
\hline\hline																															
ID	&	Date \& UT	&	Val.	&	$t_{\rm max}$	&	$m_{\rm R}$	&	$m_{\rm I}$	&	Lat.	&	Long.	&	ID	&	Date \& UT	&	Val.	&	$t_{\rm max}$	&	$m_{\rm R}$	&	$m_{\rm I}$	&	Lat.	&	Long.	\\
	&		&		&	(ms)	&	(mag)	&	(mag)	&	($\degr$)	&	($\degr$)	&		&		&		&	(ms)	&	(mag)	&	(mag)	&	($\degr$)	&	($\degr$)	\\
\hline																															
113	&	2019 08 06    18:08:34.853 	&	SC2	&	33	&	10.15(12)	&	9.81(15)	&	$	10.3	$	&	$	-69.2	$	&	155	&	2020 07 25    18:55:58.669 	&	SC2	&	33	&		&	9.38(10)	&	$	6.5	$	&	$	-78.1	$	\\
114	&	2019 08 06    18:19:16.023 	&	Val.	&	66	&	10.37(28)	&	8.71(12)	&	$	18.7	$	&	$	-26.3	$	&	156	&	2020 07 26    19:08:21.285 	&	Val.	&	33	&	10.20(12)	&	9.13(12)	&	$	14.0	$	&	$	-28.6	$	\\
115$^a$	&	2019 08 06    18:50:01.718 	&	SC2	&	66	&	8.49(11)	&	8.35(10)	&	$	18.3	$	&	$	-21.2	$	&	157	&	2020 07 26    19:10:25.428 	&	Val.	&	66	&	9.15(10)	&	7.82(6)	&	$	11.3	$	&	$	-35.2	$	\\
116	&	2019 08 06    18:56:36.990 	&	Val.	&	66	&	8.43(12)	&	7.38(7)	&	$	23.0	$	&	$	-7.3	$	&	158	&	2020 08 13     00:57:10.702 	&	Val.	&	33	&	10.16(26)	&	9.03(9)	&	$	2.3	$	&	$	40.9	$	\\
117	&	2019 08 06    18:59:15.735 	&	Val.	&	33	&	9.34(13)	&	9.00(9)	&	$	41.0	$	&	$	-56.0	$	&	159	&	2020 08 14    00:54:21.133 	&	Val.	&	33	&	9.29(15)	&	8.53(9)	&	$	-12.0	$	&	$	35.7	$	\\
118	&	2019 08 26    02:50:56.348 	&	Val.	&	66	&	10.65(27)	&	9.11(7)	&	$	9.1	$	&	$	42.7	$	&	160	&	2020 08 14    01:15:36.278 	&	Val.	&	66	&	9.95(20)	&	9.45(12)	&	$	18.0	$	&	$	20.6	$	\\
119$^a$	&	2019 08 27    02:58:28.586 	&	SC2	&	33	&	10.44(17)	&	10.29(12)	&	$	-33.5	$	&	$	40.8	$	&	161	&	2020 12 09    03:09:58.098 	&	Val.	&	33	&	9.83(18)	&	9.32(10)	&	$	0.4	$	&	$	78.1	$	\\
120	&	2019 08 28    03:03:30.931 	&	Val.	&	33	&	10.98(31)	&	9.82(16)	&	$	-17.2	$	&	$	33.9	$	&	162	&	2021 03 17    17:46:52.983 	&	Val.	&	99	&	9.48(12)	&	7.97(6)	&	$	-25.8	$	&	$	-12.3	$	\\
121	&	2019 09 05    18:11:59.277 	&	Val.	&	33	&	10.13(21)	&	9.52(11)	&	$	-22.5	$	&	$	-65.9	$	&	163	&	2021 03 17    18:03:02.205 	&	SC1	&	66	&		&	10.14(15)	&	$	-20.1	$	&	$	-69.0	$	\\
122	&	2019 09 05    18:51:32.998 	&	Val.	&	33	&	10.25(27)	&	9.53(16)	&	$	-4.8	$	&	$	-53.0	$	&	164	&	2021 03 17    18:07:18.604 	&	Val.	&	33	&	10.47(22)	&	9.21(8)	&	$	28.1	$	&	$	-28.7	$	\\
123	&	2019 09 23    01:08:44.025 	&	SC2	&	33	&		&	10.16(17)	&	$	3.0	$	&	$	66.9	$	&	165	&	2021 04 18    19:13:58.703 	&	Val.	&	99	&	9.43(15)	&	7.89(6)	&	$	17.2	$	&	$	-55.6	$	\\
124	&	2019 09 23    03:36:21.556 	&	Val.	&	66	&	10.20(28)	&	9.40(11)	&	$	-31.8	$	&	$	25.6	$	&	166	&	2021 04 18    20:19:24.000 	&	Val.	&	33	&	9.74(20)	&	8.64(10)	&	$	-19.6	$	&	$	-35.6	$	\\
125	&	2019 09 25    03:40:11.073 	&	Val.	&	99	&	8.43(8)	&	7.42(5)	&	$	12.5	$	&	$	17.8	$	&	167	&	2021 05 15    18:38:40.847 	&	Val.	&	33	&	10.12(22)	&	9.38(10)	&	$	0.3	$	&	$	-15.7	$	\\
126	&	2019 10 22    04:11:36.271 	&	Val.	&	66	&	10.00(18)	&	9.28(11)	&	$	-1.6	$	&	$	65.3	$	&	168	&	2021 05 18    20:08:21.742 	&	Val.	&	66	&	9.90(20)	&	8.66(7)	&	$	-6.0	$	&	$	-57.2	$	\\
127	&	2019 10 24    02:30:16.186 	&	Val.	&	66	&	9.49(15)	&	7.88(7)	&	$	-5.7	$	&	$	55.0	$	&	169	&	2021 06 15    19:06:12.914 	&	Val.	&	66	&	9.96(15)	&	9.19(8)	&	$	-3.6	$	&	$	-44.5	$	\\
128	&	2019 10 25    03:43:44.133 	&	SC2	&	33	&		&	10.78(20)	&	$	12.0	$	&	$	22.5	$	&	170	&	2021 06 15    19:15:27.433 	&	SC1	&	66	&		&	9.57(13)	&	$	-5.7	$	&	$	-23.2	$	\\
129	&	2019 11 02    16:20:44.065 	&	SC1	&	66	&		&	9.26(12)	&	$	9.9	$	&	$	-26.1	$	&	171	&	2021 06 15    19:23:18.620 	&	Val.	&	66	&	10.86(32)	&	10.19(17)	&	$	2.4	$	&	$	-43.7	$	\\
130	&	2019 11 02    17:19:19.668 	&	Val.	&	66	&	10.22(29)	&	9.26(12)	&	$	-33.9	$	&	$	-48.3	$	&	172	&	2021 06 15    19:38:52.236 	&	Val.	&	165	&	8.15(10)	&	6.73(6)	&	$	22.5	$	&	$	-78.8	$	\\
131	&	2019 11 03    17:49:38.486 	&	Val.	&	33	&	9.61(23)	&	8.77(12)	&	$	12.1	$	&	$	-25.3	$	&	173	&	2021 07 03    01:23:35.666 	&	SC1	&	66	&		&	9.26(12)	&	$	3.3	$	&	$	66.8	$	\\
132	&	2019 12 01    16:14:30.209 	&	Val.	&	33	&	9.17(15)	&	8.46(12)	&	$	-16.7	$	&	$	3.3	$	&	174	&	2021 07 05    01:48:25.878 	&	SC2	&	33	&		&	9.81(17)	&	$	11.3	$	&	$	26.6	$	\\
133	&	2019 12 01    16:23:13.796 	&	Val.	&	132	&	8.42(10)	&	5.57(7)	&	$	-13.1	$	&	$	-23.1	$	&	175	&	2021 07 06    02:11:11.359 	&	Val.	&	33	&	10.82(18)	&	10.33(17)	&	$	-11.3	$	&	$	82.1	$	\\
134	&	2019 12 01    16:30:43.411 	&	Val.	&	33	&	10.75(28)	&	9.12(15)	&	$	16.5	$	&	$	-40.7	$	&	176	&	2021 07 15    19:22:29.410 	&	Val.	&	33	&	10.20(27)	&	8.76(9)	&	$	13.9	$	&	$	-43.4	$	\\
135	&	2019 12 01    17:14:41.437 	&	Val.	&	33	&	11.16(39)	&	9.35(15)	&	$	-28.6	$	&	$	-32.4	$	&	177	&	2021 07 16    18:49:32.113 	&	Val.	&	66	&	9.96(34)	&	9.15(12)	&	$	3.1	$	&	$	-34.9	$	\\
136	&	2019 12 20    04:34:16.679 	&	Val.	&	99	&	9.50(20)	&	8.51(7)	&	$	-39.8	$	&	$	48.7	$	&	178$^a$	&	2021 08 02    00:48:25.336 	&	SC2	&	33	&	9.55(21)	&	9.46(15)	&	$	22.1	$	&	$	24.5	$	\\
137$^{b}$	&	2020 01 30    17:17:59.771 	&	SC2	&	165	&	10.35(21)	&	10.07(15)	&	$	23.1	$	&	$	-40.8	$	&	179	&	2021 08 02    01:26:06.111 	&	Val.	&	66	&	10.41(29)	&	9.85(14)	&	$	4.4	$	&	$	80.2	$	\\
138	&	2020 01 30    17:18:08.539 	&	Val.	&	66	&	11.03(33)	&	9.51(12)	&	$	21.9	$	&	$	-46.5	$	&	180	&	2021 08 02    01:34:28.145 	&	Val.	&	66	&	8.75(18)	&	8.28(9)	&	$	4.8	$	&	$	30.3	$	\\
139	&	2020 01 30    17:35:38.892 	&	Val.	&	33	&	10.62(22)	&	9.75(12)	&	$	-39.1	$	&	$	-40.4	$	&	181	&	2021 08 02    02:44:36.306 	&	Val.	&	66	&	10.12(26)	&	9.72(15)	&	$	-4.8	$	&	$	52.8	$	\\
140	&	2020 01 30    18:53:24.062 	&	SC2	&	33	&		&	9.58(12)	&	$	-29.9	$	&	$	-72.8	$	&	182	&	2021 08 02    02:51:02.646 	&	Val.	&	99	&	8.85(17)	&	7.74(6)	&	$	28.4	$	&	$	16.3	$	\\
141	&	2020 01 31    17:37:21.034 	&	SC2	&	33	&		&	8.82(14)	&	$	-16.2	$	&	$	-19.5	$	&	183	&	2021 10 03    03:14:17.079 	&	Val.	&	33	&	10.16(19)	&	9.37(9)	&	$	-34.7	$	&	$	33.9	$	\\
142	&	2020 03 01    16:54:23.862 	&	Val.	&	99	&	8.32(8)	&	7.15(4)	&	$	-7.8	$	&	$	-45.1	$	&	184	&	2021 10 11    16:57:00.374 	&	Val.	&	165	&	8.66(13)	&	7.73(8)	&	$	-21.2	$	&	$	-71.8	$	\\
143	&	2020 03 01    17:10:06.375 	&	Val.	&	33	&	9.92(25)	&	9.42(12)	&	$	-30.6	$	&	$	-31.3	$	&	185	&	2021 10 12    16:31:15.618 	&	Val.	&	66	&	9.33(22)	&	8.25(10)	&	$	-15.1	$	&	$	-32.5	$	\\
144	&	2020 03 27    17:40:25.320 	&	Val.	&	66	&	10.01(15)	&	8.70(8)	&	$	-1.9	$	&	$	-32.1	$	&	186	&	2021 10 12    17:42:17.599 	&	Val.	&	66	&	9.75(20)	&	8.75(10)	&	$	11.2	$	&	$	-77.3	$	\\
145	&	2020 03 29    18:14:10.875 	&	Val.	&	66	&	10.83(23)	&	9.72(9)	&	$	15.1	$	&	$	-91.3	$	&	187	&	2021 12 01    04:21:42.906 	&	SC1	&	66	&		&	10.00(13)	&	$	-30.6	$	&	$	73.8	$	\\
146	&	2020 03 29    19:07:30.900 	&	SC2	&	33	&		&	10.00(14)	&	$	0.4	$	&	$	-59.5	$	&	188	&	2021 12 08    16:15:10.805 	&	Val.	&	66	&	8.58(12)	&	7.59(7)	&	$	3.5	$	&	$	-29.5	$	\\
147	&	2020 03 29    19:16:46.509 	&	Val.	&	33	&	10.18(21)	&	9.26(9)	&	$	-22.7	$	&	$	-42.8	$	&	189	&	2021 12 08    16:34:21.521 	&	Val.	&	33	&	10.17(31)	&	8.20(8)	&	$	14.1	$	&	$	-56.0	$	\\
148	&	2020 04 28    19:19:54.525 	&	Val.	&	66	&	8.99(10)	&	8.13(5)	&	$	-26.6	$	&	$	-47.7	$	&	190	&	2022 02 05    17:41:56.145 	&	SC2	&	33	&		&	9.64(15)	&	$	-35.2	$	&	$	-19.4	$	\\
149$^a$	&	2020 06 15    02:04:12.513 	&	SC2	&	33	&	9.50(19)	&	9.34(12)	&	$	49.4	$	&	$	58.0	$	&	191	&	2022 04 05    17:30:55.889 	&	Val.	&	99	&	8.87(17)	&	7.53(6)	&	$	4.7	$	&	$	-15.4	$	\\
150	&	2020 06 16    02:09:14.848 	&	SC1	&	66	&		&	9.53(13)	&	$	-10.7	$	&	$	67.2	$	&	192	&	2022 04 05    17:54:37.532 	&	Val.	&	33	&	9.31(15)	&	7.95(7)	&	$	-7.8	$	&	$	-43.2	$	\\
151	&	2020 06 25    18:28:18.340 	&	Val.	&	132	&	7.92(10)	&	6.66(6)	&	$	10.8	$	&	$	-46.5	$	&	193	&	2022 05 07    20:47:14.061 	&	SC1	&	66	&		&	7.41(8)	&	$	8.8	$	&	$	-34.4	$	\\
152	&	2020 06 25    19:09:24.063 	&	SC1	&	66	&		&	9.76(13)	&	$	33.4	$	&	$	-33.1	$	&	194	&	2022 05 25    02:25:10.955 	&	SC2	&	33	&		&	10.15(14)	&	$	19.9	$	&	$	88.2	$	\\
153	&	2020 06 26    19:52:31.569 	&	Val.	&	66	&	9.73(11)	&	8.22(7)	&	$	12.7	$	&	$	-24.9	$	&	195	&	2022 06 03    18:21:31.378 	&	Val.	&	132	&	7.96(11)	&	6.64(6)	&	$	-4.0	$	&	$	-28.8	$	\\
154	&	2020 07 25    18:41:34.204 	&	SC2	&	33	&		&	9.61(13)	&	$	17.3	$	&	$	-30.0	$	&	196	&	2022 06 03    18:57:25.884 	&	SC1	&	66	&		&	9.17(9)	&	$	35.8	$	&	$	-29.5	$	\\

\hline																															
\end{tabular}}
\tablefoot{Val.= Validated flash, SC1/2 = Suspected flash of Class 1/2 (see Paper~III for details), $^{(a)}$ abnormal R-I index, $^{(b)}$slight displacement and abnormal temperature evolution.}
\end{sidewaystable*}
\begin{sidewaystable*}
\centering
\caption*{Table~\ref{tab:list} (cont'd)}
\scalebox{0.93}{
\begin{tabular}{cccc cccc | cccc cccc}
\hline\hline																															
ID	&	Date \& UT	&	Val.	&	$t_{\rm max}$	&	$m_{\rm R}$	&	$m_{\rm I}$	&	Lat.	&	Long.	&	ID	&	Date \& UT	&	Val.	&	$t_{\rm max}$	&	$m_{\rm R}$	&	$m_{\rm I}$	&	Lat.	&	Long.	\\
	&		&		&	(ms)	&	(mag)	&	(mag)	&	($\degr$)	&	($\degr$)	&		&		&		&	(ms)	&	(mag)	&	(mag)	&	($\degr$)	&	($\degr$)	\\
\hline																															
197	&	2022 06 03    19:21:10.176 	&	SC2	&	33	&		&	8.64(9)	&	$	-5.0	$	&	$	-22.8	$	&	239	&	2022 10 30    16:42:35.356 	&	SC2	&	33	&		&	10.16(22)	&	$	24.8	$	&	$	-53.1	$	\\
198	&	2022 06 03    19:34:16.804 	&	SC1	&	66	&		&	9.44(13)	&	$	-9.4	$	&	$	-75.9	$	&	240	&	2022 10 30    16:47:27.612 	&	Val.	&	66	&	9.68(22)	&	8.56(9)	&	$	-1.3	$	&	$	-75.6	$	\\
199	&	2022 06 04    18:20:50.612 	&	Val.	&	66	&	9.40(28)	&	8.32(9)	&	$	-7.2	$	&	$	-25.0	$	&	241	&	2022 10 30    16:54:54.929 	&	Val.	&	66	&	9.65(19)	&	8.57(9)	&	$	-27.6	$	&	$	-59.0	$	\\
200	&	2022 06 04    18:22:46.921 	&	Val.	&	33	&	10.15(38)	&	9.19(14)	&	$	16.2	$	&	$	-19.9	$	&	242	&	2022 10 30    17:13:28.165 	&	SC1	&	66	&		&	8.98(14)	&	$	-7.6	$	&	$	-21.1	$	\\
201	&	2022 06 04    19:44:16.807 	&	Val.	&	33	&	10.06(31)	&	9.45(15)	&	$	-9.4	$	&	$	-59.7	$	&	243	&	2022 10 30    17:34:04.033 	&	SC2	&	33	&		&	9.44(15)	&	$	4.1	$	&	$	-51.0	$	\\
202	&	2022 06 23    01:46:07.604 	&	Val.	&	66	&	8.73(15)	&	7.22(7)	&	$	5.3	$	&	$	84.4	$	&	244	&	2022 10 30    17:34:16.134 	&	Val.	&	132	&	8.64(14)	&	8.03(11)	&	$	-18.9	$	&	$	-45.2	$	\\
203$^c$	&	2022 07 22    00:27:39.090 	&	SC2	&	33	&	9.21(25)	&	8.84(17)	&	$	-7.3	$	&	$	24.6	$	&	245	&	2022 10 30    17:41:51.409 	&	SC1	&	66	&		&	8.98(13)	&	$	-26.8	$	&	$	-46.3	$	\\
204	&	2022 07 22    02:13:27.207 	&	Val.	&	66	&	9.97(24)	&	8.73(7)	&	$	-4.6	$	&	$	57.1	$	&	246	&	2022 10 30    18:06:20.288 	&	SC2	&	33	&		&	9.50(18)	&	$	10.5	$	&	$	-48.4	$	\\
205	&	2022 07 22    02:48:11.224 	&	Val.	&	66	&	10.27(36)	&	9.29(14)	&	$	10.7	$	&	$	51.1	$	&	247	&	2022 10 30    18:11:38.346 	&	SC2	&	33	&		&	9.02(15)	&	$	13.9	$	&	$	-57.9	$	\\
206	&	2022 07 22    02:49:51.365 	&	Val.	&	132	&	9.36(29)	&	7.51(10)	&	$	13.1	$	&	$	12.7	$	&	248	&	2022 10 30    18:13:57.723 	&	SC2	&	33	&		&	9.53(18)	&	$	17.3	$	&	$	-88.0	$	\\
207	&	2022 07 24    02:23:51.582 	&	SC2	&	33	&		&	9.95(11)	&	$	23.9	$	&	$	44.4	$	&	249	&	2022 10 31    18:58:46.938 	&	SC2	&	33	&		&	8.92(13)	&	$	-5.2	$	&	$	-79.1	$	\\
208	&	2022 08 01    18:27:23.335 	&	Val.	&	33	&	9.93(27)	&	8.66(15)	&	$	6.9	$	&	$	-45.3	$	&	250	&	2022 10 31    18:59:28.842 	&	SC2	&	33	&		&	9.35(17)	&	$	-1.5	$	&	$	-53.4	$	\\
209	&	2022 08 01    18:28:18.506 	&	Val.	&	99	&	9.42(21)	&	7.62(13)	&	$	-4.3	$	&	$	-67.7	$	&	251	&	2022 10 31    19:17:17.423 	&	Val.	&	66	&	8.60(21)	&	7.51(12)	&	$	7.1	$	&	$	-35.2	$	\\
210	&	2022 08 03    18:05:38.820 	&	SC1	&	66	&		&	8.66(9)	&	$	-4.1	$	&	$	-75.7	$	&	252	&	2022 11 19    02:39:27.785 	&	Val.	&	66	&	8.82(15)	&	7.92(8)	&	$	-13.6	$	&	$	21.6	$	\\
211	&	2022 08 03    18:17:31.174 	&	SC2	&	33	&		&	9.42(13)	&	$	10.8	$	&	$	-55.5	$	&	253	&	2022 11 19    02:45:43.608 	&	SC1	&	66	&		&	8.51(10)	&	$	-3.4	$	&	$	21.6	$	\\
212	&	2022 08 03    18:23:42.636 	&	Val.	&	66	&	9.76(12)	&	8.37(8)	&	$	28.5	$	&	$	-73.1	$	&	254	&	2022 11 19    03:51:52.914 	&	SC2	&	33	&		&	9.48(14)	&	$	-2.4	$	&	$	40.9	$	\\
213	&	2022 08 03    19:02:12.594 	&	SC1	&	66	&		&	10.04(19)	&	$	-4.1	$	&	$	-86.4	$	&	255	&	2022 11 19    04:05:36.877 	&	SC2	&	33	&		&	9.42(13)	&	$	-1.3	$	&	$	69.3	$	\\
214	&	2022 08 04    18:47:06.667 	&	Val.	&	66	&	9.26(18)	&	7.63(13)	&	$	-21.0	$	&	$	-40.7	$	&	256	&	2022 11 19    04:15:39.560 	&	SC2	&	33	&		&	8.77(9)	&	$	9.5	$	&	$	30.0	$	\\
215	&	2022 08 04    19:24:20.958 	&	SC1	&	66	&		&	8.02(12)	&	$	-4.2	$	&	$	-67.1	$	&	257	&	2022 12 18    03:37:48.471 	&	Val.	&	66	&	9.11(14)	&	7.73(6)	&	$	-2.2	$	&	$	51.7	$	\\
216	&	2022 08 04    19:25:11.729 	&	SC1	&	66	&		&	9.40(19)	&	$	6.7	$	&	$	-73.5	$	&	258	&	2022 12 18    03:41:47.156 	&	SC2	&	33	&		&	9.93(20)	&	$	2.6	$	&	$	72.5	$	\\
217	&	2022 08 22    01:33:02.291 	&	SC1	&	66	&		&	8.45(9)	&	$	-17.7	$	&	$	8.0	$	&	259	&	2022 12 26    15:46:16.570 	&	Val.	&	198	&	7.76(11)	&	6.38(8)	&	$	-15.9	$	&	$	-53.2	$	\\
218	&	2022 08 22    02:11:20.544 	&	SC2	&	33	&		&	9.86(26)	&	$	-14.1	$	&	$	89.1	$	&	260	&	2022 12 26    16:38:05.663 	&	SC2	&	33	&		&	10.45(21)	&	$	2.2	$	&	$	-29.3	$	\\
219	&	2022 09 01    18:33:36.493 	&	Val.	&	66	&	10.66(37)	&	8.04(14)	&	$	17.9	$	&	$	-62.0	$	&	261	&	2022 12 26    16:48:14.320 	&	Val.	&	66	&	10.26(20)	&	9.68(12)	&	$	0.2	$	&	$	-66.5	$	\\
220	&	2022 09 02    18:06:48.349 	&	SC2	&	33	&		&	9.36(15)	&	$	33.7	$	&	$	-34.1	$	&	262	&	2022 12 27    16:18:50.350 	&	Val.	&	33	&	9.64(22)	&	9.08(12)	&	$	-11.4	$	&	$	-8.0	$	\\
221	&	2022 09 02    19:03:02.542 	&	SC2	&	33	&		&	8.36(15)	&	$	10.2	$	&	$	-22.8	$	&	263$^a$	&	2022 12 27    17:42:53.269 	&	SC2	&	66	&	8.84(16)	&	8.71(11)	&	$	12.4	$	&	$	-11.4	$	\\
222	&	2022 10 19    01:43:11.699 	&	SC1	&	66	&		&	9.88(15)	&	$	-9.1	$	&	$	79.8	$	&	264	&	2022 12 27    17:47:37.390 	&	Val.	&	66	&	9.29(17)	&	7.98(9)	&	$	-7.2	$	&	$	-11.2	$	\\
223	&	2022 10 19    03:03:44.141 	&	Val.	&	99	&	8.47(10)	&	8.25(6)	&	$	-3.3	$	&	$	61.6	$	&	265	&	2022 12 27    17:57:38.537 	&	SC2	&	33	&		&	9.81(16)	&	$	-13.6	$	&	$	-46.8	$	\\
224	&	2022 10 20    01:12:58.151 	&	Val.	&	429	&	8.58(14)	&	7.38(8)	&	$	-14.2	$	&	$	21.4	$	&	266	&	2022 12 27    18:11:32.401 	&	Val.	&	66	&	9.73(12)	&	8.53(13)	&	$	-29.6	$	&	$	-6.6	$	\\
225	&	2022 10 20    01:23:08.098 	&	SC2	&	33	&		&	9.89(14)	&	$	-7.5	$	&	$	53.8	$	&	267	&	2023 01 16    04:11:21.224 	&	Val.	&	66	&	10.09(26)	&	9.04(9)	&	$	-36.9	$	&	$	51.7	$	\\
226	&	2022 10 20    02:05:55.914 	&	SC1	&	66	&		&	9.14(9)	&	$	6.3	$	&	$	50.0	$	&	268	&	2023 02 22    17:49:40.667 	&	Val.	&	33	&	10.37(22)	&	10.15(16)	&	$	-41.0	$	&	$	-32.0	$	\\
227	&	2022 10 20    02:56:37.878 	&	Val.	&	66	&	9.08(10)	&	8.59(6)	&	$	-7.2	$	&	$	87.3	$	&	269	&	2023 02 22    17:52:04.312 	&	Val.	&	99	&	9.97(19)	&	8.70(13)	&	$	-33.4	$	&	$	-16.7	$	\\
228	&	2022 10 20    03:32:21.841 	&	SC2	&	33	&		&	9.83(14)	&	$	0.6	$	&	$	74.8	$	&	270	&	2023 02 24    17:25:21.560 	&	SC1	&	66	&		&	10.00(13)	&	$	2.1	$	&	$	-62.2	$	\\
229	&	2022 10 20    04:11:08.419 	&	SC1	&	66	&		&	8.90(8)	&	$	-14.9	$	&	$	70.1	$	&	271	&	2023 03 24    18:36:01.406	&	SC2	&	33	&		&	10.24(15)	&	$	-14.4	$	&	$	-53.0	$	\\
230	&	2022 10 21    02:55:11.697 	&	Val.	&	66	&	10.22(24)	&	9.56(10)	&	$	-2.9	$	&	$	68.0	$	&	272	&	2023 03 26    20:25:28.546	&	Val.	&	66	&	9.88(18)	&	8.79(10)	&	$	18.4	$	&	$	-72.5	$	\\
231	&	2022 10 22    03:07:56.784 	&	Val.	&	33	&	10.92(28)	&	10.27(15)	&	$	-26.1	$	&	$	59.6	$	&	273	&	2023 03 27    17:23:31.571	&	SC2	&	33	&		&	9.70(11)	&	$	20.9	$	&	$	-55.4	$	\\
232	&	2022 10 22    03:25:17.830 	&	Val.	&	33	&	10.44(10)	&	9.55(13)	&	$	-21.8	$	&	$	21.0	$	&	274	&	2023 03 27    18:25:58.056	&	SC2	&	33	&		&	8.74(7)	&	$	-2.1	$	&	$	-38.3	$	\\
233	&	2022 10 22    03:39:55.152 	&	Val.	&	33	&	10.70(15)	&	9.91(7)	&	$	-26.4	$	&	$	22.9	$	&	275	&	2023 04 23    18:04:39.828	&	Val.	&	66	&	10.60(23)	&	9.61(10)	&	$	20.04	$	&	$	-20.5	$	\\
234	&	2022 10 22    03:55:35.517 	&	Val.	&	66	&	9.51(14)	&	8.68(7)	&	$	-25.0	$	&	$	49.2	$	&	276	&	2023 04 24    17:58:30.793	&	Val.	&	33	&	10.47(32)	&	9.10(8)	&	$	-28.9	$	&	$	-26.5	$	\\
235	&	2022 10 22    03:56:50.322 	&	Val.	&	66	&	10.31(25)	&	9.51(11)	&	$	-4.7	$	&	$	14.8	$	&	277	&	2023 04 24    19:54:33.094	&	SC2	&	33	&		&	9.71(14)	&	$	-5.5	$	&	$	-33.0	$	\\
236	&	2022 10 29    17:03:58.478 	&	Val.	&	99	&	9.25(19)	&	7.72(13)	&	$	-7.7	$	&	$	-65.2	$	&	278	&	2023 04 24    19:58:40.780	&	SC2	&	33	&		&	10.08(18)	&	$	-3.9	$	&	$	-54.8	$	\\
237	&	2022 10 29    17:13:31.322 	&	SC1	&	66	&		&	7.68(13)	&	$	4.5	$	&	$	-53.0	$	&	279	&	2023 04 24    20:02:42.098	&	Val.	&	66	&	9.63(25)	&	8.28(10)	&	$	-26.36	$	&	$	-32.3	$	\\
238	&	2022 10 30    16:41:05.648 	&	Val.	&	33	&	10.08(30)	&	9.26(14)	&	$	-16.9	$	&	$	-25.9	$	&	280	&	2023 05 23    20:06:14.792	&	Val.	&	66	&	8.94(16)	&	8.16(11)	&	$	-12.03	$	&	$	-12.9	$	\\

\hline																															
\end{tabular}}
\tablefoot{Val.= Validated flash, SC1/2 = Suspected flash of Class 1/2 (see Paper~III for details), $^{(a)}$ abnormal R-I index, $^{(b)}$slight displacement and abnormal temperature evolution, $^{(c)}$ too elongated shape in $I$ filter and too short duration for its magnitude.}
\end{sidewaystable*}

\begin{sidewaystable*}
\centering
\caption*{Table~\ref{tab:list} (cont'd)}
\scalebox{0.93}{
\begin{tabular}{cccc cccc | cccc cccc}
\hline\hline																															
ID	&	Date \& UT	&	Val.	&	$t_{\rm max}$	&	$m_{\rm R}$	&	$m_{\rm I}$	&	Lat.	&	Long.	&	ID	&	Date \& UT	&	Val.	&	$t_{\rm max}$	&	$m_{\rm R}$	&	$m_{\rm I}$	&	Lat.	&	Long.	\\
	&		&		&	(ms)	&	(mag)	&	(mag)	&	($\degr$)	&	($\degr$)	&		&		&		&	(ms)	&	(mag)	&	(mag)	&	($\degr$)	&	($\degr$)	\\
\hline	
281	&	2023 05 24    20:11:10.028	&	Val.	&	330	&	8.32(11)	&	6.41(8)	&	$	23.3	$	&	$	-67.3	$	&	289	&	2023 08 10    01:53:23.822	&	Val.	&	66	&	9.67(16)	&	8.50(6)	&	$	17.2	$	&	$	69.3	$	\\
282	&	2023 05 24    21:03:20.168	&	Val.	&	132	&	9.10(18)	&	7.53(13)	&	$	10.9	$	&	$	-36.1	$	&	290	&	2023 08 11    01:43:53.825	&	Val.	&	66	&	11.32(41)	&	9.94(14)	&	$	37.3	$	&	$	71.6	$	\\
283	&	2023 05 25    20:34:27.999	&	Val.	&	231	&	7.51(10)	&	6.28(7)	&	$	-5.9	$	&	$	-82.9	$	&	291	&	2023 08 11    02:53:30.790	&	SC1	&	66	&		&	9.28(13)	&	$	36.9	$	&	$	13.6	$	\\
284	&	2023 05 26    18:17:43.624	&	Val.	&	66	&	9.79(38)	&	8.25(9)	&	$	0.16	$	&	$	-24.5	$	&	292	&	2023 08 13    02:29:01.918	&	Val.	&	33	&	11.65(65)	&	10.07(13)	&	$	22.3	$	&	$	19.6	$	\\
285	&	2023 06 21    18:47:44.161	&	Val.	&	33	&	9.21(12)	&	9.00(9)	&	$	20.4	$	&	$	-45.4	$	&	293	&	2023 08 13    02:33:29.605	&	Val.	&	33	&	10.59(29)	&	10.04(11)	&	$	16.8	$	&	$	83.8	$	\\
286	&	2023 06 22    19:58:08.964	&	Val.	&	33	&	11.00(36)	&	10.24(17)	&	$	-3.0	$	&	$	-79.0	$	&	294	&	2023 08 13    02:57:22.550	&	SC2	&	33	&		&	11.30(33)	&	$	-6.1	$	&	$	51.0	$	\\
287	&	2023 06 22    20:03:20.188 	&	Val.	&	99	&	10.18(20)	&	8.53(12)	&	$	45.7	$	&	$	-79.2	$	&	295	&	2023 08 13    03:05:54.033	&	Val.	&	66	&	9.30(13)	&	8.76(8)	&	$	23.5	$	&	$	49.8	$	\\
288	&	2023 06 22    20:11:59.394	&	SC2	&	33	&		&	10.23(21)	&	$	-24.6	$	&	$	-44.7	$	&		&		&		&		&		&		&				&				\\

\hline																															
\end{tabular}}
\tablefoot{Val.= Validated flash, SC1/2 = Suspected flash of Class 1/2 (see Paper~III for details), $^{(a)}$ abnormal $R-I$ index, $^{(b)}$slight displacement and abnormal temperature evolution, $^{(c)}$ too elongated shape in $I$ filter and too short duration for its magnitude.}
\end{sidewaystable*}

\begin{sidewaystable*}
\centering
\caption{Results for the validated flashes detected by the NELIOTA campaign between August 2019 and mid-August 2023 as well as for the corresponding projectiles and craters for various $\eta$ values. Errors are included in parentheses alongside values and correspond to the last digit(s).}
\label{tab:ResultsReal}
\scalebox{0.92}{
\begin{tabular}{ccccc|cccc|cccc|cccc}
\hline\hline																																	
	&		&		&		&		&	\multicolumn{4}{c}{$\eta=5\times10^{-3}$}							&	\multicolumn{4}{c}{$\eta=1.5\times10^{-3}$}							&	\multicolumn{4}{c}{$\eta=5\times10^{-4}$}							\\
\hline																																	
ID	&	Stream	&	$LE_{\rm R}$	&	$LE_{\rm I}$	&	$T$	&	$KE_{\rm p}$	&	$m_{\rm p}$	&	$r_{\rm p}$	&	$d_{\rm c}$	&	$KE_{\rm p}$	&	$m_{\rm p}$	&	$r_{\rm p}$	&	$d_{\rm c}$	&	$KE_{\rm p}$	&	$m_{\rm p}$	&	$r_{\rm p}$	&	$d_{\rm c}$	\\
	&		&	($\times 10^4$~J)	&	($\times 10^4$~J)	&	(K)	&	($\times 10^6$~J)	&	(g)	&	(cm)	&	(m)	&	($\times 10^6$~J)	&	(g)	&	(cm)	&	(m)	&	($\times 10^6$~J)	&	(g)	&	(cm)	&	(m)	\\
\hline																																														
114	&	PER	&	0.5(1)	&	1.6(1)	&	2237(317)	&	4.2(4)	&	2.3(2)	&	0.8(1)	&	1.39(5)	&	14(1)	&	8(1)	&	1.1(1)	&	2.0(1)	&	42(4)	&	23(2)	&	1.6(0.2)	&	2.7(1)	\\
116	&	PER	&	2.9(3)	&	5.3(3)	&	3151(305)	&	16.3(8)	&	8.8(4)	&	1.2(1)	&	2.07(6)	&	54(3)	&	29(1)	&	1.8(2)	&	2.9(1)	&	163(8)	&	88(4)	&	2.6(0.3)	&	4.0(1)	\\
117	&	PER	&	1.3(2)	&	0.9(1)	&	6149(1410)	&	4.2(3)	&	2.3(2)	&	0.8(1)	&	1.40(5)	&	14(1)	&	8(1)	&	1.1(1)	&	2.0(1)	&	42(3)	&	23(2)	&	1.6(0.2)	&	2.7(1)	\\
118	&	spo	&	0.4(1)	&	1.5(1)	&	2933(600)	&	3.8(2)	&	26.3(1.5)	&	1.5(5)	&	1.45(5)	&	13(1)	&	88(5)	&	2.2(7)	&	2.1(1)	&	38(2)	&	263(15)	&	3.2(1.1)	&	2.8(1)	\\
120	&	spo	&	0.3(1)	&	0.4(1)	&	2933(600)	&	1.3(2)	&	9.2(1.4)	&	1.1(3)	&	1.07(5)	&	4(1)	&	31(5)	&	1.6(5)	&	1.5(1)	&	13(2)	&	92(14)	&	2.3(0.8)	&	2.1(1)	\\
121	&	spo	&	0.6(1)	&	0.6(1)	&	4443(1207)	&	2.3(3)	&	16.2(1.8)	&	1.3(4)	&	1.26(5)	&	8(1)	&	54(6)	&	1.9(6)	&	1.8(1)	&	23(3)	&	162(18)	&	2.8(0.9)	&	2.5(1)	\\
122	&	spo	&	0.5(1)	&	0.6(1)	&	4016(1214)	&	2.2(3)	&	15.2(2.2)	&	1.3(4)	&	1.24(6)	&	7(1)	&	51(7)	&	1.9(6)	&	1.8(1)	&	22(3)	&	152(22)	&	2.7(0.9)	&	2.4(1)	\\
124	&	spo	&	0.6(1)	&	1.2(1)	&	3777(1042)	&	3.5(3)	&	24.4(2.4)	&	1.5(5)	&	1.42(6)	&	12(1)	&	81(8)	&	2.2(7)	&	2.0(1)	&	35(3)	&	244(24)	&	3.2(1.0)	&	2.8(1)	\\
125	&	spo	&	4.1(2)	&	5.7(2)	&	3212(207)	&	19.6(6)	&	135.8(3.8)	&	2.6(8)	&	2.33(7)	&	65(2)	&	453(13)	&	4(1)	&	3.3(1)	&	196(6)	&	1358(38)	&	5.6(1.8)	&	4.5(1)	\\
126	&	spo	&	0.7(1)	&	1.1(1)	&	4036(810)	&	3.6(3)	&	25.1(2.0)	&	1.5(5)	&	1.43(5)	&	12(1)	&	84(7)	&	2.2(7)	&	2.0(1)	&	36(3)	&	251(20)	&	3.2(1.0)	&	2.8(1)	\\
127	&	spo	&	1.1(1)	&	3.4(2)	&	2288(160)	&	9.0(4)	&	62.6(3.0)	&	2.0(7)	&	1.86(6)	&	30(1)	&	209(10)	&	3(1)	&	2.6(1)	&	90(4)	&	626(30)	&	4.3(1.4)	&	3.6(1)	\\
130	&	spo	&	0.5(1)	&	1.3(1)	&	3345(770)	&	3.7(4)	&	25.5(2.5)	&	1.5(5)	&	1.43(6)	&	12(1)	&	85(8)	&	2.2(7)	&	2.0(1)	&	37(4)	&	255(25)	&	3.2(1.1)	&	2.8(1)	\\
131	&	spo	&	0.9(2)	&	1.2(1)	&	3641(770)	&	4.3(5)	&	29.7(3.3)	&	1.6(5)	&	1.50(6)	&	14(2)	&	99(11)	&	2.3(8)	&	2.1(1)	&	43(5)	&	297(33)	&	3.4(1.1)	&	2.9(1)	\\
132	&	spo	&	1.4(2)	&	1.6(2)	&	4080(647)	&	6.0(5)	&	41.3(3.7)	&	1.8(6)	&	1.65(6)	&	20(2)	&	138(12)	&	2.6(9)	&	2.3(1)	&	60(5)	&	413(37)	&	3.7(1.2)	&	3.2(1)	\\
133	&	spo	&	4.6(3)	&	34.8(1.6)	&	1438(53)	&	79(3)	&	544.8(22.7)	&	4(1)	&	3.49(11)	&	262(11)	&	1816(76)	&	6(2)	&	4.9(1)	&	787(33)	&	5448(227)	&	8.8(2.9)	&	6.8(2)	\\
134	&	spo	&	0.4(1)	&	0.9(1)	&	2277(353)	&	2.5(3)	&	17.3(2.1)	&	1.3(4)	&	1.28(6)	&	8(1)	&	58(7)	&	2.0(6)	&	1.8(1)	&	25(3)	&	173(21)	&	2.8(0.9)	&	2.5(1)	\\
135	&	spo	&	0.3(1)	&	0.7(1)	&	2093(363)	&	2.0(3)	&	13.5(1.9)	&	1.2(4)	&	1.19(6)	&	7(1)	&	45(6)	&	1.8(6)	&	1.7(1)	&	20(3)	&	135(19)	&	2.6(0.9)	&	2.3(1)	\\
136	&	spo	&	1.2(2)	&	3.1(1)	&	3254(477)	&	8.7(5)	&	60.2(3.6)	&	2.0(6)	&	1.84(6)	&	29(2)	&	201(12)	&	3(1)	&	2.6(1)	&	87(5)	&	602(36)	&	4.2(1.4)	&	3.6(1)	\\
138	&	spo	&	0.3(1)	&	1.4(1)	&	2393(427)	&	3.4(3)	&	23.5(1.7)	&	1.5(5)	&	1.40(5)	&	11(1)	&	78(6)	&	2.2(7)	&	2.0(1)	&	34(3)	&	235(17)	&	3.1(1.0)	&	2.7(1)	\\
139	&	spo	&	0.4(1)	&	0.5(1)	&	3558(727)	&	1.9(2)	&	13.2(1.5)	&	1.2(4)	&	1.19(5)	&	6(1)	&	44(5)	&	1.8(6)	&	1.7(1)	&	19(2)	&	132(15)	&	2.6(0.8)	&	2.3(1)	\\
142	&	spo	&	3.7(3)	&	8.9(2)	&	2919(161)	&	25.3(7)	&	175.4(4.9)	&	2.8(9)	&	2.51(7)	&	84(2)	&	585(16)	&	4(1)	&	3.6(1)	&	253(7)	&	1754(49)	&	6.0(2.0)	&	4.9(1)	\\
143	&	spo	&	0.9(2)	&	0.7(1)	&	4991(2002)	&	3.0(4)	&	20.9(3.0)	&	1.4(5)	&	1.35(7)	&	10(1)	&	70(10)	&	2.1(7)	&	1.9(1)	&	30(4)	&	209(30)	&	3.0(1.0)	&	2.6(1)	\\
144	&	spo	&	0.8(1)	&	1.7(1)	&	2677(263)	&	5.1(3)	&	35.1(2.1)	&	1.7(5)	&	1.57(5)	&	17(1)	&	117(7)	&	2.5(8)	&	2.2(1)	&	51(3)	&	351(21)	&	3.6(1.2)	&	3.1(1)	\\
145	&	spo	&	0.4(1)	&	0.7(1)	&	3030(507)	&	2.1(2)	&	14.5(1.2)	&	1.2(4)	&	1.22(4)	&	7(1)	&	48(4)	&	1.8(6)	&	1.7(1)	&	21(2)	&	145(12)	&	2.7(0.9)	&	2.4(1)	\\
147	&	spo	&	0.6(1)	&	0.8(1)	&	3430(605)	&	2.8(3)	&	19.5(1.9)	&	1.4(4)	&	1.33(5)	&	9(1)	&	65(6)	&	2.0(7)	&	1.9(1)	&	28(3)	&	195(19)	&	2.9(1.0)	&	2.6(1)	\\
148	&	spo	&	2.0(2)	&	3.5(1)	&	3587(323)	&	11.0(4)	&	76.4(3.0)	&	2.1(7)	&	1.97(6)	&	37(1)	&	255(10)	&	3(1)	&	2.8(1)	&	110(4)	&	764(30)	&	4.6(1.5)	&	3.8(1)	\\
151	&	spo	&	6.2(5)	&	11.9(4)	&	2763(192)	&	36(1)	&	250.9(8.6)	&	3(1)	&	2.78(8)	&	121(4)	&	836(29)	&	5(2)	&	3.9(1)	&	363(12)	&	2509(86)	&	6.8(2.2)	&	5.4(2)	\\
153	&	spo	&	0.9(1)	&	2.6(1)	&	2408(160)	&	7.0(3)	&	48.3(2.0)	&	1.8(6)	&	1.73(5)	&	23(1)	&	161(7)	&	2.7(9)	&	2.4(1)	&	70(3)	&	483(20)	&	3.9(1.3)	&	3.4(1)	\\
156	&	SDA	&	0.6(1)	&	0.7(1)	&	2788(365)	&	2.6(2)	&	3.1(2)	&	0.7(8)	&	1.37(6)	&	9(1)	&	10(1)	&	1.0(2)	&	1.9(1)	&	26(2)	&	31(2)	&	1.5(1.7)	&	2.7(1)	\\
157	&	SDA	&	1.5(1)	&	3.7(2)	&	2649(171)	&	10.4(4)	&	12.3(5)	&	1.1(5)	&	2.03(9)	&	35(1)	&	41(2)	&	1.6(8)	&	2.9(1)	&	104(4)	&	123(5)	&	2.3(2.7)	&	4.0(2)	\\
158	&	spo	&	0.6(1)	&	0.9(1)	&	2991(540)	&	3.1(3)	&	21.2(2.2)	&	1.4(5)	&	1.36(6)	&	10(1)	&	71(7)	&	2.1(7)	&	1.9(1)	&	31(3)	&	212(22)	&	3.0(1.0)	&	2.7(1)	\\
159	&	spo	&	1.3(2)	&	1.5(1)	&	3910(623)	&	5.6(4)	&	38.5(3.0)	&	1.7(6)	&	1.62(6)	&	19(1)	&	128(10)	&	2.5(8)	&	2.3(1)	&	56(4)	&	385(30)	&	3.7(1.2)	&	3.2(1)	\\
160	&	spo	&	0.7(1)	&	1.4(1)	&	4961(1518)	&	4.1(3)	&	28.3(2.2)	&	1.5(5)	&	1.48(5)	&	14(1)	&	94(7)	&	2.3(8)	&	2.1(1)	&	41(3)	&	283(22)	&	3.3(1.1)	&	2.9(1)	\\
161	&	spo	&	0.8(1)	&	0.6(1)	&	4967(1227)	&	2.8(3)	&	19.5(2.0)	&	1.4(4)	&	1.33(5)	&	9(1)	&	65(7)	&	2.0(7)	&	1.9(1)	&	28(3)	&	195(20)	&	2.9(1.0)	&	2.6(1)	\\
162	&	spo	&	1.3(1)	&	4.3(2)	&	2410(168)	&	11.1(4)	&	77.0(3.0)	&	2.2(7)	&	1.98(6)	&	37(1)	&	257(10)	&	3(1)	&	2.8(1)	&	111(4)	&	770(30)	&	4.6(1.5)	&	3.9(1)	\\
164	&	spo	&	0.4(1)	&	0.8(1)	&	2754(391)	&	2.5(2)	&	17.6(1.5)	&	1.3(4)	&	1.29(5)	&	8(1)	&	59(5)	&	2.0(6)	&	1.8(1)	&	25(2)	&	176(15)	&	2.8(0.9)	&	2.5(1)	\\
165	&	LYR	&	1.3(2)	&	5.9(2)	&	2370(184)	&	14.4(5)	&	12.0(4)	&	1.5(1)	&	1.86(4)	&	48(2)	&	40(1)	&	2.3(1)	&	2.6(1)	&	144(5)	&	120(4)	&	3.3(0.1)	&	3.6(1)	\\
166	&	spo	&	1.0(2)	&	1.4(1)	&	3038(460)	&	4.8(5)	&	33.0(3.1)	&	1.6(5)	&	1.55(6)	&	16(2)	&	110(10)	&	2.4(8)	&	2.2(1)	&	48(5)	&	330(31)	&	3.5(1.1)	&	3.0(1)	\\
167	&	spo	&	0.7(1)	&	0.7(1)	&	3971(926)	&	2.9(3)	&	19.8(2.2)	&	1.4(4)	&	1.33(6)	&	10(1)	&	66(7)	&	2.0(7)	&	1.9(1)	&	29(3)	&	198(22)	&	2.9(1.0)	&	2.6(1)	\\
168	&	spo	&	0.9(2)	&	2.0(1)	&	2788(365)	&	5.8(4)	&	40.1(2.7)	&	1.7(6)	&	1.64(6)	&	19(1)	&	134(9)	&	2.6(8)	&	2.3(1)	&	58(4)	&	401(27)	&	3.7(1.2)	&	3.2(1)	\\
169	&	spo	&	0.8(1)	&	1.3(1)	&	3863(596)	&	4.2(3)	&	29.1(1.8)	&	1.6(5)	&	1.49(5)	&	14(1)	&	97(6)	&	2.3(8)	&	2.1(1)	&	42(3)	&	291(18)	&	3.3(1.1)	&	2.9(1)	\\
171	&	spo	&	0.3(1)	&	0.5(1)	&	4457(1492)	&	1.7(2)	&	12.1(1.5)	&	1.2(4)	&	1.16(5)	&	6(1)	&	40(5)	&	1.7(6)	&	1.6(1)	&	17(2)	&	121(15)	&	2.5(0.8)	&	2.3(1)	\\
172	&	spo	&	7.0(4)	&	15.5(5)	&	2520(158)	&	45(1)	&	311.4(8.7)	&	3(1)	&	2.96(9)	&	150(4)	&	1038(29)	&	5.1(1.7)	&	4.2(1)	&	450(13)	&	3114(87)	&	7.3(2.4)	&	5.8(2)	\\
175	&	spo	&	0.4(1)	&	0.3(1)	&	5070(1586)	&	1.4(2)	&	9.4(1.1)	&	1.1(4)	&	1.07(5)	&	5(1)	&	31(4)	&	1.6(5)	&	1.5(1)	&	14(2)	&	94(11)	&	2.3(0.8)	&	2.1(1)	\\
176	&	spo	&	0.6(1)	&	1.1(1)	&	2497(385)	&	3.4(3)	&	23.4(2.4)	&	1.5(5)	&	1.40(6)	&	11(1)	&	78(8)	&	2.2(7)	&	2.0(1)	&	34(3)	&	234(24)	&	3.1(1.0)	&	2.7(1)	\\
177	&	spo	&	0.7(2)	&	1.7(1)	&	3746(1274)	&	4.8(5)	&	33.4(3.6)	&	1.6(5)	&	1.55(6)	&	16(2)	&	111(12)	&	2.4(8)	&	2.2(1)	&	48(5)	&	334(36)	&	3.5(1.2)	&	3.0(1)	\\
\hline														
\end{tabular}}
\end{sidewaystable*}
\begin{sidewaystable*}
\centering
\caption*{Table~\ref{tab:ResultsReal} (cont'd)}
\scalebox{0.92}{
\begin{tabular}{ccccc|cccc|cccc|cccc}
\hline\hline																																	
	&		&		&		&		&	\multicolumn{4}{c}{$\eta=5\times10^{-3}$}							&	\multicolumn{4}{c}{$\eta=1.5\times10^{-3}$}							&	\multicolumn{4}{c}{$\eta=5\times10^{-4}$}							\\
\hline																																	
ID	&	Stream	&	$LE_{\rm R}$	&	$LE_{\rm I}$	&	$T$	&	$KE_{\rm p}$	&	$m_{\rm p}$	&	$r_{\rm p}$	&	$d_{\rm c}$	&	$KE_{\rm p}$	&	$m_{\rm p}$	&	$r_{\rm p}$	&	$d_{\rm c}$	&	$KE_{\rm p}$	&	$m_{\rm p}$	&	$r_{\rm p}$	&	$d_{\rm c}$	\\
	&		&	($\times 10^4$~J)	&	($\times 10^4$~J)	&	(K)	&	($\times 10^6$~J)	&	(g)	&	(cm)	&	(m)	&	($\times 10^6$~J)	&	(g)	&	(cm)	&	(m)	&	($\times 10^6$~J)	&	(g)	&	(cm)	&	(m)	\\
\hline																															
179	&	SDA	&	0.5(1)	&	1.2(1)	&	4646(1967)	&	3.4(3)	&	4.1(4)	&	0.7(9)	&	1.47(7)	&	11(1)	&	14(1)	&	1.1(3)	&	2.1(1)	&	34(3)	&	41(4)	&	1.6(1.8)	&	2.9(1)	\\
180	&	spo	&	2.5(4)	&	3.3(2)	&	5171(1433)	&	11.6(9)	&	80.1(6.4)	&	2.2(7)	&	2.00(7)	&	39(3)	&	267(21)	&	3(1)	&	2.8(1)	&	116(9)	&	801(64)	&	4.7(1.5)	&	3.9(1)	\\
181	&	SDA	&	0.7(2)	&	1.2(1)	&	5634(2105)	&	3.8(4)	&	4.5(5)	&	0.8(9)	&	1.51(8)	&	13(1)	&	15(2)	&	1.1(3)	&	2.1(1)	&	38(4)	&	45(5)	&	1.6(1.9)	&	3.0(2)	\\
182	&	spo	&	2.3(3)	&	7.4(2)	&	3022(349)	&	19.3(8)	&	133.6(5.9)	&	2.6(8)	&	2.32(7)	&	64(3)	&	445(20)	&	4(1)	&	3.3(1)	&	193(8)	&	1336(59)	&	5.5(1.8)	&	4.5(1)	\\
183	&	spo	&	0.6(1)	&	0.6(1)	&	3791(726)	&	2.5(2)	&	17.5(1.7)	&	1.3(4)	&	1.29(5)	&	8(1)	&	58(6)	&	2.0(6)	&	1.8(1)	&	25(2)	&	175(17)	&	2.8(0.9)	&	2.5(1)	\\
184	&	spo	&	3.5(3)	&	7.1(3)	&	3404(385)	&	21.2(8)	&	146.9(5.3)	&	2.7(9)	&	2.38(7)	&	71(3)	&	490(18)	&	4(1)	&	3.4(1)	&	212(8)	&	1469(53)	&	5.7(1.9)	&	4.6(1)	\\
185	&	spo	&	1.3(3)	&	2.5(2)	&	3066(516)	&	7.6(6)	&	52.3(4.4)	&	1.9(6)	&	1.77(6)	&	25(2)	&	174(15)	&	2.8(9)	&	2.5(1)	&	76(6)	&	523(44)	&	4.1(1.3)	&	3.4(1)	\\
186	&	spo	&	0.9(2)	&	1.6(1)	&	3248(528)	&	4.9(4)	&	33.9(2.7)	&	1.6(5)	&	1.56(6)	&	16(1)	&	113(9)	&	2.4(8)	&	2.2(1)	&	49(4)	&	339(27)	&	3.5(1.2)	&	3.0(1)	\\
188	&	spo	&	2.5(2)	&	5.3(1)	&	3261(322)	&	15.5(5)	&	107.6(3.5)	&	2.4(8)	&	2.18(6)	&	52(2)	&	359(12)	&	4(1)	&	3.1(1)	&	155(5)	&	1076(35)	&	5.1(1.7)	&	4.2(1)	\\
189	&	spo	&	0.6(2)	&	1.8(1)	&	1953(252)	&	4.7(4)	&	32.8(3.0)	&	1.6(5)	&	1.54(6)	&	16(1)	&	109(10)	&	2.4(8)	&	2.2(1)	&	47(4)	&	328(30)	&	3.5(1.1)	&	3.0(1)	\\
191	&	spo	&	3.4(4)	&	9.5(3)	&	2635(265)	&	26(1)	&	178.1(7.1)	&	2.8(9)	&	2.52(8)	&	86(3)	&	594(24)	&	4(1)	&	3.6(1)	&	257(10)	&	1781(71)	&	6.1(2.0)	&	4.9(1)	\\
192	&	spo	&	1.5(2)	&	2.6(2)	&	2600(232)	&	8.2(5)	&	56.9(3.7)	&	1.9(6)	&	1.81(6)	&	27(2)	&	190(12)	&	3(1)	&	2.6(1)	&	82(5)	&	569(37)	&	4.2(1.4)	&	3.5(1)	\\
195	&	spo	&	7.2(6)	&	16.4(5)	&	2666(192)	&	47.3(1.5)	&	327.1(10.5)	&	3.5(1.1)	&	3.01(9)	&	158(5)	&	1090(35)	&	5(2)	&	4.3(1)	&	473(15)	&	3271(105)	&	7.4(2.5)	&	5.9(2)	\\
199	&	spo	&	1.4(3)	&	3.4(2)	&	3081(624)	&	9.5(8)	&	65.4(5.4)	&	2.0(7)	&	1.89(7)	&	32(3)	&	218(18)	&	3(1)	&	2.7(1)	&	95(8)	&	654(54)	&	4.4(1.4)	&	3.7(1)	\\
200	&	spo	&	0.7(2)	&	0.8(1)	&	3326(1077)	&	3.1(5)	&	21.2(3.7)	&	1.4(5)	&	1.36(8)	&	10(2)	&	71(12)	&	2.1(7)	&	1.9(1)	&	31(5)	&	212(37)	&	3.0(1.0)	&	2.7(2)	\\
201	&	spo	&	0.8(2)	&	0.7(1)	&	4477(1909)	&	2.8(5)	&	19.6(3.2)	&	1.4(5)	&	1.33(7)	&	9(2)	&	65(11)	&	2.0(7)	&	1.9(1)	&	28(5)	&	196(32)	&	2.9(1.0)	&	2.6(1)	\\
202	&	spo	&	2.4(3)	&	7.4(3)	&	2409(207)	&	19.8(9)	&	136.8(6.1)	&	2.6(9)	&	2.34(7)	&	66(3)	&	456(20)	&	4(1)	&	3.3(1)	&	198(9)	&	1368(61)	&	5.6(1.8)	&	4.6(1)	\\
204	&	spo	&	0.8(2)	&	2.0(1)	&	2788(425)	&	5.6(4)	&	39.0(2.6)	&	1.7(6)	&	1.62(6)	&	19(1)	&	130(9)	&	2.6(8)	&	2.3(1)	&	56(4)	&	390(26)	&	3.7(1.2)	&	3.2(1)	\\
205	&	spo	&	0.6(2)	&	1.5(1)	&	3297(993)	&	4.2(5)	&	29.3(3.1)	&	1.6(5)	&	1.49(6)	&	14(2)	&	98(10)	&	2.3(8)	&	2.1(1)	&	42(5)	&	293(31)	&	3.3(1.1)	&	2.9(1)	\\
206	&	spo	&	1.3(4)	&	10.7(5)	&	2056(257)	&	24(1)	&	166.5(8.0)	&	2.8(9)	&	2.47(8)	&	80(4)	&	555(27)	&	4(1)	&	3.5(1)	&	241(12)	&	1665(80)	&	5.9(2.0)	&	4.8(1)	\\
208	&	SDA	&	0.8(2)	&	1.3(2)	&	2744(516)	&	4.3(6)	&	5.1(7)	&	0.8(9)	&	1.57(9)	&	14(2)	&	17(2)	&	1.2(4)	&	2.2(1)	&	43(6)	&	51(7)	&	1.7(2.0)	&	3.1(2)	\\
209	&	SDA	&	2.6(3)	&	7.2(5)	&	2435(410)	&	20(1)	&	23.4(1.3)	&	1.3(1.5)	&	2.45(11)	&	66(4)	&	78(4)	&	2(1)	&	3.5(2)	&	197(11)	&	234(13)	&	2.8(3.3)	&	4.8(2)	\\
212	&	SDA	&	0.9(1)	&	2.7(1)	&	2575(204)	&	7.2(3)	&	8.6(4)	&	1.0(1.1)	&	1.83(8)	&	24(1)	&	29(1)	&	1.4(6)	&	2.6(1)	&	72(3)	&	86(4)	&	2.0(2.4)	&	3.6(2)	\\
214	&	SDA	&	1.4(2)	&	4.9(4)	&	2271(242)	&	12.6(9)	&	15.0(1.1)	&	1.1(1.3)	&	2.15(10)	&	42(3)	&	50(4)	&	1.7(5)	&	3.1(1)	&	126(9)	&	150(11)	&	2.4(2.8)	&	4.2(2)	\\
219	&	spo	&	0.4(1)	&	3.5(3)	&	1543(186)	&	7.7(6)	&	53.3(4.3)	&	1.9(6)	&	1.78(6)	&	26(2)	&	178(14)	&	2.8(9)	&	2.5(1)	&	77(6)	&	533(43)	&	4.1(1.3)	&	3.5(1)	\\
223	&	spo	&	3.2(3)	&	3.1(1)	&	7426(1907)	&	12.7(6)	&	88.1(4.4)	&	2.3(7)	&	2.06(6)	&	42(2)	&	294(15)	&	3(1)	&	2.9(1)	&	127(6)	&	881(44)	&	4.8(1.6)	&	4.0(1)	\\
224	&	spo	&	14.6(4)	&	28.6(3)	&	2858(285)	&	86(1)	&	597.4(7.0)	&	4.2(1.4)	&	3.58(10)	&	288(3)	&	1991(23)	&	6(2)	&	5.1(1)	&	863(10)	&	5974(70)	&	9.1(3.0)	&	7.0(2)	\\
227	&	ORI	&	1.8(1)	&	3.4(1)	&	5084(755)	&	10.4(3)	&	4.6(1)	&	1.1(1)	&	1.73(16)	&	35(1)	&	15(1)	&	1.6(2)	&	2.4(2)	&	104(3)	&	46(1)	&	2.3(0.3)	&	3.4(3)	\\
230	&	ORI	&	0.6(1)	&	1.2(1)	&	4254(1213)	&	3.7(3)	&	1.6(1)	&	0.8(1)	&	1.28(12)	&	12(1)	&	5.4(4)	&	1.1(2)	&	1.8(2)	&	37(3)	&	16(1)	&	1.6(0.2)	&	2.5(2)	\\
231	&	ORI	&	0.3(1)	&	0.3(1)	&	4285(1532)	&	1.2(2)	&	0.5(1)	&	0.5(1)	&	0.93(9)	&	4(1)	&	1.8(2)	&	0.8(1)	&	1.3(1)	&	12(2)	&	5(1)	&	1.1(0.2)	&	1.8(2)	\\
232	&	ORI	&	0.5(1)	&	0.6(1)	&	3512(466)	&	2.1(2)	&	1.0(1)	&	0.6(1)	&	1.09(10)	&	7(1)	&	3.2(2)	&	0.9(1)	&	1.5(1)	&	21(2)	&	10(1)	&	1.4(0.2)	&	2.1(2)	\\
233	&	ORI	&	0.4(1)	&	0.4(1)	&	3800(543)	&	1.6(1)	&	0.7(1)	&	0.6(1)	&	1.00(9)	&	5(1)	&	2.4(1)	&	0.9(1)	&	1.4(1)	&	16(1)	&	7(1)	&	1.2(0.2)	&	2.0(2)	\\
234	&	ORI	&	1.2(1)	&	2.0(1)	&	3677(488)	&	6.4(3)	&	2.8(1)	&	0.9(1)	&	1.50(14)	&	21(1)	&	10(1)	&	1.4(2)	&	2.1(2)	&	64(3)	&	28(1)	&	1.9(0.3)	&	2.9(3)	\\
235	&	spo	&	0.6(1)	&	1.5(1)	&	3778(935)	&	4.1(3)	&	28.2(2.0)	&	1.5(5)	&	1.48(5)	&	14(1)	&	94(7)	&	2.3(8)	&	2.1(1)	&	41(3)	&	282(20)	&	3.3(1.1)	&	2.9(1)	\\
236	&	TAU	&	1.3(1)	&	6.8(4)	&	2382(279)	&	16.3(8)	&	41.7(2.0)	&	1.5(8)	&	2.36(4)	&	54(3)	&	139(7)	&	2(1)	&	3.4(1)	&	163(8)	&	417(20)	&	3.3(1.6)	&	4.6(1)	\\
238	&	TAU	&	0.6(2)	&	0.7(1)	&	3711(1109)	&	2.6(4)	&	6.6(9)	&	0.8(4)	&	1.38(6)	&	9(1)	&	22(3)	&	1.2(6)	&	2.0(1)	&	26(4)	&	66(9)	&	1.8(0.9)	&	2.7(1)	\\
240	&	TAU	&	0.9(2)	&	2.2(1)	&	3005(472)	&	6.2(4)	&	15.7(1.1)	&	1.1(5)	&	1.78(4)	&	21(1)	&	52(4)	&	1.7(8)	&	2.5(1)	&	62(4)	&	157(11)	&	2.4(1.2)	&	3.5(1)	\\
241	&	TAU	&	0.9(1)	&	2.3(1)	&	3080(453)	&	6.4(4)	&	16.3(1.0)	&	1.1(5)	&	1.80(4)	&	21(1)	&	54(3)	&	1.7(8)	&	2.6(1)	&	64(4)	&	163(10)	&	2.4(1.2)	&	3.5(1)	\\
244	&	TAU	&	5.2(4)	&	9.4(3)	&	4475(883)	&	29(1)	&	74.0(2.5)	&	1.9(9)	&	2.79(4)	&	97(3)	&	247(8)	&	3(1)	&	4.0(1)	&	290(10)	&	740(25)	&	4.0(2.0)	&	5.4(1)	\\
251	&	TAU	&	2.4(4)	&	8.1(4)	&	3065(517)	&	21(1)	&	54.0(2.8)	&	1.7(8)	&	2.55(5)	&	71(4)	&	180(9)	&	3(1)	&	3.6(1)	&	212(11)	&	540(28)	&	3.6(1.8)	&	5.0(1)	\\
252	&	spo	&	2.2(3)	&	6.8(2)	&	3492(462)	&	18.0(7)	&	124.8(4.7)	&	2.5(8)	&	2.27(7)	&	60(2)	&	416(16)	&	4(1)	&	3.2(1)	&	180(7)	&	1248(47)	&	5.4(1.8)	&	4.4(1)	\\
257	&	spo	&	1.7(2)	&	4.5(2)	&	2577(225)	&	12.3(5)	&	85.3(3.8)	&	2.2(7)	&	2.04(6)	&	41(2)	&	284(13)	&	3(1)	&	2.9(1)	&	123(5)	&	853(38)	&	4.8(1.6)	&	4.0(1)	\\
259	&	spo	&	10.3(5)	&	23.3(7)	&	2579(197)	&	67(2)	&	465.1(12.0)	&	3.9(1.3)	&	3.33(10)	&	224(6)	&	1550(40)	&	66(2)	&	4.7(1)	&	672(17)	&	4651(120)	&	8.3(2.8)	&	6.5(2)	\\
261	&	spo	&	0.5(1)	&	1.0(1)	&	4580(1240)	&	3.0(2)	&	20.5(1.4)	&	1.4(5)	&	1.35(5)	&	10(1)	&	68(5)	&	2.1(7)	&	1.9(1)	&	30(2)	&	205(14)	&	3.0(1.0)	&	2.6(1)	\\
262	&	spo	&	0.9(2)	&	0.8(1)	&	4669(1310)	&	3.4(4)	&	23.4(2.7)	&	1.5(5)	&	1.40(6)	&	11(1)	&	78(9)	&	2.2(7)	&	2.0(1)	&	34(4)	&	234(27)	&	3.1(1.0)	&	2.7(1)	\\
\hline														
\end{tabular}}
\end{sidewaystable*}
\begin{sidewaystable*}
\centering
\caption*{Table~\ref{tab:ResultsReal} (cont'd)}
\scalebox{0.92}{
\begin{tabular}{ccccc|cccc|cccc|cccc}
\hline\hline																																	
	&		&		&		&		&	\multicolumn{4}{c}{$\eta=5\times10^{-3}$}							&	\multicolumn{4}{c}{$\eta=1.5\times10^{-3}$}							&	\multicolumn{4}{c}{$\eta=5\times10^{-4}$}							\\
\hline																																	
ID	&	Stream	&	$LE_{\rm R}$	&	$LE_{\rm I}$	&	$T$	&	$KE_{\rm p}$	&	$m_{\rm p}$	&	$r_{\rm p}$	&	$d_{\rm c}$	&	$KE_{\rm p}$	&	$m_{\rm p}$	&	$r_{\rm p}$	&	$d_{\rm c}$	&	$KE_{\rm p}$	&	$m_{\rm p}$	&	$r_{\rm p}$	&	$d_{\rm c}$	\\
	&		&	($\times 10^4$~J)	&	($\times 10^4$~J)	&	(K)	&	($\times 10^6$~J)	&	(g)	&	(cm)	&	(m)	&	($\times 10^6$~J)	&	(g)	&	(cm)	&	(m)	&	($\times 10^6$~J)	&	(g)	&	(cm)	&	(m)	\\
\hline																																	
264	&	spo	&	1.3(2)	&	3.1(2)	&	2675(298)	&	8.7(5)	&	60.6(3.5)	&	2.0(7)	&	1.84(6)	&	29(2)	&	202(12)	&	3(1)	&	2.6(1)	&	87(5)	&	606(35)	&	4.3(1.4)	&	3.6(1)	\\
266	&	spo	&	0.8(1)	&	2.0(2)	&	2856(307)	&	5.6(4)	&	38.9(2.5)	&	1.7(6)	&	1.62(5)	&	19(1)	&	130(8)	&	2.6(8)	&	2.3(1)	&	56(4)	&	389(25)	&	3.7(1.2)	&	3.2(1)	\\
267	&	spo	&	0.6(1)	&	1.6(1)	&	3144(623)	&	4.4(4)	&	30.4(2.4)	&	1.6(5)	&	1.51(5)	&	15(1)	&	101(8)	&	2.4(8)	&	2.1(1)	&	44(4)	&	304(24)	&	3.4(1.1)	&	2.9(1)	\\
268	&	spo	&	0.5(1)	&	0.3(1)	&	7336(2244)	&	1.5(2)	&	10.6(1.1)	&	1.1(4)	&	1.11(5)	&	5(1)	&	35(4)	&	1.7(5)	&	1.6(1)	&	15(2)	&	106(11)	&	2.4(0.8)	&	2.2(1)	\\
269	&	spo	&	0.7(1)	&	2.7(1)	&	2745(381)	&	6.7(3)	&	46.3(2.3)	&	1.8(6)	&	1.71(5)	&	22(1)	&	154(8)	&	2.7(9)	&	2.4(1)	&	67(3)	&	463(23)	&	3.9(1.3)	&	3.3(1)	\\
272	&	spo	&	0.8(1)	&	1.9(1)	&	3054(428)	&	5.4(3)	&	38(2)	&	1.7(6)	&	1.61(5)	&	18(1)	&	125(8)	&	2.5(8)	&	2.3(1)	&	54(3)	&	375(23)	&	3.6(1.2)	&	3.1(1)	\\
275	&	LYR	&	0.4(1)	&	0.8(1)	&	3260(600)	&	2.5(2)	&	2.1(2)	&	0.9(1)	&	1.12(3)	&	8(1)	&	7(1)	&	1.3(1)	&	1.6(1)	&	25(2)	&	21(2)	&	1.8(0.1)	&	2.2(1)	\\
276	&	spo	&	0.5(1)	&	0.9(1)	&	2597(485)	&	2.8(3)	&	19(2)	&	1.4(4)	&	1.32(6)	&	9(1)	&	64(7)	&	2.0(7)	&	1.9(1)	&	28(3)	&	191(22)	&	2.9(1.0)	&	2.6(1)	\\
279	&	spo	&	1.1(2)	&	2.9(2)	&	2630(400)	&	8.0(6)	&	56(4)	&	1.9(6)	&	1.80(6)	&	27(2)	&	185(14)	&	2.9(9)	&	2.5(1)	&	80(6)	&	555(41)	&	4.1(1.4)	&	3.5(1)	\\
280	&	spo	&	2.1(2)	&	3.9(2)	&	3860(681)	&	12.0(6)	&	83(4)	&	2.2(7)	&	2.02(6)	&	40(2)	&	277(14)	&	3.3(1.1)	&	2.9(1)	&	120(6)	&	830(41)	&	4.7(1.6)	&	3.9(1)	\\
281	&	spo	&	13.9(2)	&	37.5(9)	&	2824(237)	&	103(2)	&	711(13)	&	4.5(1.5)	&	3.77(11)	&	342(6)	&	2370(43)	&	6.7(2.2)	&	5.3(2)	&	1027(19)	&	7111(129)	&	9.6(3.2)	&	7.3(2)	\\
282	&	spo	&	3.3(4)	&	9.1(3)	&	2335(262)	&	25(1)	&	171(7)	&	2.8(9)	&	2.49(8)	&	82(3)	&	569(23)	&	4.2(1.4)	&	3.5(1)	&	247(10)	&	1706(70)	&	6.0(2.0)	&	4.9(1)	\\
283	&	spo	&	11.6(7)	&	19.9(8)	&	2935(465)	&	63(2)	&	435(15)	&	3.8(1.3)	&	3.27(10)	&	210(7)	&	1451(50)	&	5.7(1.9)	&	4.6(1)	&	629(22)	&	4354(150)	&	8.2(2.7)	&	6.4(2)	\\
284	&	spo	&	1.0(3)	&	2.9(2)	&	2376(469)	&	7.7(8)	&	53(5)	&	1.9(6)	&	1.78(7)	&	26(3)	&	177(18)	&	2.8(9)	&	2.5(1)	&	77(8)	&	531(54)	&	4.1(1.3)	&	3.5(1)	\\
285	&	spo	&	1.7(1)	&	1.0(1)	&	7520(2682)	&	5.4(3)	&	37.2(1.8)	&	1.7(6)	&	1.60(5)	&	18(1)	&	124(6)	&	2.5(8)	&	2.3(1)	&	54(3)	&	372(18)	&	3.6(1.2)	&	3.1(1)	\\
286	&	spo	&	0.3(1)	&	0.3(1)	&	3911(1643)	&	1.3(2)	&	9.0(1.4)	&	1.1(3)	&	1.06(6)	&	4(1)	&	30(5)	&	1.6(5)	&	1.5(1)	&	13(2)	&	90(14)	&	2.3(8)	&	2.1(1)	\\
287	&	spo	&	1.2(1)	&	3.4(2)	&	2372(375)	&	9.3(4)	&	64.3(3.1)	&	2.0(7)	&	1.88(6)	&	31(1)	&	214(10)	&	3(1)	&	2.7(1)	&	93(4)	&	643(31)	&	4.3(1.4)	&	3.7(1)	\\
289	&	spo	&	1.0(1)	&	2.4(1)	&	2905(316)	&	6.8(3)	&	47.1(2.4)	&	1.8(6)	&	1.71(5)	&	23(1)	&	157(8)	&	2.7(9)	&	2.4(1)	&	68(3)	&	471(24)	&	3.9(1.3)	&	3.3(1)	\\
290	&	PER	&	0.2(1)	&	0.6(1)	&	2581(641)	&	1.7(2)	&	0.9(1)	&	0.6(1)	&	1.06(5)	&	6(1)	&	3(1)	&	0.8(1)	&	1.5(1)	&	17(2)	&	9(1)	&	1.2(1)	&	2.1(1)	\\
292	&	spo	&	0.2(1)	&	0.4(1)	&	2326(791)	&	1.1(2)	&	7.6(1.6)	&	1.0(3)	&	1.01(7)	&	4(1)	&	25(5)	&	1.5(5)	&	1.4(1)	&	11(2)	&	76(16)	&	2.1(7)	&	2.0(1)	\\
293	&	PER	&	0.5(1)	&	0.4(1)	&	4747(1153)	&	1.7(2)	&	0.9(1)	&	0.6(1)	&	1.07(5)	&	6(1)	&	3(1)	&	0.8(1)	&	1.5(1)	&	17(2)	&	9(1)	&	1.2(1)	&	2.1(1)	\\
295	&	PER	&	1.5(1)	&	1.9(1)	&	4820(876)	&	6.8(4)	&	3.7(2)	&	0.9(1)	&	1.60(5)	&	23(1)	&	12(1)	&	1.3(1)	&	2.3(1)	&	68(4)	&	37(2)	&	1.9(2)	&	3.1(1)	\\
\hline														
\end{tabular}}
\end{sidewaystable*}

\begin{table*}
\centering
\caption{Rough estimation of parameters of the suspected LIFs (based on the whole NELIOTA sample), projectiles and impact craters for $\eta=1.5\times10^{-3}$. Errors are included in parentheses alongside values and correspond to the last digit(s).}
\label{tab:ResultsSusp}
\begin{tabular}{ccccccccccc}
\hline\hline																					
ID	&	Stream	&	$m_{\rm I}$	&	$m_{\rm R, ~est}$	&	$LE_{\rm I}$	&	$LE_{\rm R, ~est}$	&	$T_{\rm est}$	&	$KE_{\rm p, ~est}$	&	$m_{\rm p, ~est}$	&	$r_{\rm p, ~est}$	&	$d_{\rm c, ~est}$	\\
	&		&	(mag)	&	(mag)	&	($\times 10^3$~J)	&	($\times 10^3$~J)	&	(K)	&	($\times 10^6$~J)	&	(g)	&	(cm)	&	(m)	\\
\hline																					
7	&	spo	&	10.9(3)	&	12.7(4)	&	1.5(5)	&	1.3(5)	&	2141(475)	&	1.9(3)	&	13(2)	&	1.2(4)	&	1.2(4)	\\
8	&	spo	&	9.8(1)	&	11.7(3)	&	4.1(5)	&	3.1(9)	&	2012(276)	&	4.8(6)	&	33(4)	&	1.6(5)	&	1.6(5)	\\
9	&	spo	&	9.6(1)	&	11.5(3)	&	5.0(5)	&	4(1)	&	1991(267)	&	5.7(7)	&	40(5)	&	1.7(6)	&	1.7(6)	\\
10	&	spo	&	11.0(4)	&	12.7(4)	&	1.3(5)	&	1.2(5)	&	2158(510)	&	1.7(3)	&	12(2)	&	1.1(4)	&	1.2(4)	\\
11	&	spo	&	11.0(4)	&	12.7(5)	&	1.4(5)	&	1.2(5)	&	2156(557)	&	1.7(3)	&	12(2)	&	1.1(4)	&	1.2(4)	\\
15	&	SDA	&	9.3(1)	&	11.3(3)	&	6.7(7)	&	5(1)	&	1960(260)	&	7.6(9)	&	9(1)	&	1.0(5)	&	1.0(5)	\\
25	&	spo	&	10.2(2)	&	12.1(4)	&	3.1(7)	&	2.4(8)	&	2061(360)	&	3.7(5)	&	26(4)	&	1.5(5)	&	1.5(5)	\\
34	&	GEM	&	9.6(1)	&	11.6(3)	&	5.0(5)	&	4(1)	&	1990(261)	&	5.8(7)	&	9(1)	&	0.9(4)	&	0.9(5)	\\
35	&	GEM	&	8.9(1)	&	10.7(3)	&	14(1)	&	9(3)	&	2163(321)	&	15(2)	&	23(3)	&	1.2(1)	&	1.2(6)	\\
36	&	GEM	&	9.6(1)	&	11.5(3)	&	5.3(5)	&	4(1)	&	1985(261)	&	6.1(7)	&	9(1)	&	0.9(5)	&	0.9(5)	\\
38	&	GEM	&	9.6(1)	&	11.5(3)	&	5.4(8)	&	4(1)	&	1984(281)	&	6.2(8)	&	10(1)	&	0.9(5)	&	0.9(6)	\\
40	&	GEM	&	10.1(2)	&	12.0(3)	&	3.3(6)	&	2.6(8)	&	2054(331)	&	3.9(5)	&	6(1)	&	0.8(3)	&	0.8(3)	\\
41	&	GEM	&	10.3(2)	&	12.1(3)	&	2.9(5)	&	2.3(7)	&	2069(316)	&	3.4(5)	&	5(1)	&	0.7(3)	&	0.8(3)	\\
43	&	GEM	&	9.9(2)	&	11.7(3)	&	4.3(7)	&	3(1)	&	2006(292)	&	5.0(6)	&	8(1)	&	0.9(4)	&	0.9(5)	\\
44	&	GEM	&	9.7(2)	&	11.6(3)	&	5.3(8)	&	4(1)	&	1985(285)	&	6.1(8)	&	9(1)	&	0.9(5)	&	0.9(5)	\\
45	&	GEM	&	9.3(1)	&	11.2(3)	&	8(1)	&	5(2)	&	1949(268)	&	9(1)	&	13(2)	&	1.0(7)	&	1.0(7)	\\
48	&	spo	&	9.6(1)	&	11.6(3)	&	5.6(7)	&	4(1)	&	1980(268)	&	6.4(8)	&	44(5)	&	1.8(6)	&	1.8(6)	\\
57	&	spo	&	10.5(3)	&	12.3(4)	&	2.2(7)	&	1.8(7)	&	2098(433)	&	2.7(4)	&	19(3)	&	1.3(4)	&	1.4(4)	\\
60	&	spo	&	9.8(2)	&	11.7(4)	&	4(1)	&	3(1)	&	2013(355)	&	4.8(7)	&	33(5)	&	1.6(5)	&	1.6(5)	\\
77	&	spo	&	11.6(5)	&	13.2(5)	&	0.7(3)	&	0.7(3)	&	2232(718)	&	1.0(2)	&	7(2)	&	0.9(3)	&	1.0(3)	\\
79	&	spo	&	9.7(1)	&	11.6(3)	&	5.1(9)	&	4(1)	&	1988(288)	&	5.9(8)	&	41(5)	&	1.7(6)	&	1.7(6)	\\
80	&	spo	&	8.7(1)	&	10.7(3)	&	13.3(3)	&	8(2)	&	1894(233)	&	15(2)	&	100(11)	&	2.3(8)	&	2.3(8)	\\
82	&	GEM	&	10.1(2)	&	12.0(3)	&	3.7(7)	&	2.8(9)	&	2043(324)	&	4.3(6)	&	7(1)	&	0.8(4)	&	0.8(4)	\\
83	&	GEM	&	8.9(1)	&	11.0(3)	&	11(1)	&	7(2)	&	1914(256)	&	12(1)	&	18(2)	&	1.1(4)	&	1.1(6)	\\
86	&	spo	&	9.0(2)	&	11.0(3)	&	10(1)	&	6(2)	&	1925(273)	&	11(1)	&	74(9)	&	2.1(7)	&	2.1(7)	\\
87	&	spo	&	9.7(2)	&	11.6(3)	&	5.1(7)	&	4(1)	&	1990(283)	&	5.8(7)	&	40(5)	&	1.7(6)	&	1.7(6)	\\
91	&	spo	&	9.6(2)	&	11.5(4)	&	5(1)	&	4(1)	&	1993(346)	&	5.7(8)	&	39(6)	&	1.7(6)	&	1.7(6)	\\
92	&	spo	&	9.6(2)	&	11.5(4)	&	6(1)	&	4(1)	&	1974(331)	&	6.7(9)	&	47(6)	&	1.8(6)	&	1.8(6)	\\
99	&	spo	&	9.5(2)	&	11.4(3)	&	6(1)	&	4(1)	&	1980(304)	&	6.3(8)	&	44(6)	&	1.8(6)	&	1.8(6)	\\
113	&	SDA	&	9.8(2)	&	11.7(3)	&	4.1(6)	&	3.0(9)	&	2013(288)	&	4.7(6)	&	6(1)	&	0.8(9)	&	0.8(4)	\\
123	&	spo	&	10.2(2)	&	12.0(3)	&	3.0(5)	&	2.3(7)	&	2066(328)	&	3.6(5)	&	25(3)	&	1.9(6)	&	1.5(5)	\\
128	&	spo	&	10.8(2)	&	12.5(3)	&	1.6(3)	&	1.4(4)	&	2139(352)	&	2.0(3)	&	14(2)	&	1.5(5)	&	1.2(4)	\\
129	&	spo	&	9.3(1)	&	10.7(3)	&	14(1)	&	9(3)	&	2548(447)	&	15(2)	&	102(11)	&	1.2(4)	&	2.4(8)	\\
140	&	spo	&	9.6(1)	&	11.5(3)	&	5.9(6)	&	4(1)	&	1944(292)	&	6.7(8)	&	47(6)	&	2.3(8)	&	1.8(6)	\\
141	&	spo	&	8.8(1)	&	10.9(3)	&	12(2)	&	8(2)	&	1906(257)	&	13(2)	&	90(11)	&	1.8(6)	&	2.3(7)	\\
146	&	spo	&	10.0(1)	&	11.9(3)	&	3.9(5)	&	2.9(8)	&	2038(286)	&	4.5(6)	&	31(4)	&	2.2(7)	&	1.6(5)	\\
150	&	spo	&	9.5(1)	&	10.8(3)	&	12.3(9)	&	8(2)	&	2679(502)	&	13(2)	&	93(10)	&	1.6(5)	&	2.3(8)	\\
152	&	spo	&	9.8(1)	&	10.9(3)	&	9.4(6)	&	6(2)	&	2950(595)	&	10(1)	&	72(8)	&	2.6(9)	&	2.1(7)	\\
154	&	SDA	&	9.6(1)	&	11.5(3)	&	4.8(6)	&	4(1)	&	1995(273)	&	5.6(7)	&	7(1)	&	2.3(7)	&	0.9(5)	\\
155	&	SDA	&	9.4(1)	&	11.3(3)	&	5.9(6)	&	4(1)	&	1974(255)	&	6.8(8)	&	8(1)	&	2.1(7)	&	0.9(5)	\\
163	&	spo	&	10.1(2)	&	11.3(3)	&	7(1)	&	5(2)	&	2920(626)	&	8(1)	&	58(7)	&	0.9(5)	&	2.0(6)	\\
170	&	spo	&	9.6(1)	&	11.0(3)	&	9.6(9)	&	6(2)	&	2551(451)	&	11(1)	&	74(8)	&	0.9(5)	&	2.1(7)	\\
173	&	spo	&	9.3(1)	&	10.6(3)	&	15(1)	&	10(3)	&	2632(484)	&	17(2)	&	115(13)	&	1.9(6)	&	2.5(8)	\\
174	&	spo	&	9.8(2)	&	11.7(3)	&	4.8(8)	&	4(1)	&	1994(299)	&	5.6(7)	&	39(5)	&	2.1(7)	&	1.7(6)	\\
187	&	spo	&	10.0(1)	&	11.3(3)	&	6.2(5)	&	4(1)	&	2743(523)	&	7.1(8)	&	49(6)	&	2.4(8)	&	1.9(6)	\\
190	&	spo	&	9.6(1)	&	11.5(3)	&	5.2(7)	&	4(1)	&	1988(280)	&	5.9(7)	&	41(5)	&	1.7(6)	&	1.8(6)	\\
193	&	ETA	&	7.4(1)	&	8.9(3)	&	98(3)	&	47(15)	&	2522(448)	&	96(10)	&	667(67)	&	1.8(6)	&	4(2)	\\
194	&	spo	&	10.1(1)	&	12.0(3)	&	3.2(4)	&	2.5(7)	&	2059(286)	&	3.8(5)	&	27(3)	&	1.7(6)	&	1.5(5)	\\
196	&	spo	&	9.2(1)	&	10.6(3)	&	16(1)	&	9.7(3)	&	2521(424)	&	17(2)	&	117(13)	&	4(1)	&	2.5(8)	\\
197	&	spo	&	8.6(1)	&	10.7(3)	&	14(1)	&	8.9(3)	&	1888(235)	&	15(2)	&	107(12)	&	1.5(5)	&	2.4(8)	\\
198	&	spo	&	9.4(1)	&	10.8(3)	&	13(1)	&	8.5(2)	&	2659(500)	&	15(2)	&	100(11)	&	2.5(8)	&	2.4(8)	\\
207	&	SDA	&	9.9(1)	&	11.8(3)	&	4.2(4)	&	3.2(9)	&	2008(267)	&	4.9(6)	&	34(4)	&	2.4(8)	&	1.6(5)	\\
210	&	SDA	&	8.7(1)	&	10.2(3)	&	22(1)	&	13(4)	&	2441(397)	&	24(3)	&	28(3)	&	2.3(8)	&	1.4(6)	\\
211	&	SDA	&	9.4(1)	&	11.4(3)	&	6.2(8)	&	4(1)	&	1968(270)	&	7.1(9)	&	8(1)	&	1.6(5)	&	0.9(5)	\\
213	&	SDA	&	10.0(2)	&	11.1(3)	&	8.2(8)	&	6(2)	&	3104(745)	&	9(1)	&	11(1)	&	1.4(6)	&	1.0(5)	\\
215	&	SDA	&	8.0(1)	&	9.6(3)	&	39(3)	&	21(6)	&	2347(382)	&	40(4)	&	48(5)	&	0.9(5)	&	1.7(5)	\\
216	&	SDA	&	9.4(2)	&	10.5(3)	&	15(2)	&	9(3)	&	2988(697)	&	16(2)	&	19(2)	&	1.0(5)	&	1.2(5)	\\
217	&	spo	&	8.5(1)	&	9.9(3)	&	34(2)	&	19(6)	&	2531(440)	&	35(4)	&	42(4)	&	1.7(7)	&	1.6(5)	\\
\hline																																										
\end{tabular}
\end{table*}
\begin{table*}
\centering
\caption*{Table~\ref{tab:ResultsSusp} (cont'd)}
\begin{tabular}{ccccccccccc}
\hline\hline																					
ID	&	Stream	&	$m_{\rm I}$	&	$m_{\rm R, ~est}$	&	$LE_{\rm I}$	&	$LE_{\rm R, ~est}$	&	$T_{\rm est}$	&	$KE_{\rm p, ~est}$	&	$m_{\rm p, ~est}$	&	$r_{\rm p, ~est}$	&	$d_{\rm c, ~est}$	\\
	&		&	(mag)	&	(mag)	&	($\times 10^3$~J)	&	($\times 10^3$~J)	&	(K)	&	($\times 10^6$~J)	&	(g)	&	(cm)	&	(m)	\\
\hline																					
218	&	spo	&	9.9(3)	&	11.8(4)	&	5(1)	&	3.4(1)	&	2000(359)	&	5.3(8)	&	6(1)	&	1.2(5)	&	0.9(5)	\\
220	&	spo	&	9.4(1)	&	11.3(3)	&	6.3(9)	&	4.4(1)	&	1968(276)	&	7.1(9)	&	49(6)	&	1.6(8)	&	1.9(6)	\\
221	&	spo	&	8.4(1)	&	10.4(3)	&	16(2)	&	9.9(3)	&	1878(257)	&	17(2)	&	119(14)	&	0.8(4)	&	2.5(8)	\\
222	&	ORI	&	9.9(2)	&	11.1(3)	&	9.5(8)	&	6.3(2)	&	2922(609)	&	11(1)	&	73(8)	&	1.8(6)	&	2.1(7)	\\
225	&	spo	&	9.9(1)	&	11.8(3)	&	4.4(6)	&	3.3(9)	&	2005(279)	&	5.1(6)	&	35(4)	&	2.5(8)	&	1.7(5)	\\
226	&	spo	&	9.1(1)	&	10.7(3)	&	13.5(8)	&	9(2)	&	2357(360)	&	15(2)	&	101(11)	&	2.1(7)	&	2.4(8)	\\
228	&	ORI	&	9.8(1)	&	11.7(3)	&	4.6(6)	&	3(1)	&	2000(278)	&	5.3(6)	&	37(4)	&	1.6(5)	&	1.7(6)	\\
229	&	ORI	&	8.9(1)	&	10.3(3)	&	20.6(9)	&	12(4)	&	2554(434)	&	22(2)	&	10(1)	&	2.3(8)	&	1.4(2)	\\
237	&	TAU	&	7.7(1)	&	9.5(3)	&	43(4)	&	23(7)	&	2131(326)	&	44(5)	&	303(32)	&	1.7(5)	&	3(1)	\\
239	&	TAU	&	10.2(2)	&	12.0(3)	&	2.9(6)	&	2.3(7)	&	2069(342)	&	3.5(5)	&	9(1)	&	1.4(2)	&	0.9(4)	\\
242	&	TAU	&	9.0(1)	&	10.3(3)	&	17(1)	&	10(3)	&	2611(487)	&	18(2)	&	46(5)	&	3(1)	&	1.6(8)	\\
243	&	TAU	&	9.4(2)	&	11.4(3)	&	5.7(8)	&	4(1)	&	1978(281)	&	6.5(8)	&	17(2)	&	0.9(4)	&	1.1(6)	\\
245	&	TAU	&	9.0(1)	&	10.4(3)	&	16(1)	&	10(3)	&	2567(461)	&	18(2)	&	45(5)	&	1.6(8)	&	1.6(8)	\\
246	&	TAU	&	9.5(2)	&	11.4(3)	&	5.4(9)	&	4(1)	&	1984(301)	&	6.2(8)	&	16(2)	&	1.1(5)	&	1.1(5)	\\
247	&	TAU	&	9.0(2)	&	11.0(3)	&	8(1)	&	6(2)	&	1940(272)	&	9(1)	&	24(3)	&	1.6(8)	&	1.3(6)	\\
248	&	TAU	&	9.5(2)	&	11.4(3)	&	5.2(9)	&	4(1)	&	1987(301)	&	6.0(8)	&	15(2)	&	1.1(5)	&	1.1(5)	\\
249	&	TAU	&	8.9(1)	&	10.9(3)	&	9(1)	&	6(2)	&	1930(262)	&	10(1)	&	26(3)	&	1.3(6)	&	1.3(6)	\\
250	&	TAU	&	9.4(2)	&	11.3(3)	&	6(1)	&	4(1)	&	1970(288)	&	7.0(9)	&	18(2)	&	1.1(5)	&	1.2(6)	\\
253	&	spo	&	8.5(1)	&	10.0(3)	&	28(2)	&	16(5)	&	2475(418)	&	29(3)	&	74(8)	&	1.3(6)	&	1.9(9)	\\
254	&	spo	&	9.5(1)	&	11.4(3)	&	6.1(8)	&	4(1)	&	1971(275)	&	6.9(8)	&	48(6)	&	1.2(6)	&	1.8(6)	\\
255	&	spo	&	9.4(1)	&	11.4(3)	&	6.4(8)	&	5(1)	&	1966(270)	&	7.2(9)	&	50(6)	&	1.8(9)	&	1.9(6)	\\
256	&	spo	&	8.8(1)	&	10.8(3)	&	12(1)	&	8(2)	&	1908(238)	&	13(1)	&	88(10)	&	1.8(6)	&	2.2(7)	\\
258	&	spo	&	9.9(2)	&	11.8(3)	&	3.9(7)	&	3.0(9)	&	2016(320)	&	4.6(6)	&	32(4)	&	1.8(6)	&	1.6(5)	\\
260	&	spo	&	10.4(2)	&	12.2(3)	&	2.2(4)	&	1.8(5)	&	2103(347)	&	2.6(4)	&	18(3)	&	2.2(7)	&	1.3(4)	\\
265	&	spo	&	9.8(2)	&	11.7(3)	&	4.0(6)	&	3.0(9)	&	2014(297)	&	4.7(6)	&	32(4)	&	1.6(5)	&	1.6(5)	\\
270	&	spo	&	10.0(1)	&	11.3(3)	&	6.6(6)	&	5(1)	&	2714(513)	&	7.5(9)	&	52(6)	&	1.3(4)	&	1.9(6)	\\
271	&	spo	&	10.2(1)	&	12.1(3)	&	2.9(4)	&	2.3(7)	&	2070(295)	&	3.4(4)	&	24(3)	&	1.6(5)	&	1.5(5)	\\
273	&	spo	&	9.7(1)	&	11.6(3)	&	5.0(6)	&	4(1)	&	1990(265)	&	5.8(7)	&	40(5)	&	1.9(6)	&	1.7(6)	\\
274	&	spo	&	8.7(1)	&	10.8(3)	&	12.3(9)	&	8(2)	&	1902(234)	&	13(2)	&	93(10)	&	1.4(5)	&	2.3(8)	\\
277	&	LYR	&	9.7(1)	&	11.6(3)	&	5.1(7)	&	4(1)	&	1989(279)	&	5.9(7)	&	5(1)	&	1.7(6)	&	1.1(1)	\\
278	&	LYR	&	10.1(2)	&	11.9(3)	&	3.6(6)	&	2.8(8)	&	2045(309)	&	4.3(6)	&	30(4)	&	2.3(7)	&	1.6(5)	\\
288	&	spo	&	10.2(2)	&	12.1(3)	&	3.3(7)	&	2.6(8)	&	2055(337)	&	3.9(5)	&	27(4)	&	1.1(1)	&	1.5(5)	\\
291	&	spo	&	9.3(1)	&	10.7(3)	&	13(1)	&	8(3)	&	2513(443)	&	14(2)	&	99(11)	&	1.5(5)	&	2.3(8)	\\
294	&	spo	&	11.3(3)	&	13.0(4)	&	1.2(4)	&	1.1(4)	&	2170(486)	&	1.5(3)	&	11(2)	&	1.5(5)	&	1.1(4)	\\
\hline																																										
\end{tabular}
\end{table*}

\section{Multiframe flashes}
\label{sec:App1}
\begin{figure*}[h]
\begin{tabular}{ccc}
\includegraphics[width=5.6cm]{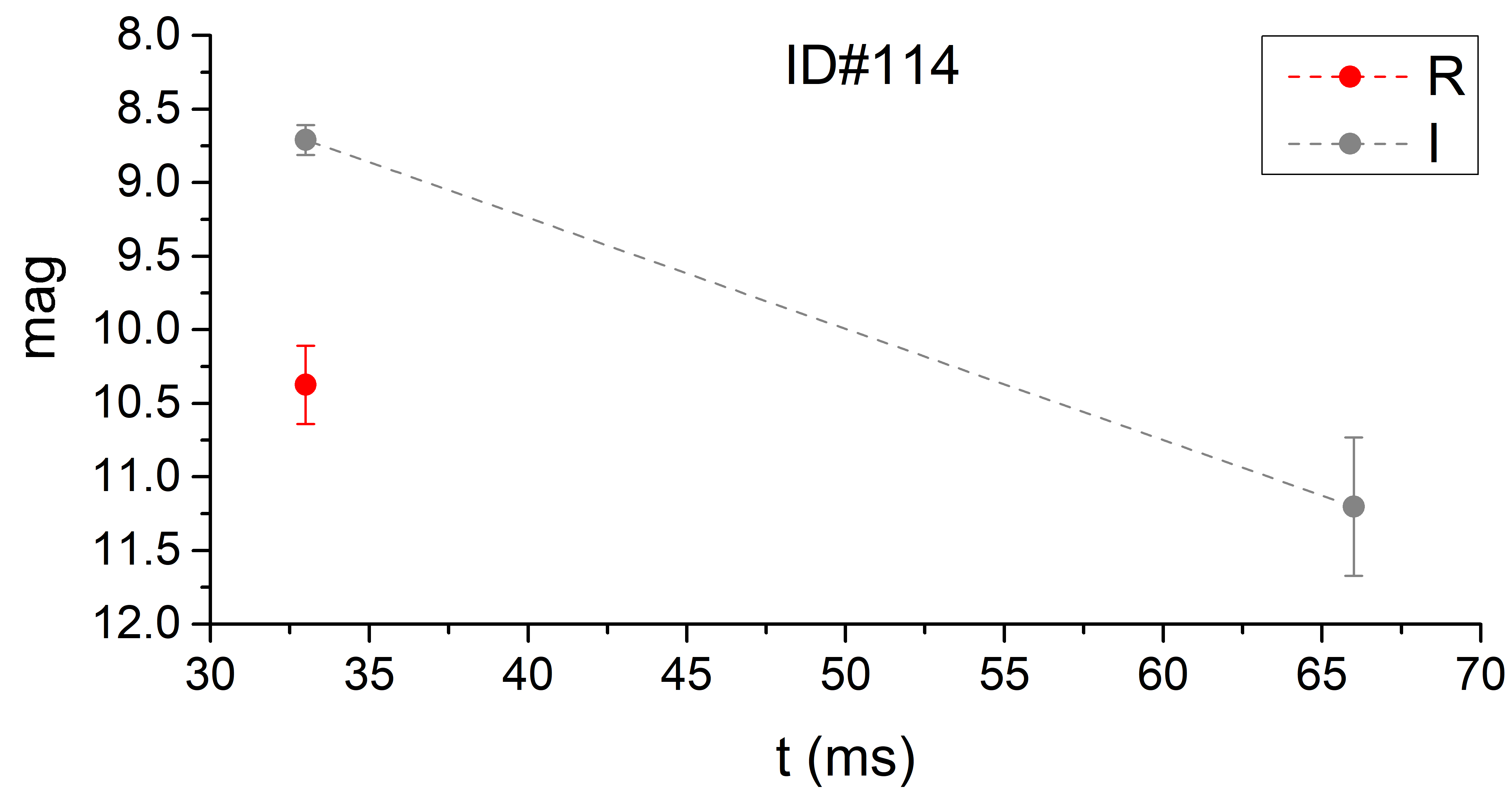}&\includegraphics[width=5.6cm]{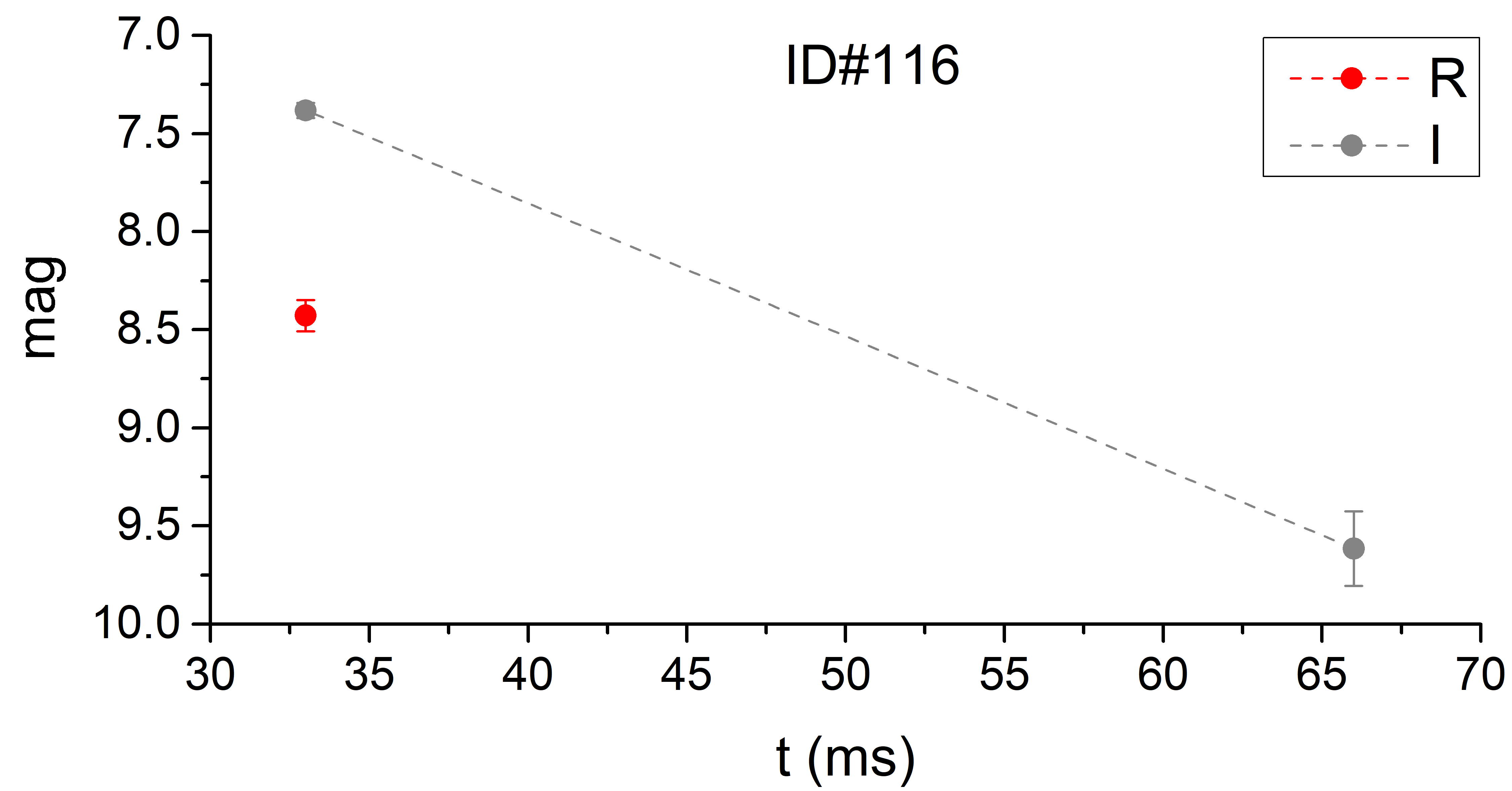}&\includegraphics[width=5.6cm]{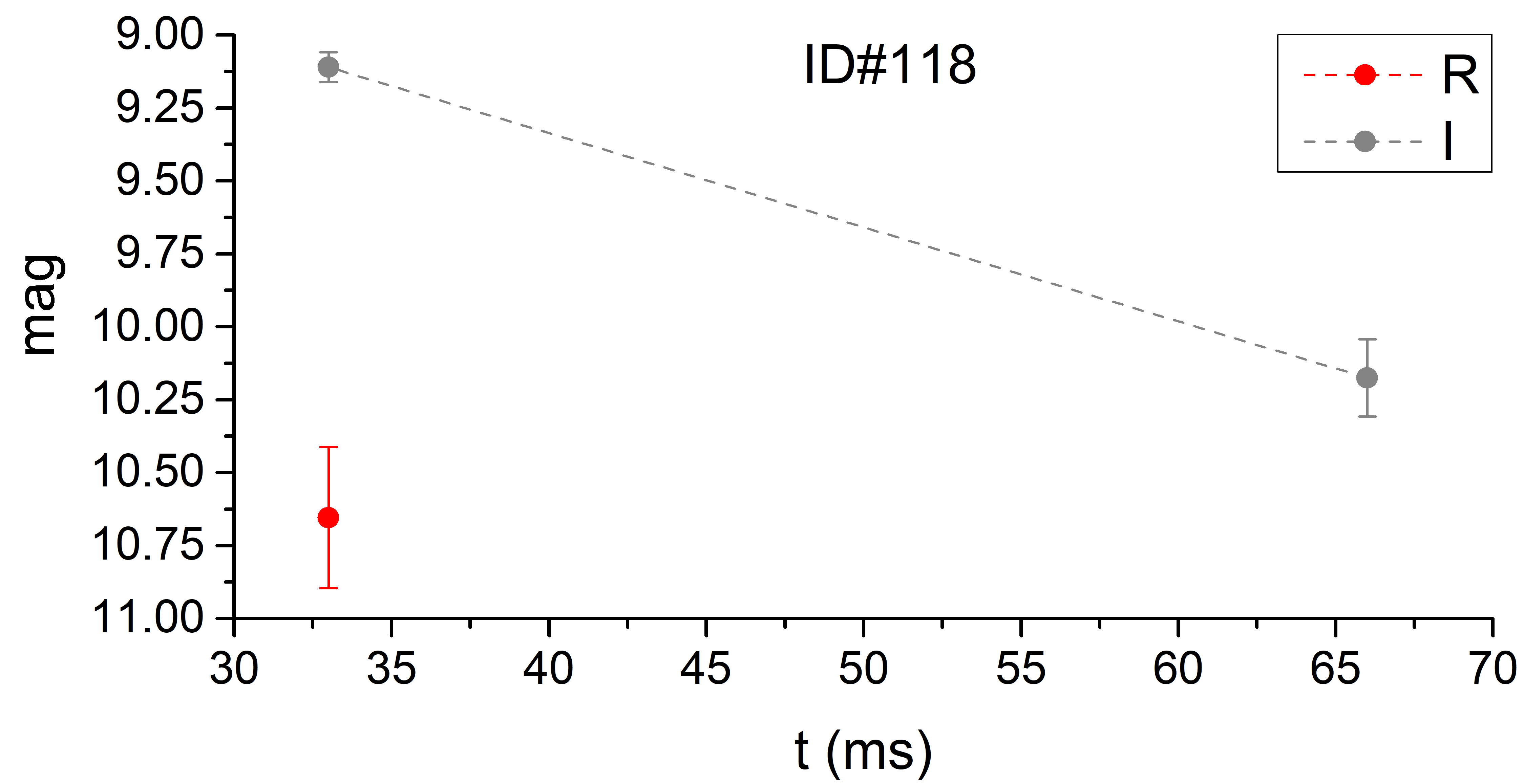}\\
\includegraphics[width=5.6cm]{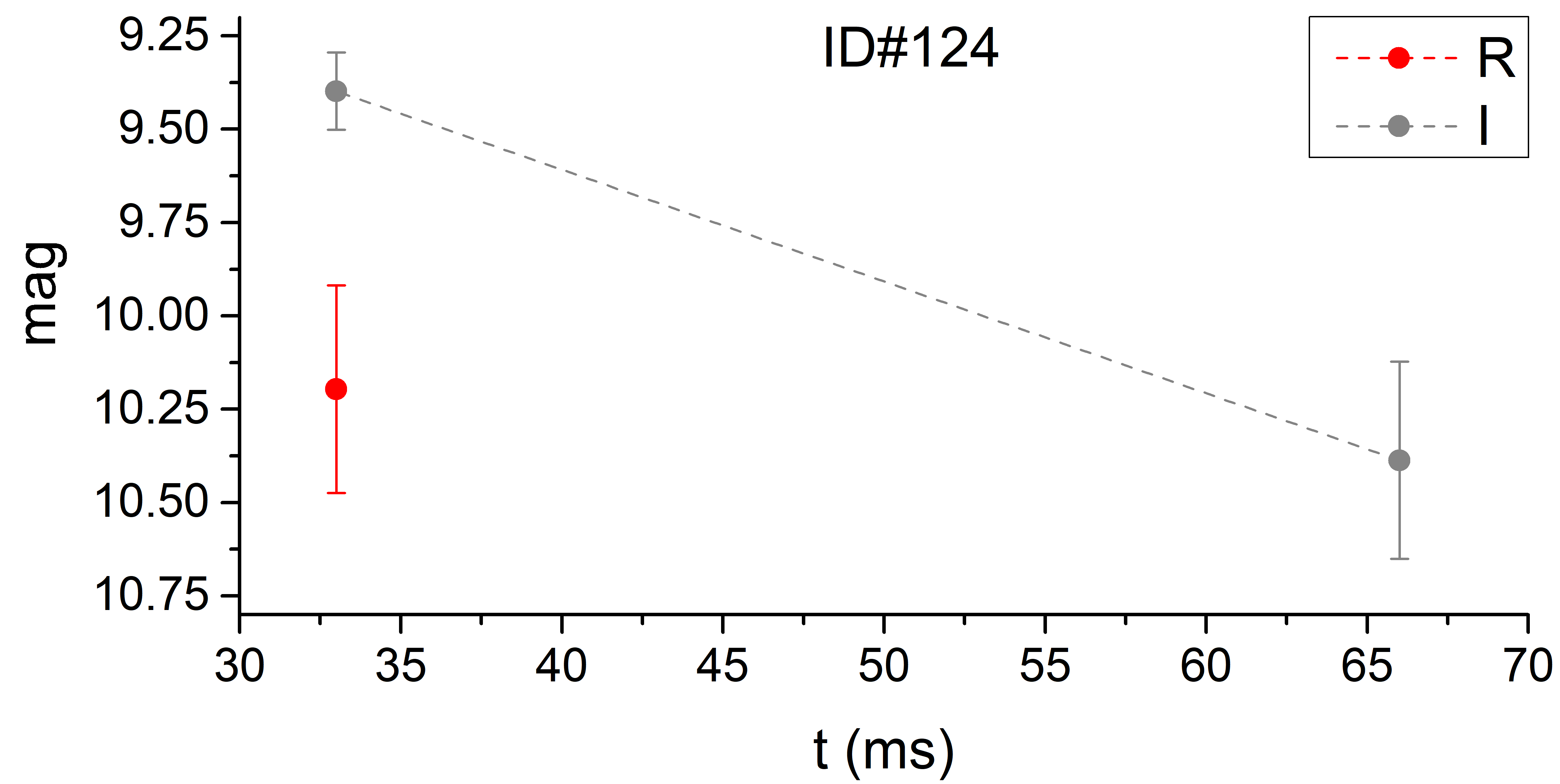}&\includegraphics[width=5.6cm]{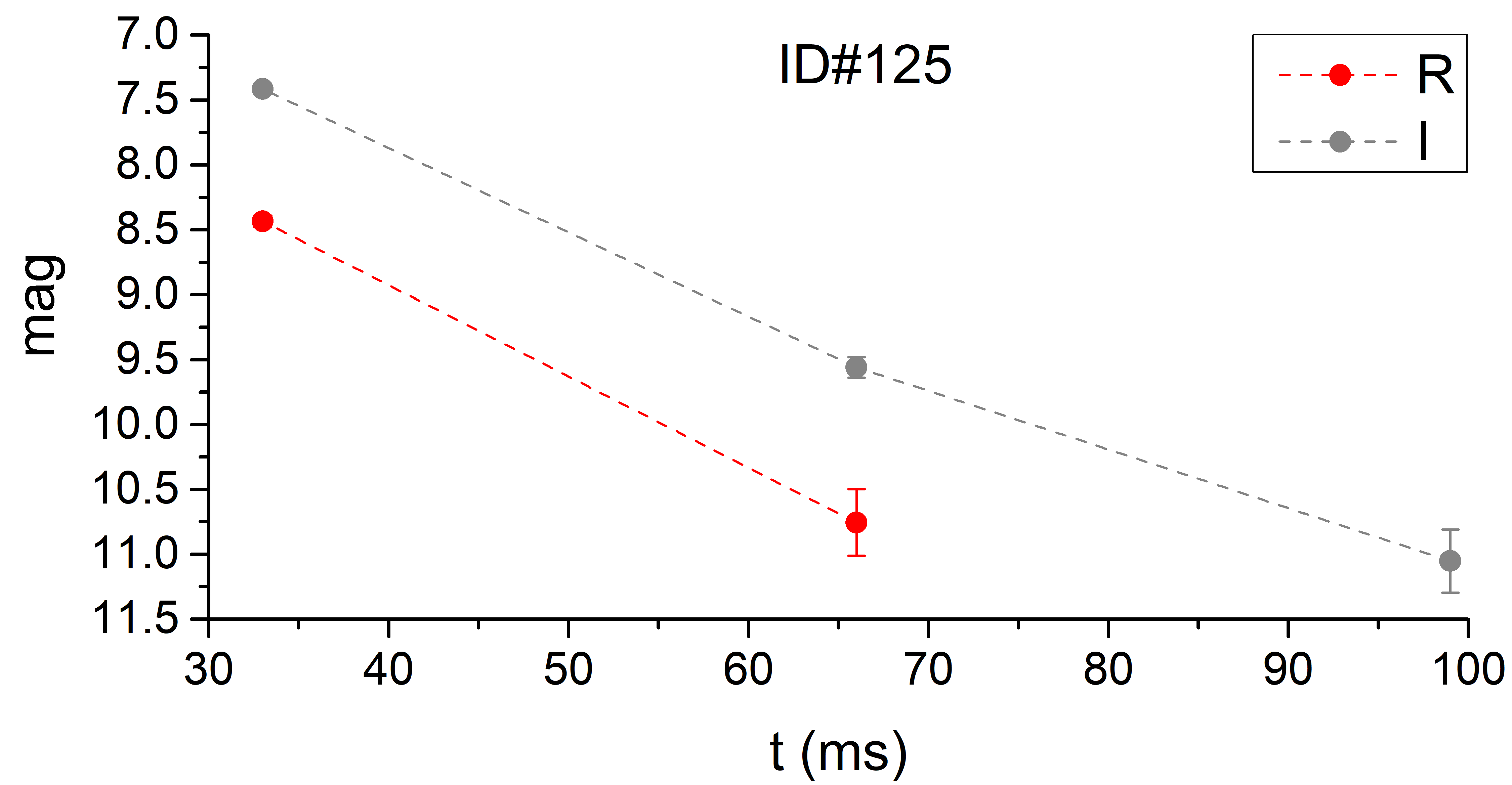}&\includegraphics[width=5.6cm]{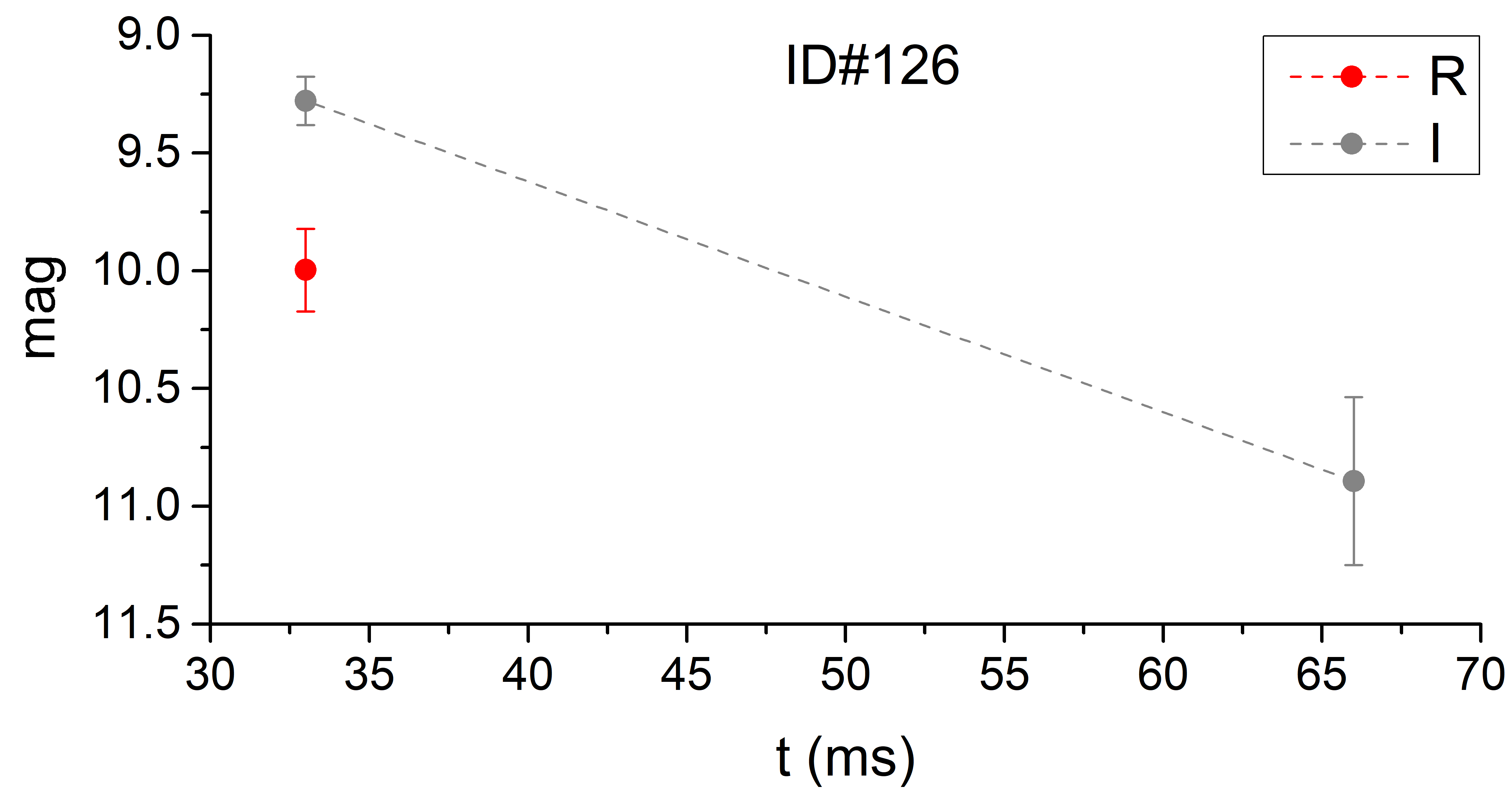}\\
\includegraphics[width=5.6cm]{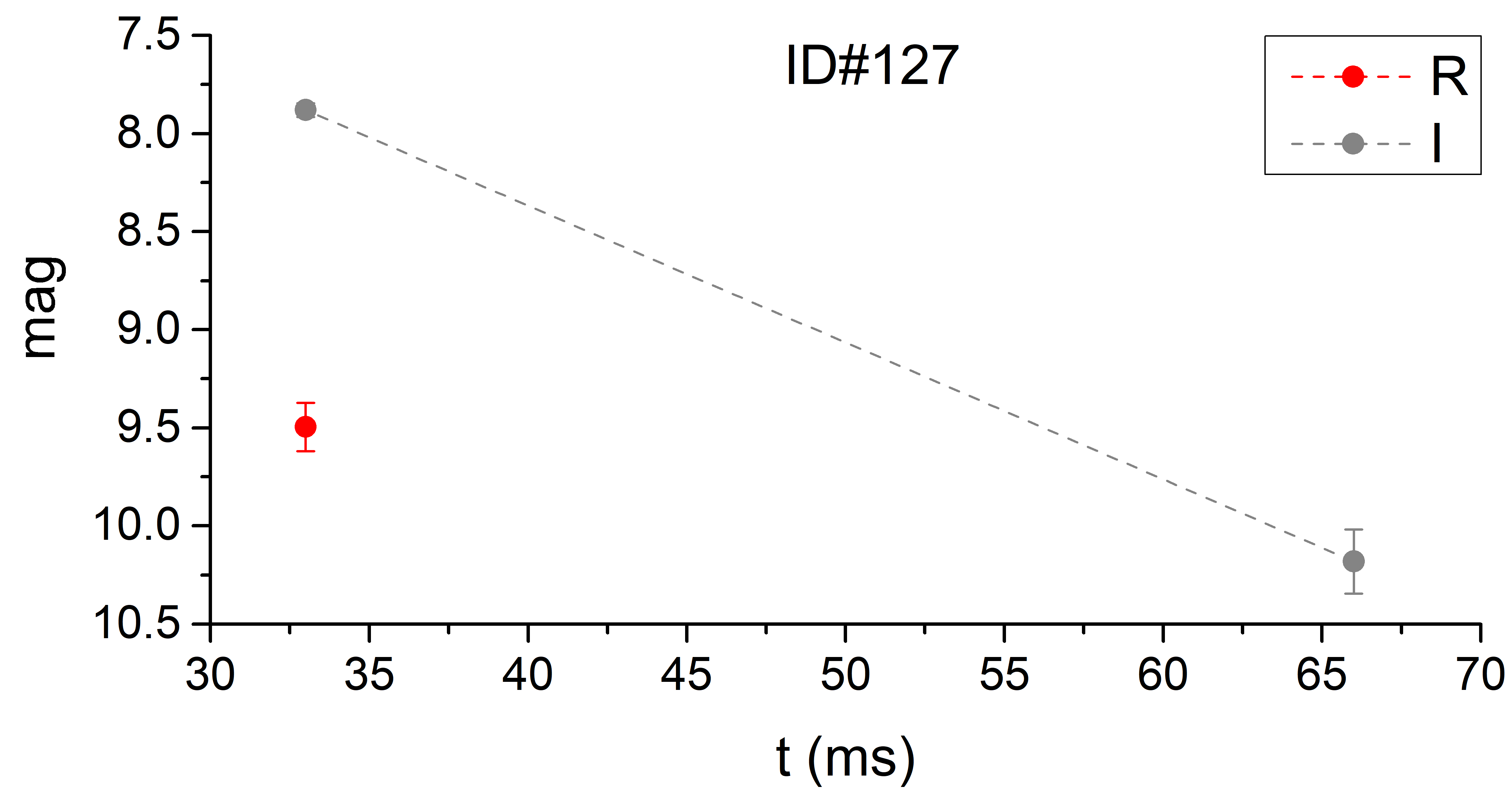}&\includegraphics[width=5.6cm]{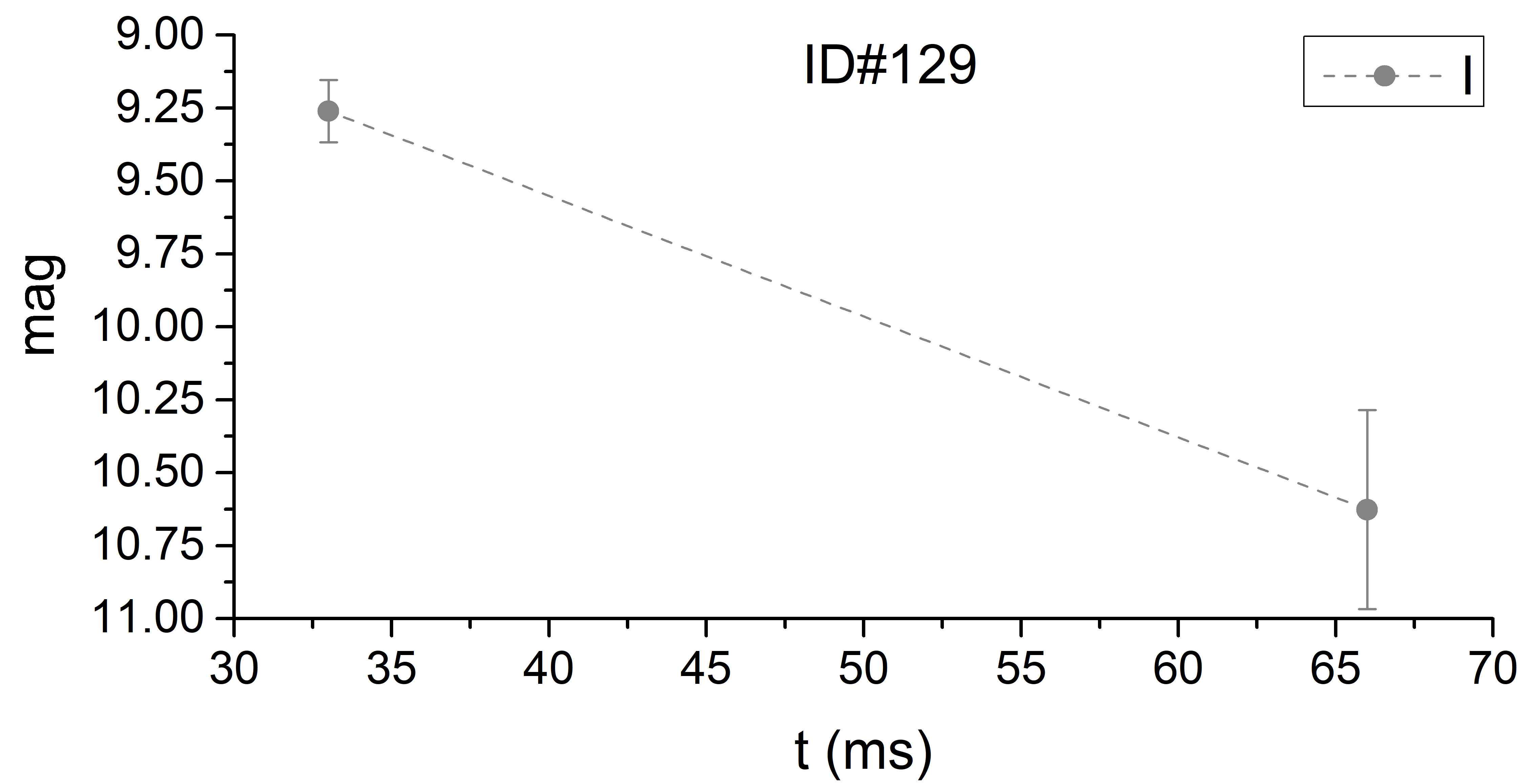}&\includegraphics[width=5.6cm]{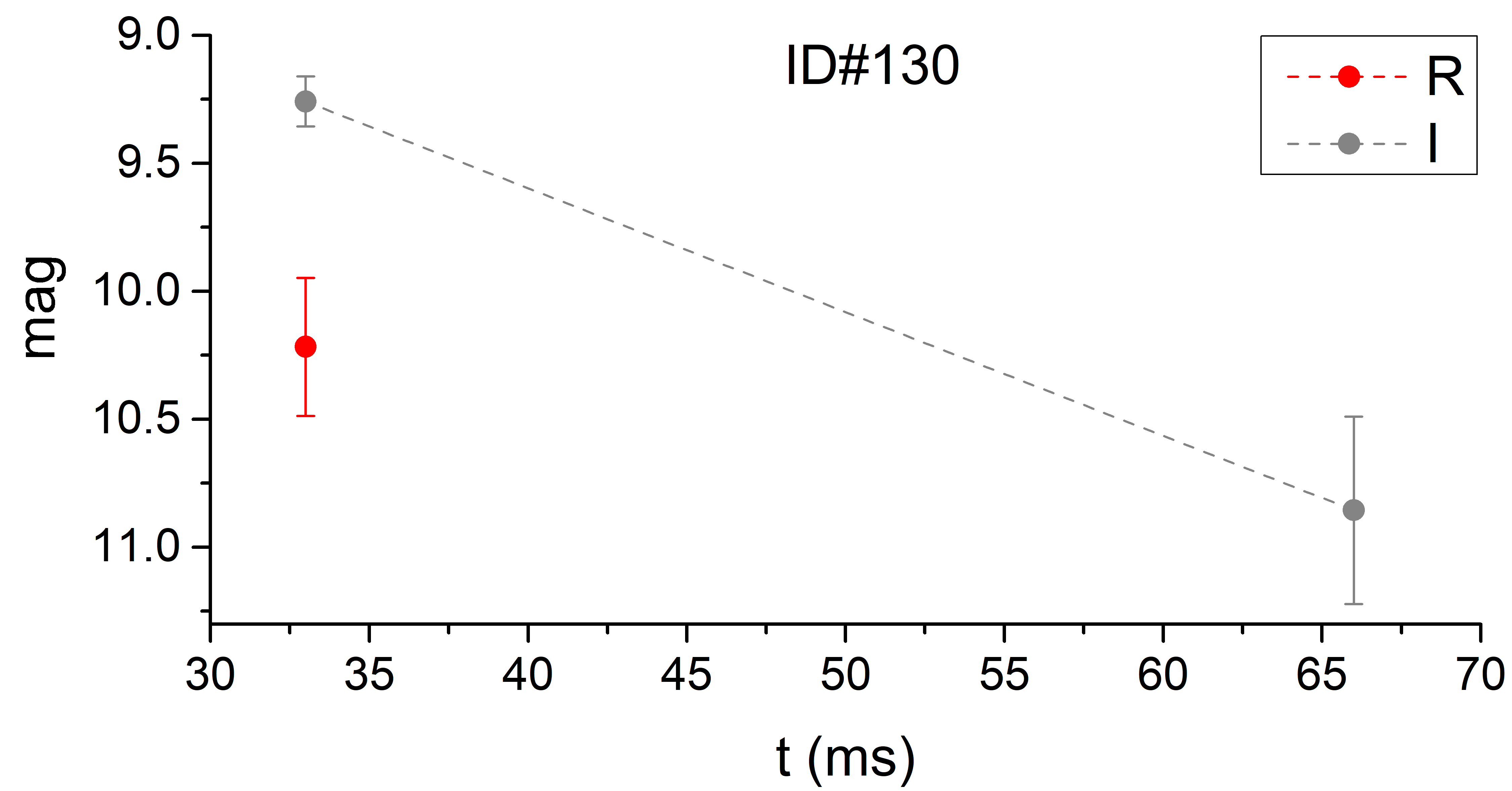}\\
\includegraphics[width=5.6cm]{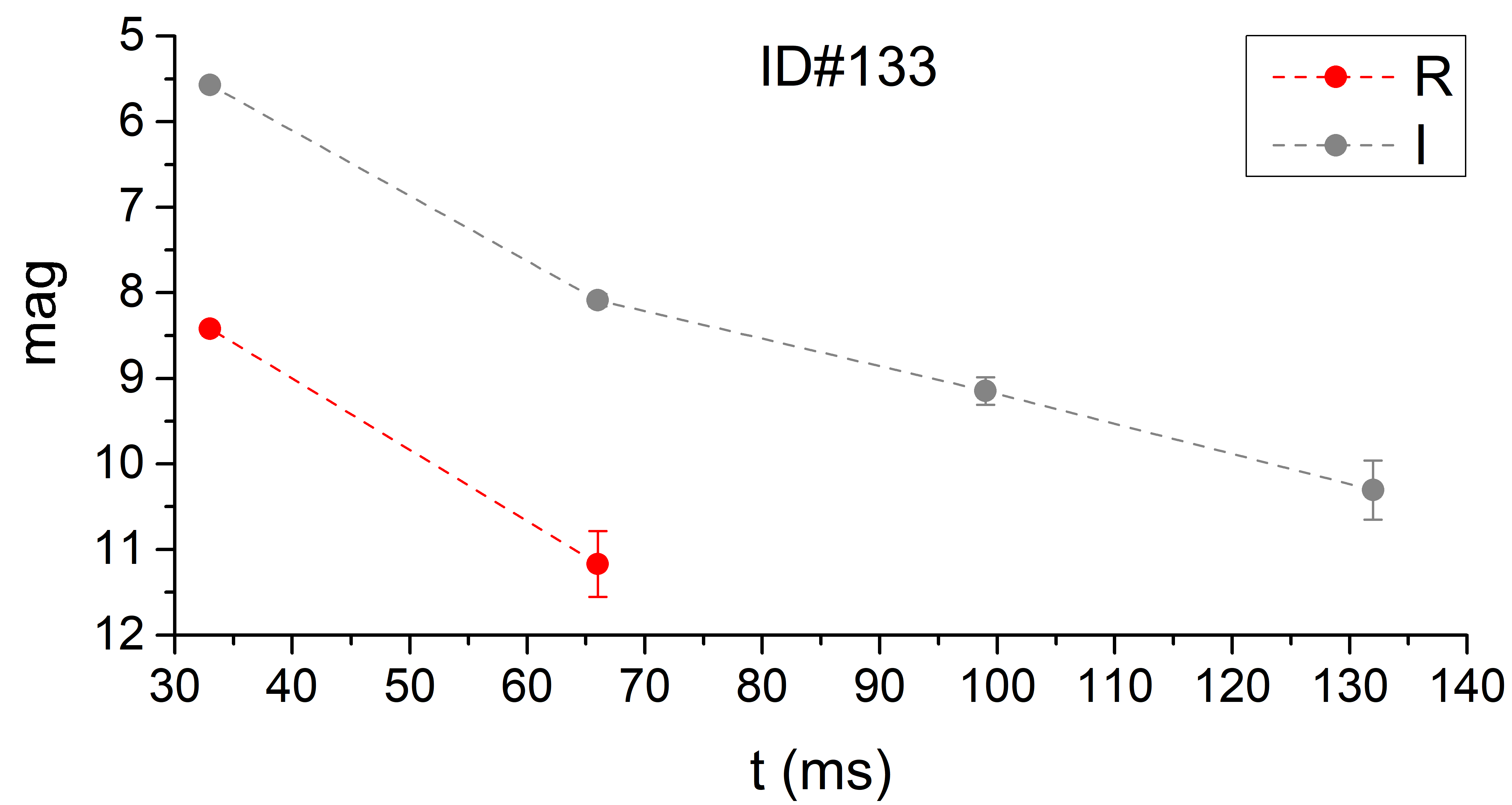}&\includegraphics[width=5.6cm]{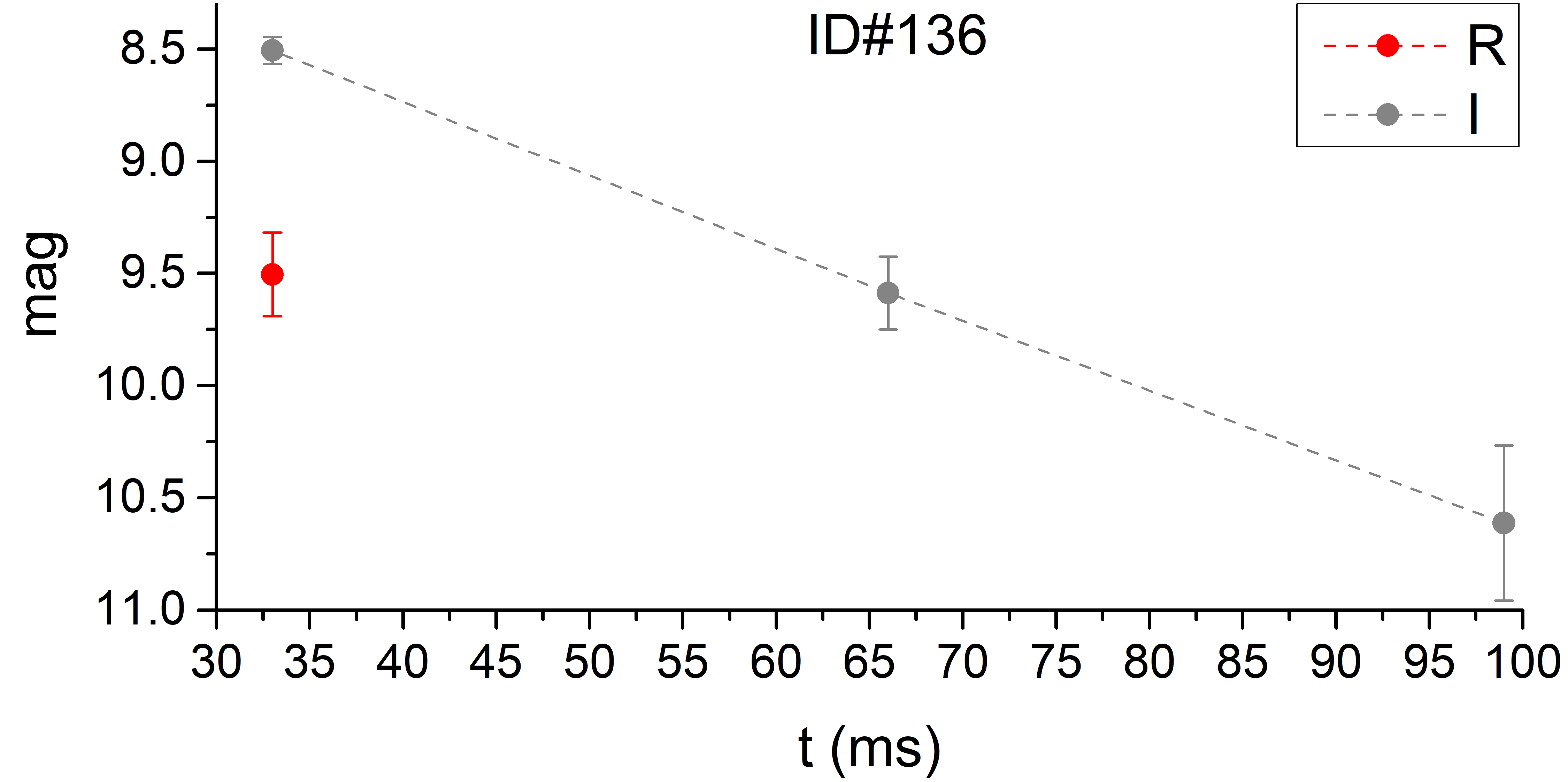}&\includegraphics[width=5.6cm]{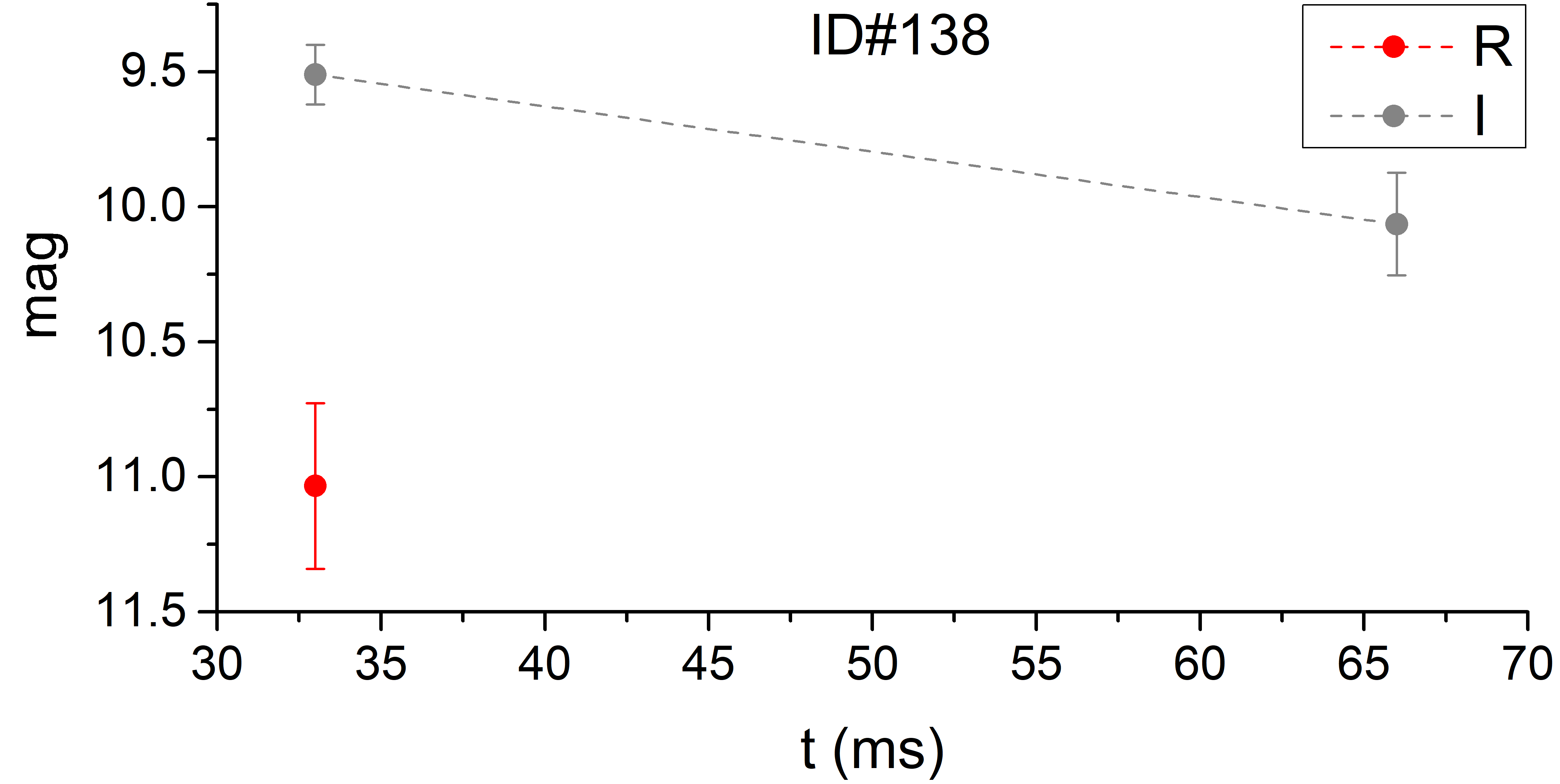}\\
\includegraphics[width=5.6cm]{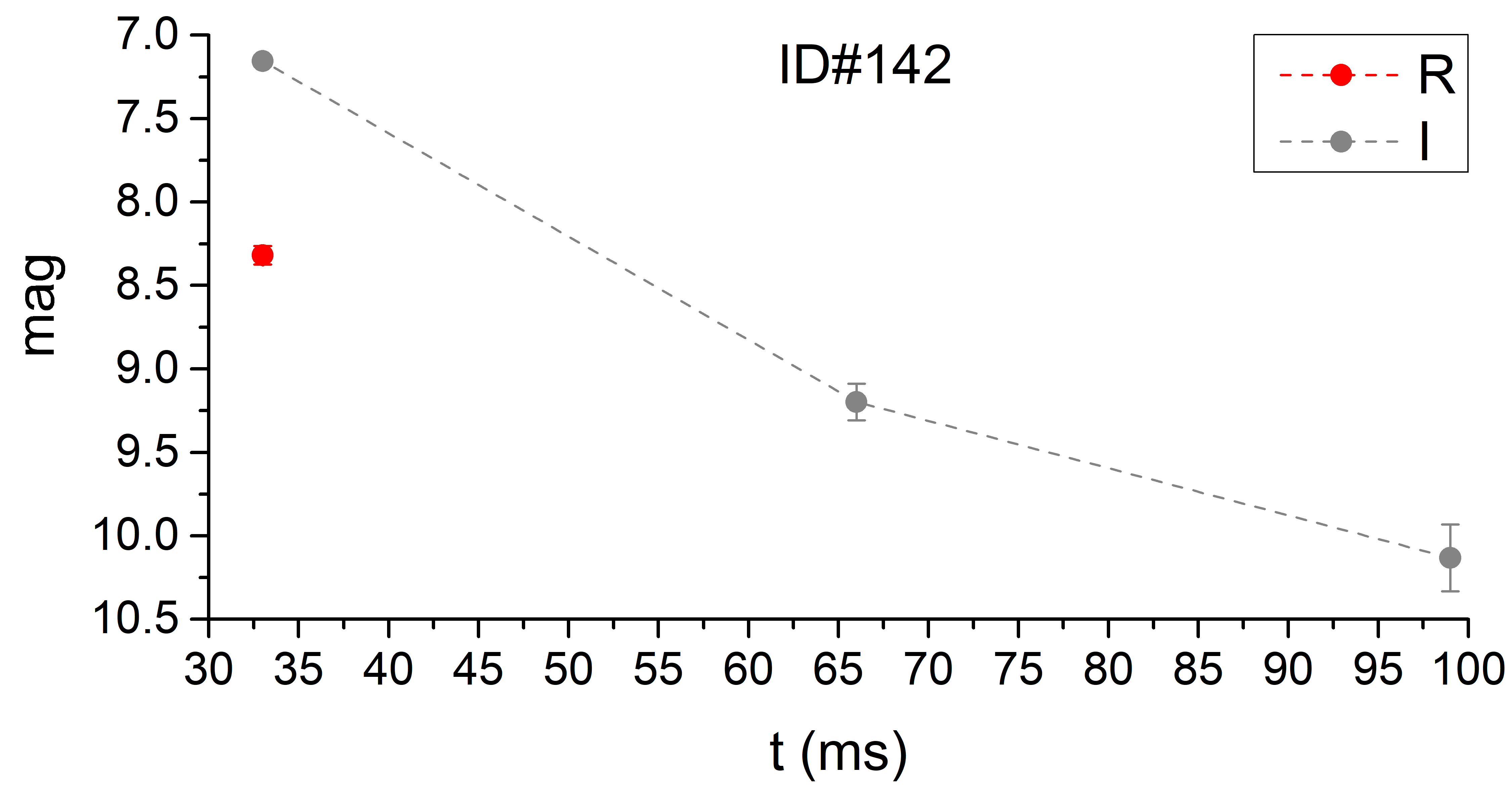}&\includegraphics[width=5.6cm]{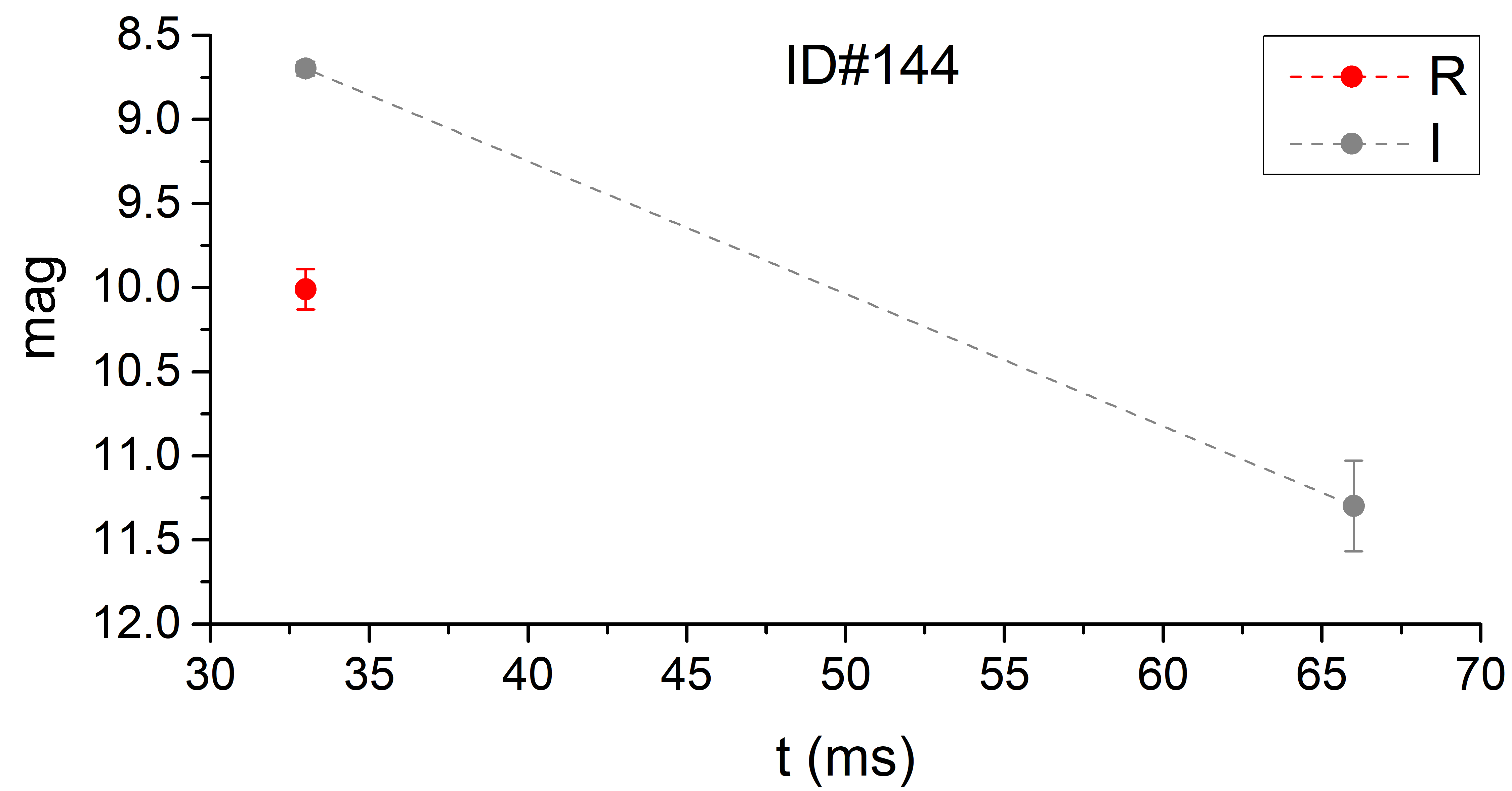}&\includegraphics[width=5.6cm]{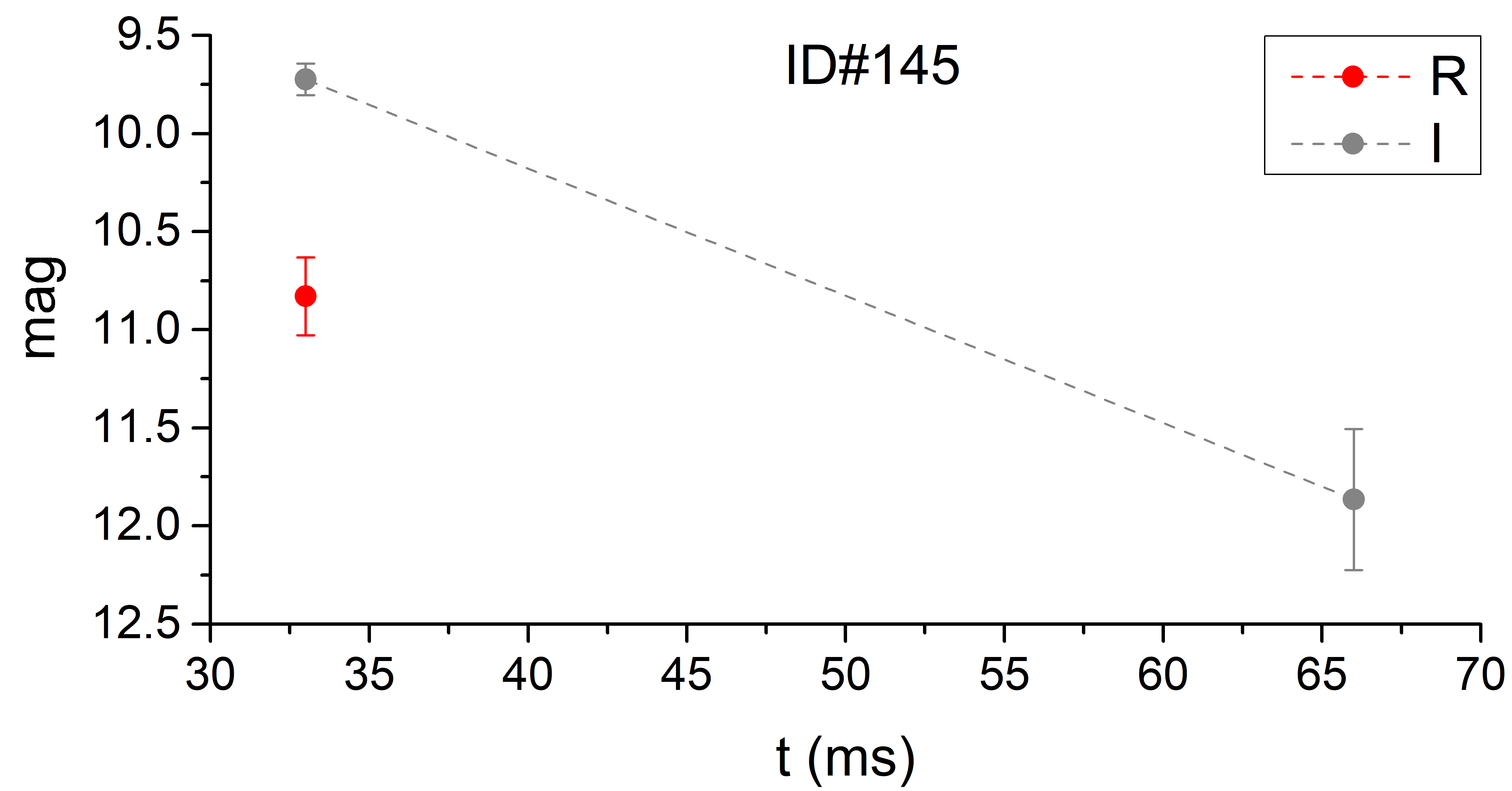}\\
\includegraphics[width=5.6cm]{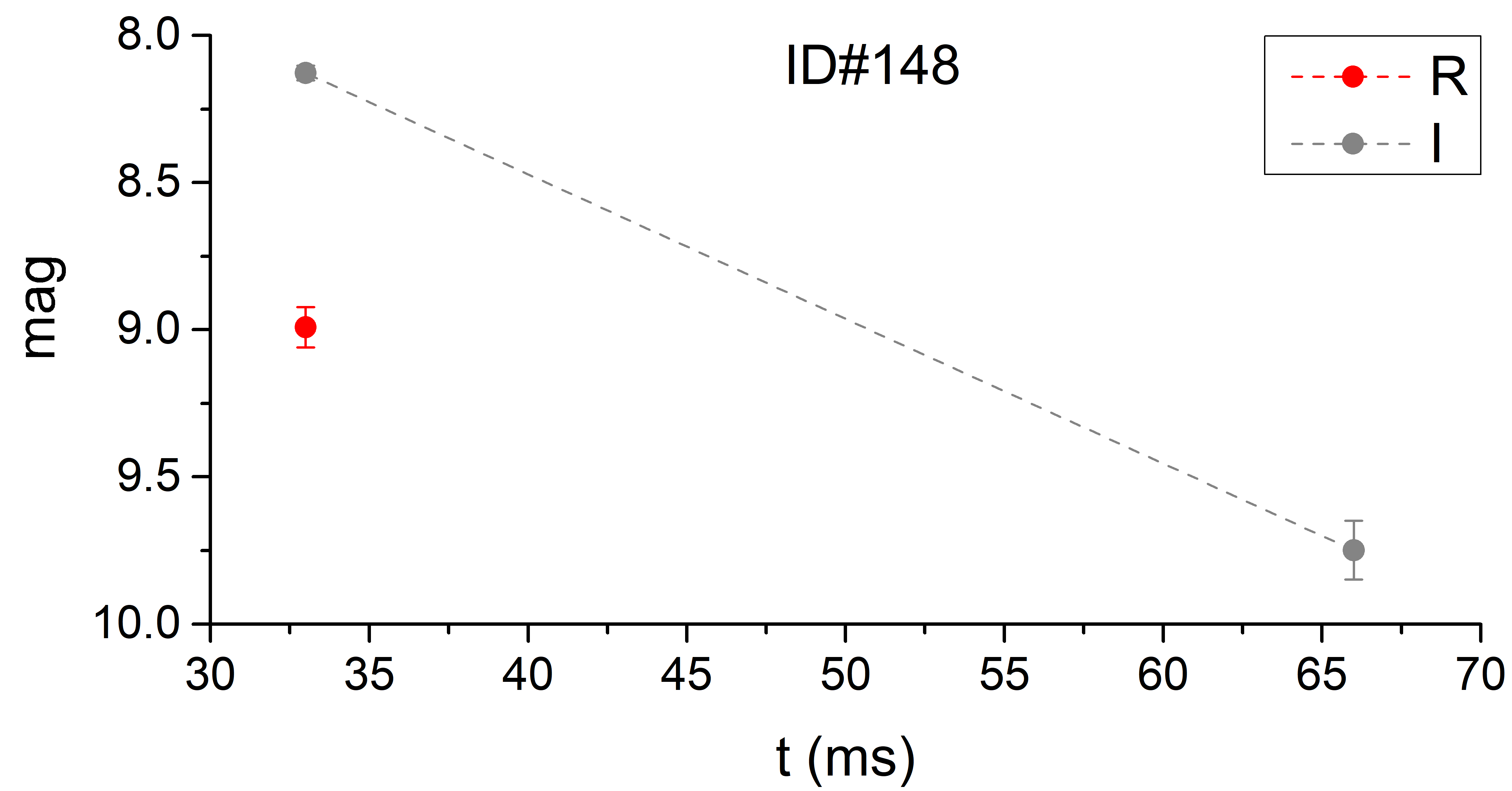}&\includegraphics[width=5.6cm]{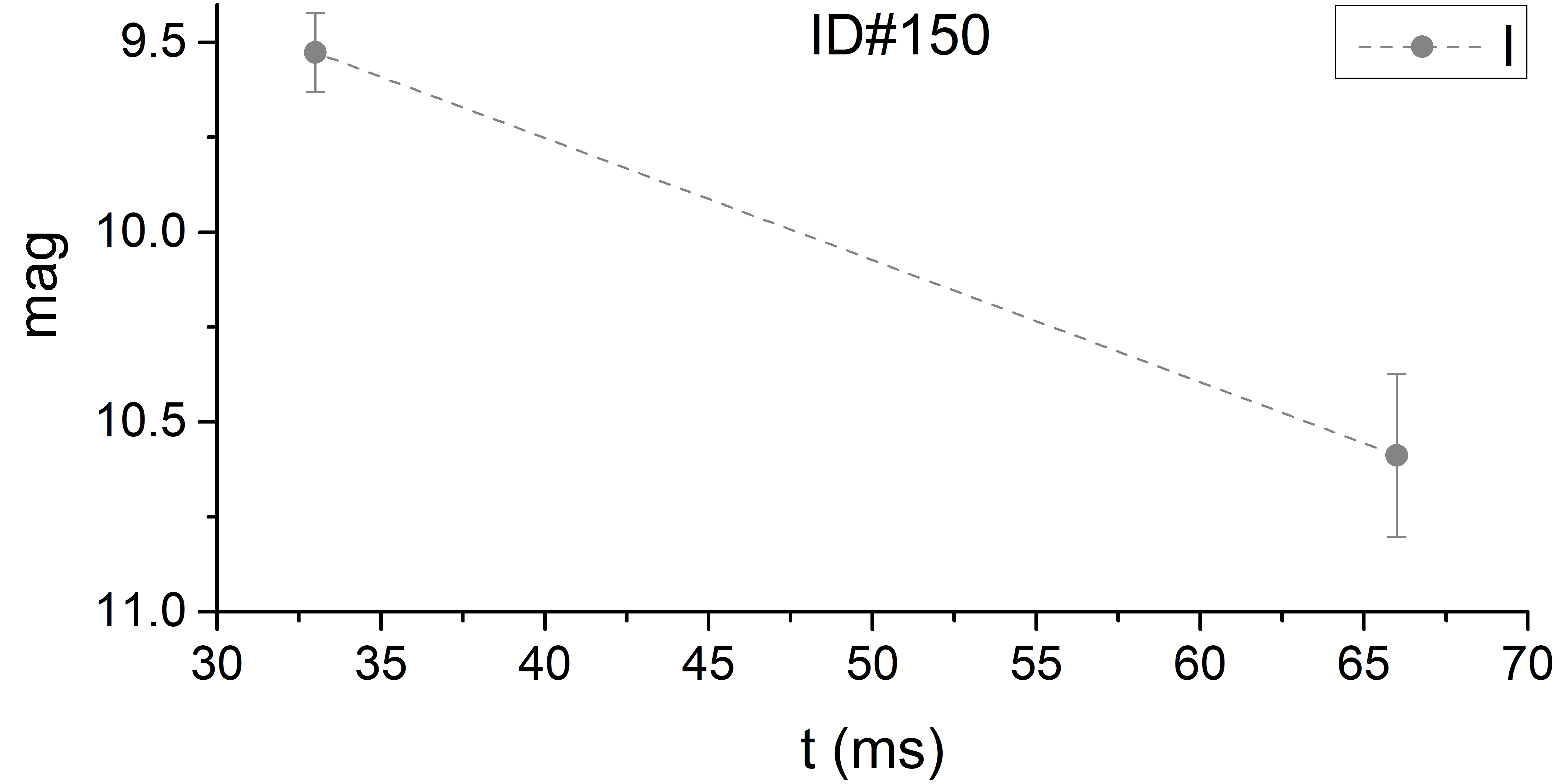}&\includegraphics[width=5.6cm]{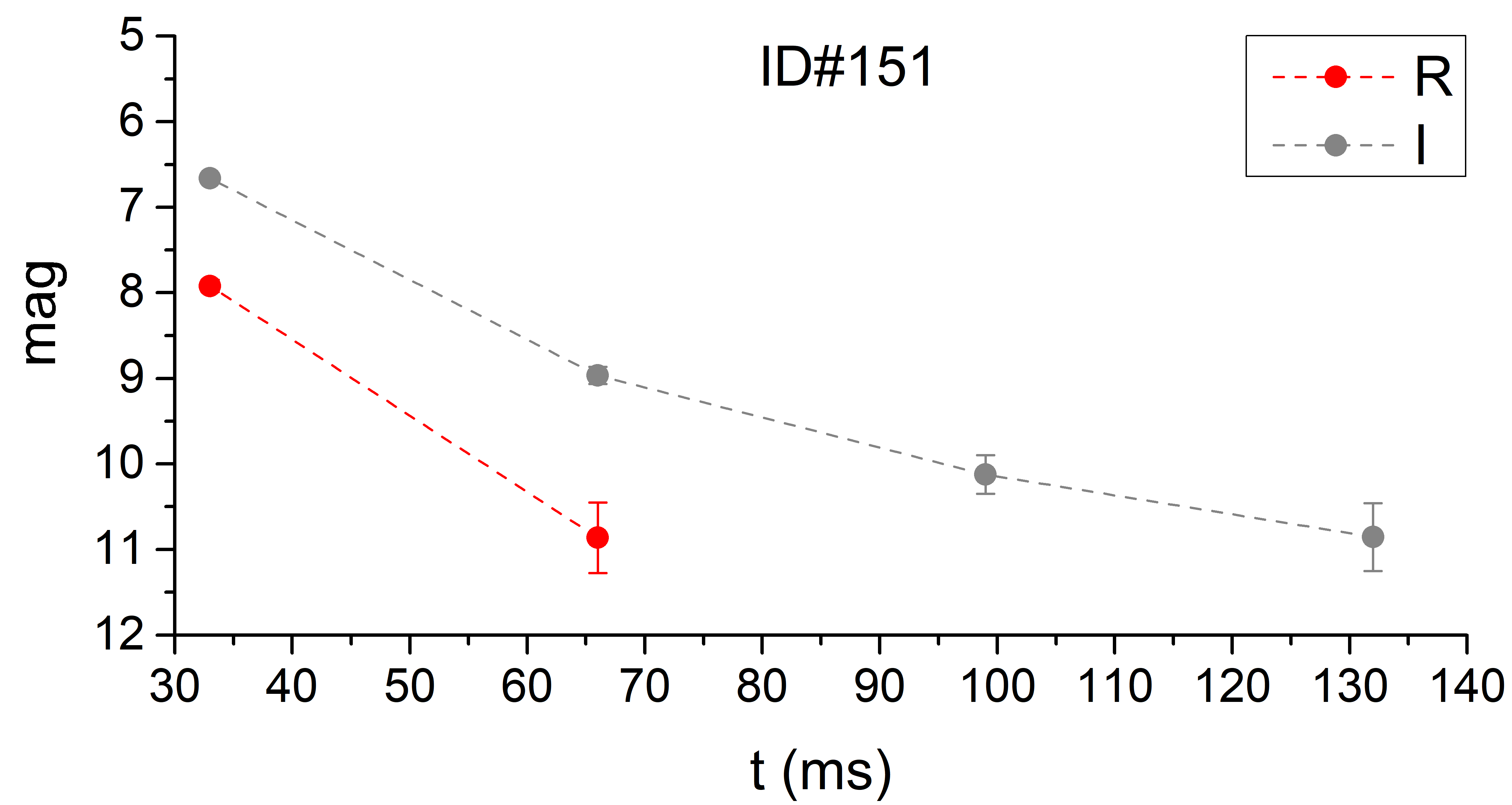}\\
\includegraphics[width=5.6cm]{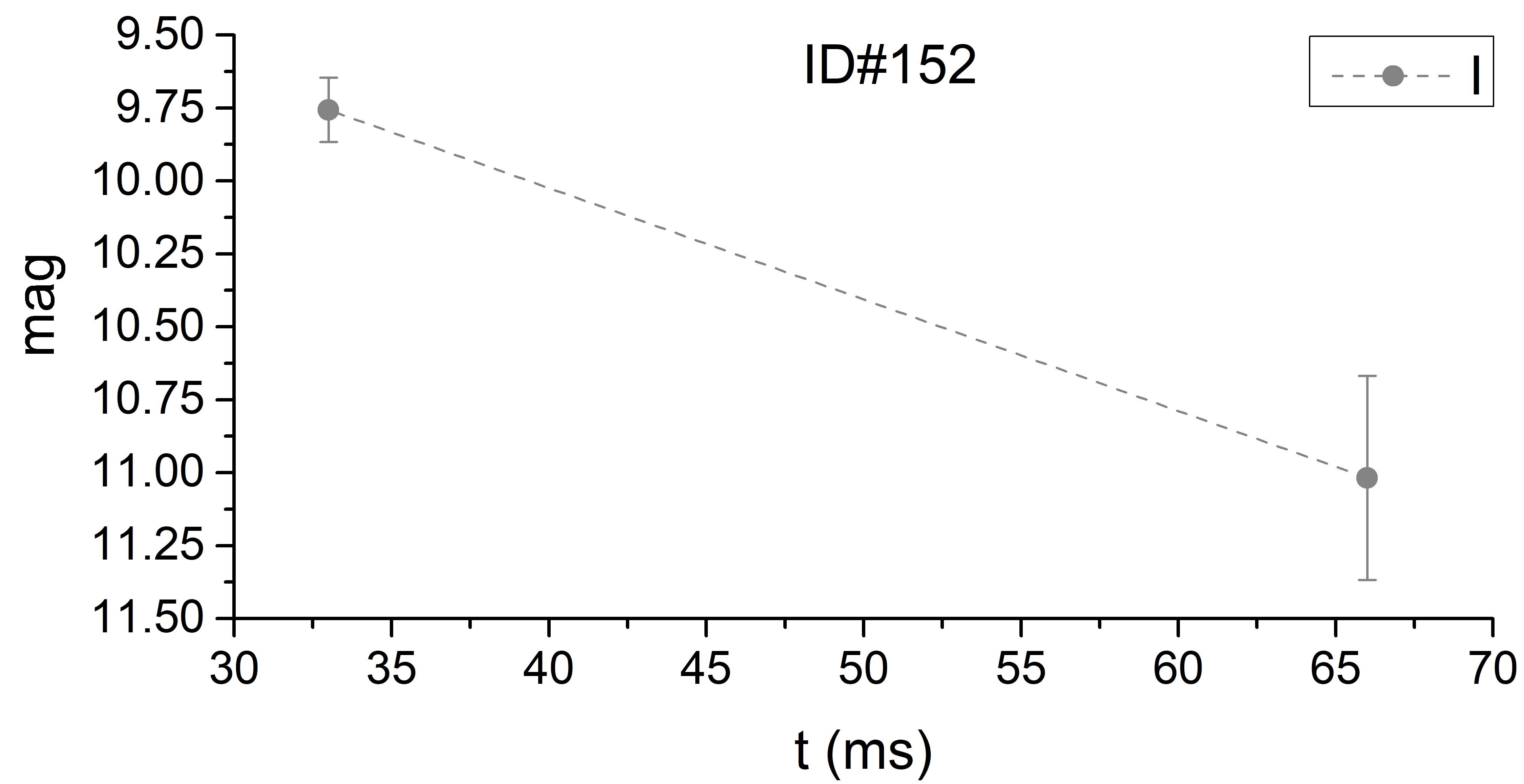}&\includegraphics[width=5.6cm]{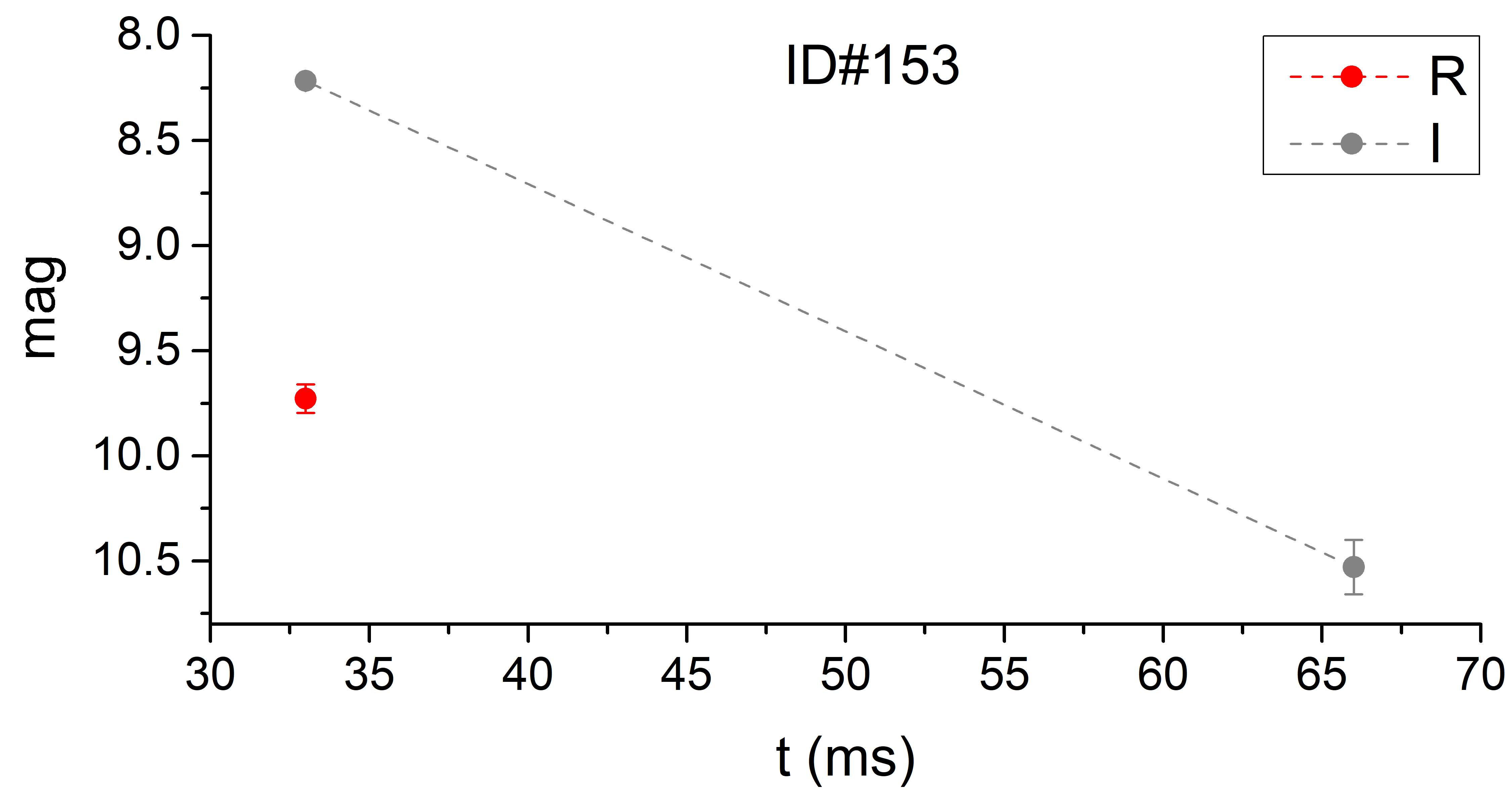}&\includegraphics[width=5.6cm]{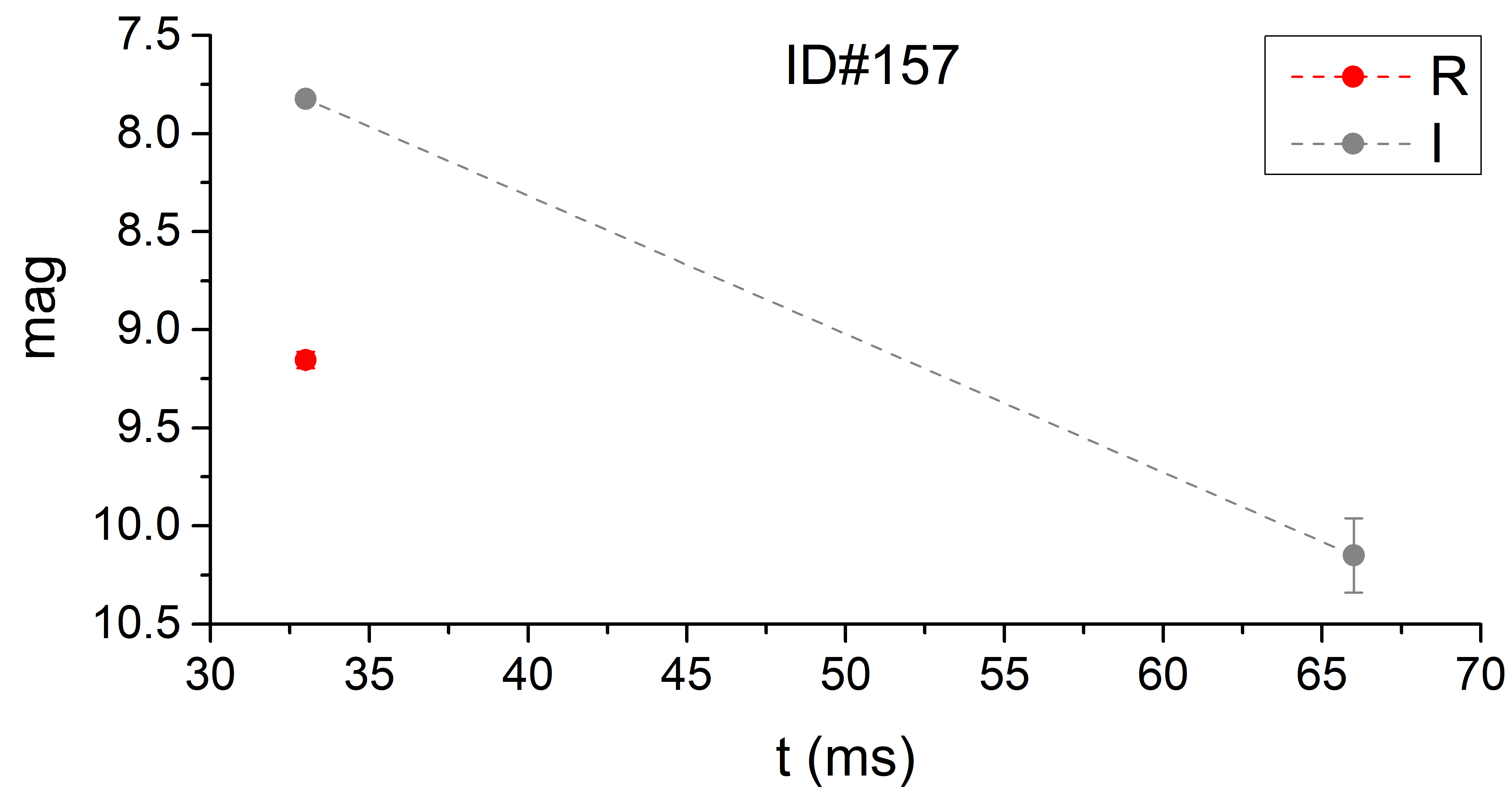}\\
\includegraphics[width=5.6cm]{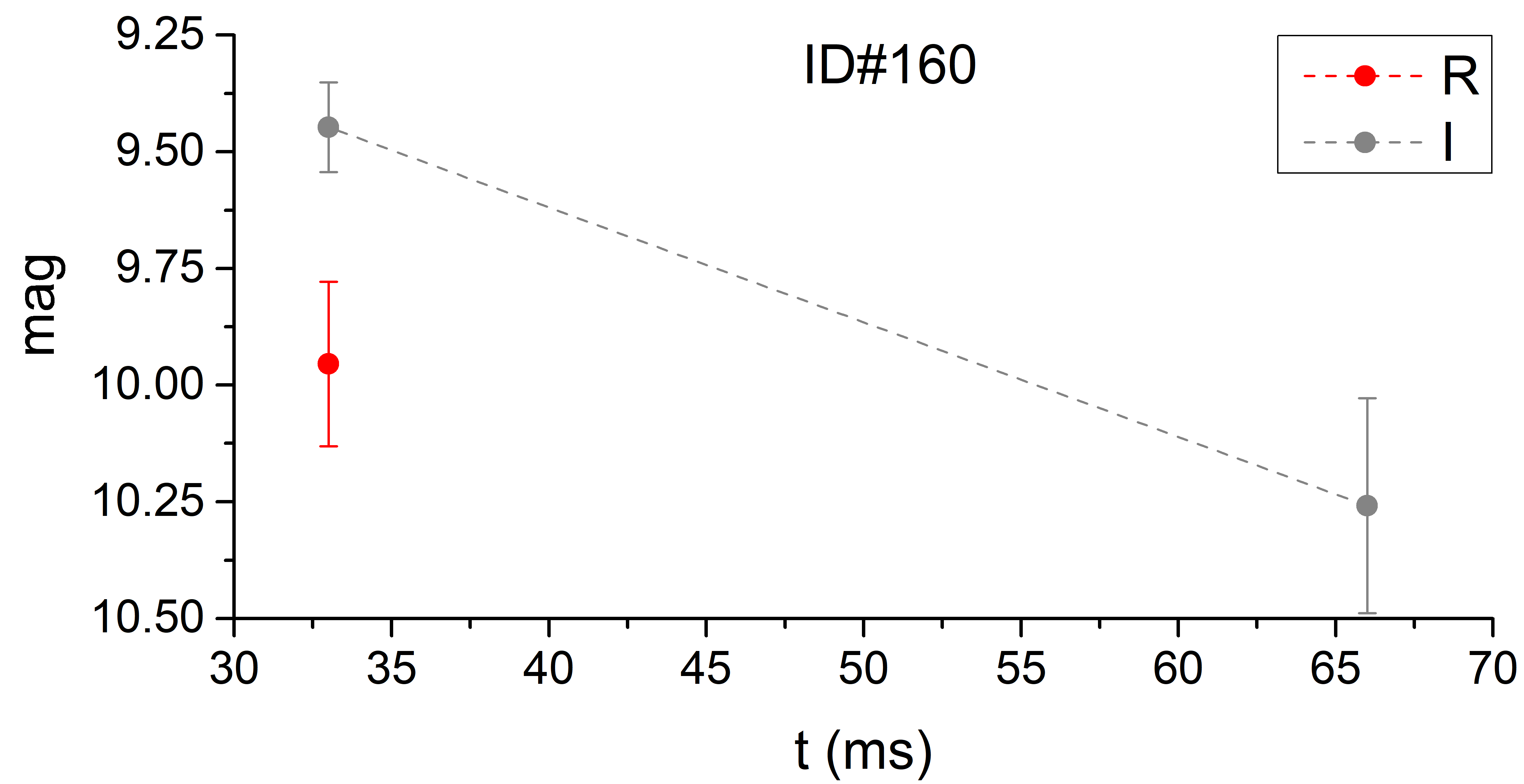}&\includegraphics[width=5.6cm]{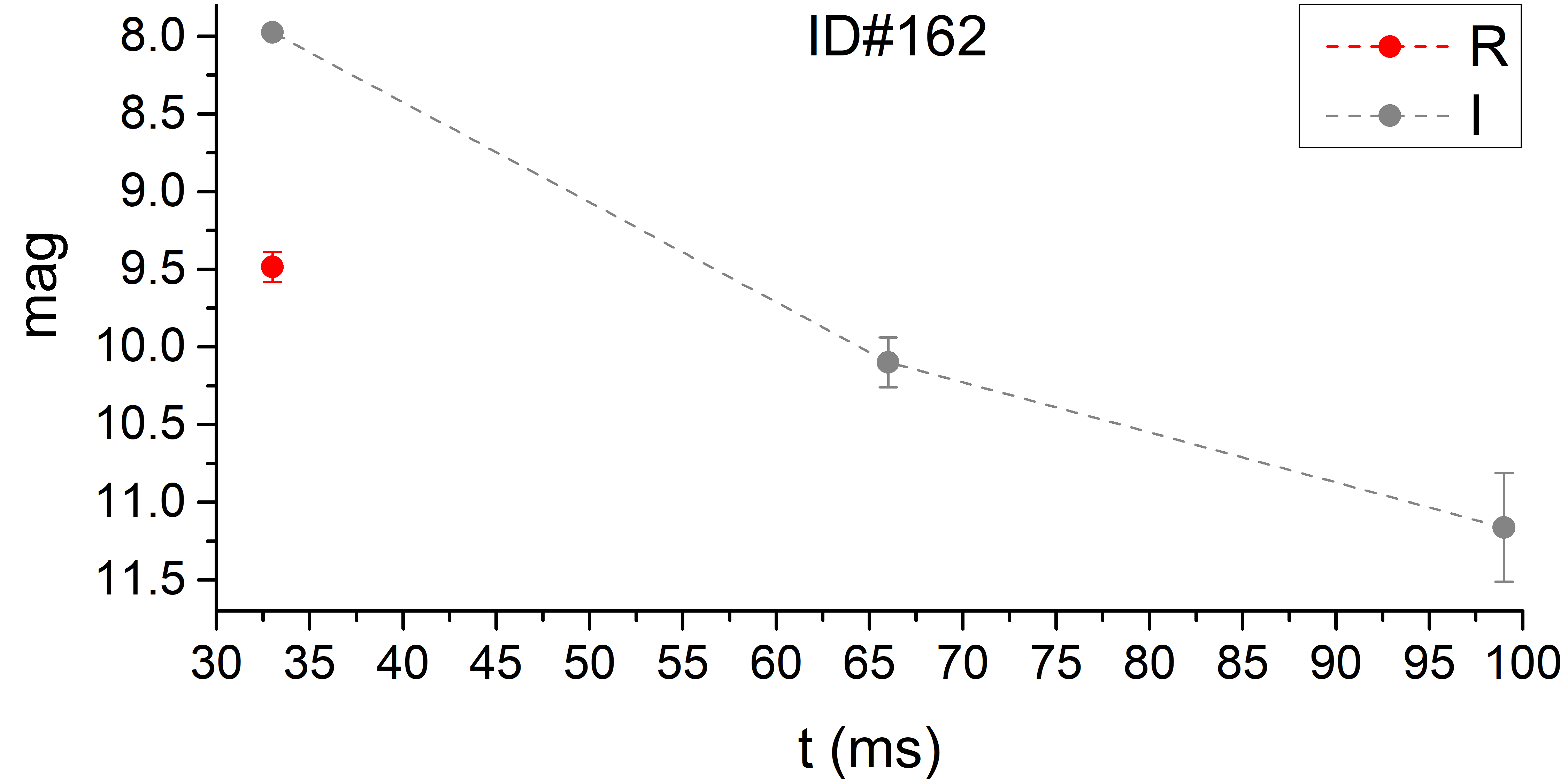}&\includegraphics[width=5.6cm]{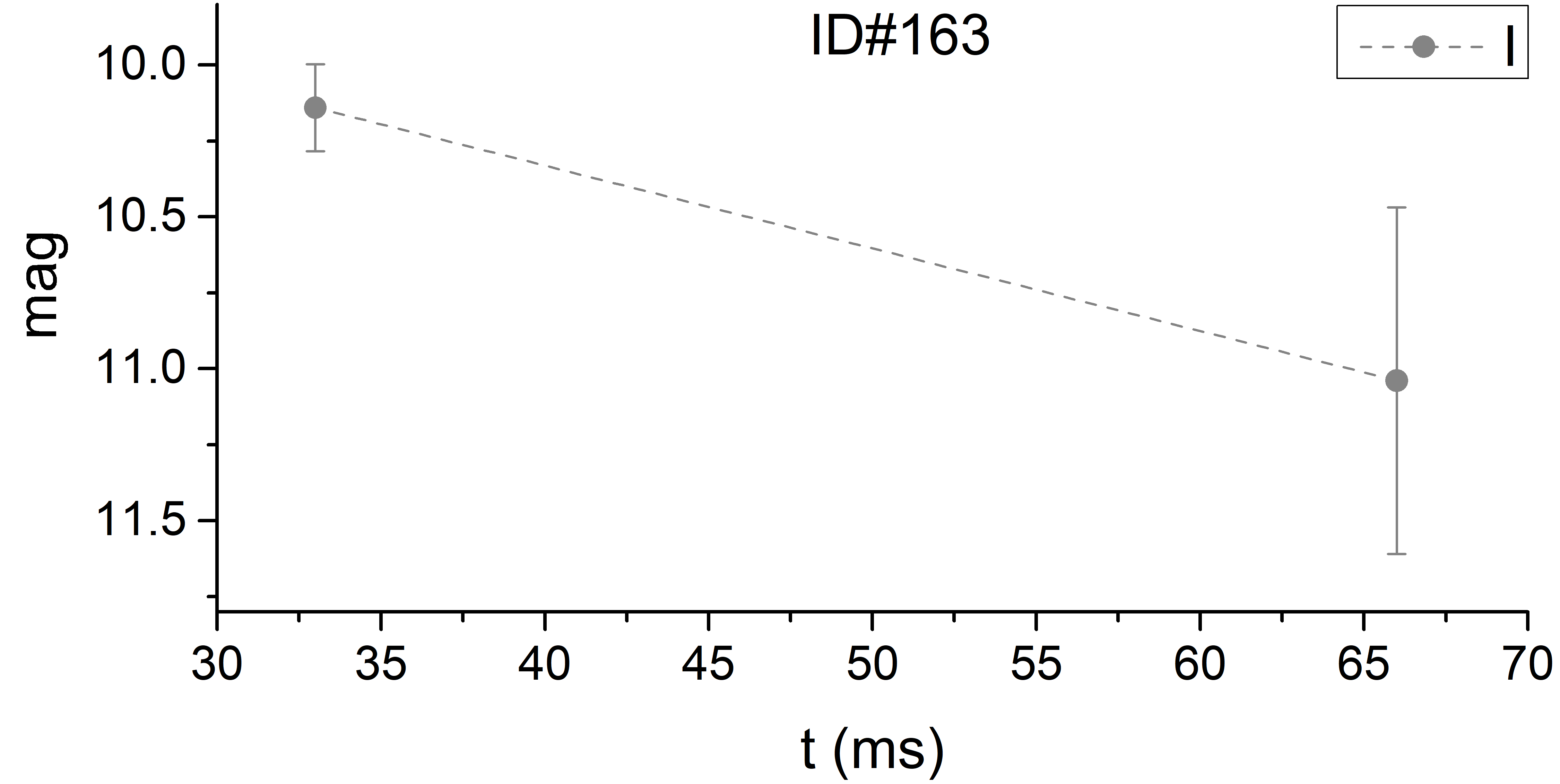}\\
\end{tabular}
\caption{Light curves of the multiframe flashes.}
\label{fig:LCs1}
\end{figure*}

In this appendix, we provide the magnitude values for all the multiframe flashes (Table~\ref{tab:multi}). For flashes with more than two set of frames in $R$ and $I$ bands, the corresponding temperature of each set of frames is also given. Moreover, the light curves of these flashes are plotted in Fig.~\ref{fig:LCs1} and the temperature evolution curves in Fig.~\ref{fig:TCs1}.

\begin{table*}
\centering
\caption{List of multiframe flashes after July 2019. Errors are included in parentheses alongside values and correspond to the last digit(s).}
\label{tab:multi}
\scalebox{0.85}{
\begin{tabular}{lcccc | lcccc }
\hline\hline
ID	&	Date \& UT	&	$m_{\rm R}$	&	$m_{\rm I}$	&	T	&	ID	&	Date \& UT	&	$m_{\rm R}$	&	$m_{\rm I}$	&	T	\\
	&		&	(mag)	&	(mag)	&	(K)	&		&		&	(mag)	&	(mag)	&	(K)	\\
\hline																			
114i	&	2019 08 06    18:19:16.023	&	10.37(28)	&	8.71(12)	&	2237(317)	&	172iv	&		&		&	10.22(18)	&		\\
114ii	&		&		&	11.20(51)	&		&	172v	&		&		&	11.56(58)	&		\\
116i	&	2019 08 06    18:56:36.990	&	8.43(12)	&	7.38(7)	&	3151(305)	&	173i	&	2021 07 03    01:23:35.666	&		&	9.26(12)	&		\\
116ii	&		&		&	9.62(21)	&		&	173ii	&		&		&	10.37(26)	&		\\
118i	&	2019 08 26    02:50:56.348	&	10.65(27)	&	9.11(7)	&	2368(308)	&	177i	&	2021 07 16    18:49:32.113	&	9.96(34)	&	9.15(12)	&	3746(1274)	\\
118ii	&		&		&	10.18(17)	&		&	177ii	&		&		&	9.77(25)	&		\\
124i	&	2019 09 23    03:36:21.556	&	10.20(28)	&	9.40(11)	&	3777(1042)	&	179i	&	2021 08 02    01:26:06.111	&	10.41(29)	&	9.85(14)	&	4646(1967)	\\
124ii	&		&		&	10.39(32)	&		&	179ii	&		&		&	10.18(19)	&		\\
125i	&	2019 09 25    03:40:11.073	&	8.43(8)	&	7.42(5)	&	3212(207)	&	180i	&	2021 08 02    01:34:28.145	&	8.75(18)	&	8.28(9)	&	5171(1433)	\\
125ii	&		&	10.75(26)	&	9.56(9)	&	2864(504)	&	180ii	&		&		&	10.02(36)	&		\\
125iii	&		&		&	11.05(26)	&		&	181i	&	2021 08 02    02:44:36.306	&	10.12(26)	&	9.72(15)	&	5634(2105)	\\
126i	&	2019 10 22    04:11:36.271	&	10.00(18)	&	9.28(11)	&	4036(810)	&	181ii	&		&		&	10.37(20)	&		\\
126ii	&		&		&	10.89(40)	&		&	182i	&	2021 08 02    02:51:02.646	&	8.85(17)	&	7.74(6)	&	3022(349)	\\
127i	&	2019 10 24    02:30:16.186	&	9.49(15)	&	7.88(7)	&	2288(160)	&	182ii	&		&		&	8.89(13)	&		\\
127ii	&		&		&	10.18(21)	&		&	182iii	&		&		&	9.81(27)	&		\\
129i	&	2019 11 02    16:20:44.065	&		&	9.26(12)	&		&	184i	&	2021 10 11    16:57:00.374	&	8.66(13)	&	7.73(8)	&	3404(385)	\\
129ii	&		&		&	10.63(40)	&		&	184ii	&		&	10.79(35)	&	8.86(10)	&	1986(301)	\\
130i	&	2019 11 02    17:19:19.668	&	10.22(29)	&	9.26(12)	&	3345(770)	&	184iii	&		&		&	9.69(16)	&		\\
130ii	&		&		&	10.86(40)	&		&	184iv	&		&		&	10.37(26)	&		\\
133i	&	2019 12 01    16:23:13.796	&	8.42(10)	&	5.57(7)	&	1438(53)	&	184e	&		&		&	10.83(32)	&		\\
133ii	&		&	11.17(41)	&	8.09(9)	&	1343(149)	&	185i	&	2021 10 12    16:31:15.618	&	9.33(22)	&	8.25(10)	&	3066(516)	\\
133iii	&		&		&	9.15(17)	&		&	185ii	&		&		&	10.61(41)	&		\\
133iv	&		&		&	10.31(35)	&		&	186i	&	2021 10 12    17:42:17.599	&	9.75(20)	&	8.75(10)	&	3248(528)	\\
136i	&	2019 12 20    04:34:16.679	&	9.50(20)	&	8.51(7)	&	3254(477)	&	186ii	&		&		&	11.09(41)	&		\\
136ii	&		&		&	9.59(17)	&		&	187i	&	2021 12 01    04:21:42.906	&		&	10.00(13)	&		\\
136iii	&		&		&	10.61(35)	&		&	187ii	&		&		&	11.26(36)	&		\\
138i	&	2020 01 30    17:18:08.539	&	11.03(33)	&	9.51(12)	&	2393(427)	&	188i	&	2021 12 08    16:15:10.805	&	8.58(12)	&	7.59(7)	&	3261(322)	\\
138ii	&		&		&	10.06(19)	&		&	188ii	&		&		&	9.34(18)	&		\\
142i	&	2020 03 01    16:54:23.862	&	8.32(8)	&	7.15(4)	&	2919(161)	&	191i	&	2022 04 05    17:30:55.889	&	8.87(17)	&	7.53(6)	&	2635(265)	\\
142ii	&		&		&	9.20(11)	&		&	191ii	&		&	9.55(22)	&	8.51(1)	&	3158(512)	\\
142iii	&		&		&	10.13(20)	&		&	191iii	&		&		&	9.56(26)	&		\\
144i	&	2020 03 27    17:40:25.320	&	10.01(15)	&	8.70(8)	&	2677(263)	&	193i	&	2022 05 07    20:47:14.061	&		&	7.41(8)	&		\\
144ii	&		&		&	11.30(28)	&		&	193ii	&		&		&	8.05(10)	&		\\
145i	&	2020 03 29    18:14:10.875	&	10.83(23)	&	9.72(9)	&	3030(507)	&	195i	&	2022 06 03    18:21:31.378	&	7.96(11)	&	6.64(6)	&	2666(192)	\\
145ii	&		&		&	11.87(36)	&		&	195ii	&		&	10.73(44)	&	8.43(7)	&	1725(270)	\\
148i	&	2020 04 28    19:19:54.525 	&	8.99(10)	&	8.13(5)	&	3587(323)	&	195iii	&		&		&	10.04(19)	&		\\
148ii	&		&		&	9.75(10)	&		&	195iv	&		&		&	10.04(19)	&		\\
150i	&	2020 06 16    02:09:14.848	&		&	9.53(13)	&		&	196i	&	2022 06 03    18:57:25.884	&		&	9.17(9)	&		\\
150ii	&		&		&	10.59(23)	&		&	196ii	&		&		&	10.58(25)	&		\\
151i	&	2020 06 25    18:28:18.340	&	7.92(7)	&	6.66(6)	&	2763(192)	&	198i	&	2022 06 03    19:34:16.804	&		&	10.55(23)	&		\\
151ii	&		&	10.86(41)	&	8.96(9)	&	2016(357)	&	198ii	&		&	9.44(13)	&	9.44(13)	&		\\
151iii	&		&		&	10.12(23)	&		&	199i	&	2022 06 04    18:20:50.612	&	9.40(28)	&	8.32(9)	&	3081(624)	\\
151iv	&		&		&	10.85(40)	&		&	199ii	&		&		&	9.76(23)	&		\\
152i	&	2020 06 25    19:09:24.063	&		&	9.76(13)	&		&	202i	&	2022 06 23    01:46:07.604	&	8.73(15)	&	7.22(7)	&	2409(207)	\\
152ii	&		&		&	11.02(35)	&		&	202ii	&		&		&	9.44(11)	&		\\
153i	&	2020 06 26    19:52:31.569	&	9.73(11)	&	8.22(7)	&	2408(160)	&	204i	&	2022 07 22    02:13:27.207	&	9.97(24)	&	8.73(7)	&	2788(425)	\\
153ii	&		&		&	10.53(13)	&		&	204ii	&		&		&	10.45(22)	&		\\
157i	&	2020 07 26    19:10:25.428	&	9.15(10)	&	7.82(6)	&	2649(171)	&	205i	&	2022 07 22    02:48:11.224	&	10.27(36)	&	9.29(14)	&	3297(993)	\\
157ii	&		&		&	10.15(19)	&		&	205ii	&		&		&	10.11(25)	&		\\
160i	&	2020 08 14    01:15:36.278	&	9.95(20)	&	9.45(12)	&	4967(1227)	&	206i	&	2022 07 22    02:49:51.365	&	9.36(29)	&	7.51(10)	&	2056(257)	\\
160ii	&		&		&	10.26(23)	&		&	206ii	&		&		&	8.15(12)	&		\\
162i	&	2021 03 17    17:46:52.983	&	9.48(12)	&	7.97(6)	&	2410(168)	&	206iii	&		&		&	9.65(25)	&		\\
162ii	&		&		&	10.10(16)	&		&	206iv	&		&		&	10.30(44)	&		\\
162iii	&		&		&	11.16(35)	&		&	209i	&	2022 08 01    18:28:18.506	&	9.42(21)	&	7.62(13)	&	2103(230)	\\
163i	&	2021 03 17    18:03:02.205	&		&	10.14(15)	&		&	209ii	&		&	10.45(28)	&	8.96(15)	&	2435(410)	\\
163ii	&		&		&	11.04(57)	&		&	209iii	&		&		&	10.13(23)	&		\\
165i	&	2021 04 18    19:13:58.703	&	9.43(15)	&	7.89(6)	&	2370(184)	&	210i	&	2022 08 03    18:05:38.820	&		&	8.66(9)	&		\\
165ii	&		&		&	9.08(11)	&		&	210ii	&		&		&	10.16(23)	&		\\
165iii	&		&		&	10.55(25)	&		&	212i	&	2022 08 03    18:23:42.636	&	9.76(12)	&	8.37(8)	&	2575(204)	\\
168i	&	2021 05 18    20:08:21.742	&	9.90(20)	&	8.66(7)	&	2788(365)	&	212ii	&		&		&	10.16(17)	&		\\
168ii	&		&		&	10.66(28)	&		&	213i	&	2022 08 03    19:02:12.594	&		&	10.04(19)	&		\\
169i	&	2021 06 15    19:06:12.914	&	9.96(15)	&	9.19(8)	&	3863(596)	&	213ii	&		&		&	10.59(27)	&		\\
169ii	&		&		&	10.94(25)	&		&	214i	&	2022 08 04    18:47:06.667	&	9.26(18)	&	7.63(13)	&	2271(242)	\\
170i	&	2021 06 15    19:15:27.433	&		&	9.57(13)	&		&	214ii	&		&		&	9.71(33)	&		\\
170ii	&		&		&	11.11(48)	&		&	215i	&	2022 08 04    19:24:20.958	&		&	8.02(12)	&		\\
171i	&	2021 06 15    19:23:18.620	&	10.86(32)	&	10.19(16)	&	4457(1492)	&	215ii	&		&		&	9.55(18)	&		\\
171ii	&		&		&	10.70(26)	&		&	216i	&	2022 08 04    19:25:11.729	&		&	9.40(19)	&		\\
172i	&	2021 06 15    19:38:52.236	&	8.15(10)	&	6.73(6)	&	2520(158)	&	216ii	&		&		&	9.93(29)	&		\\
172ii	&		&	9.70(16)	&	8.33(8)	&	2595(250)	&	217i	&	2022 08 22    01:33:02.291	&		&	8.45(9)	&		\\
172iii	&		&		&	9.44(13)	&		&	217ii	&		&		&	9.48(19)	&		\\
\hline
\end{tabular}}
\end{table*}

\begin{table*}
\centering
\caption*{Table~\ref{tab:multi} (cont'd)}
\scalebox{0.85}{
\begin{tabular}{lcccc | lcccc }
\hline\hline
ID	&	Date \& UT	&	$m_{\rm R}$	&	$m_{\rm I}$	&	T	&	ID	&	Date \& UT	&	$m_{\rm R}$	&	$m_{\rm I}$	&	T	\\
	&		&	(mag)	&	(mag)	&	(K)	&		&		&	(mag)	&	(mag)	&	(K)	\\
\hline	
219i	&	2022 09 01    18:33:36.493	&	10.66(37)	&	8.04(14)	&	1543(186)	&	259iii	&		&	9.91(34)	&	7.78(10)	&	1839(244)	\\
219ii	&		&		&	9.99(23)	&		&	259iv	&		&	10.56(56)	&	8.58(13)	&	1945(455)	\\
222i	&	2022 10 19    01:43:11.699	&		&	9.88(15)	&		&	259v	&		&		&	9.24(17)	&		\\
222ii	&		&		&	10.68(26)	&		&	259vi	&		&		&	9.62(26)	&		\\
223i	&	2022 10 19    03:03:44.141	&	8.47(10)	&	8.25(6)	&	7426(1907)	&	261i	&	2022 12 26    16:48:14.320	&	10.26(20)	&	9.68(12)	&	4580(1240)	\\
223ii	&		&		&	9.93(20)	&		&	261ii	&		&		&	10.37(20)	&		\\
223iii	&		&		&	10.75(37)	&		&	264i	&	2022 12 27    17:47:37.390	&	9.29(17)	&	7.98(9)	&	2675(298)	\\
224i	&	2022 10 20    01:12:58.151	&	9.80(27)	&	7.94(8)	&	2053(245)	&	264ii	&		&		&	10.43(34)	&		\\
224ii	&		&	8.58(14)	&	7.38(8)	&	2858(285)	&	266i	&	2022 12 27    18:11:32.401	&	9.73(12)	&	8.53(13)	&	2856(307)	\\
224iii	&		&	9.25(19)	&	7.94(8)	&	2680(321)	&	266ii	&		&		&	10.73(52)	&		\\
224iv	&		&	9.73(24)	&	8.51(10)	&	2829(446)	&	267i	&	2023 01 16    04:11:21.224	&	10.09(26)	&	9.04(9)	&	3144(623)	\\
224v	&		&	10.04(26)	&	8.99(12)	&	3132(633)	&	267ii	&		&		&	10.34(26)	&		\\
224vi	&		&	10.36(33)	&	9.35(14)	&	3223(879)	&	269i	&	2023 02 22    17:52:04.312	&	9.97(19)	&	8.70(13)	&	2745(381)	\\
224vii	&		&		&	9.70(19)	&		&	269ii	&		&		&	9.88(16)	&		\\
224viii	&		&		&	9.89(18)	&		&	269iii	&		&		&	10.59(23)	&	2991(540)	\\
224ix	&		&		&	10.23(21)	&		&	270i	&	2023 02 24    17:25:21.560	&		&	10.00(13)	&		\\
224x	&		&		&	10.38(24)	&		&	270ii	&		&		&	11.30(34)	&		\\
224xi	&		&		&	10.45(28)	&		&	272i	&	2023 03 26    20:25:28.546	&	9.88(15)	&	8.79(5)	&	3054(428)	\\
224xii	&		&		&	10.72(36)	&		&	272ii	&		&		&	10.72(23)	&		\\
224xiii	&		&		&	11.61(80)	&		&	275i	&	2023 04 23    18:04:39.828	&	10.60(18)	&	9.61(9)	&	3260(600)	\\
226i	&	2022 10 20    02:05:55.914	&		&	9.14(9)	&		&	275ii	&		&		&	11.99(54)	&		\\
226ii	&		&		&	10.64(24)	&		&	279i	&	2023 04 24    20:02:42.098	&	9.6(2)	&	8.28(6)	&	2630(400)	\\
227i	&	2022 10 20    02:56:37.878	&	9.08(10)	&	8.59(6)	&	5084(755)	&	279ii	&		&		&	10.43(32)	&		\\
227ii	&		&		&	10.48(20)	&		&	280i	&	2023 05 23    20:06:14.792	&	8.94(9)	&	8.16(3)	&	3860(681)	\\
229i	&	2022 10 20    04:11:08.419	&		&	8.90(8)	&		&	280ii	&		&		&	9.62(15)	&		\\
229ii	&		&		&	10.08(16)	&		&	281i	&	2023 05 24    20:11:10.028	&	8.50(11)	&	7.28(8)	&	2824(237)	\\
230i	&	2022 10 21    02:55:11.697	&	10.22(24)	&	9.56(10)	&	4254(1213)	&	281ii	&		&	8.32(11)	&	6.41(8)	&	2003(114)	\\
230ii	&		&		&	12.44(98)	&		&	281iii	&		&	10.05(18)	&	8.11(8)	&	1980(167)	\\
234i	&	2022 10 22    03:55:35.517	&	9.51(14)	&	8.68(7)	&	3677(488)	&	281iv	&		&	10.29(25)	&	8.96(9)	&	2511(365)	\\
234ii	&		&		&	10.69(26)	&		&	281v	&		&		&	9.69(12)	&		\\
235i	&	2022 10 22    03:56:50.322	&	10.31(25)	&	9.51(11)	&	3778(935)	&	281vi	&		&		&	10.10(17)	&		\\
235ii	&		&		&	11.04(32)	&		&	281vii	&		&		&	10.40(22)	&		\\
236i	&	2022 10 29    17:03:58.478	&	9.25(19)	&	7.72(13)	&	2382(279)	&	281viii	&		&		&	10.76(30)	&		\\
236ii	&		&		&	8.72(14)	&		&	281ix	&		&		&	11.13(35)	&		\\
236iii	&		&		&	9.60(19)	&		&	281x	&		&		&	11.40(44)	&		\\
237i	&	2022 10 29    17:13:31.322	&		&	7.68(13)	&		&	282i	&	2023 05 24    21:03:20.168	&	9.10(18)	&	7.53(13)	&	2335(262)	\\
237ii	&		&		&	9.96(24)	&		&	282ii	&		&	10.58(44)	&	8.78(14)	&	2102(427)	\\
240i	&	2022 10 30    16:47:27.612	&	9.68(22)	&	8.56(9)	&	3005(472)	&	282iii	&		&		&	9.86(20)	&		\\
240ii	&		&		&	10.17(28)	&		&	282iv	&		&		&	10.79(32)	&		\\
241i	&	2022 10 30    16:54:54.929	&	9.65(19)	&	8.57(9)	&	3080(453)	&	283i	&	2023 05 25    20:34:27.999	&	7.51(10)	&	6.28(7)	&	2811(200)	\\
241ii	&		&		&	9.87(18)	&		&	283ii	&		&	10.01(23)	&	8.86(9)	&	2935(465)	\\
242i	&	2022 10 30    17:13:28.165	&		&	8.98(14)	&		&	283iii	&		&	10.83(46	&	9.44(13)	&	2569(694)	\\
242ii	&		&		&	10.11(29)	&		&	283iv	&		&		&	10.30(20)	&		\\
244i	&	2022 10 30    17:34:16.134	&	8.64(14)	&	8.03(11)	&	4475(883)	&	283v	&		&		&	10.64(29)	&		\\
244ii	&		&	9.36(19)	&	8.05(10)	&	2676(334)	&	283vi	&		&		&	11.40(53)	&		\\
244iii	&		&		&	9.01(13)	&		&	283vii	&		&		&	11.71(63)	&		\\
244iv	&		&		&	9.74(19)	&		&	284i	&	2023 05 26    18:17:43.624	&	9.79(37)	&	8.28(8)	&	2376(469)	\\
245i	&	2022 10 30    17:41:51.409	&		&	8.98(13)	&		&	284ii	&		&		&	10.72(64)	&		\\
245ii	&		&		&	10.24(28)	&		&	287i	&	2023 06 22    20:03:20.188	&	10.18(20)	&	8.53(12)	&	2253(254)	\\
251i	&	2022 10 31    19:17:17.423	&	8.60(21)	&	7.51(12)	&	3065(517)	&	287ii	&		&	10.51(28)	&	8.97(13)	&	2372(375)	\\
251ii	&		&		&	9.19(22)	&		&	287iii	&		&		&	10.16(17)	&		\\
252i	&	2022 11 19    02:39:27.785	&	8.82(15)	&	7.92(8)	&	3492(462)	&	289i	&	2023 08 10    01:53:23.822	&	9.67(16)	&	8.50(6)	&	2905(316)	\\
252ii	&		&		&	9.14(15)	&		&	289ii	&		&		&	10.44(28)	&		\\
253i	&	2022 11 19    02:45:43.608	&		&	8.51(10)	&		&	290i	&	2023 08 11    01:43:53.825	&	11.32(41)	&	9.94(14)	&	2581(641)	\\
253ii	&		&		&	9.79(25)	&		&	290ii	&		&		&	12.29(77)	&		\\
257i	&	2022 12 18    03:37:48.471	&	9.11(14)	&	7.73(6)	&	2577(225)	&	291i	&	2023 08 11    02:53:30.790	&		&	9.28(13)	&		\\
257ii	&		&		&	9.98(19)	&		&	291ii	&		&		&	10.81(58)	&		\\
259i	&	2022 12 26    15:46:16.570	&	7.76(11)	&	6.38(8)	&	2579(197)	&	295i	&	2023 08 13    03:05:54.033	&	9.30(13)	&	8.76(8)	&	4820(876)	\\
259ii	&		&	8.62(14)	&	6.67(8)	&	1969(131)	&	295ii	&		&		&	11.02(43)	&		\\
\hline
\end{tabular}}
\end{table*}

\newpage
\begin{figure*}[h]
\begin{tabular}{ccc}
\includegraphics[width=5.6cm]{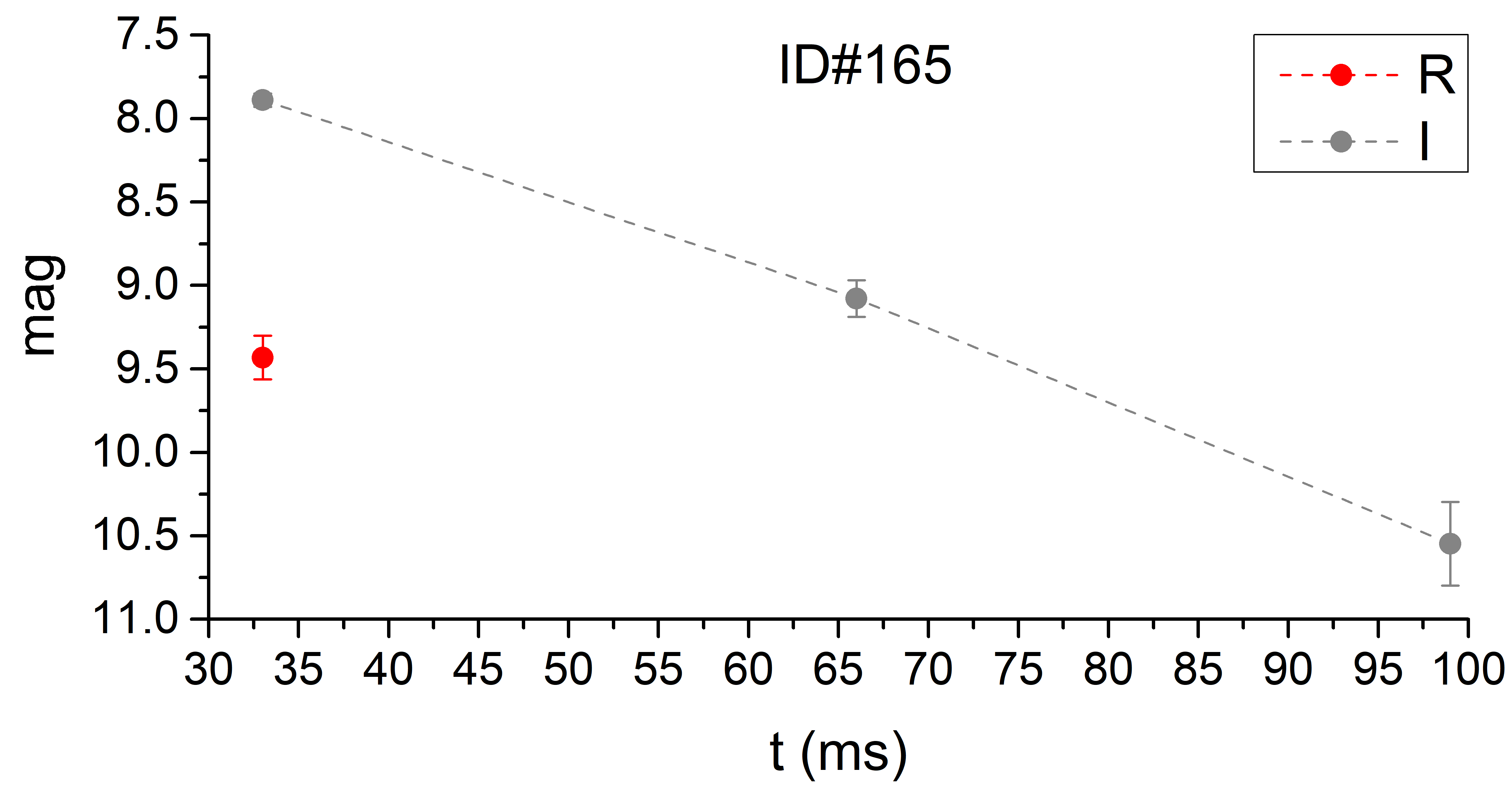}&\includegraphics[width=5.6cm]{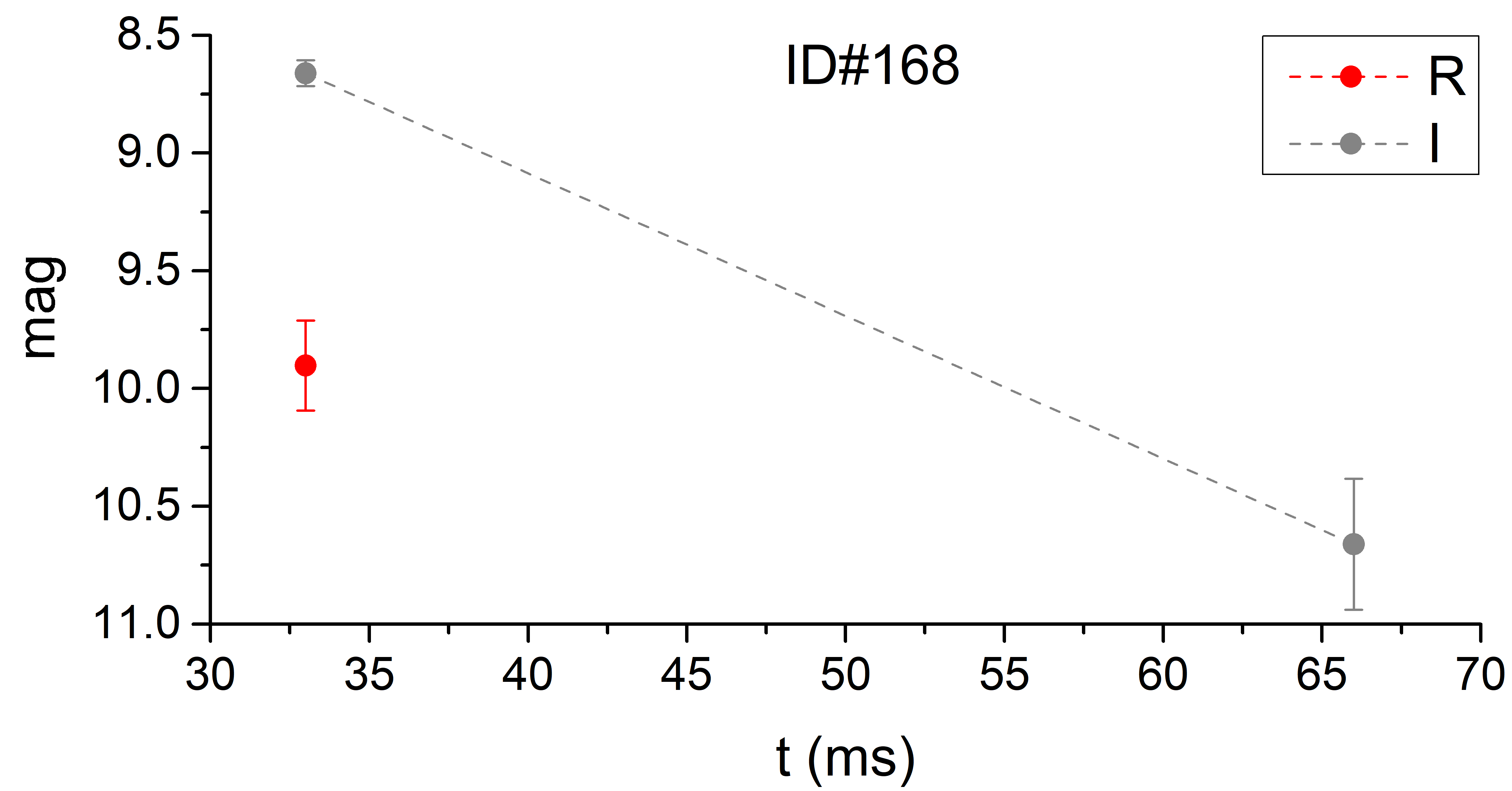}&\includegraphics[width=5.6cm]{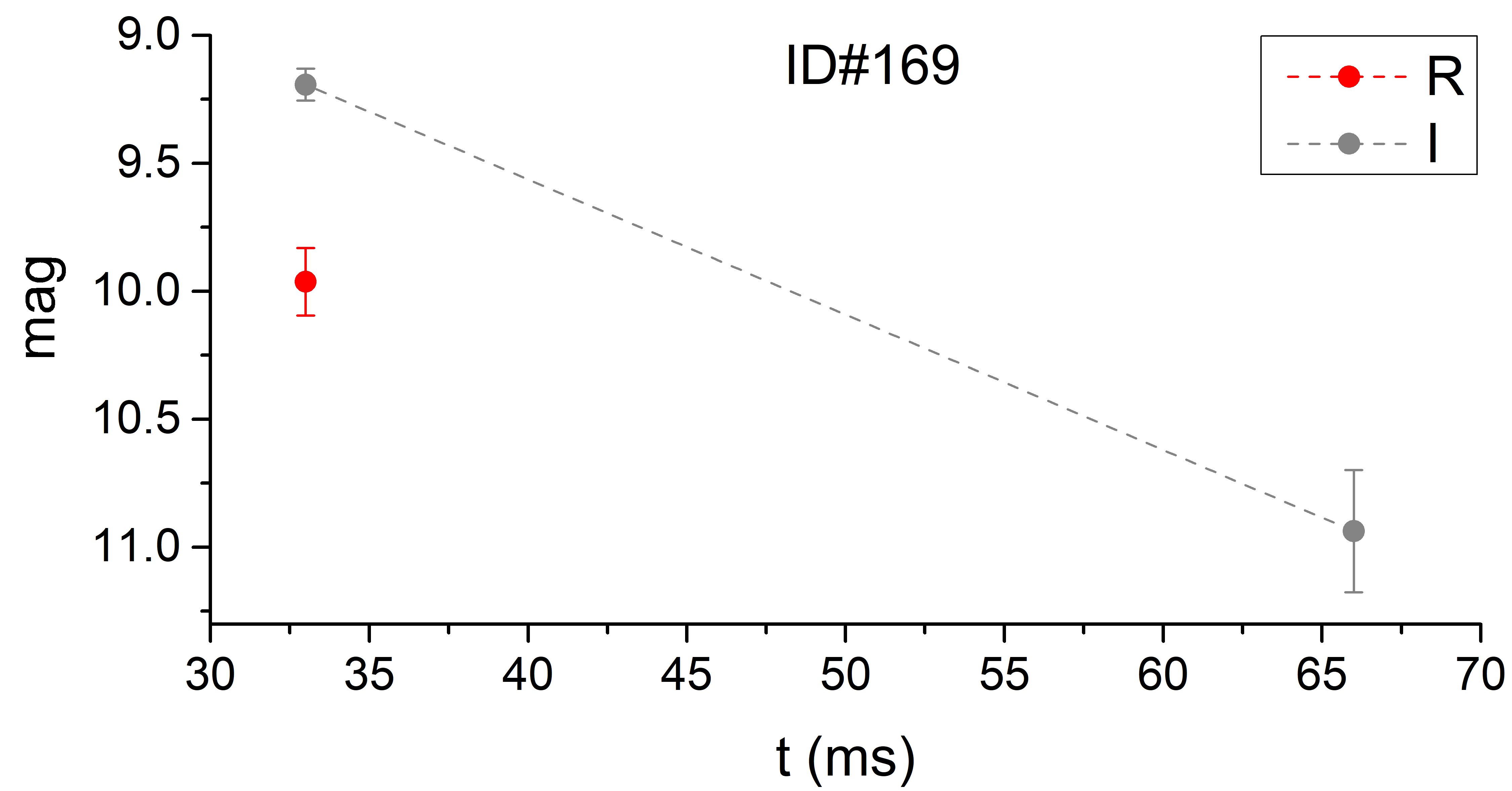}\\
\includegraphics[width=5.6cm]{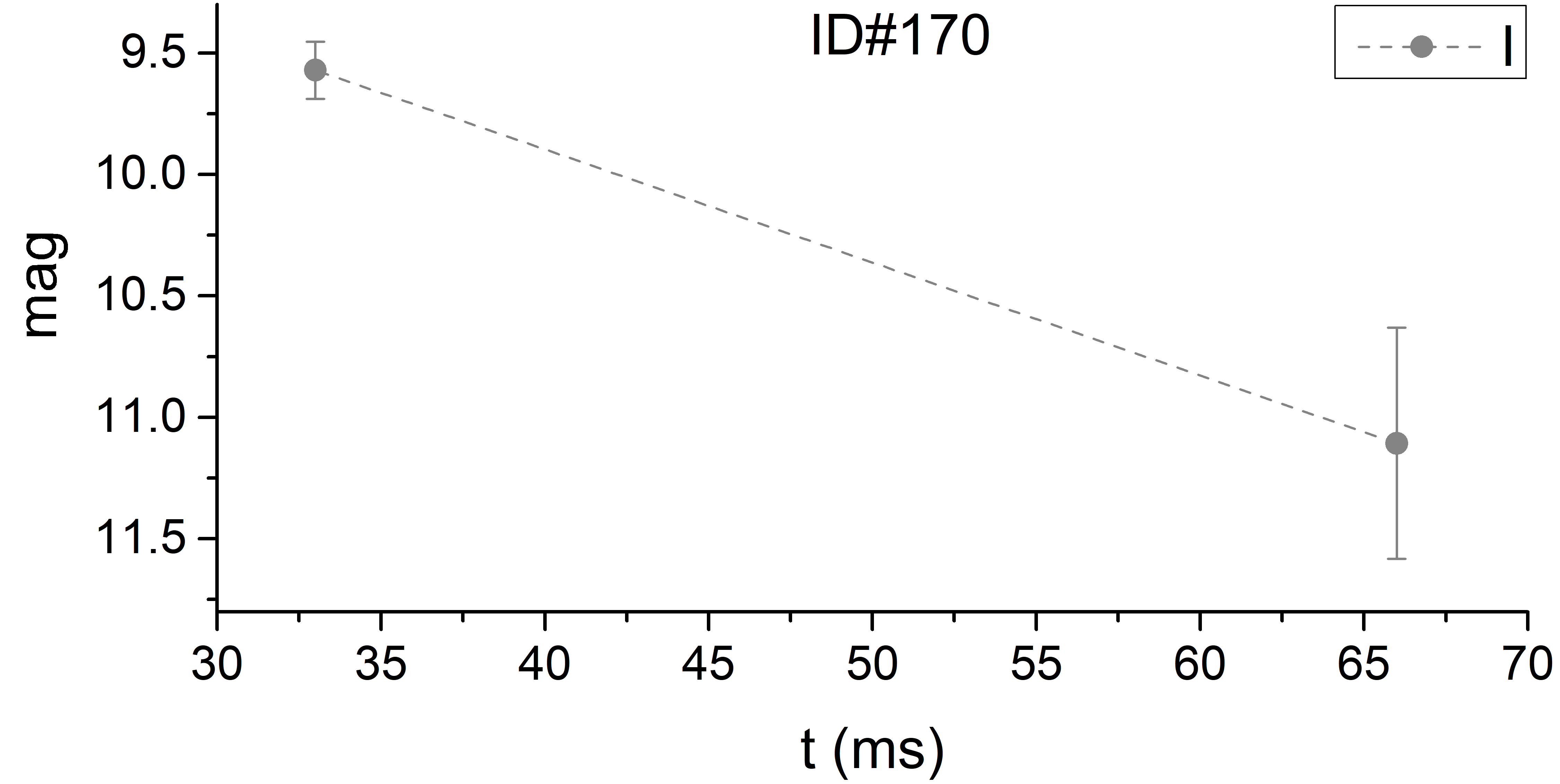}&\includegraphics[width=5.6cm]{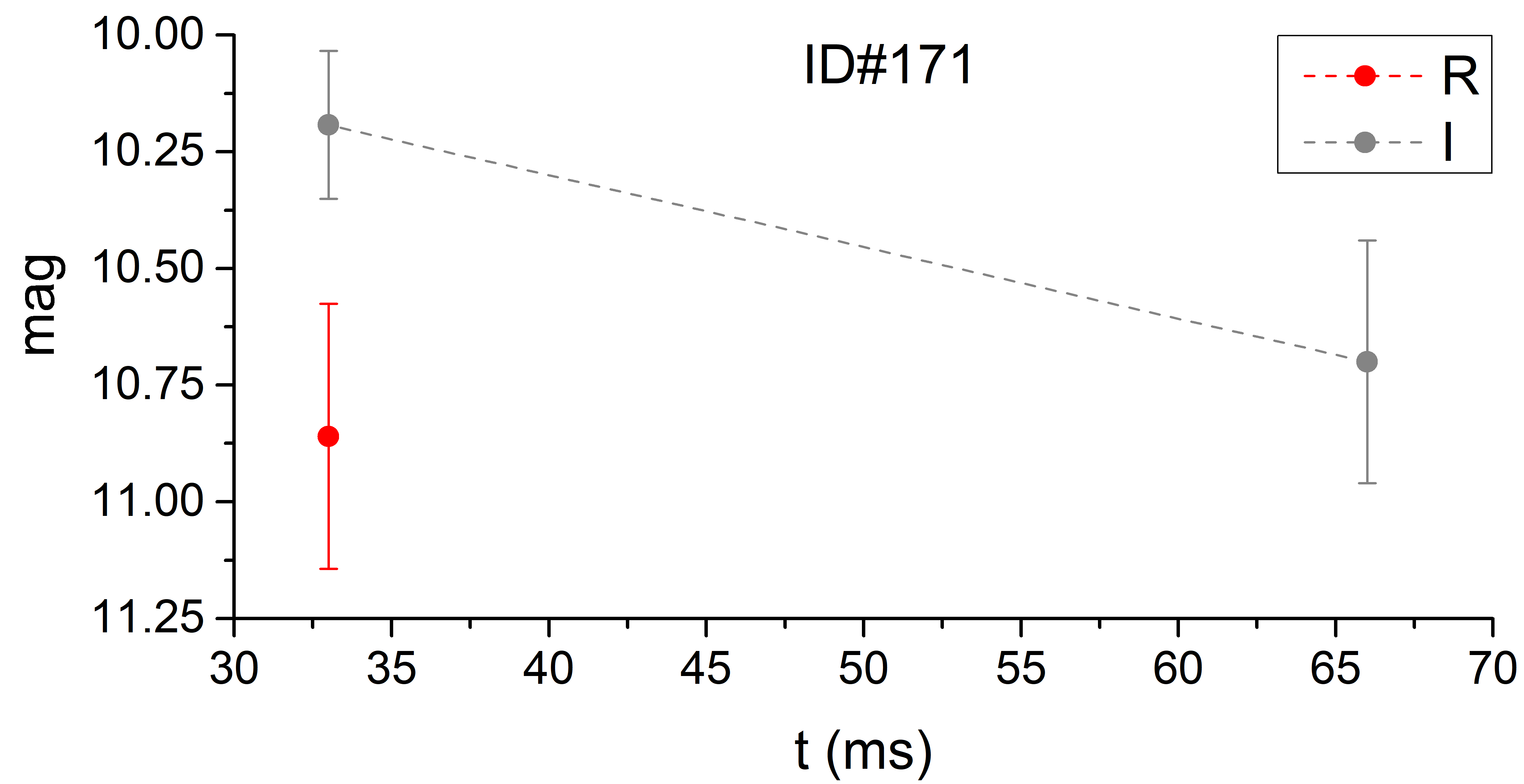}&\includegraphics[width=5.6cm]{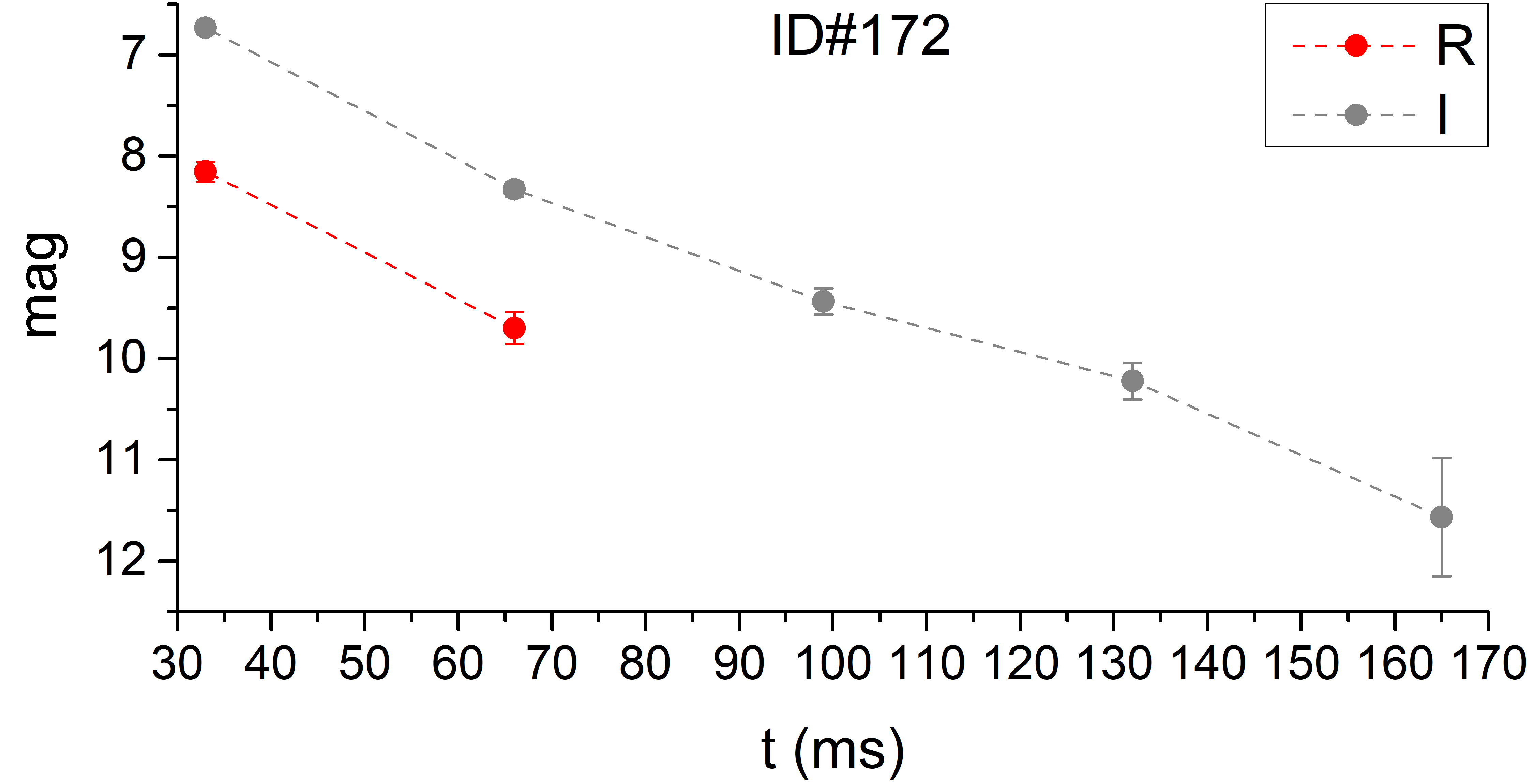}\\
\includegraphics[width=5.6cm]{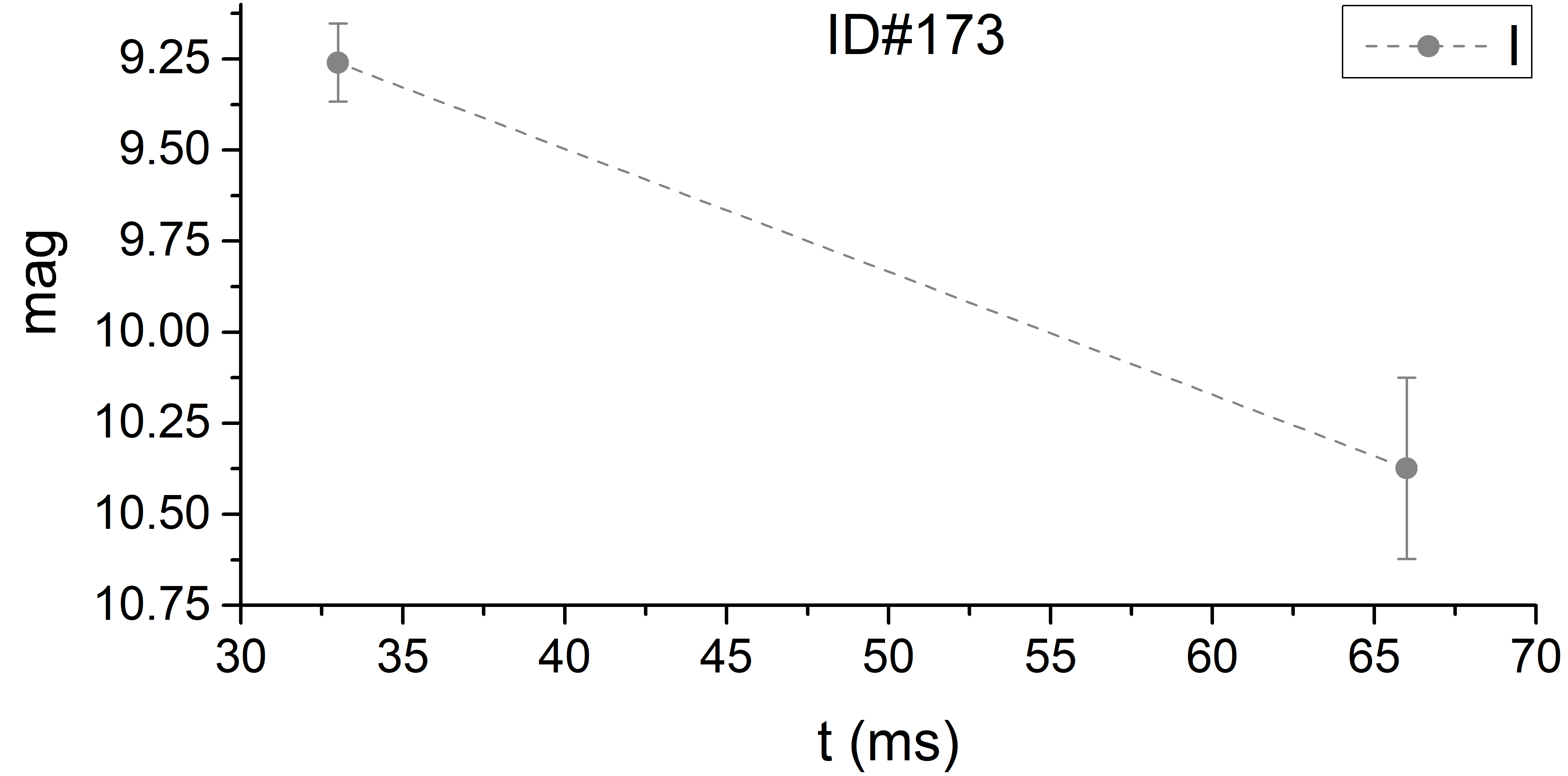}&\includegraphics[width=5.6cm]{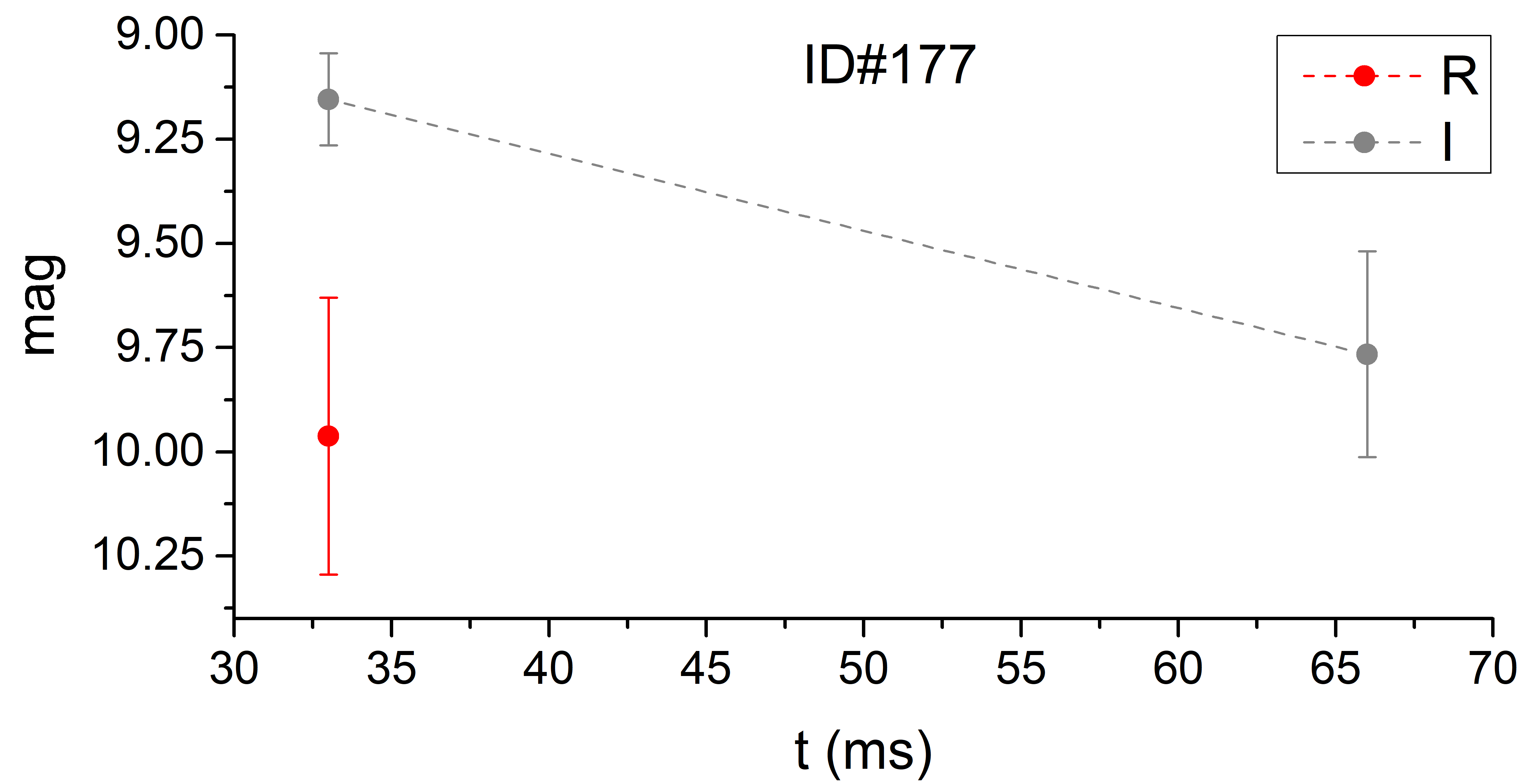}&\includegraphics[width=5.6cm]{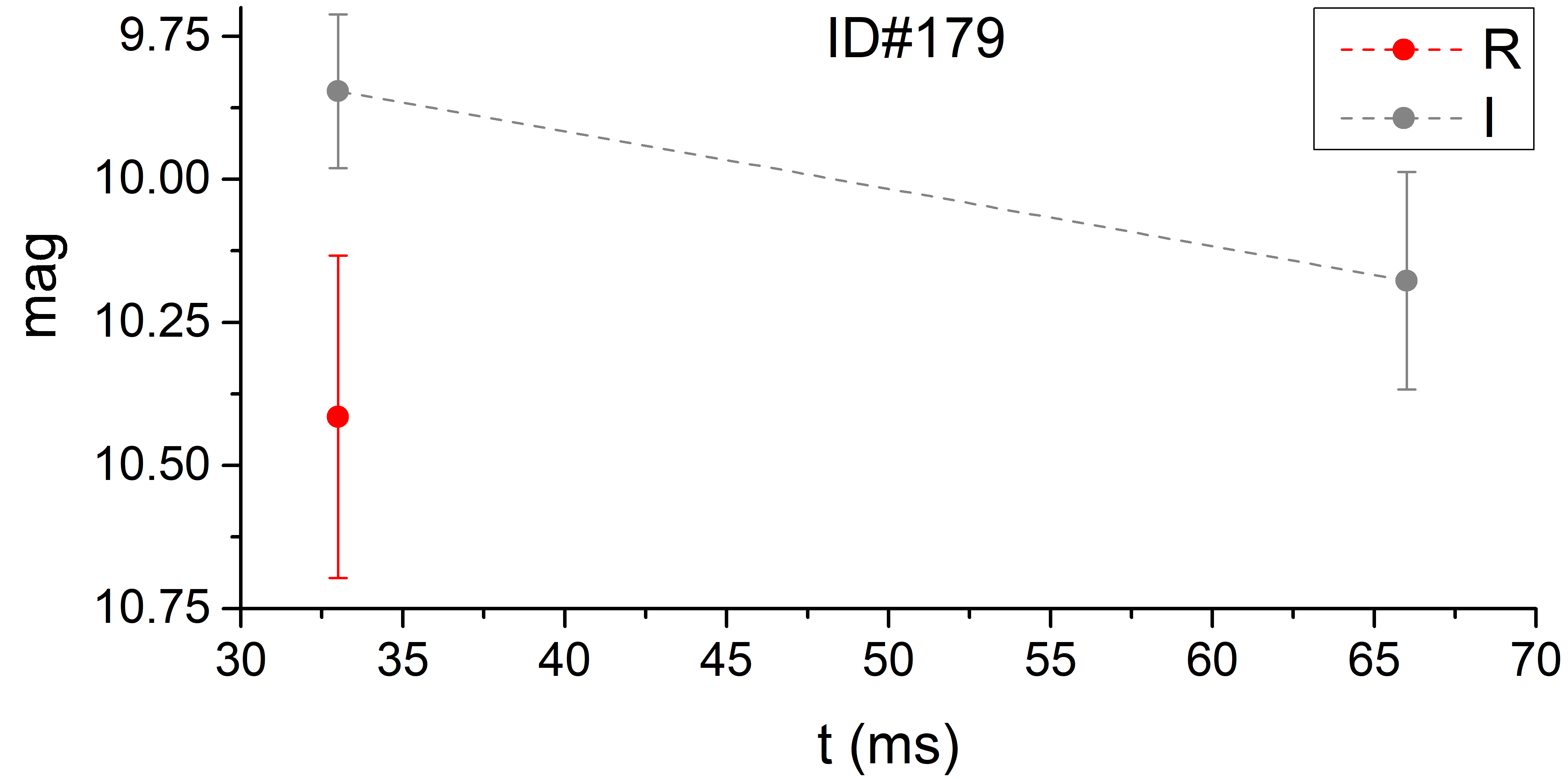}\\
\includegraphics[width=5.6cm]{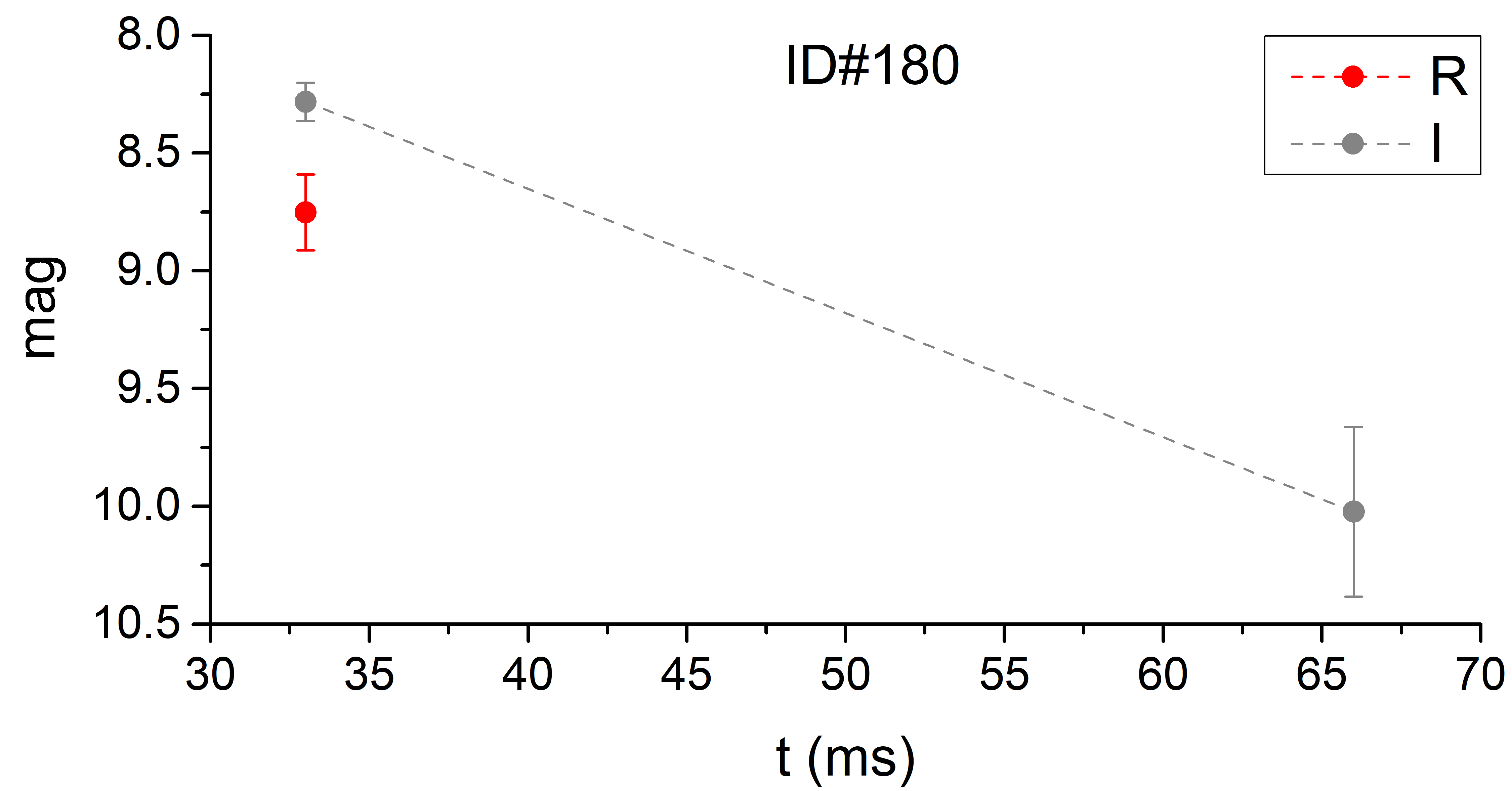}&\includegraphics[width=5.6cm]{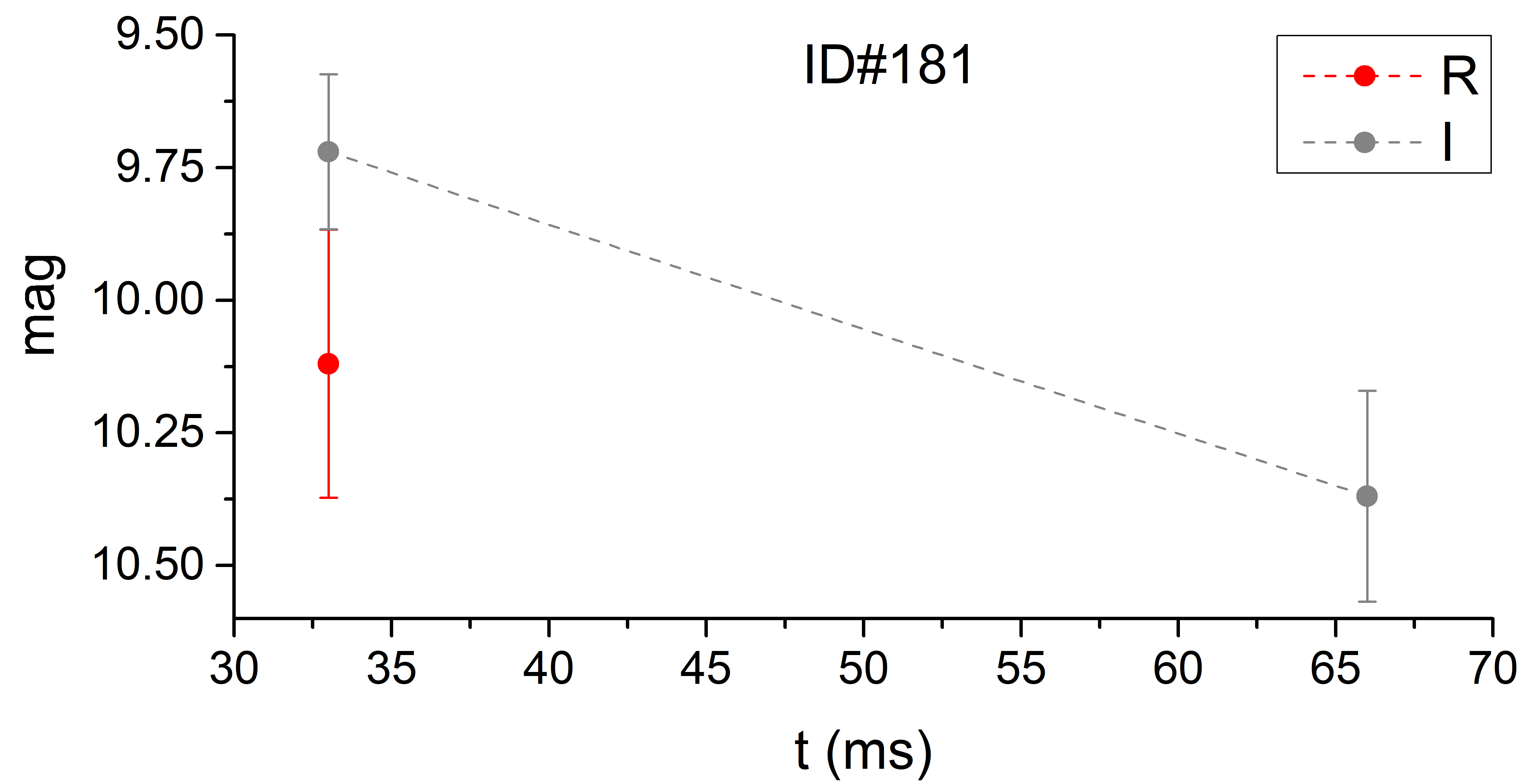}&\includegraphics[width=5.6cm]{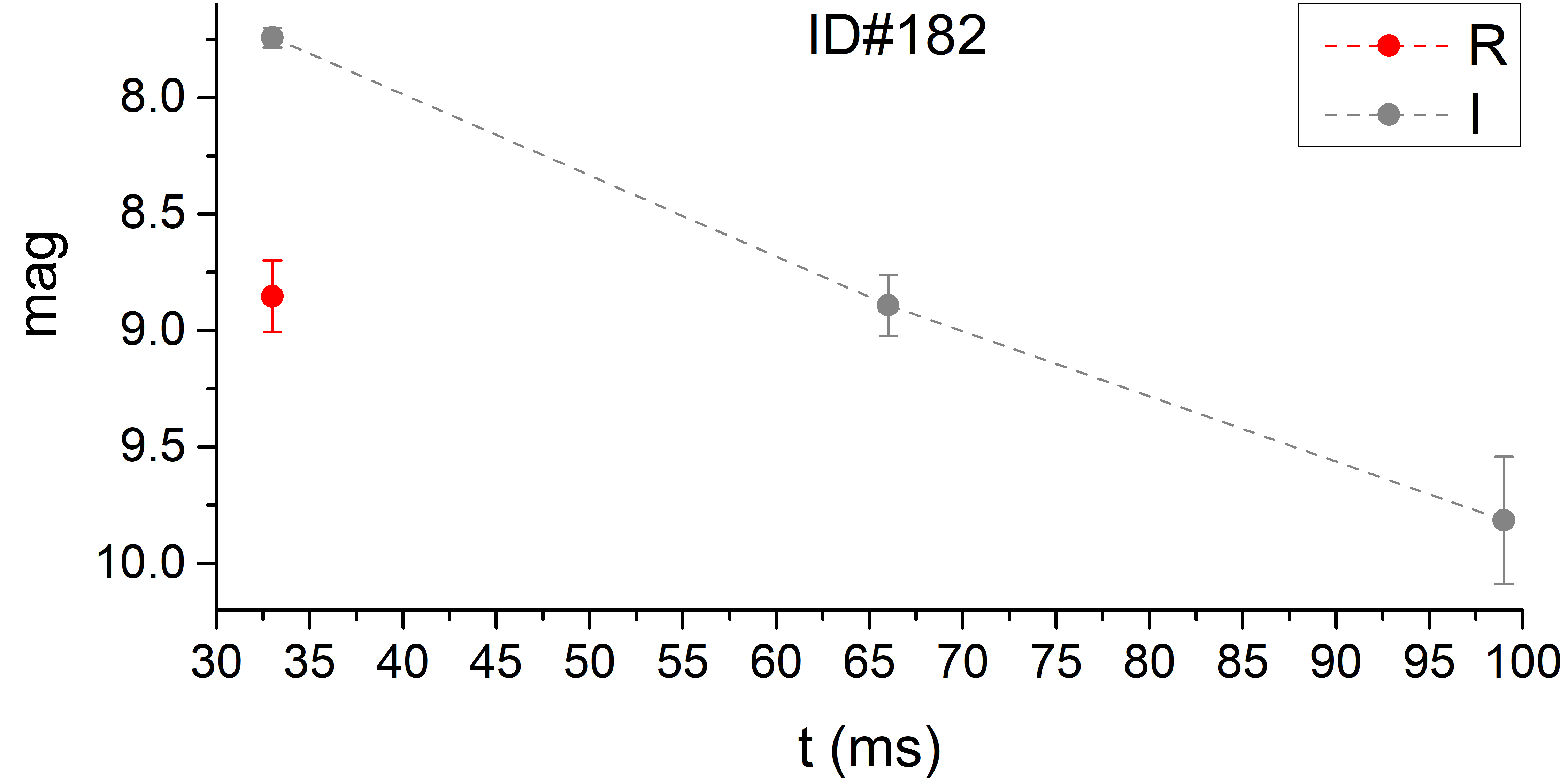}\\
\includegraphics[width=5.6cm]{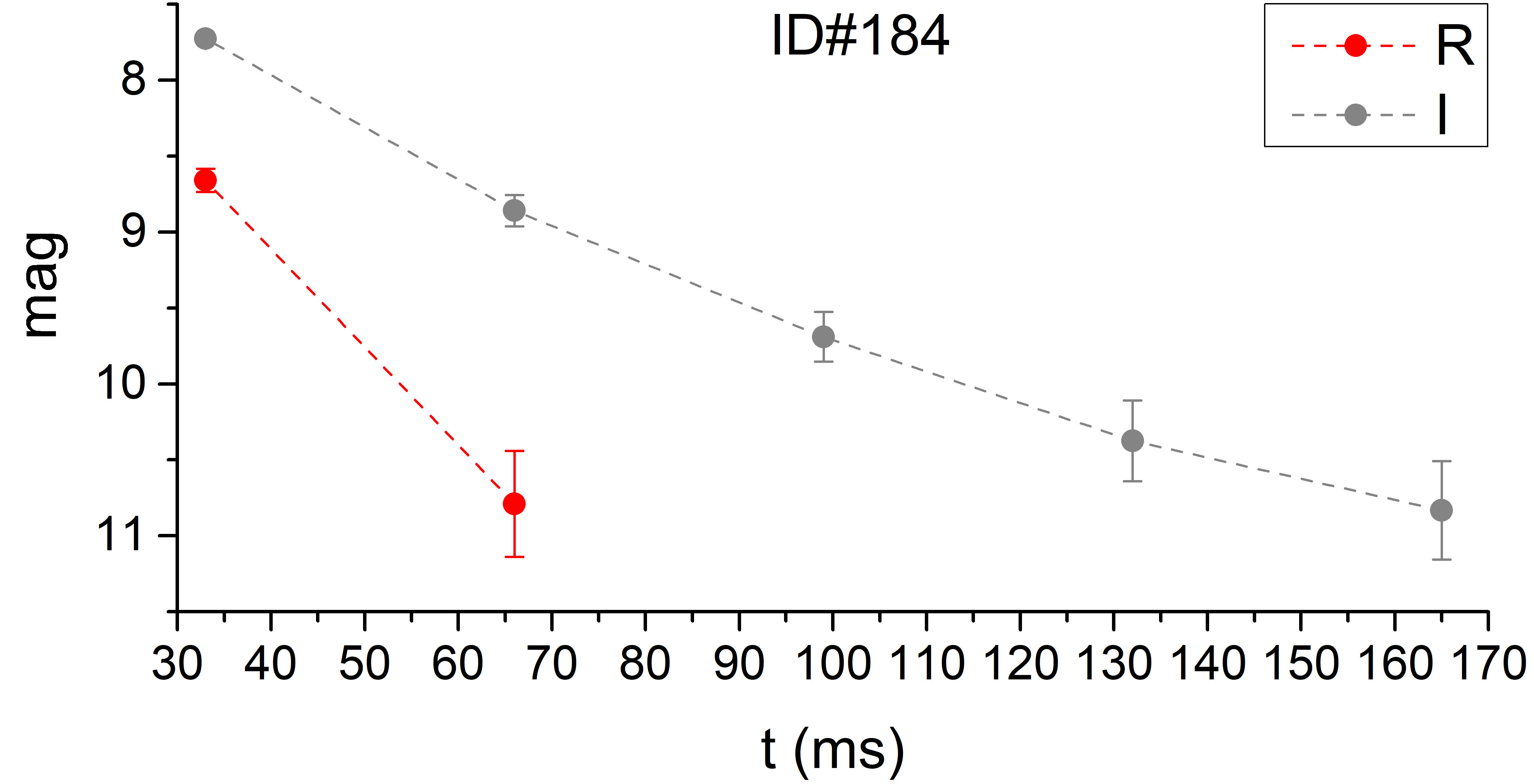}&\includegraphics[width=5.6cm]{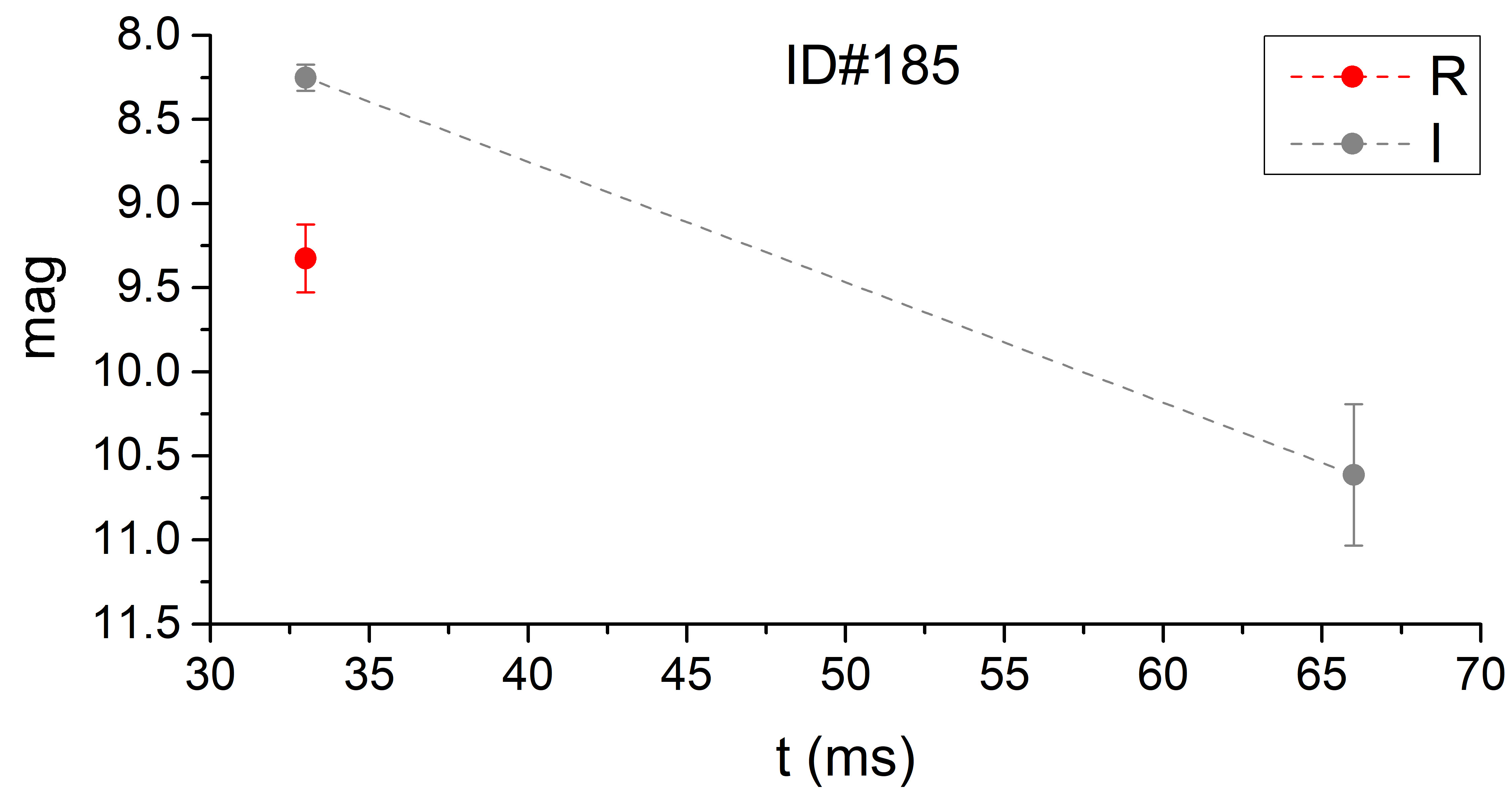}&\includegraphics[width=5.6cm]{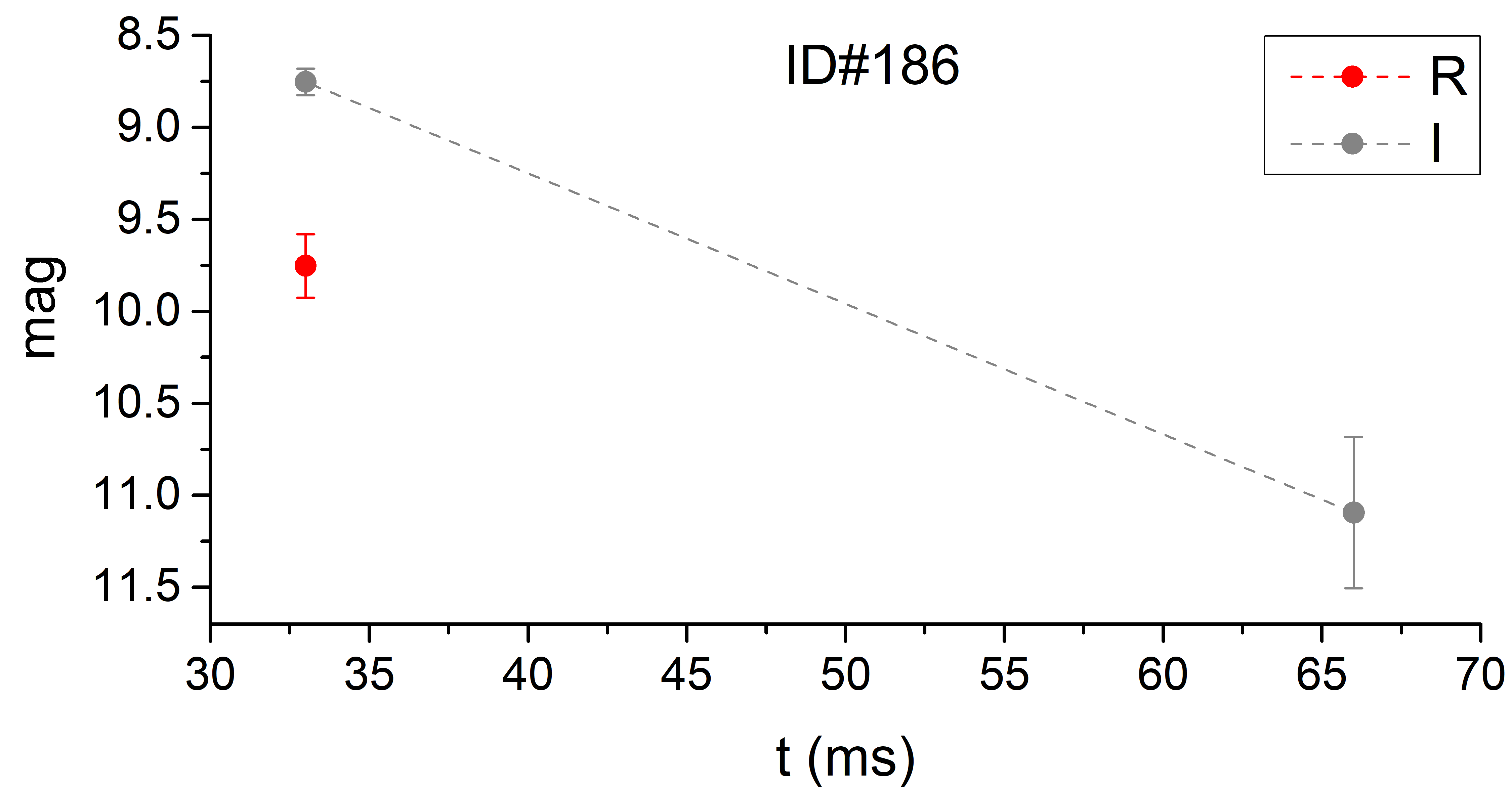}\\
\includegraphics[width=5.6cm]{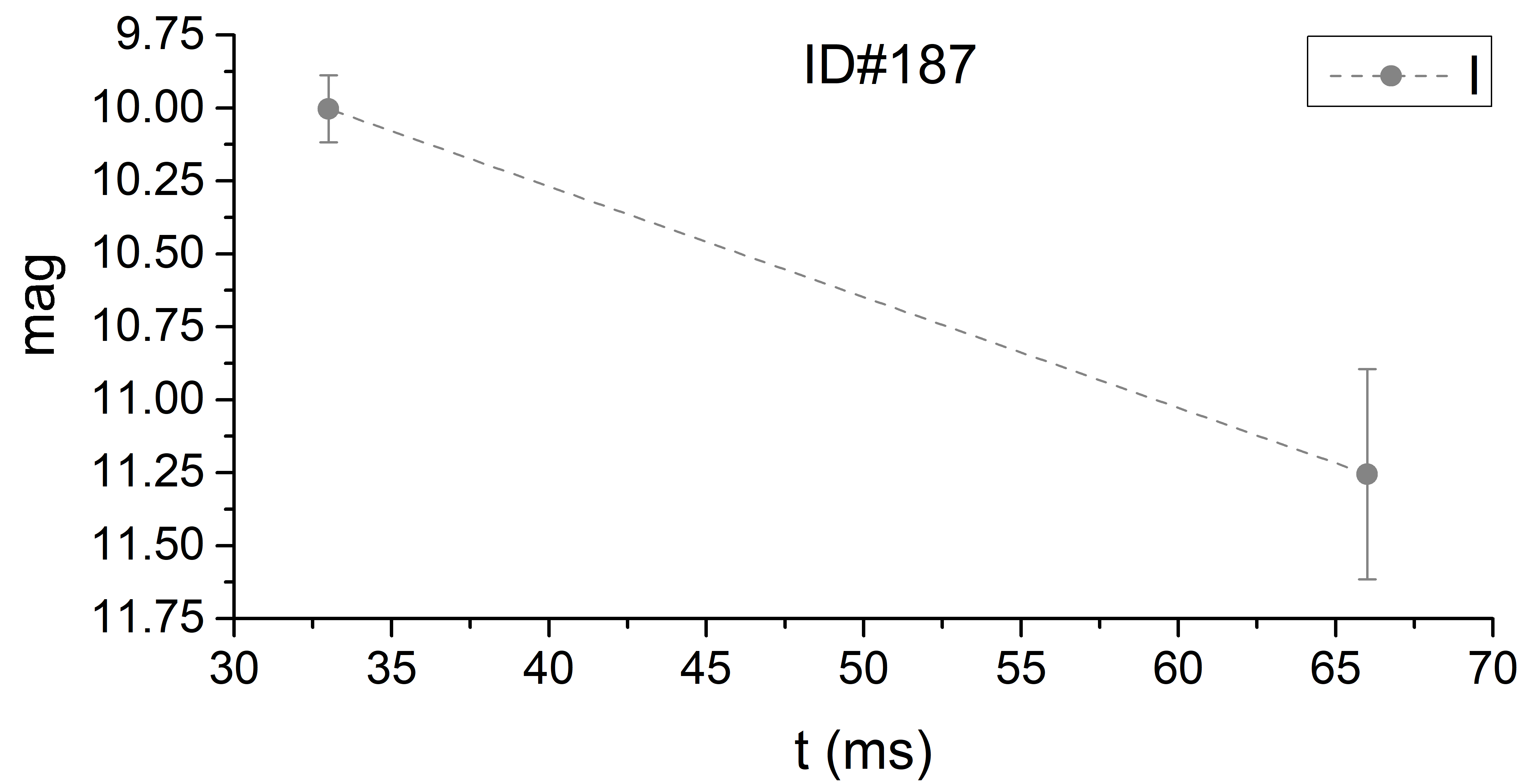}&\includegraphics[width=5.6cm]{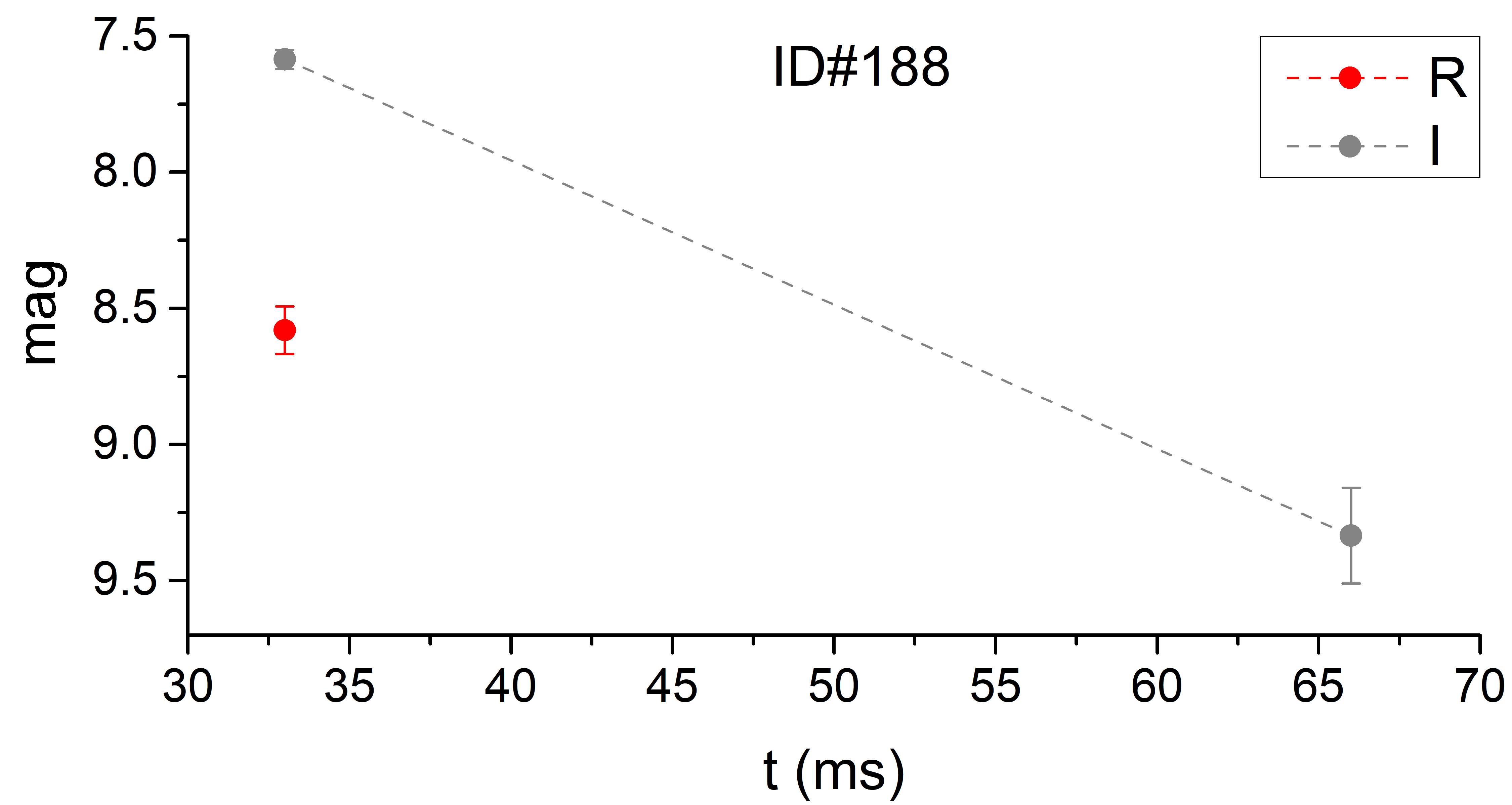}&\includegraphics[width=5.6cm]{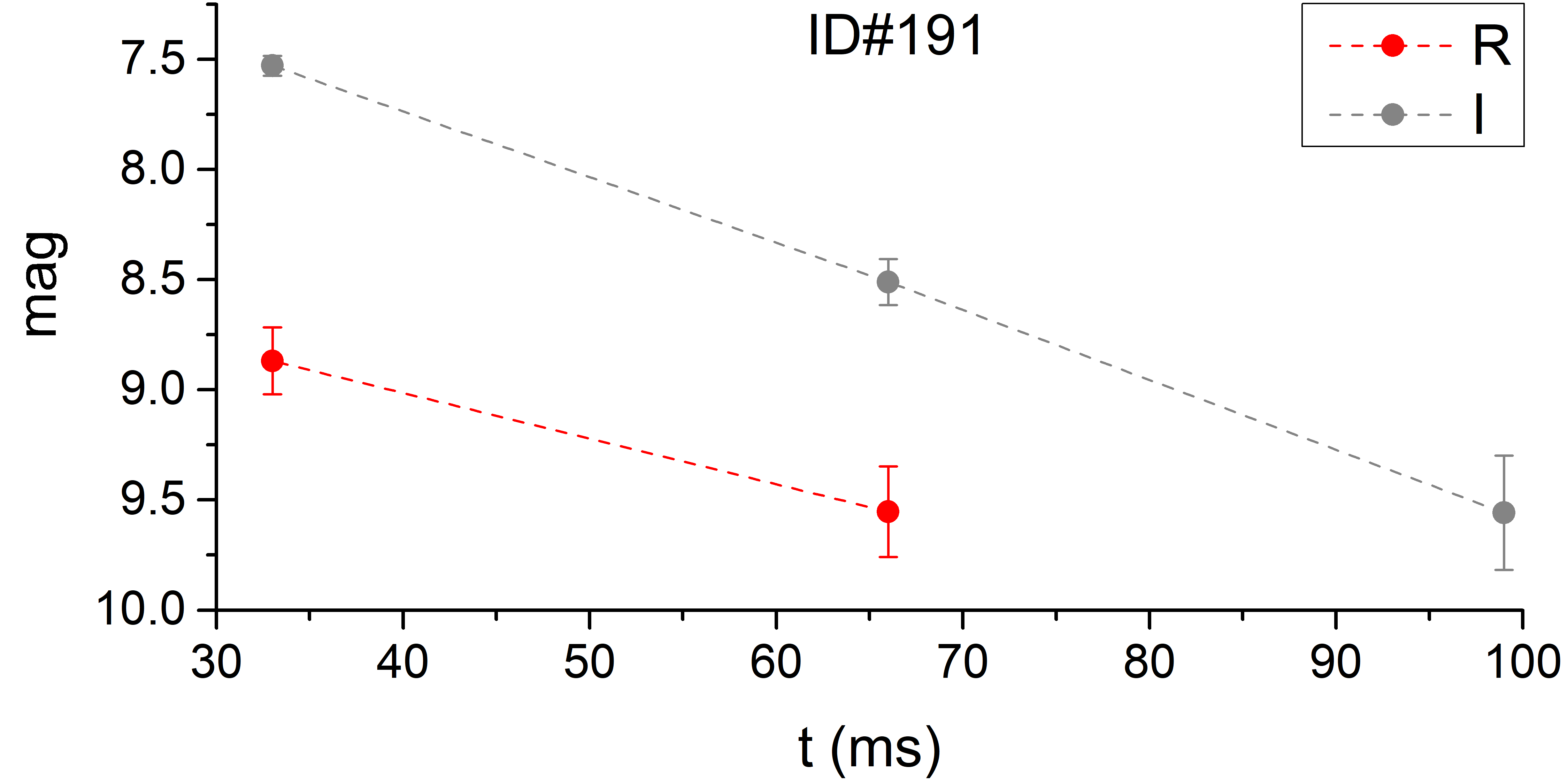}\\
\includegraphics[width=5.6cm]{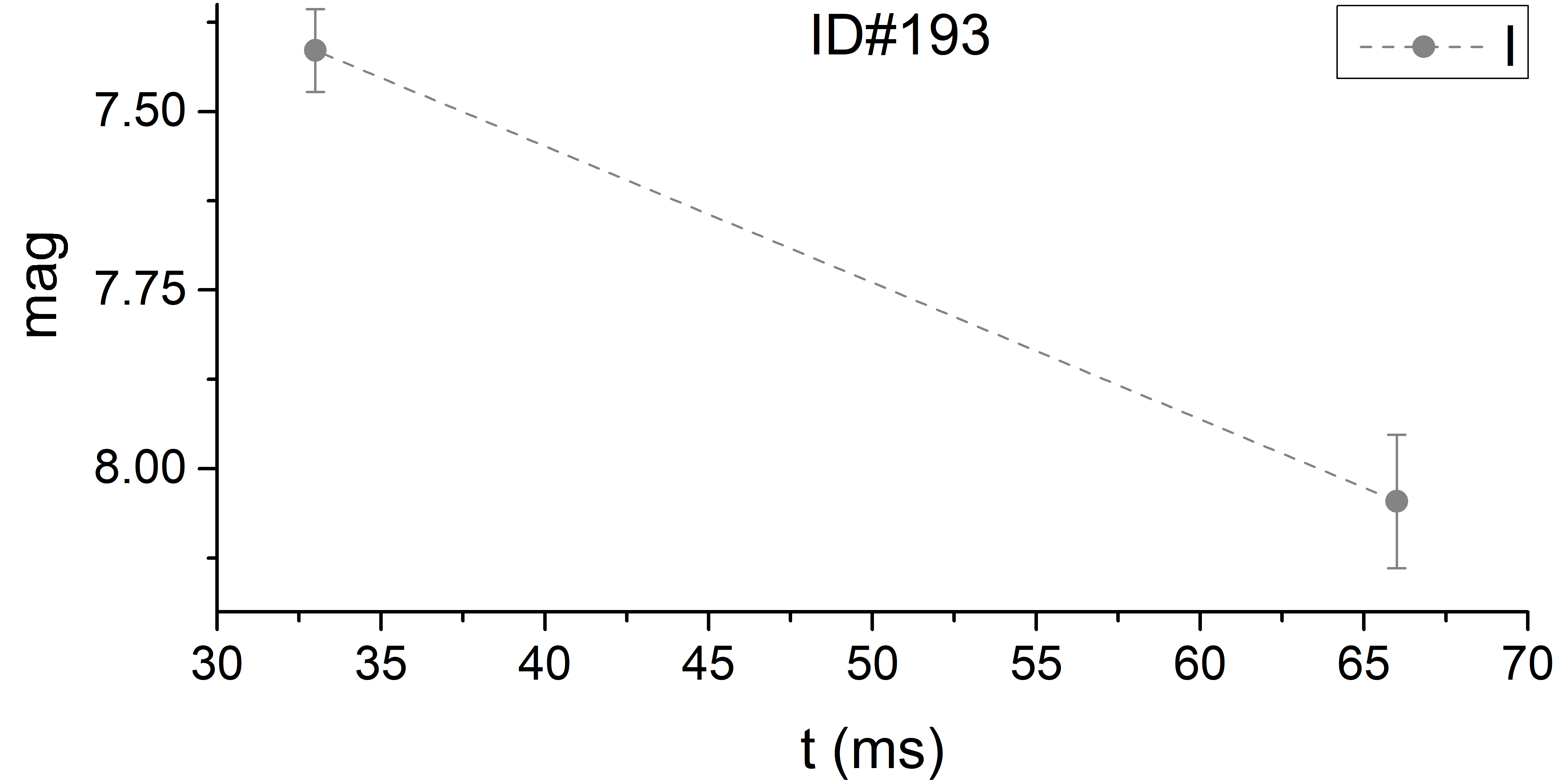}&\includegraphics[width=5.6cm]{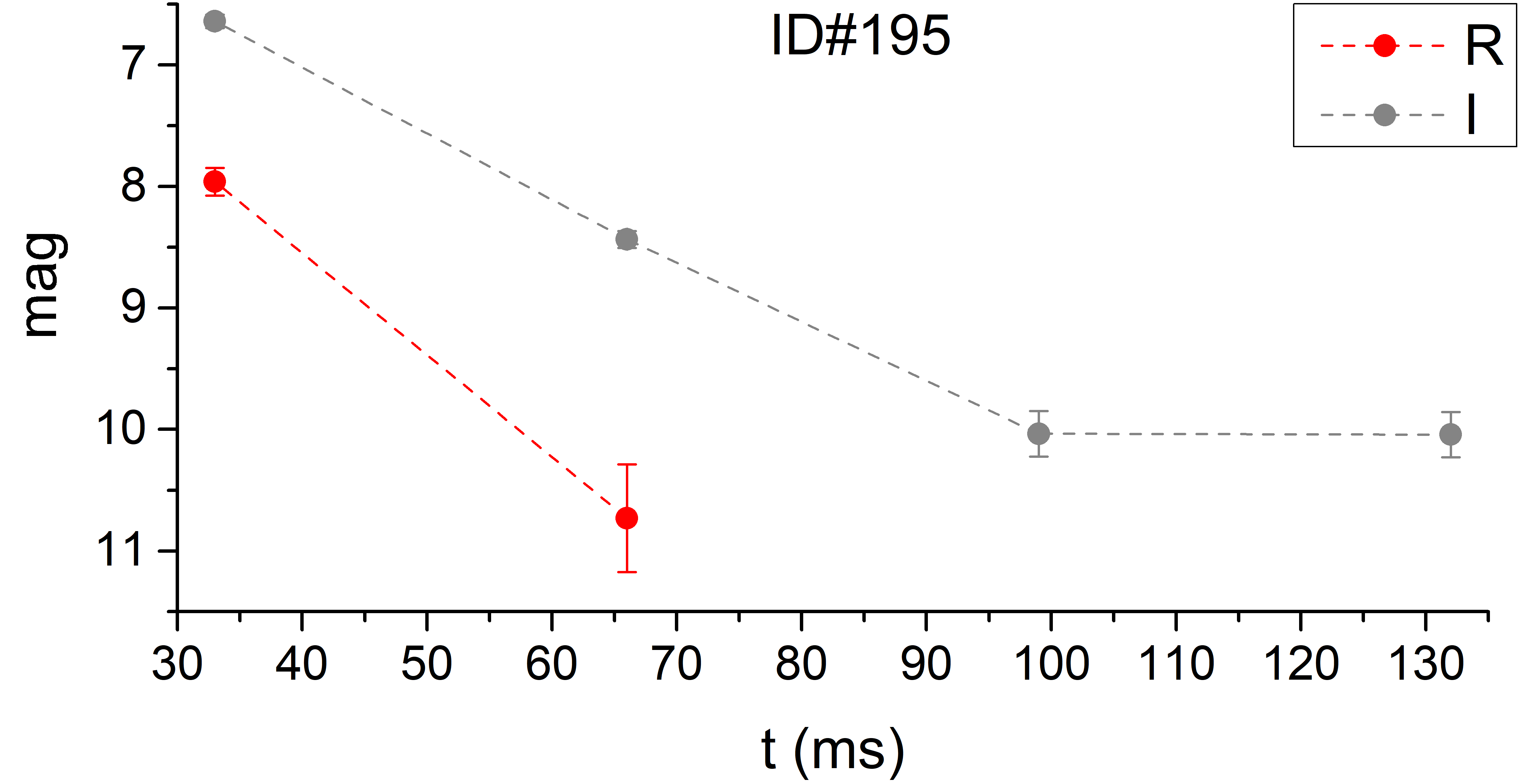}&\includegraphics[width=5.6cm]{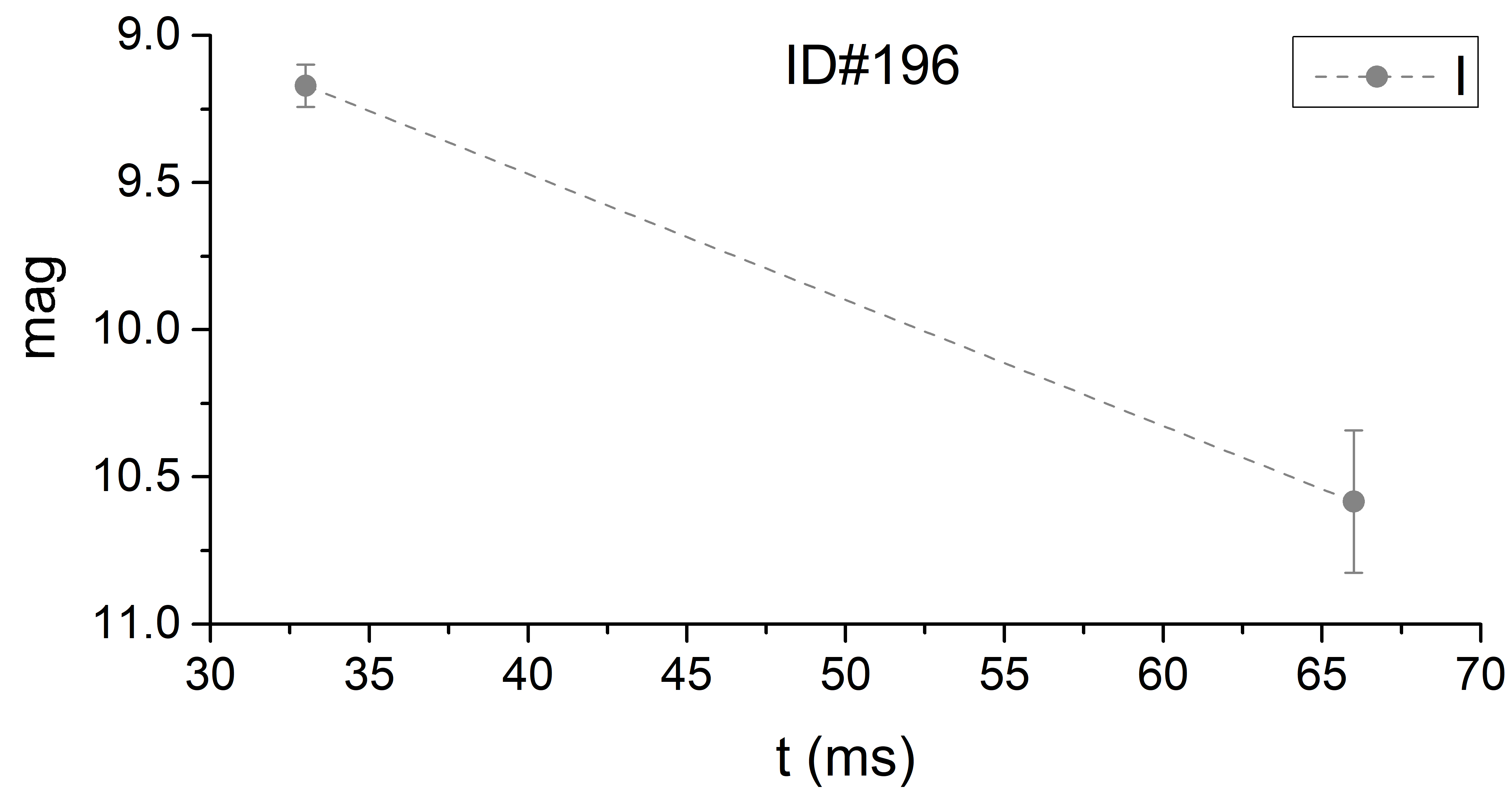}\\
\includegraphics[width=5.6cm]{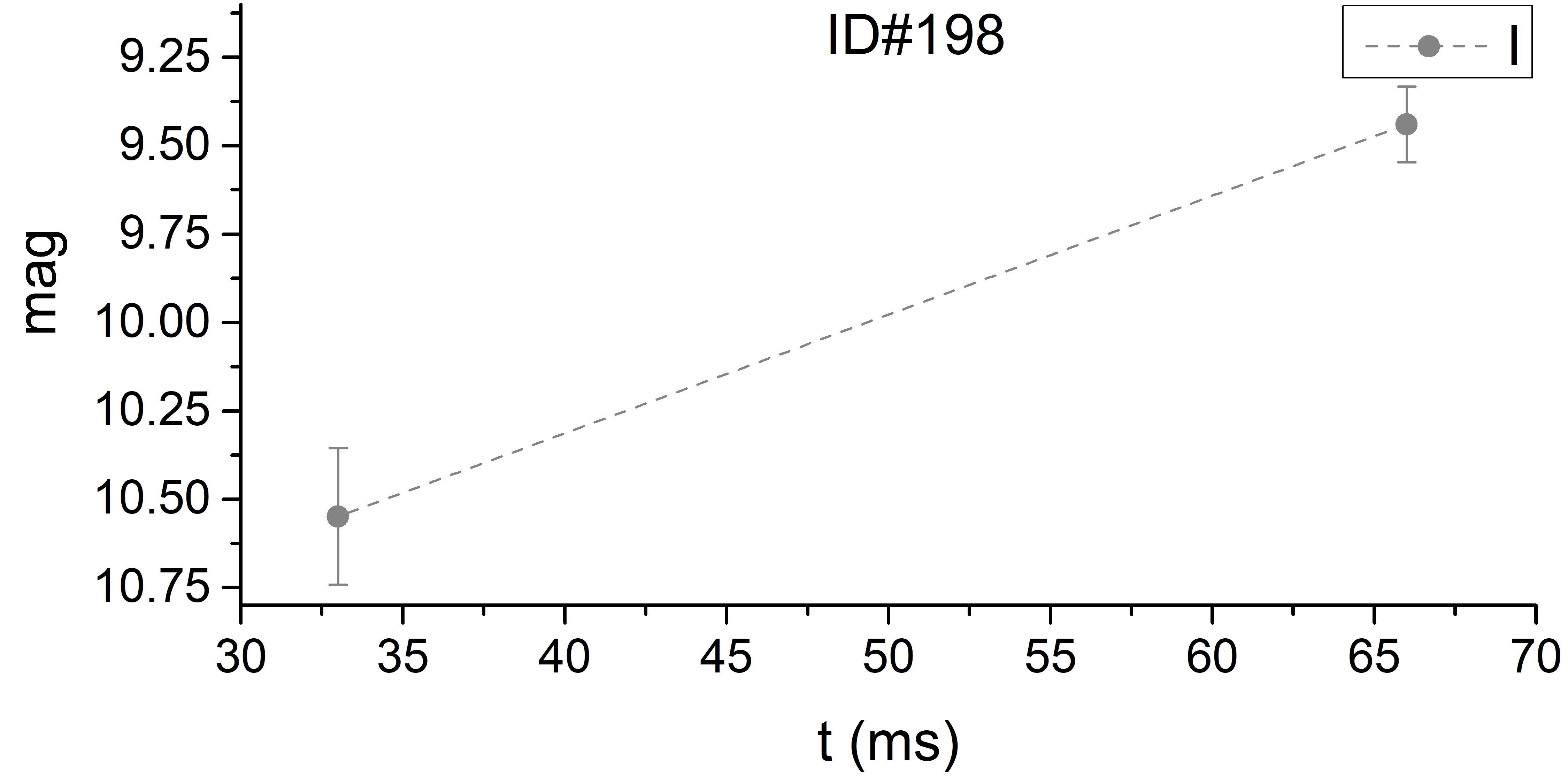}&\includegraphics[width=5.6cm]{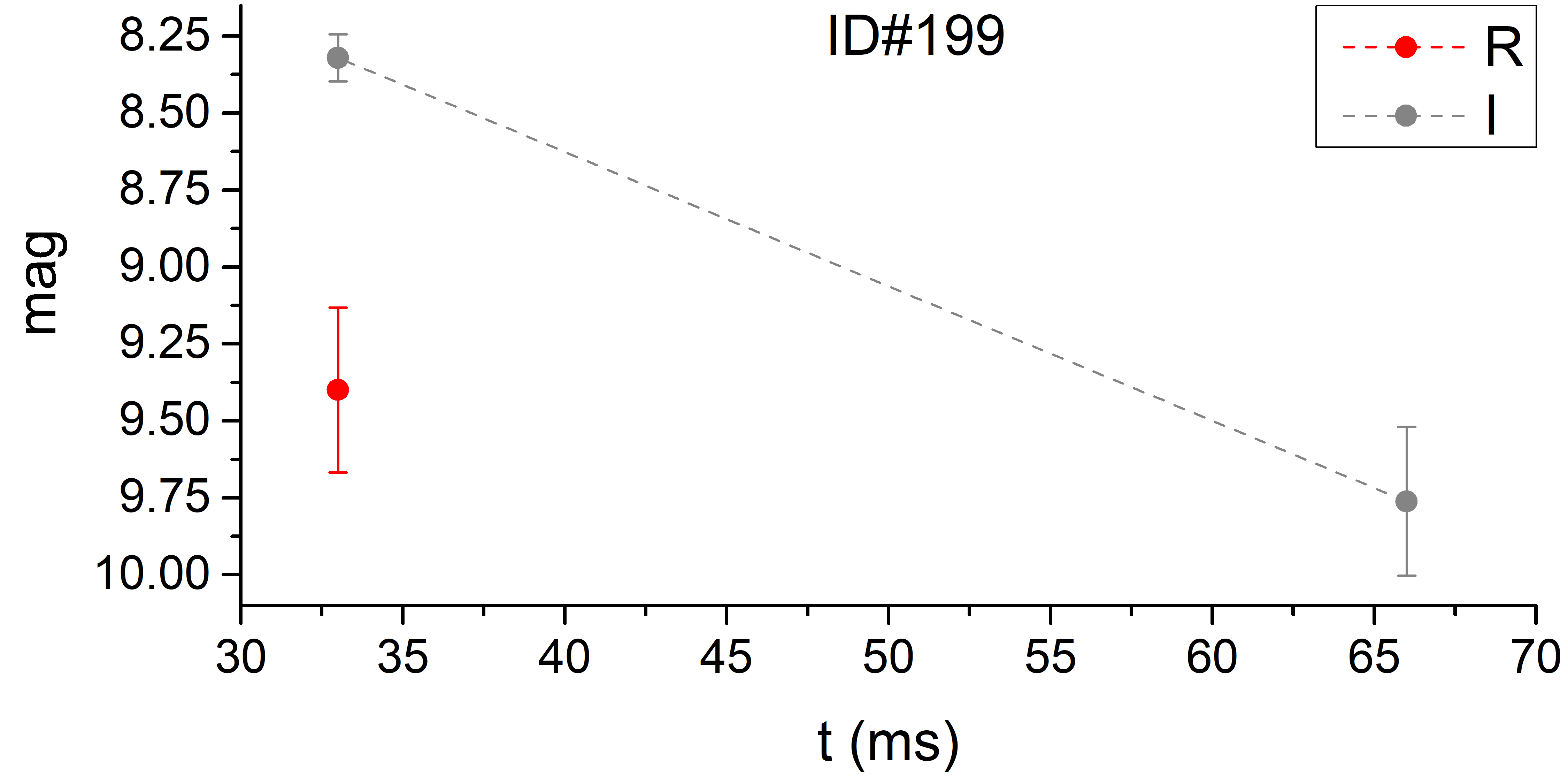}&\includegraphics[width=5.6cm]{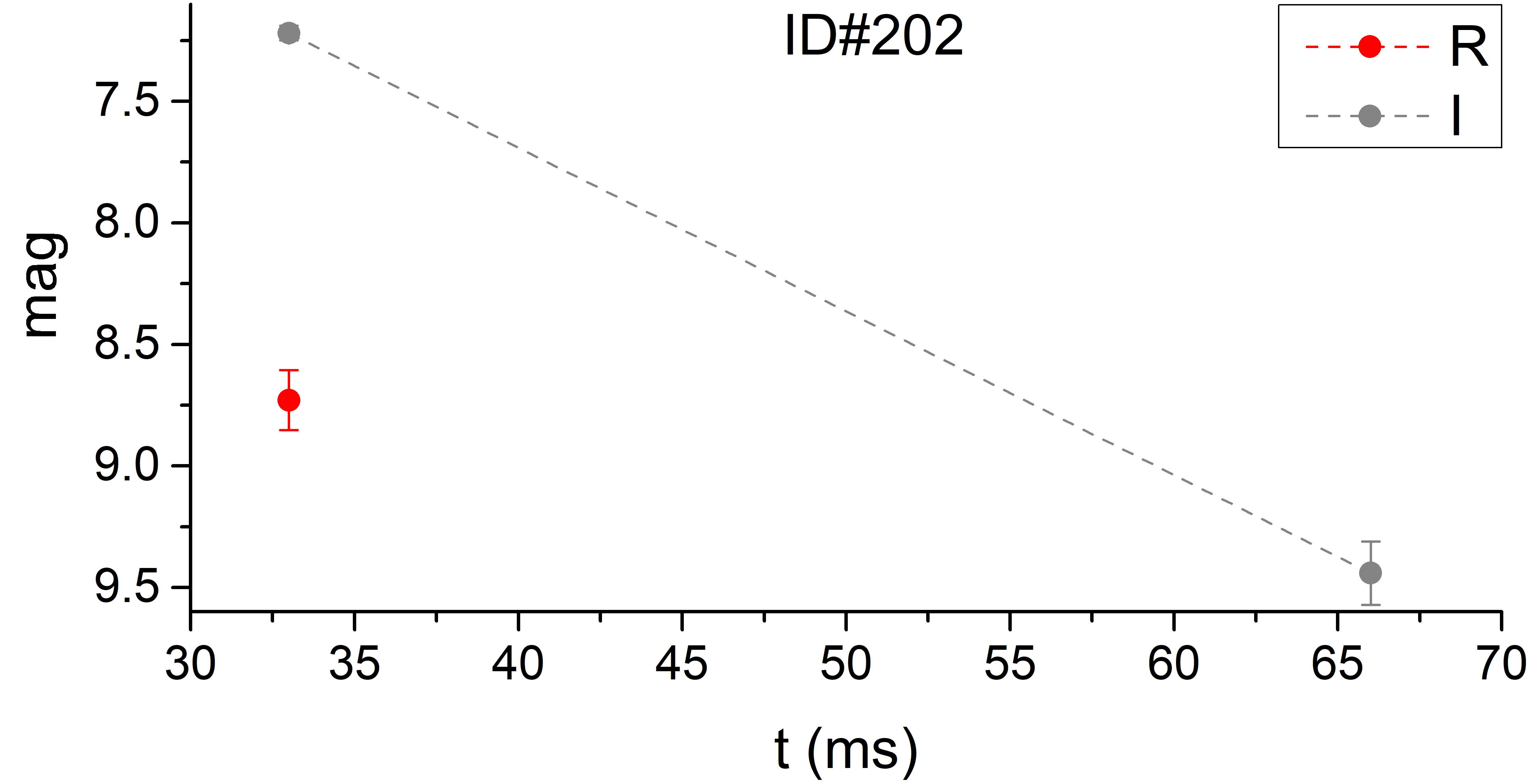}\\
\end{tabular}
\caption*{Fig.~\ref{fig:LCs1}~(cont'd)}
\end{figure*}
\begin{figure*}[h]
\begin{tabular}{ccc}
\includegraphics[width=5.6cm]{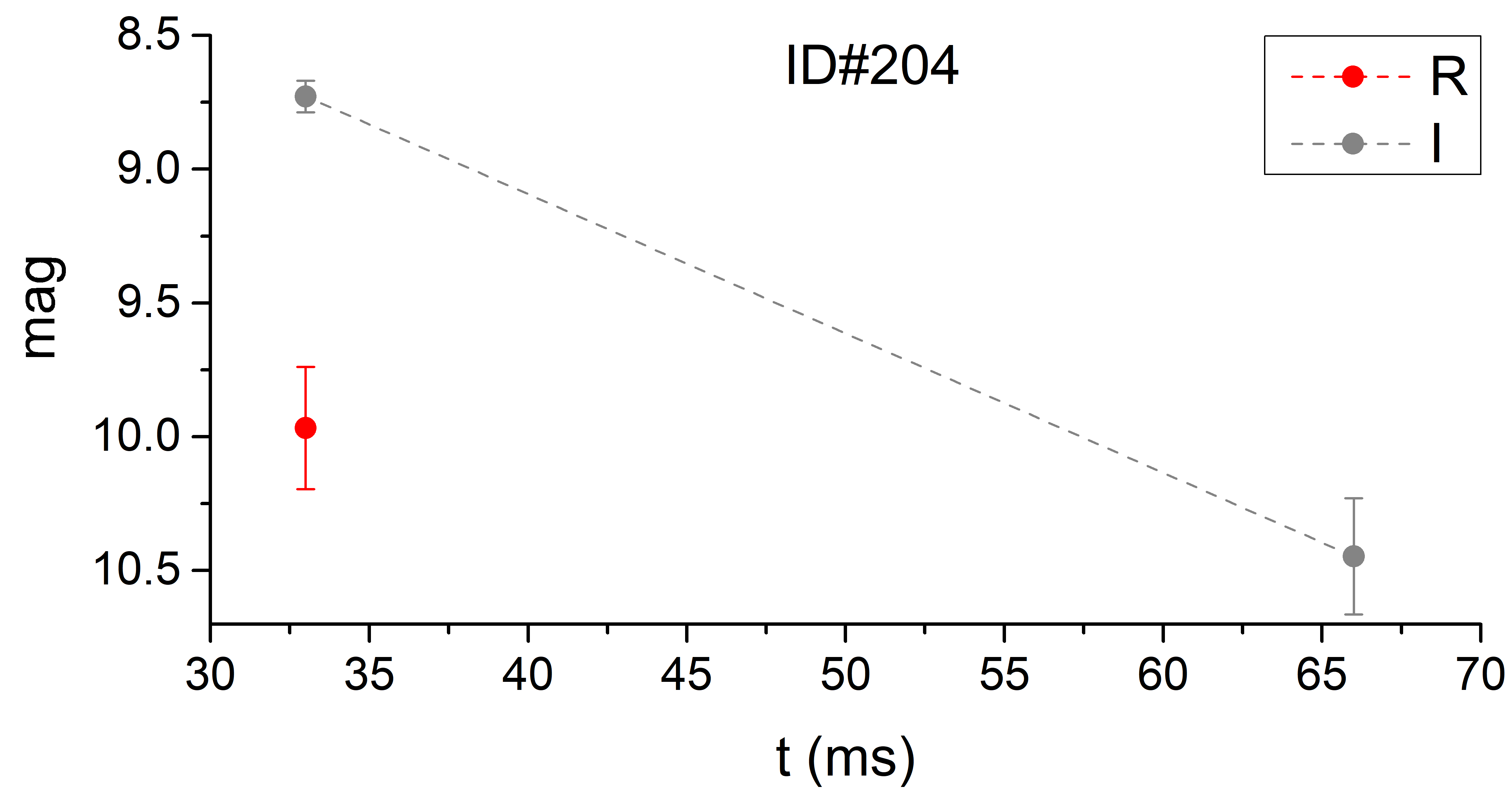}&\includegraphics[width=5.6cm]{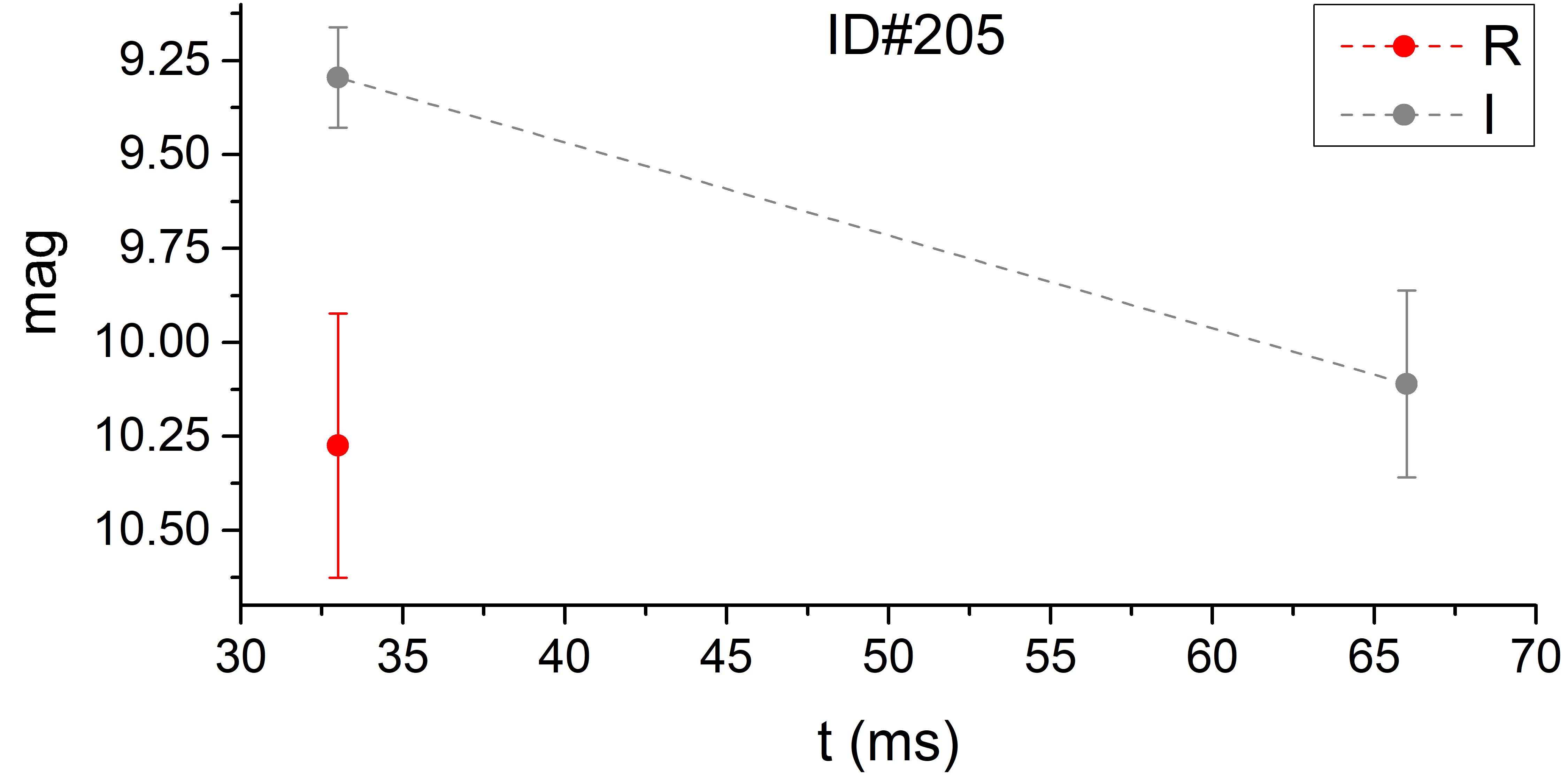}&\includegraphics[width=5.6cm]{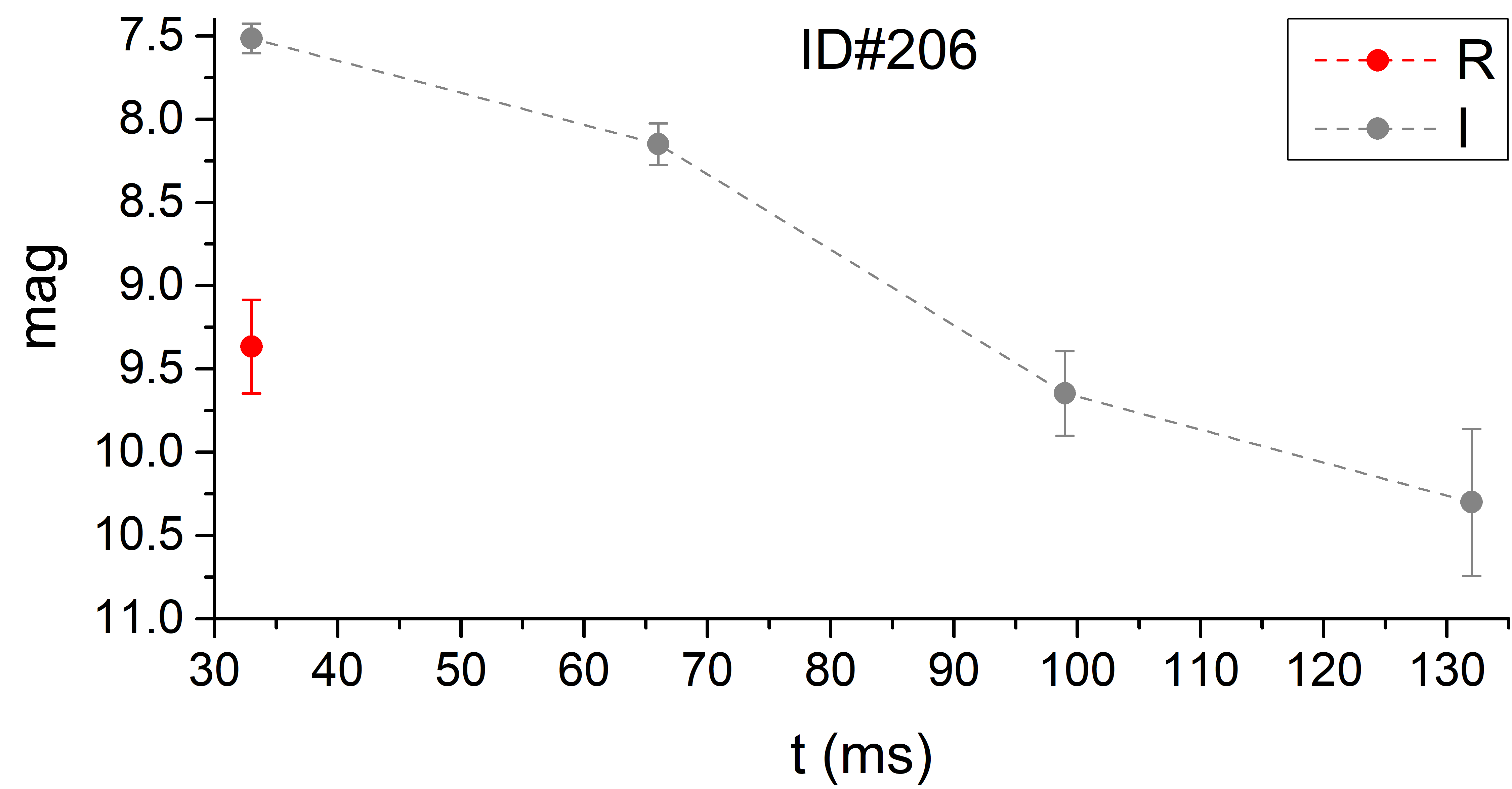}\\
\includegraphics[width=5.6cm]{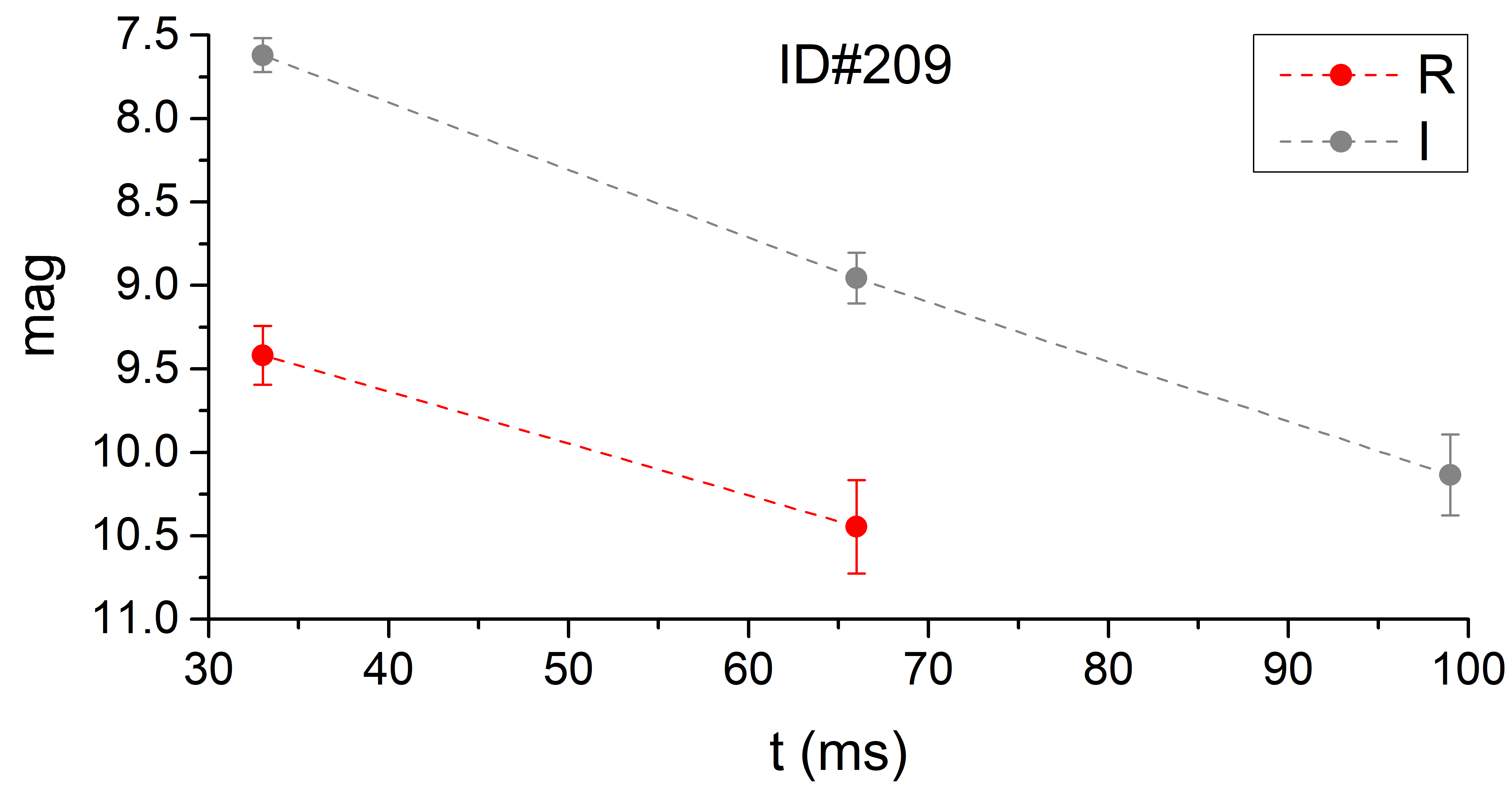}&\includegraphics[width=5.6cm]{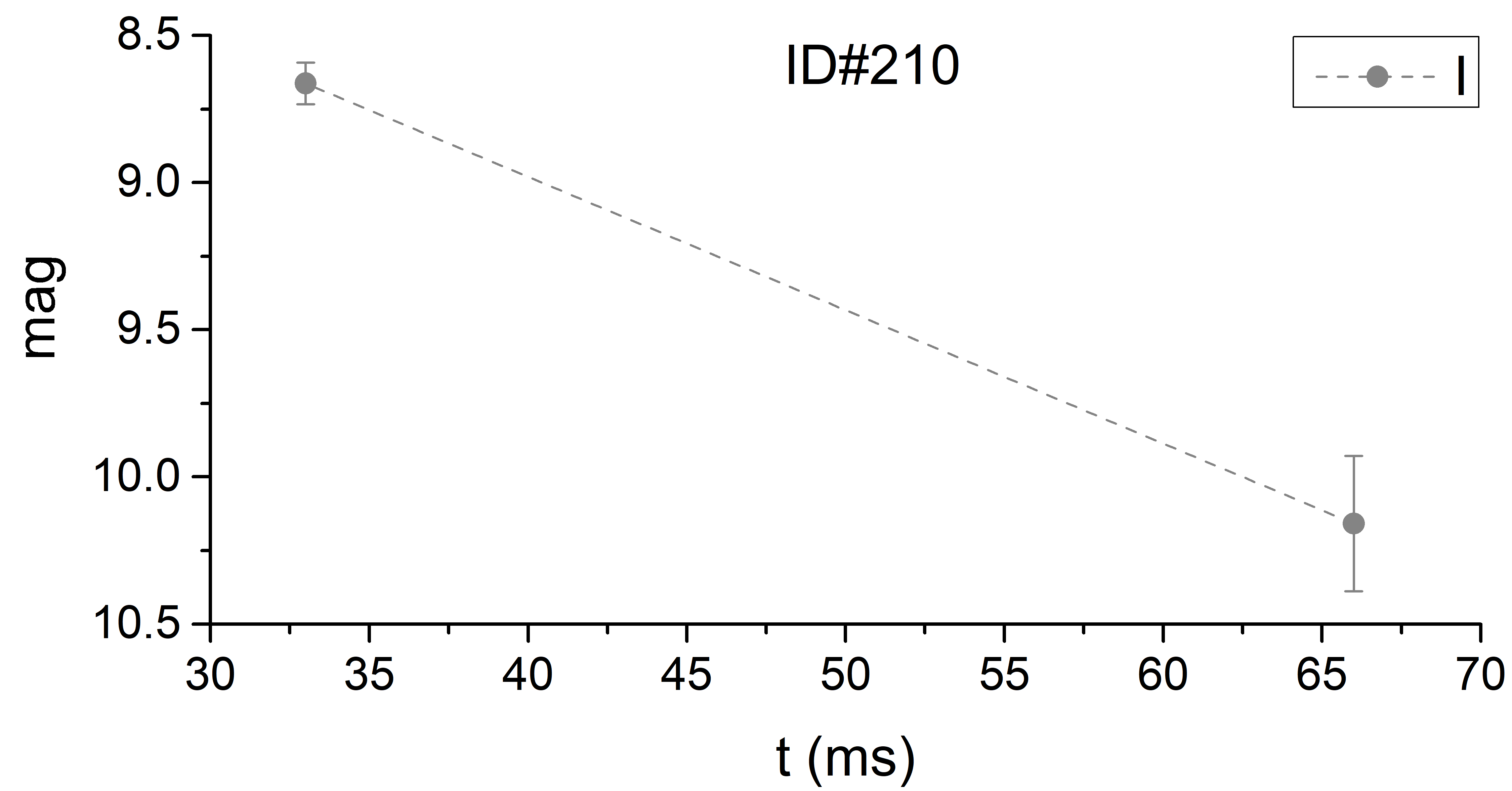}&\includegraphics[width=5.6cm]{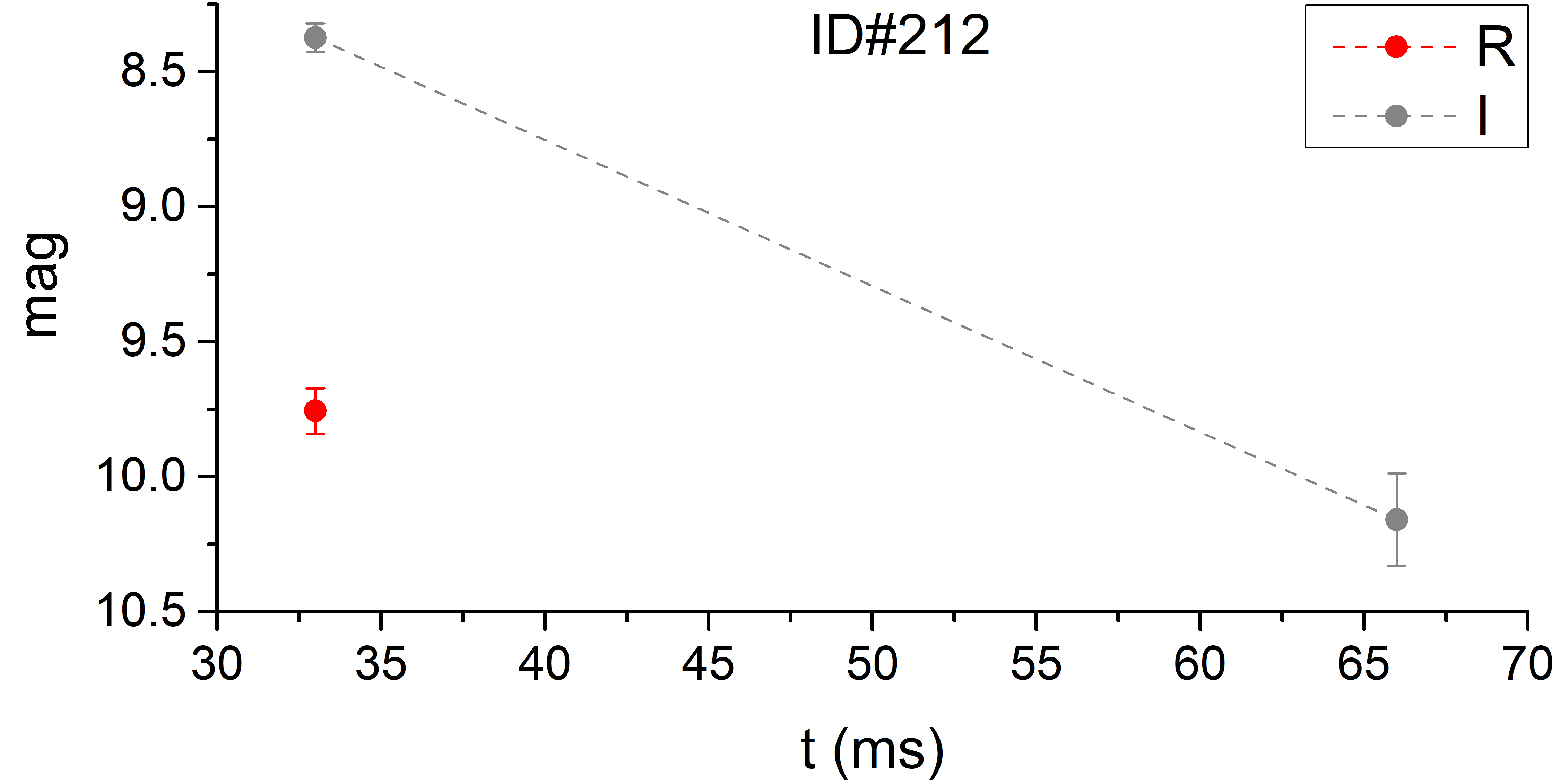}\\
\includegraphics[width=5.6cm]{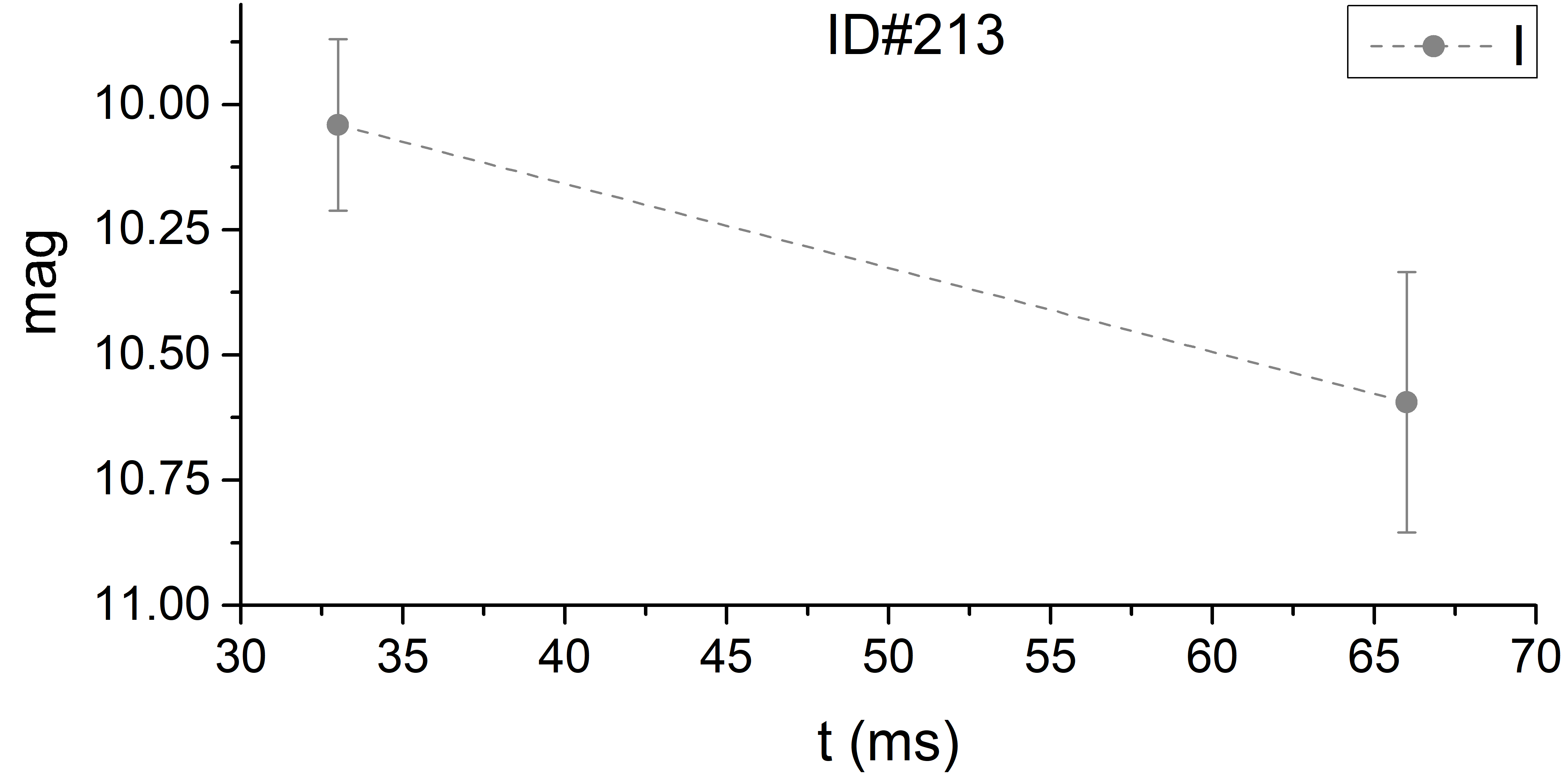}&\includegraphics[width=5.6cm]{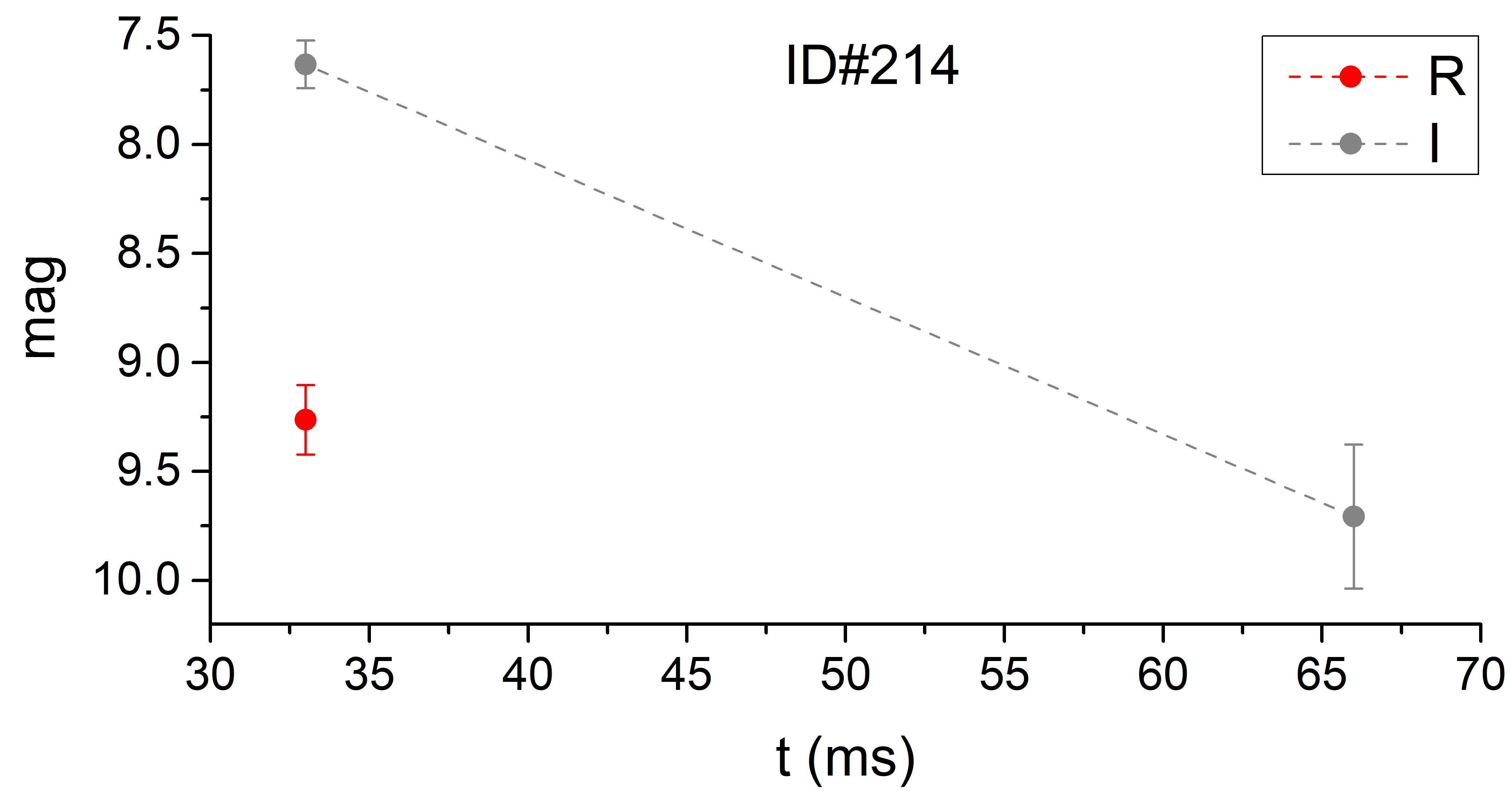}&\includegraphics[width=5.6cm]{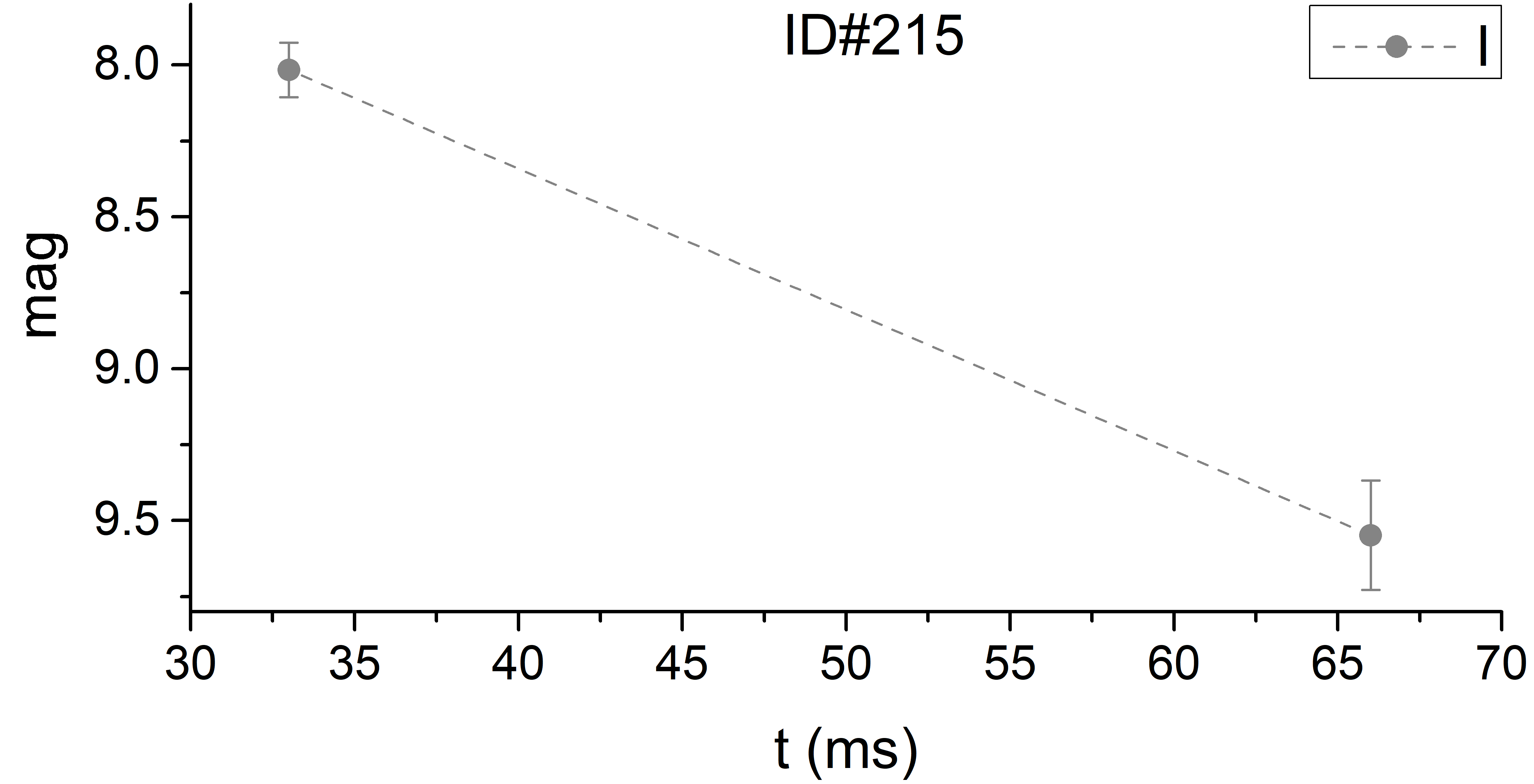}\\
\includegraphics[width=5.6cm]{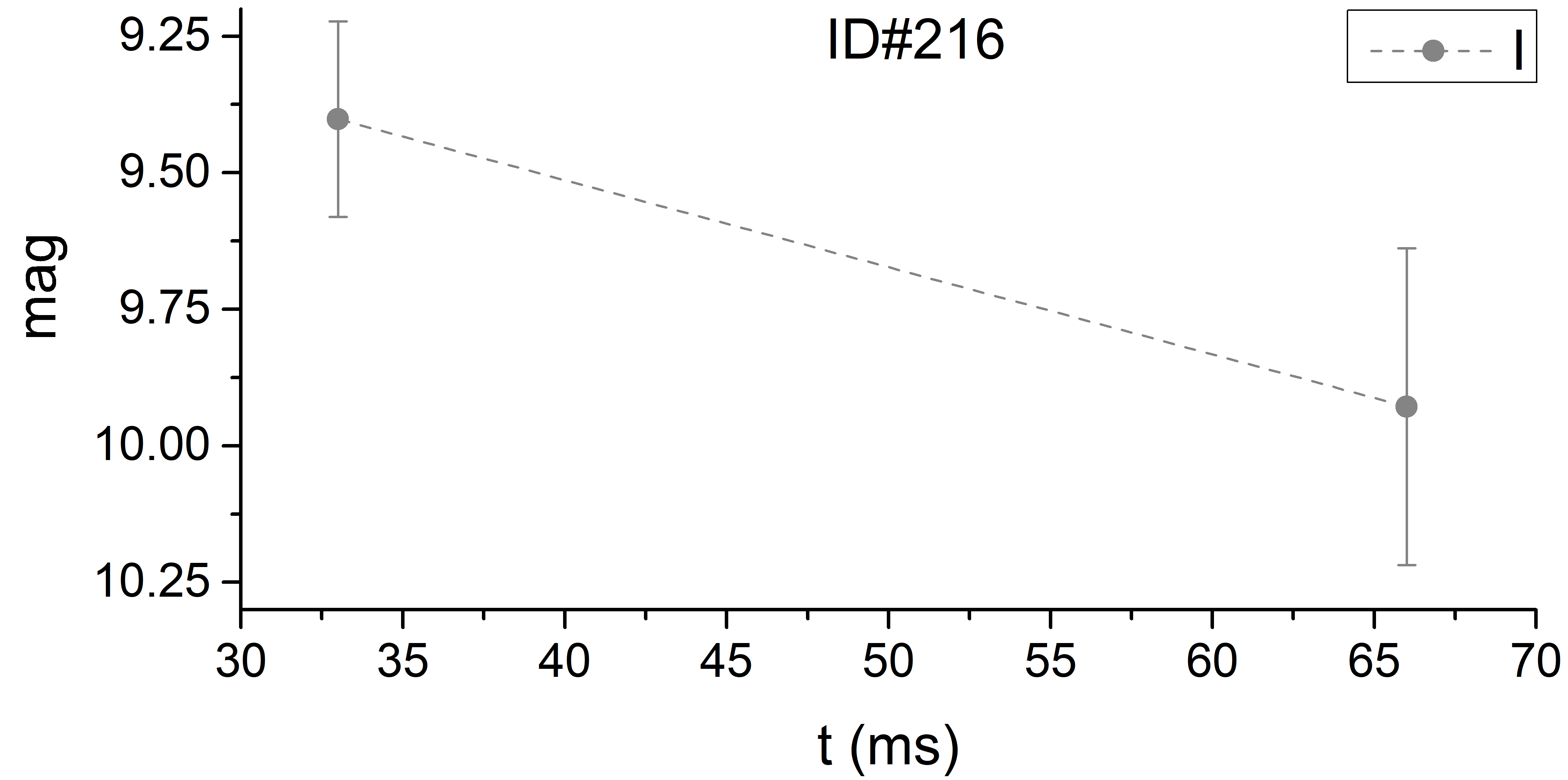}&\includegraphics[width=5.6cm]{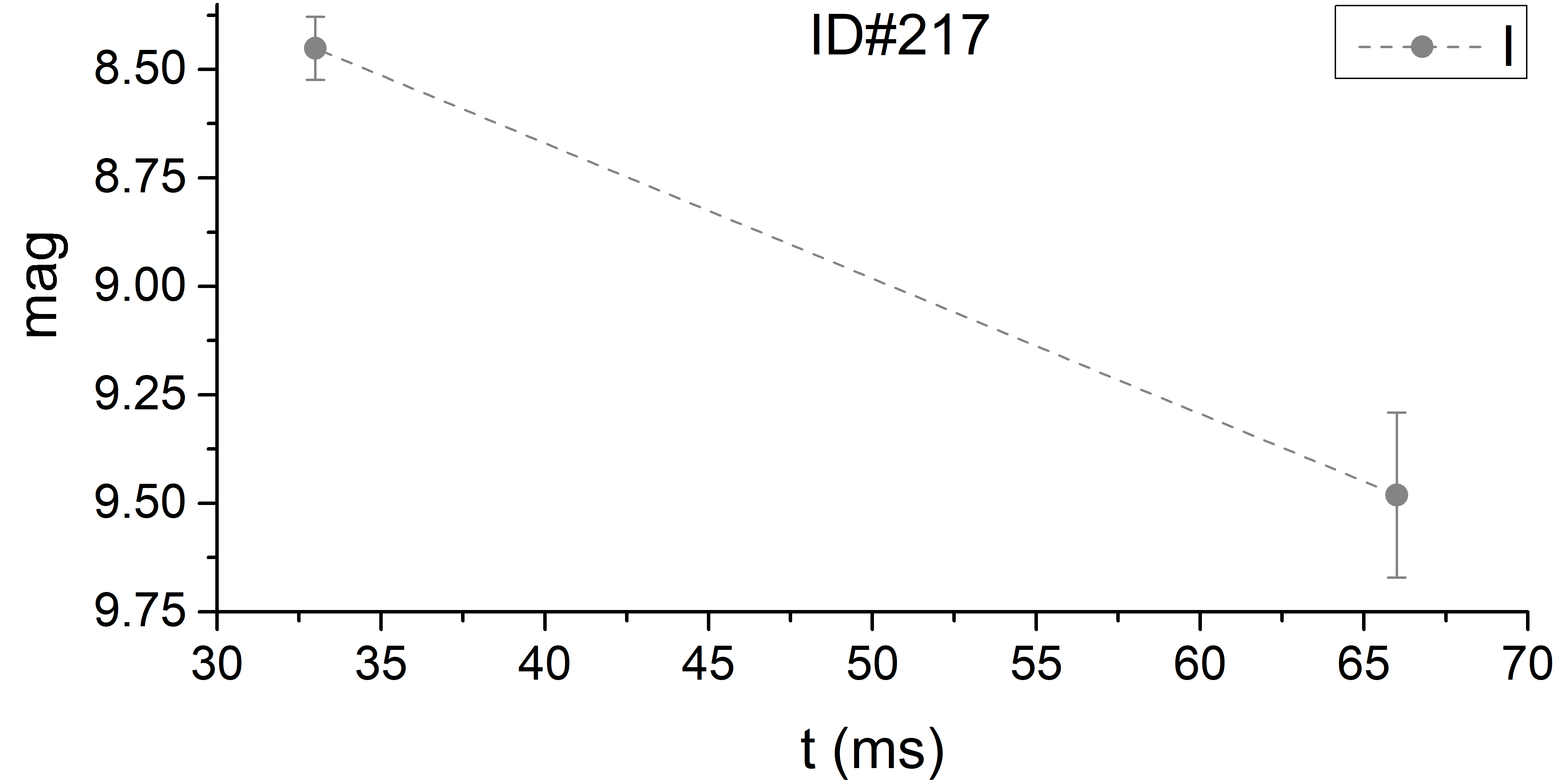}&\includegraphics[width=5.6cm]{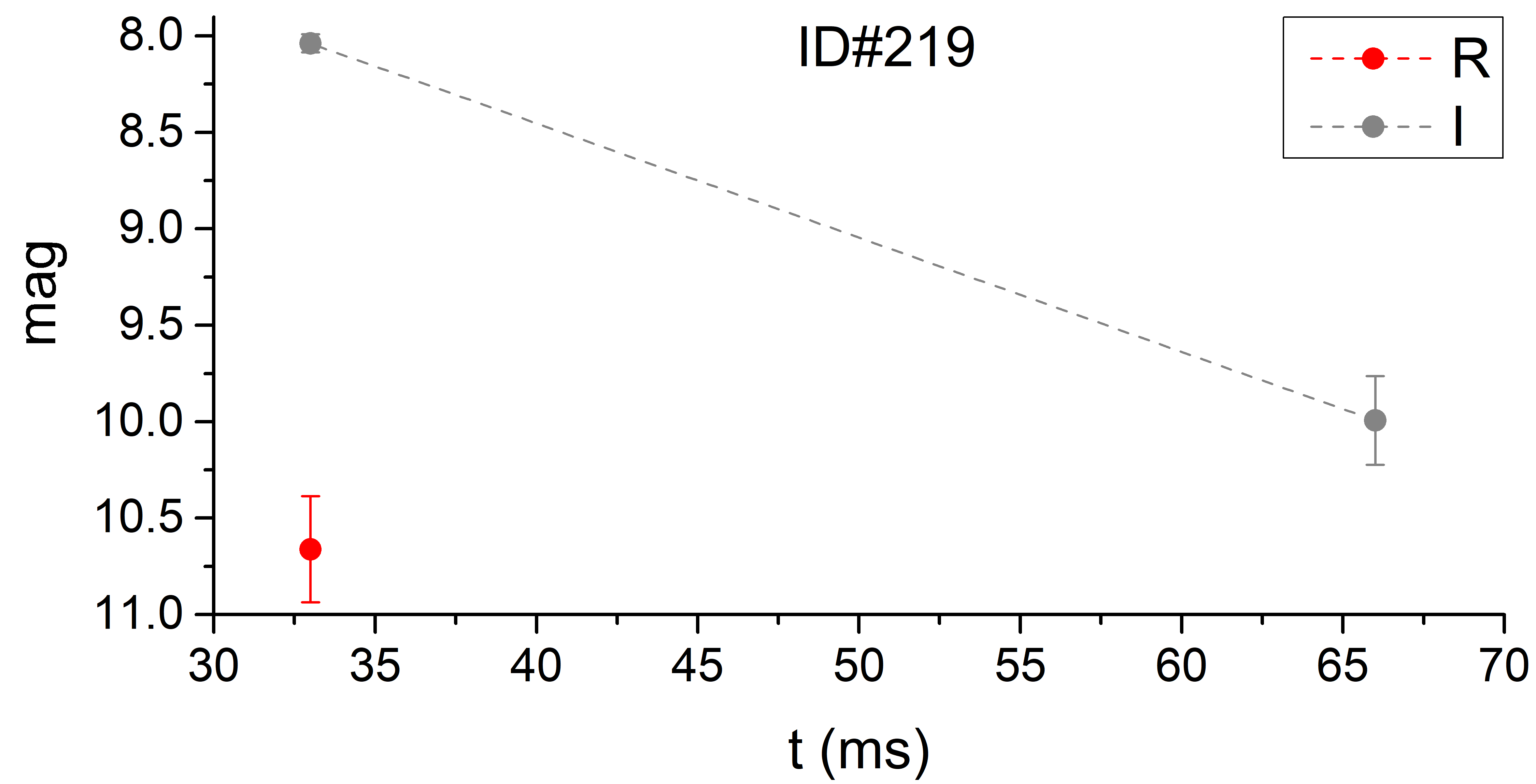}\\
\includegraphics[width=5.6cm]{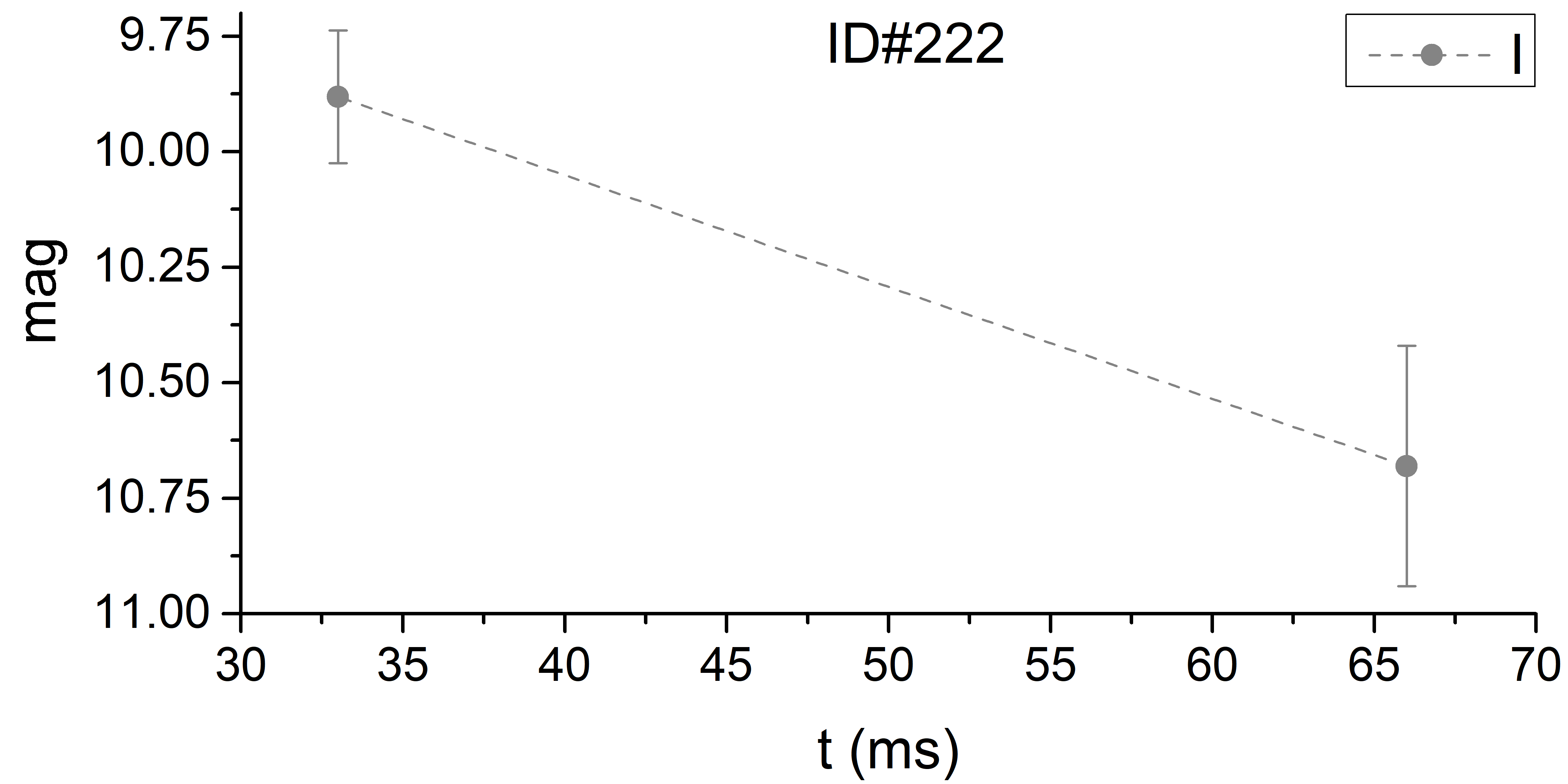}&\includegraphics[width=5.6cm]{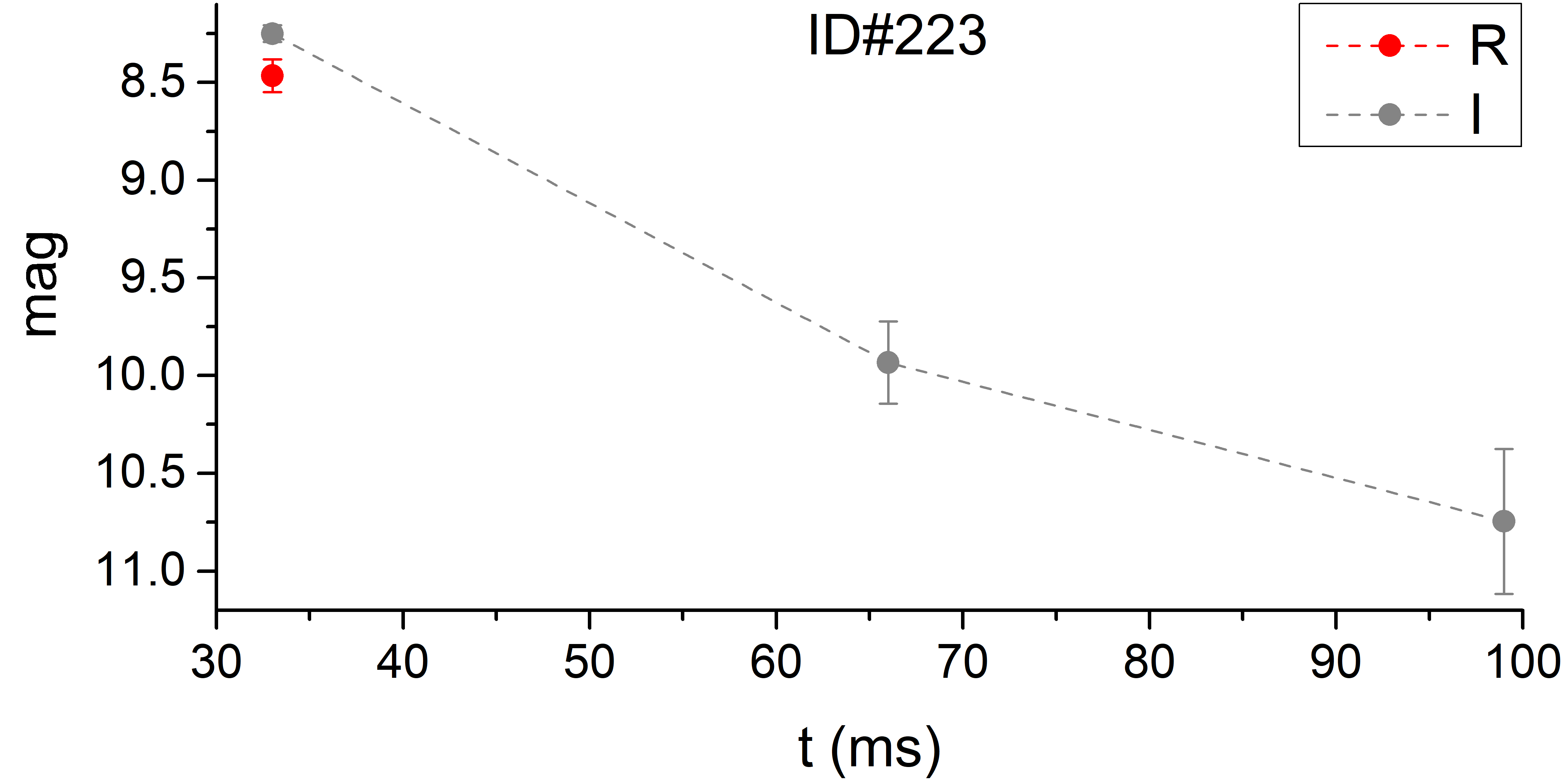}&\includegraphics[width=5.6cm]{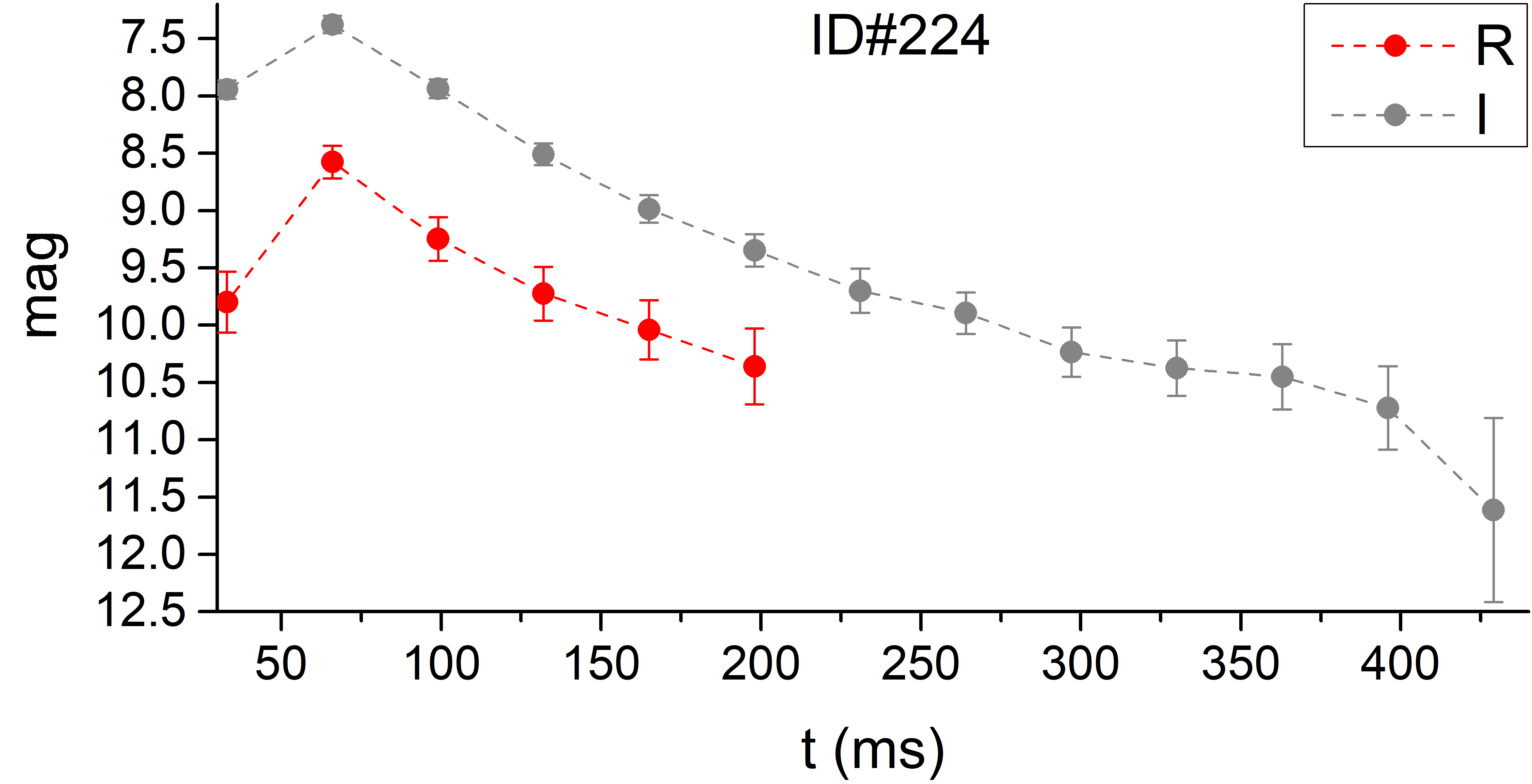}\\
\includegraphics[width=5.6cm]{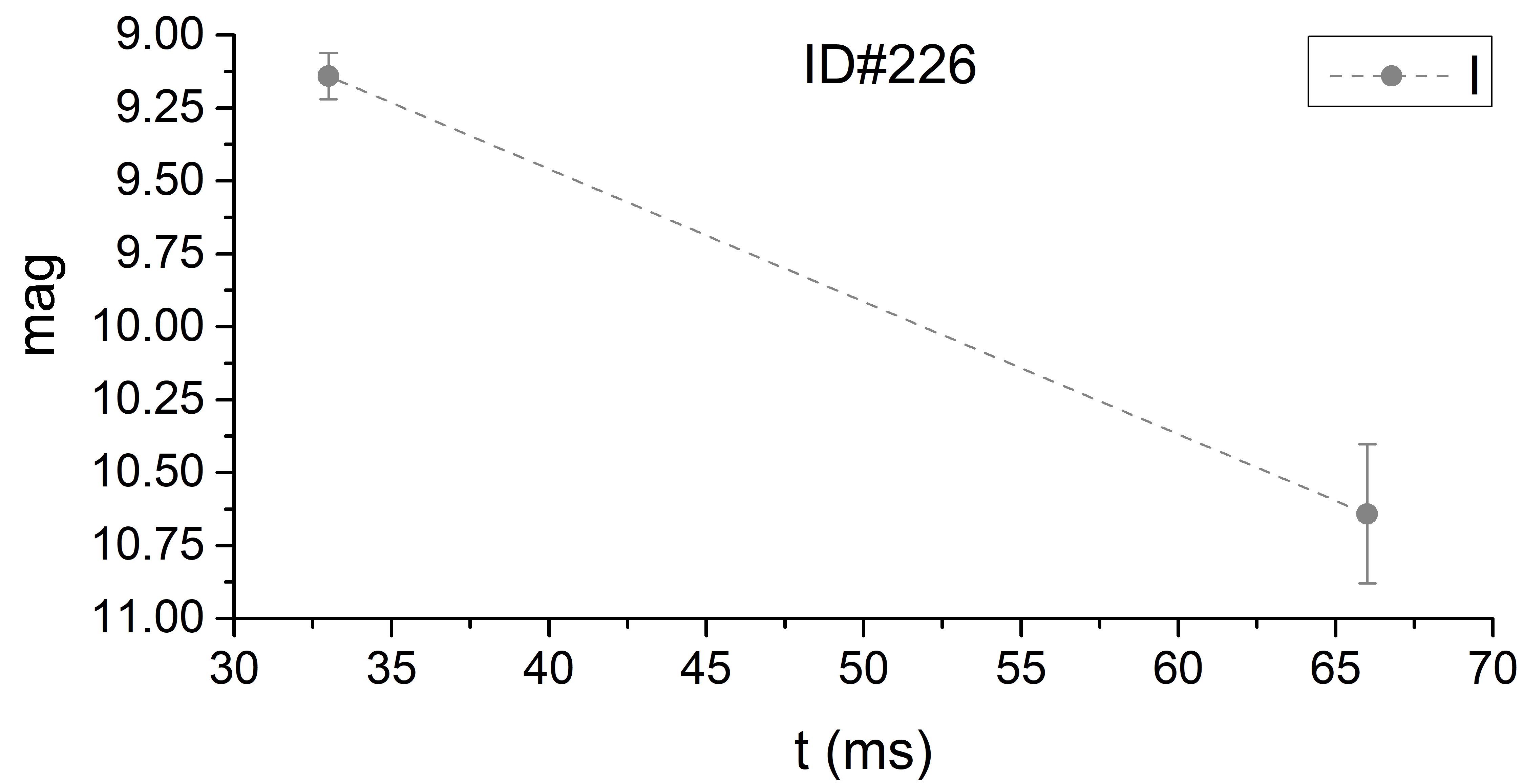}&\includegraphics[width=5.6cm]{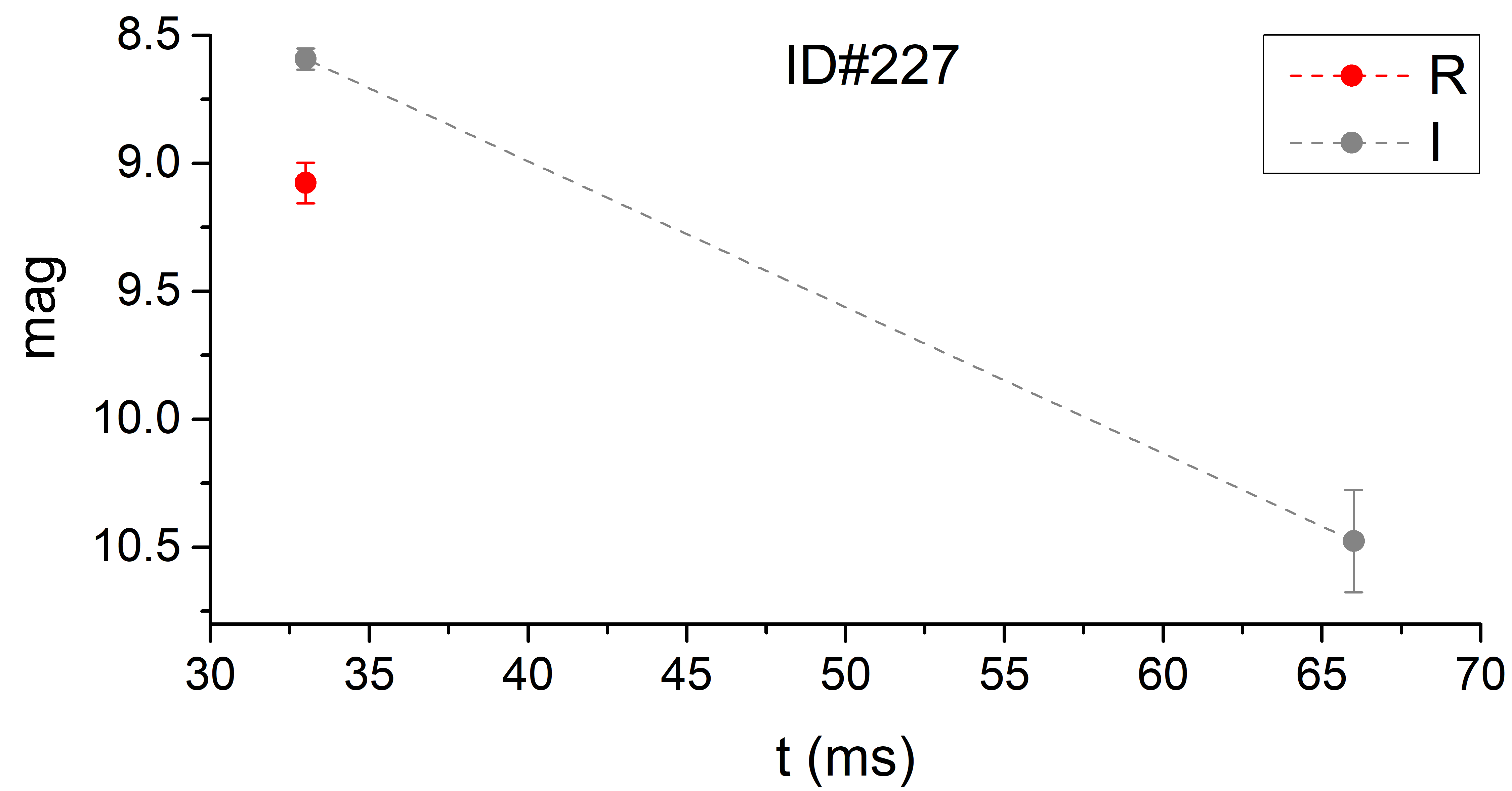}&\includegraphics[width=5.6cm]{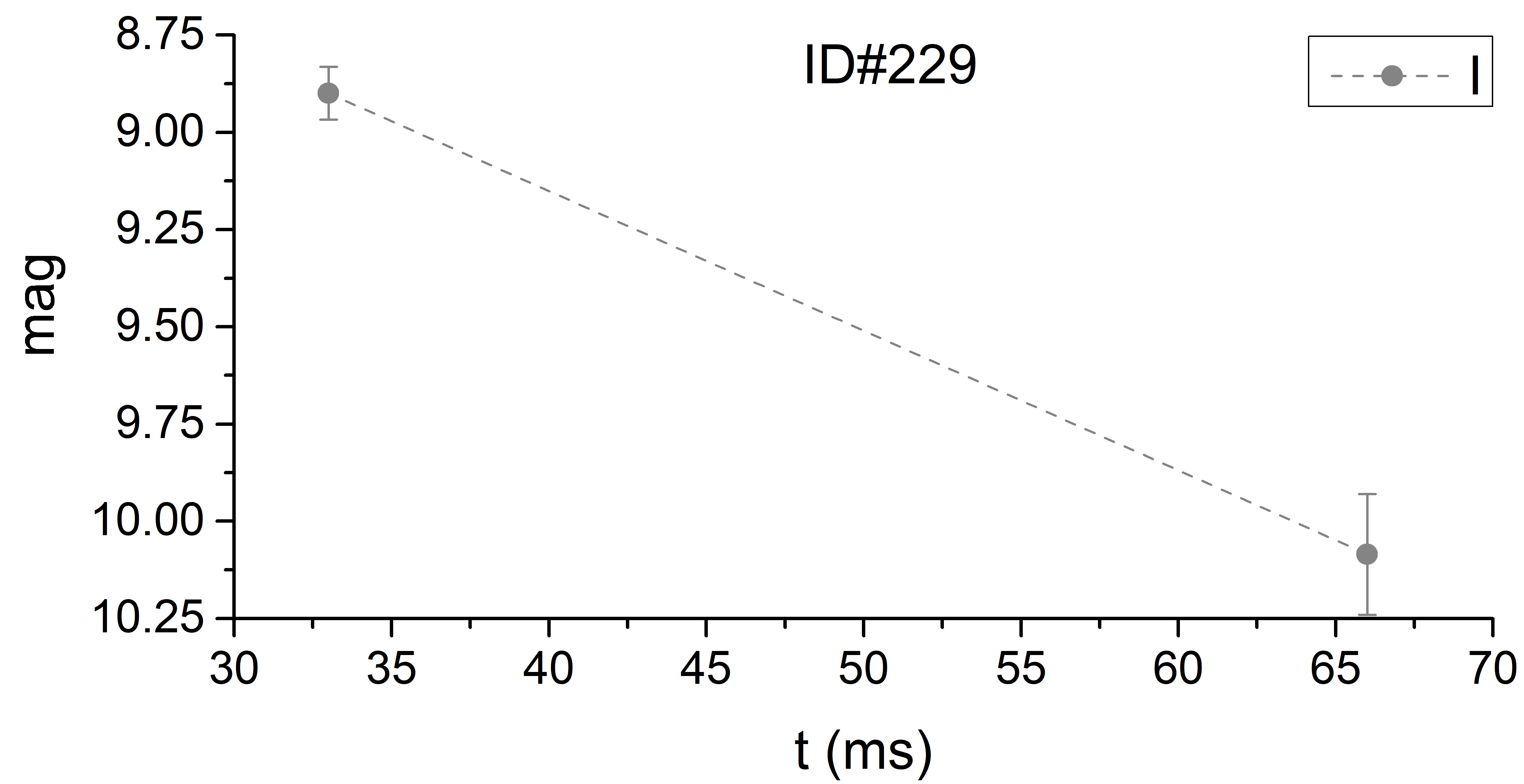}\\
\includegraphics[width=5.6cm]{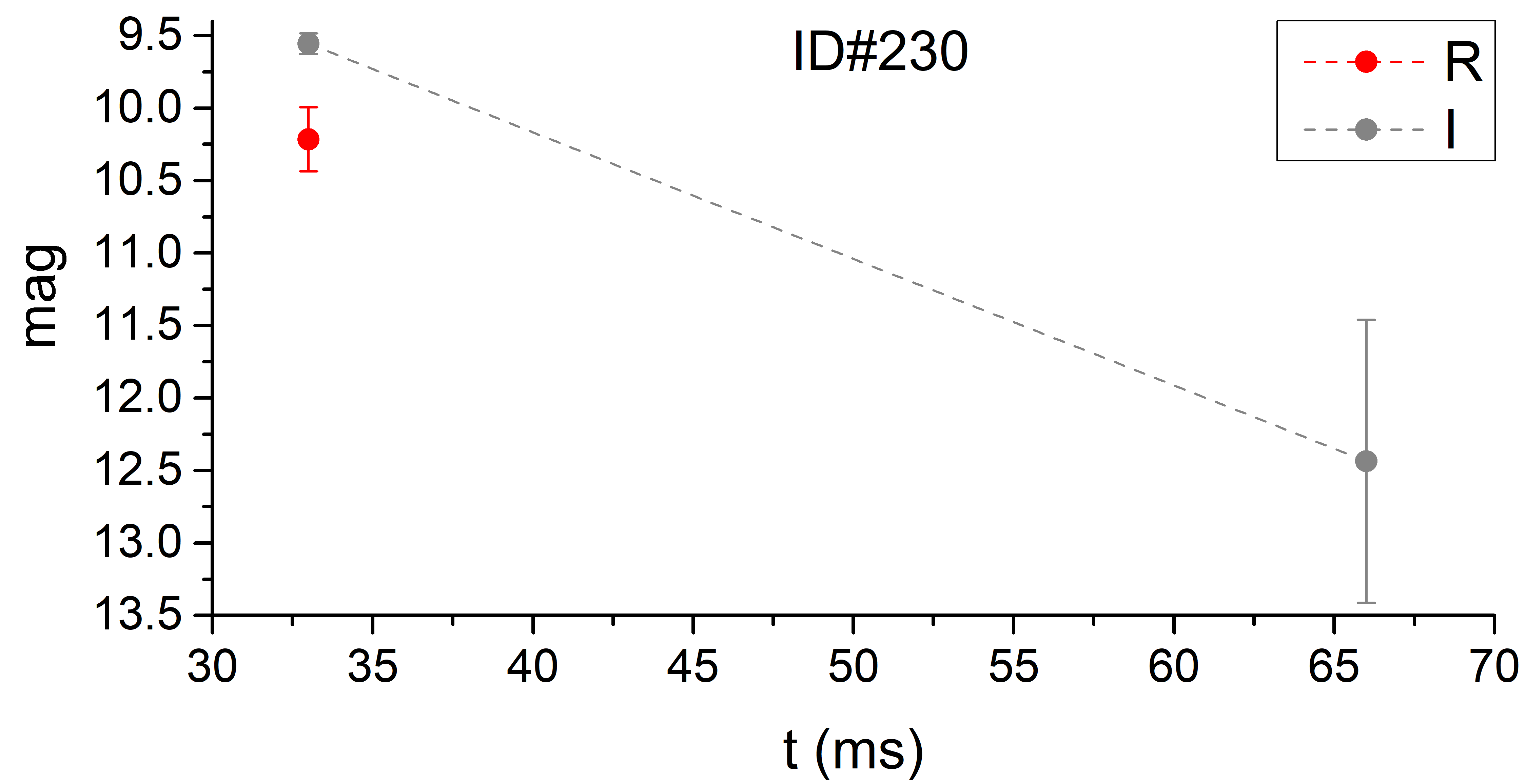}&\includegraphics[width=5.6cm]{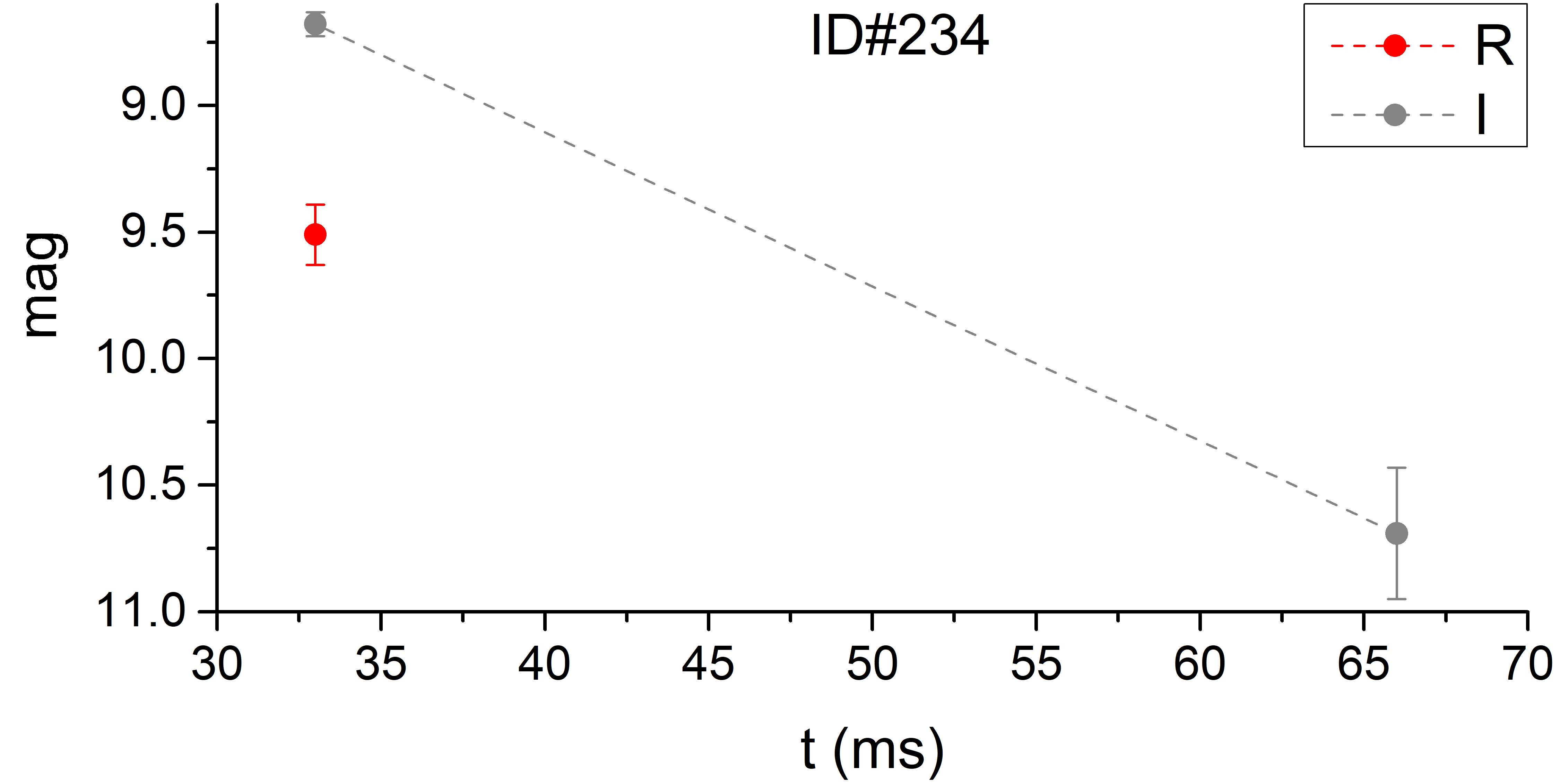}&\includegraphics[width=5.6cm]{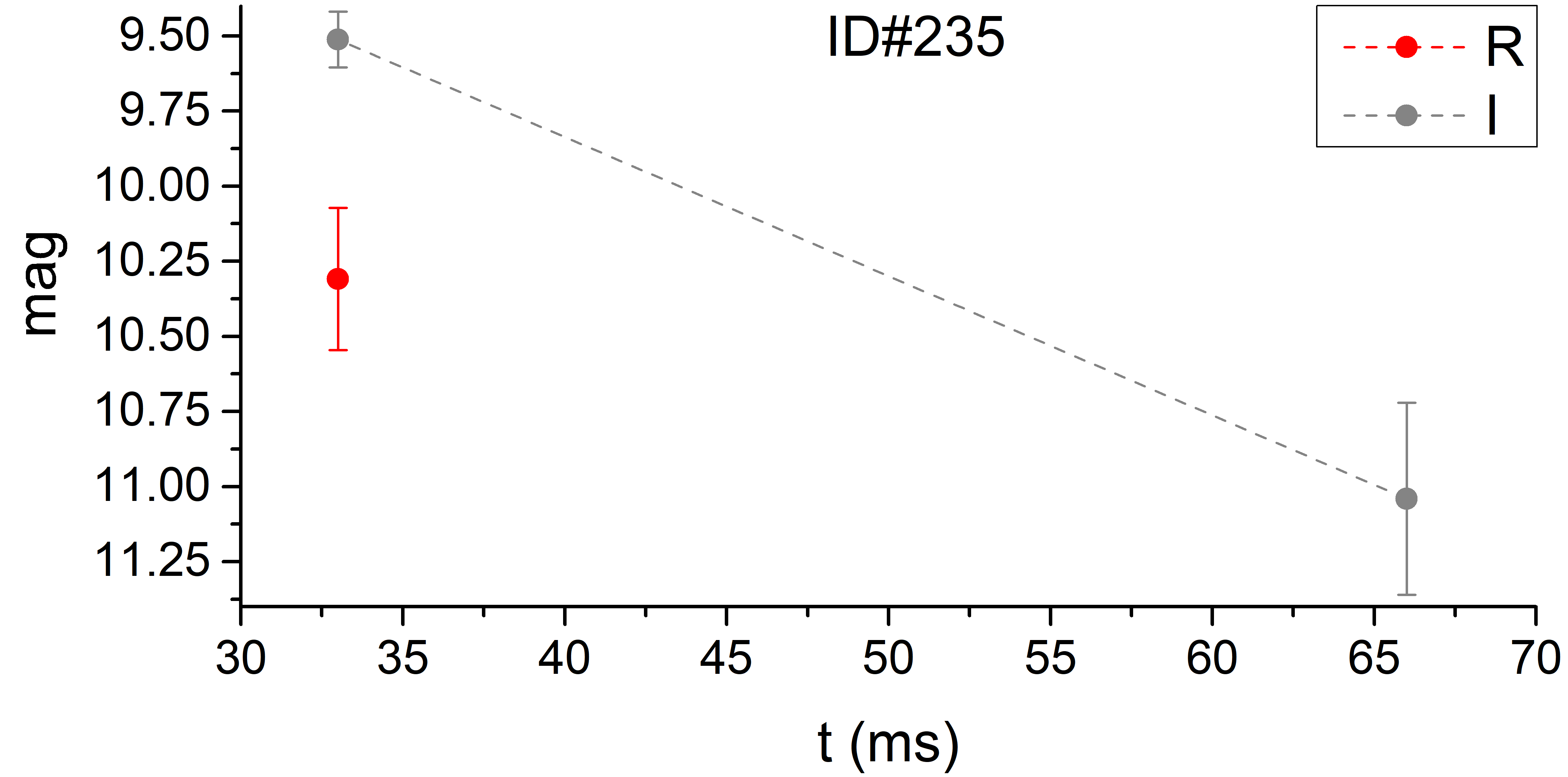}\\
\includegraphics[width=5.6cm]{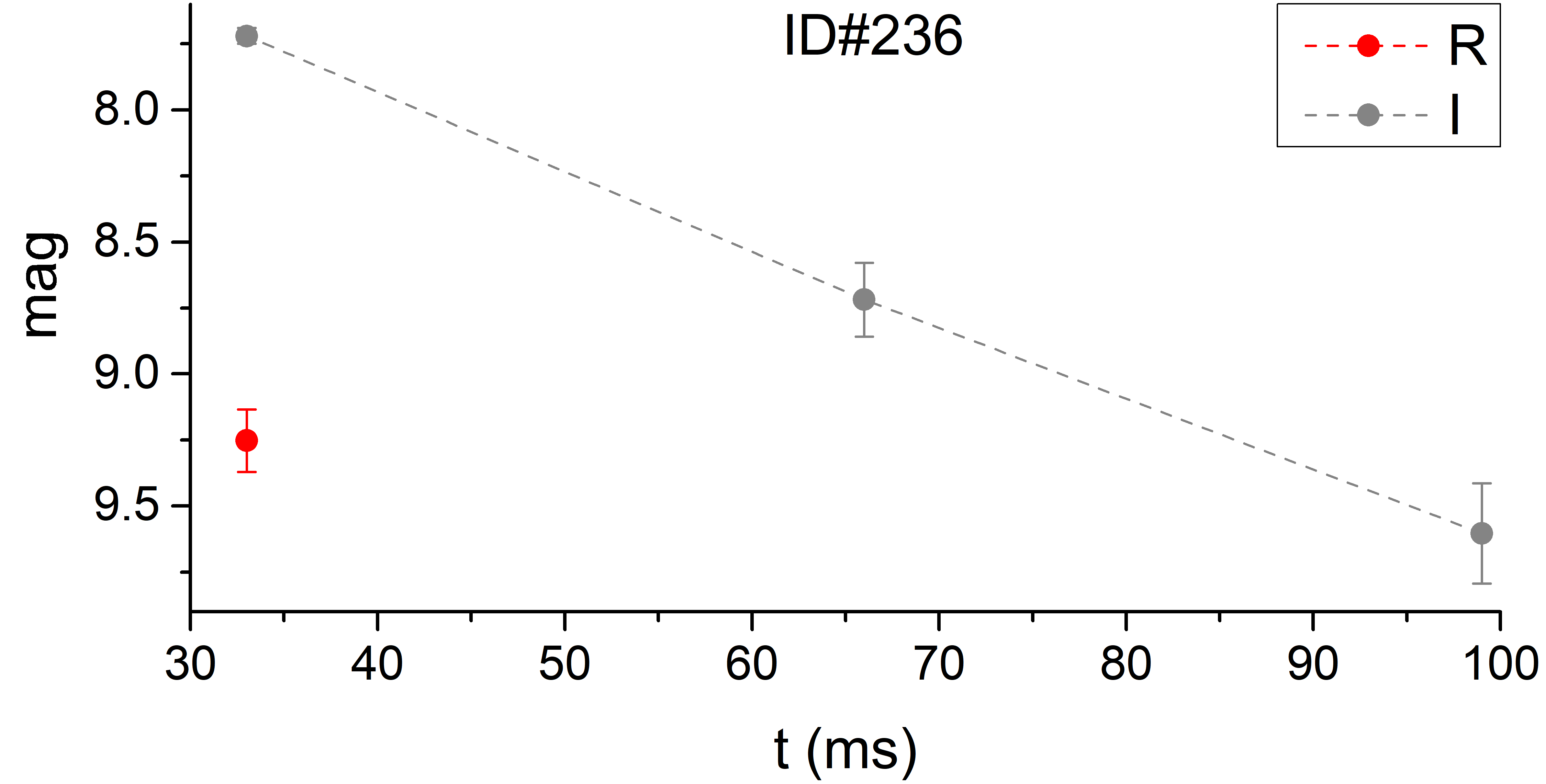}&\includegraphics[width=5.6cm]{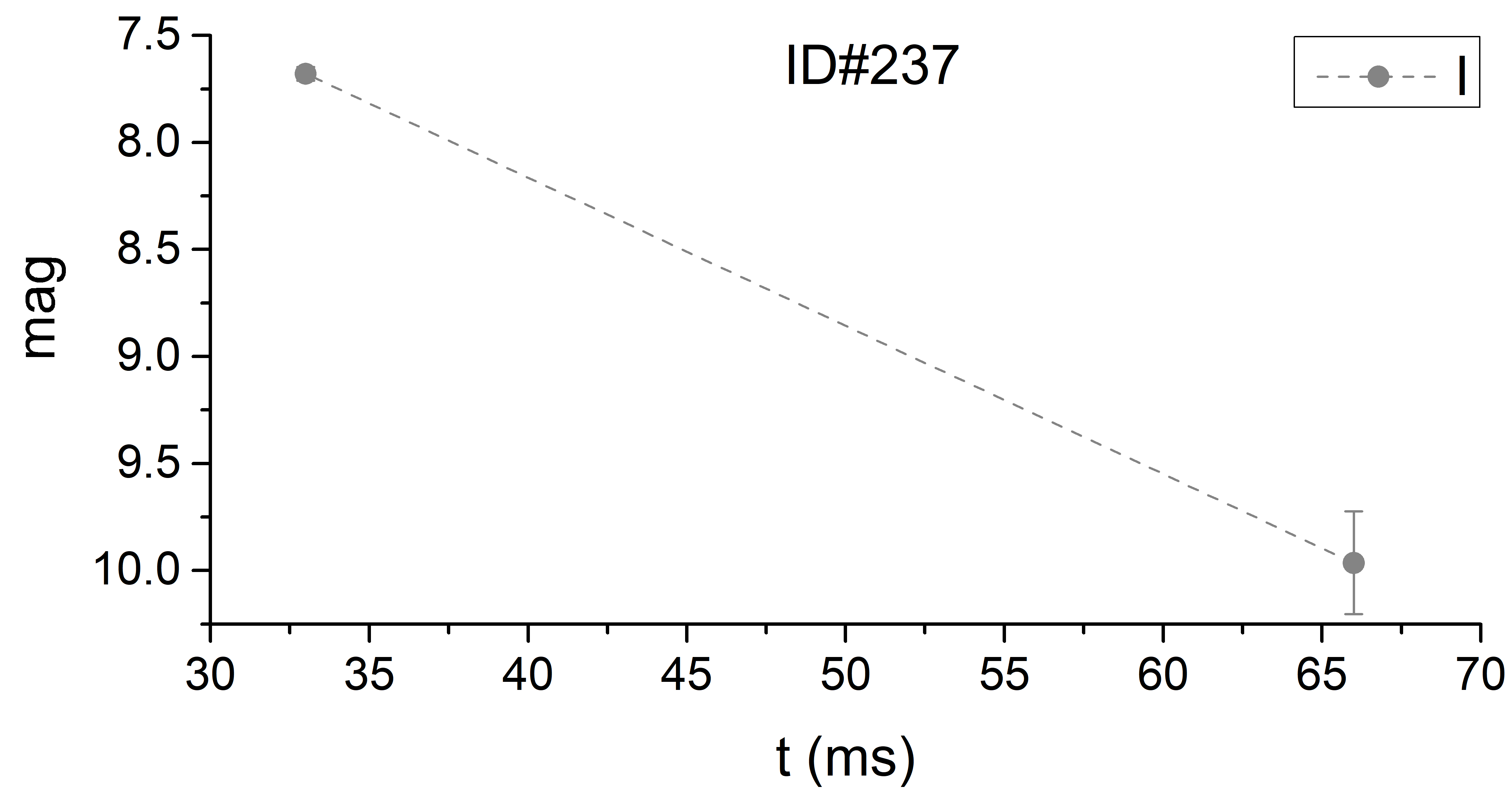}&\includegraphics[width=5.6cm]{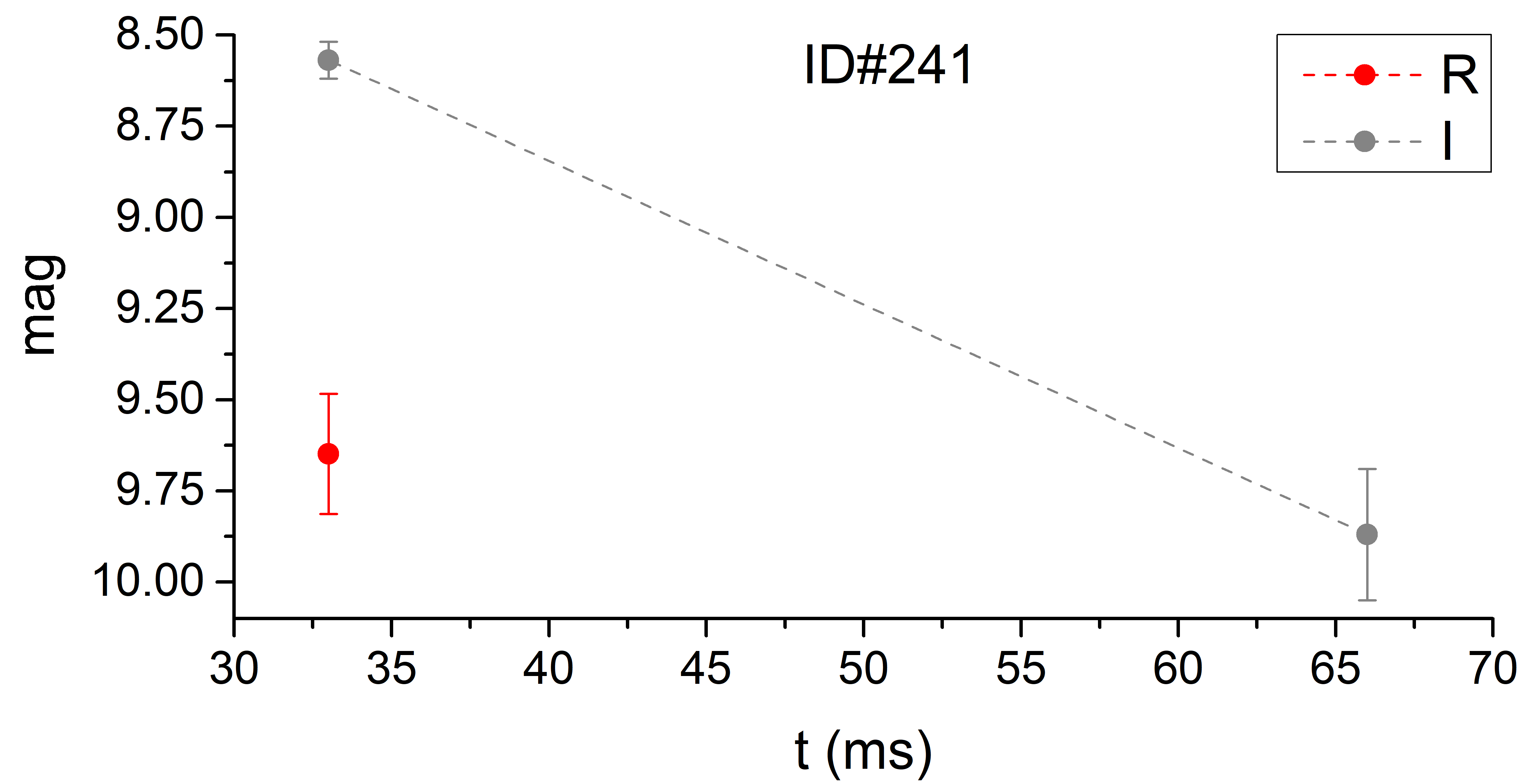}\\
\end{tabular}
\caption*{Fig.~\ref{fig:LCs1}~(cont'd)}
\end{figure*}
\begin{figure*}[h]
\begin{tabular}{ccc}
\includegraphics[width=5.6cm]{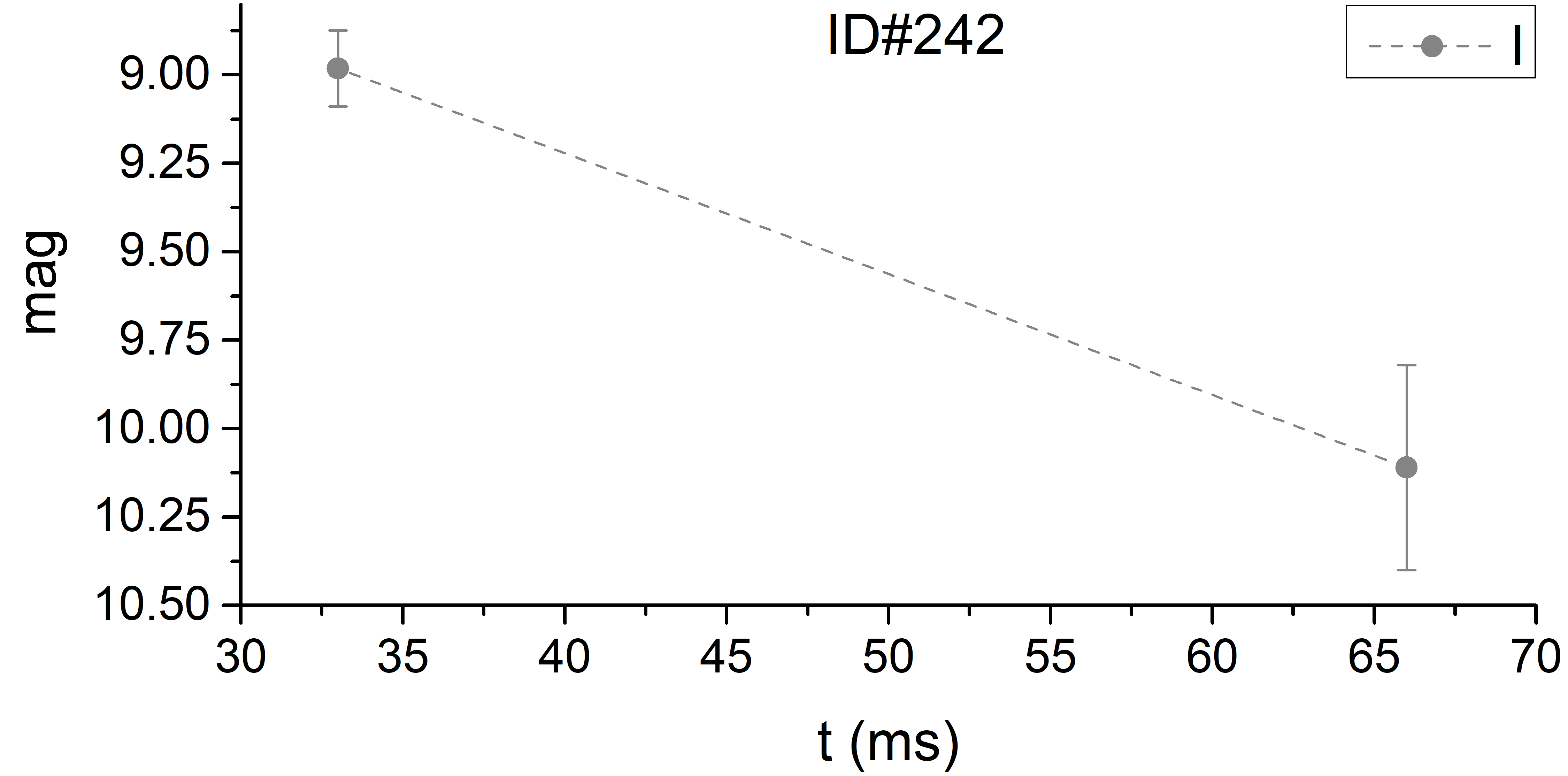}&\includegraphics[width=5.6cm]{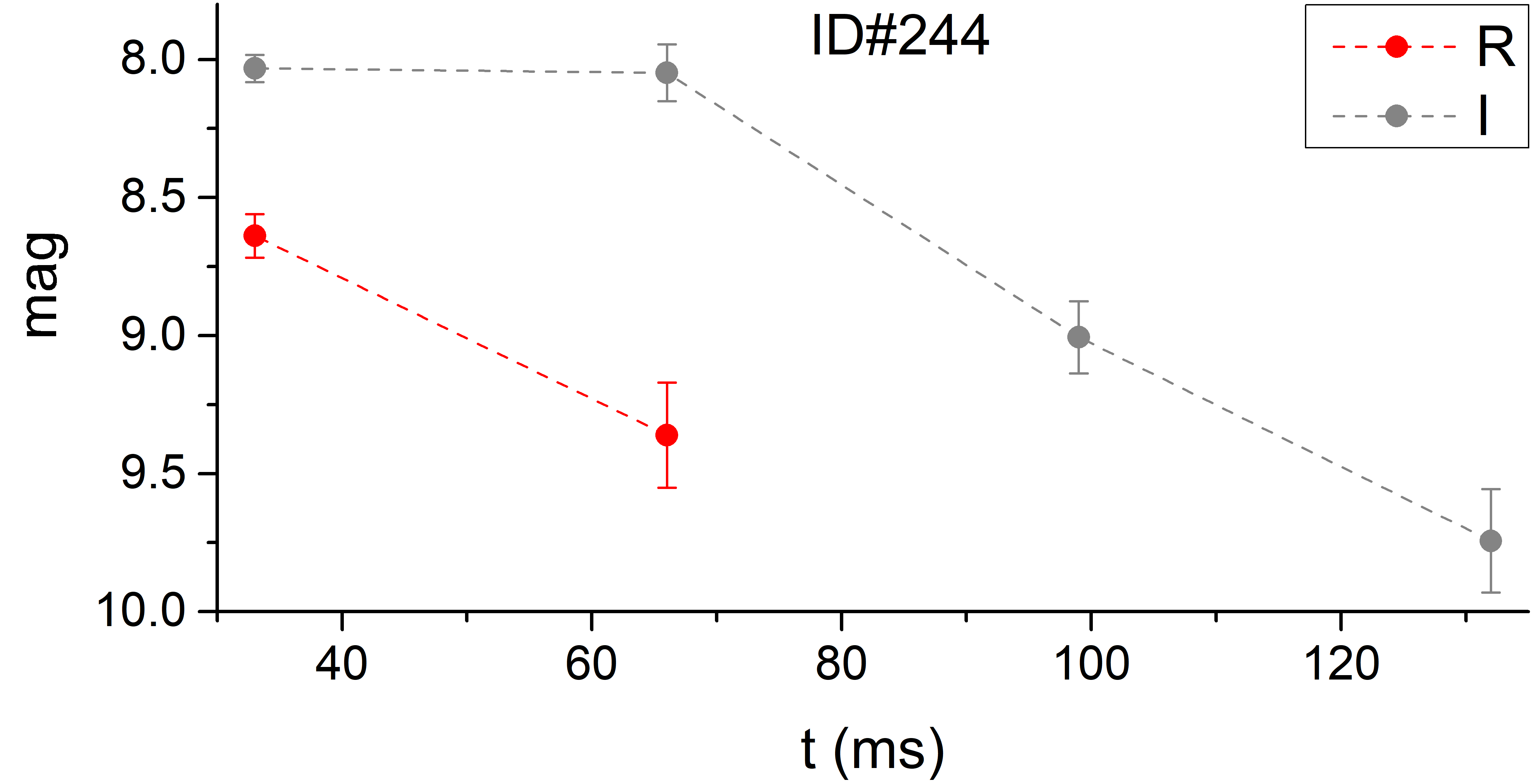}&\includegraphics[width=5.6cm]{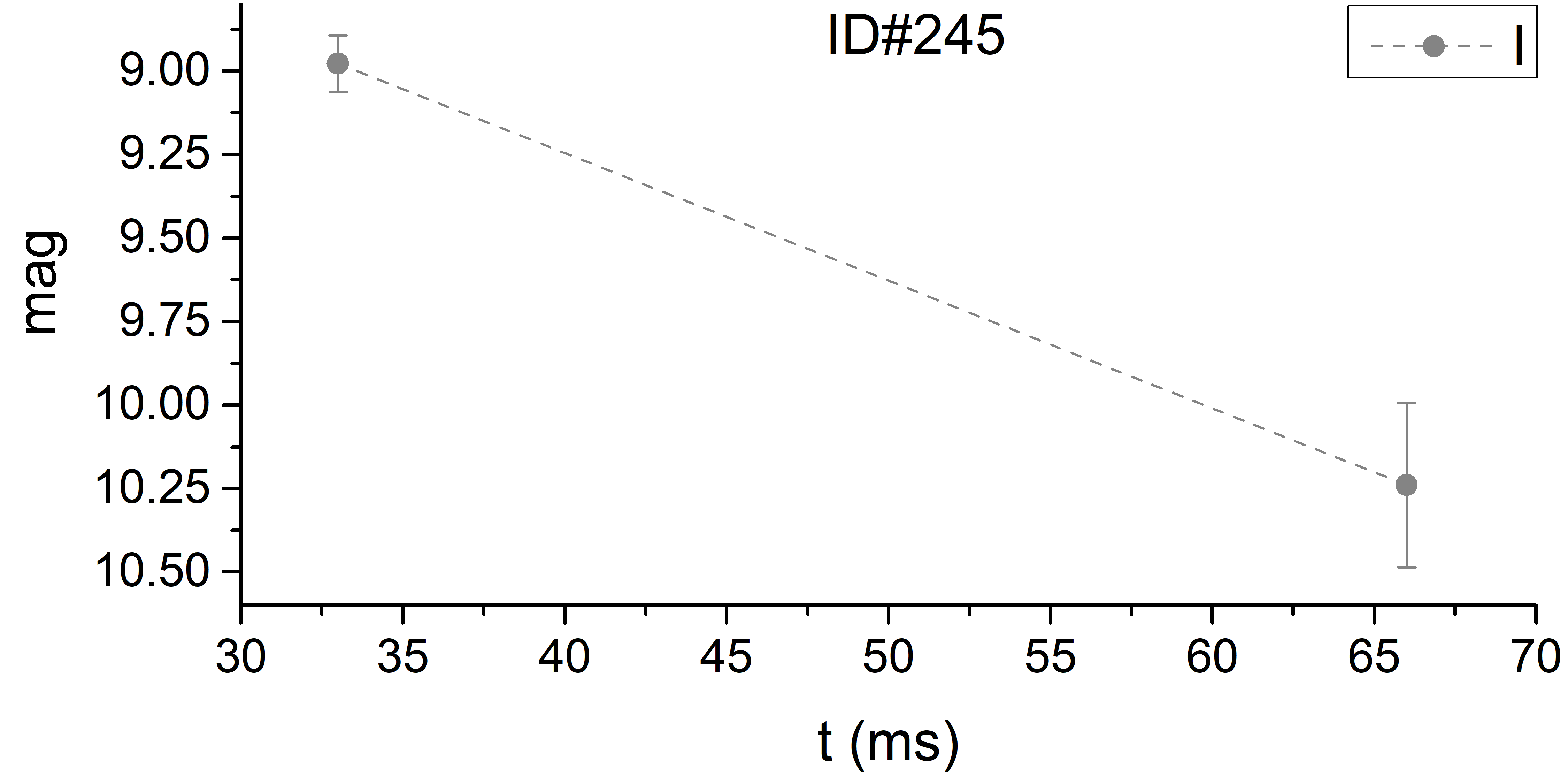}\\
\includegraphics[width=5.6cm]{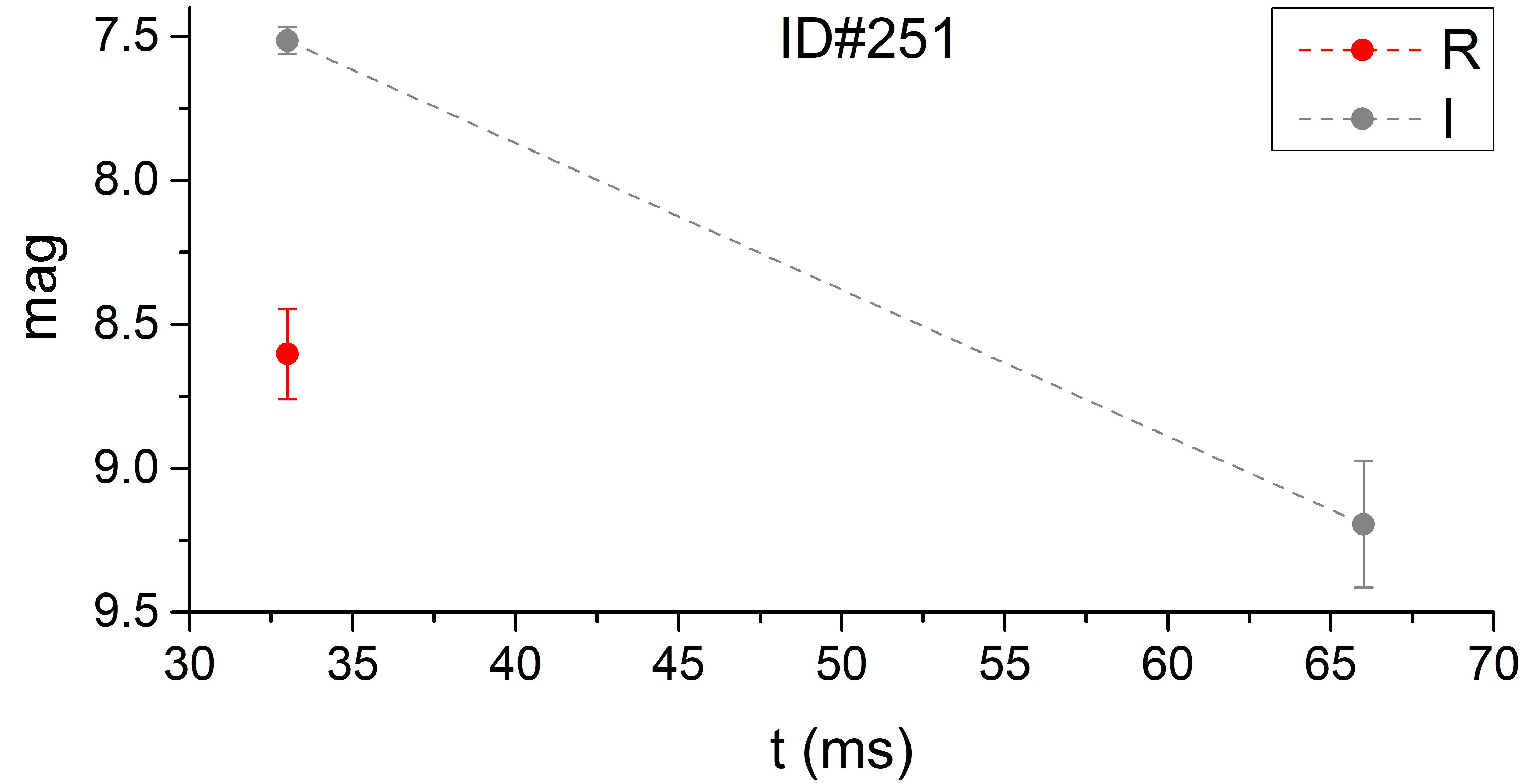}&\includegraphics[width=5.6cm]{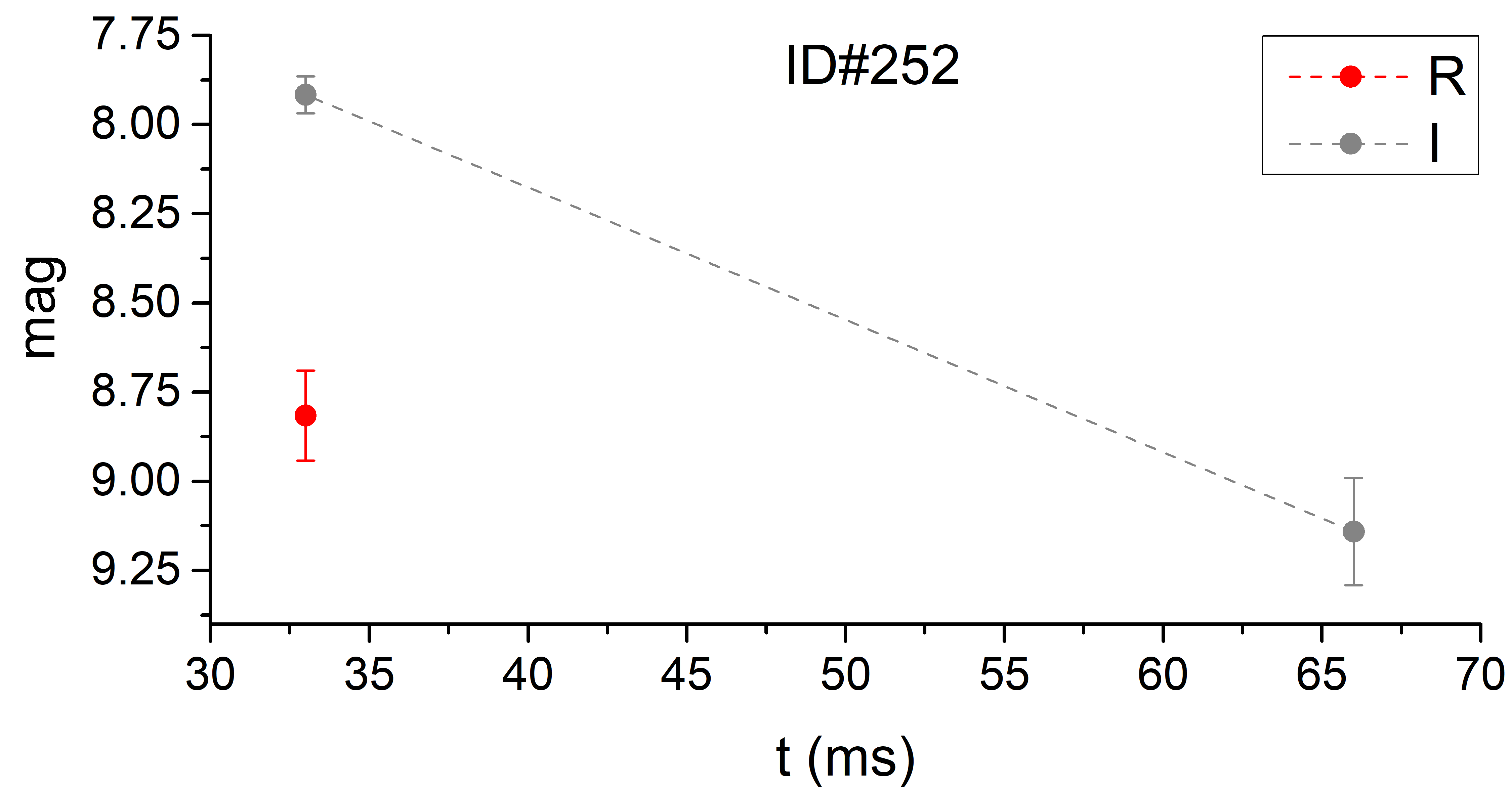}&\includegraphics[width=5.6cm]{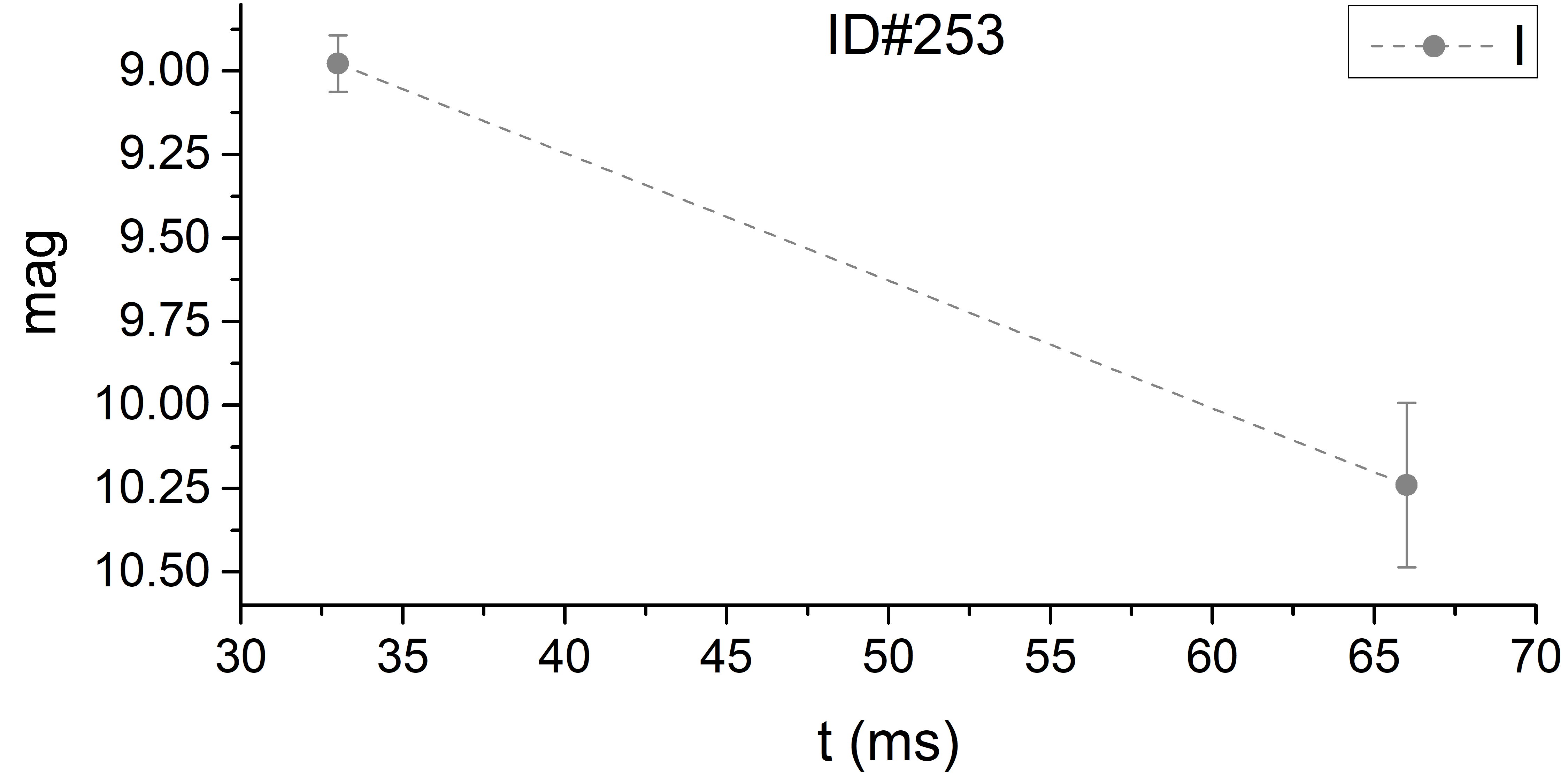}\\
\includegraphics[width=5.6cm]{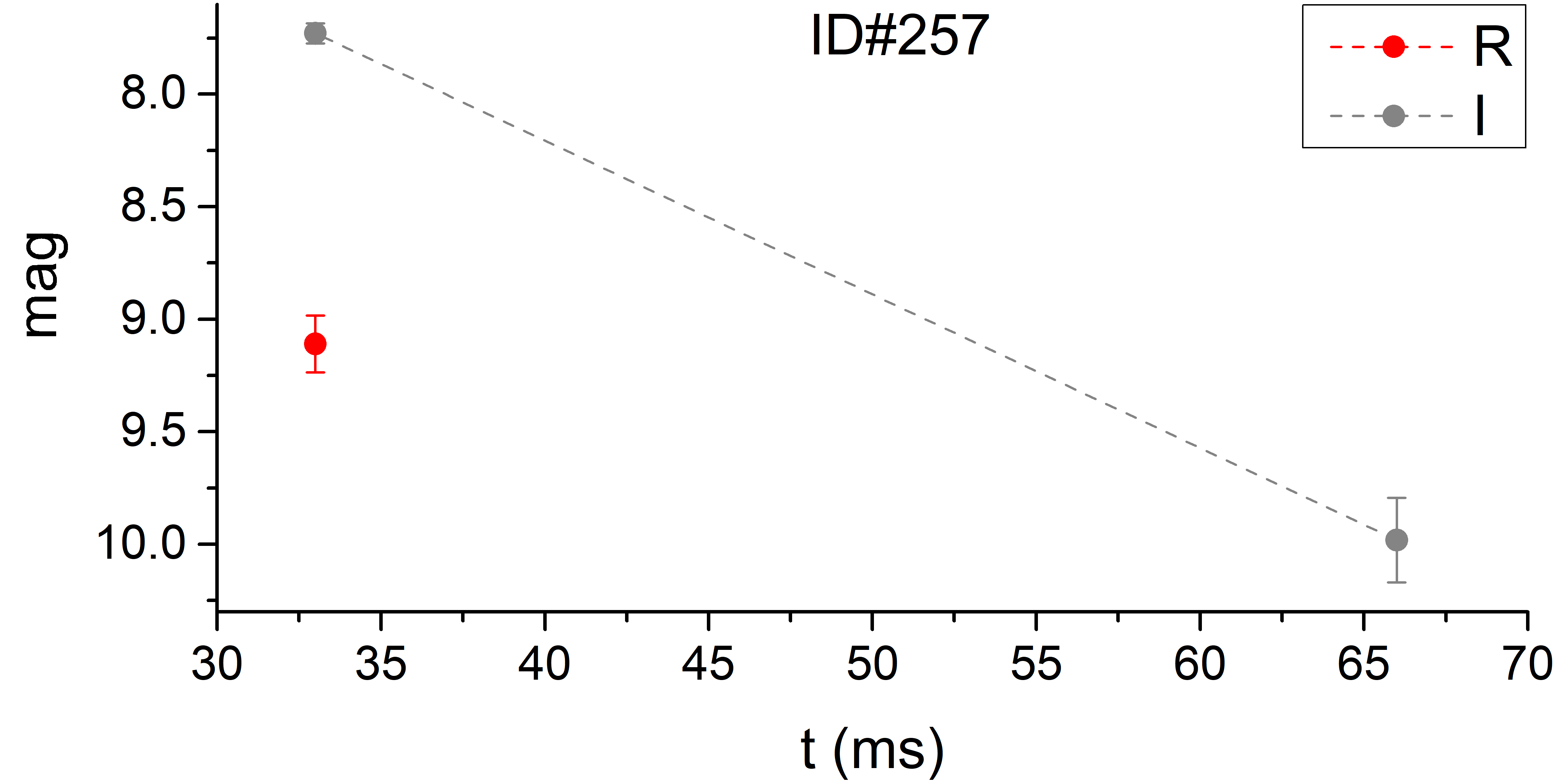}&\includegraphics[width=5.6cm]{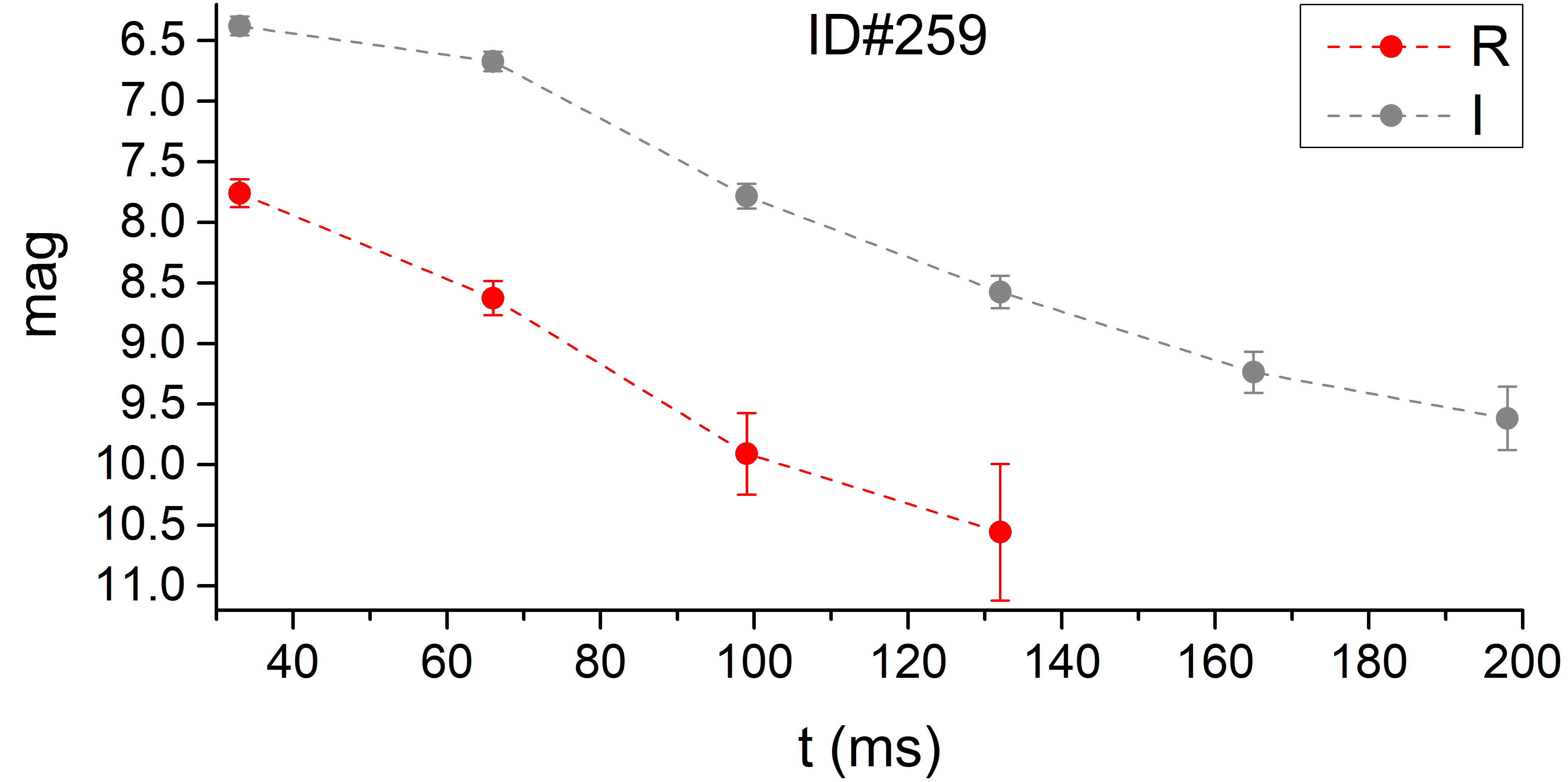}&\includegraphics[width=5.6cm]{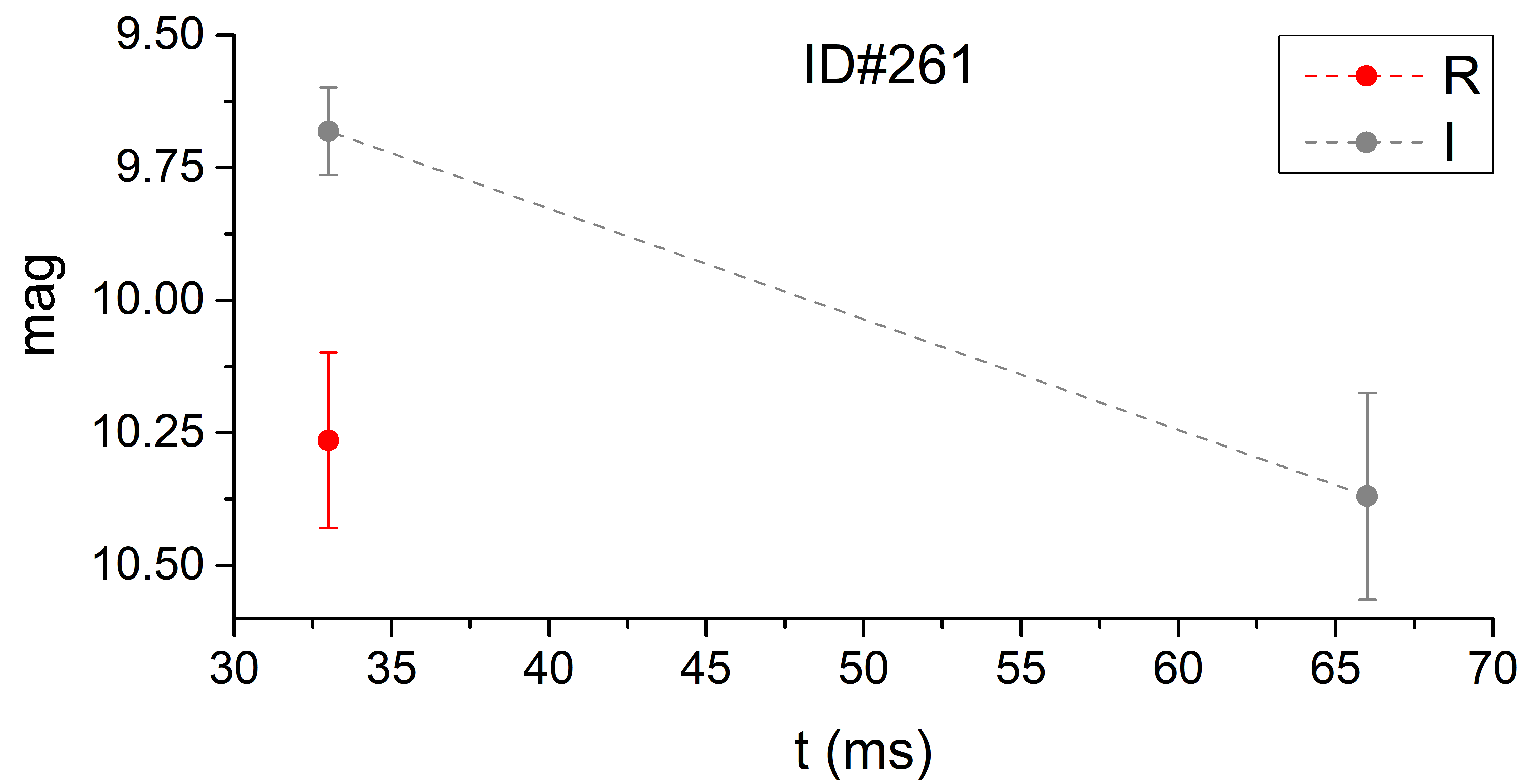}\\
\includegraphics[width=5.6cm]{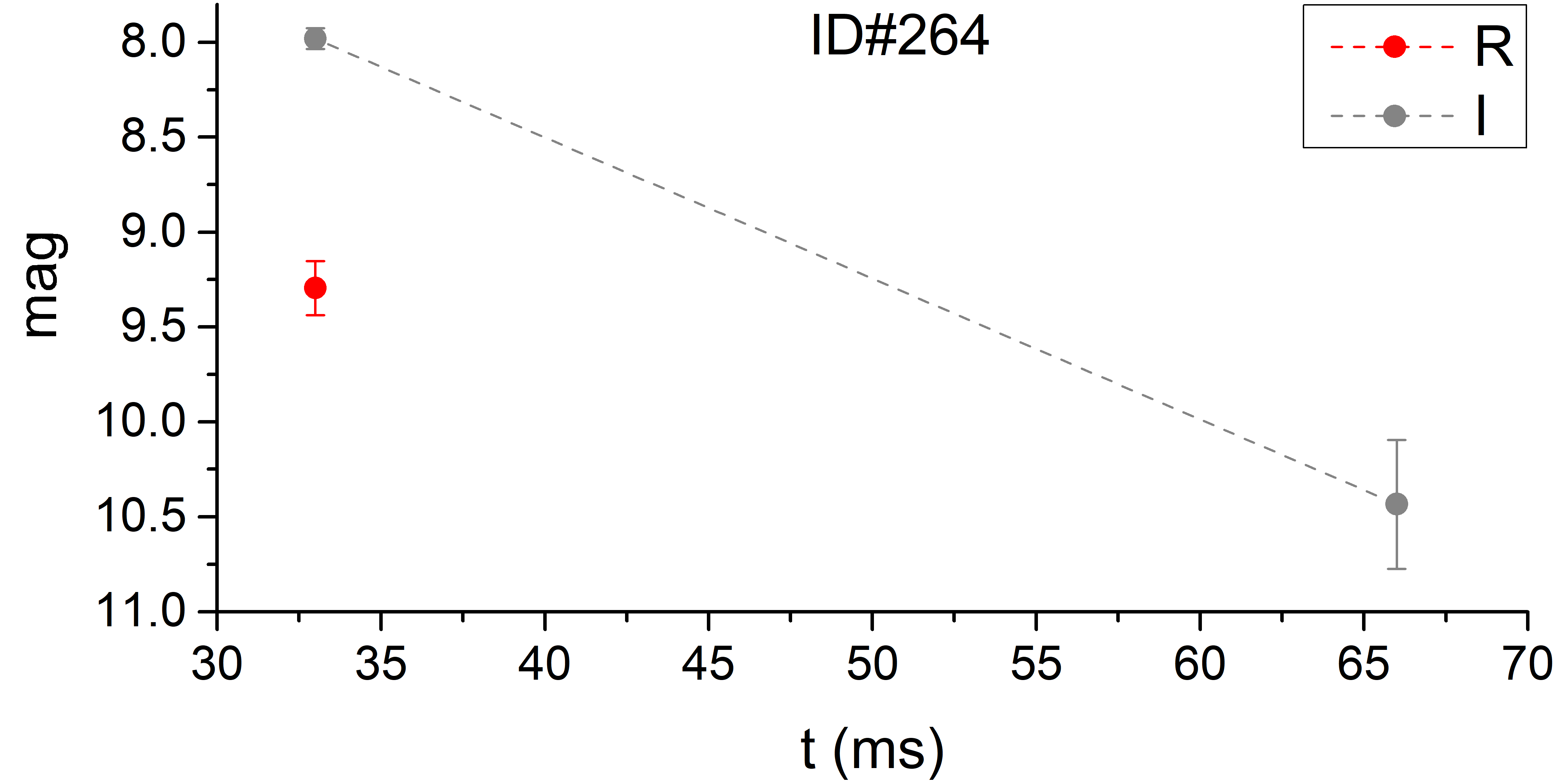}&\includegraphics[width=5.6cm]{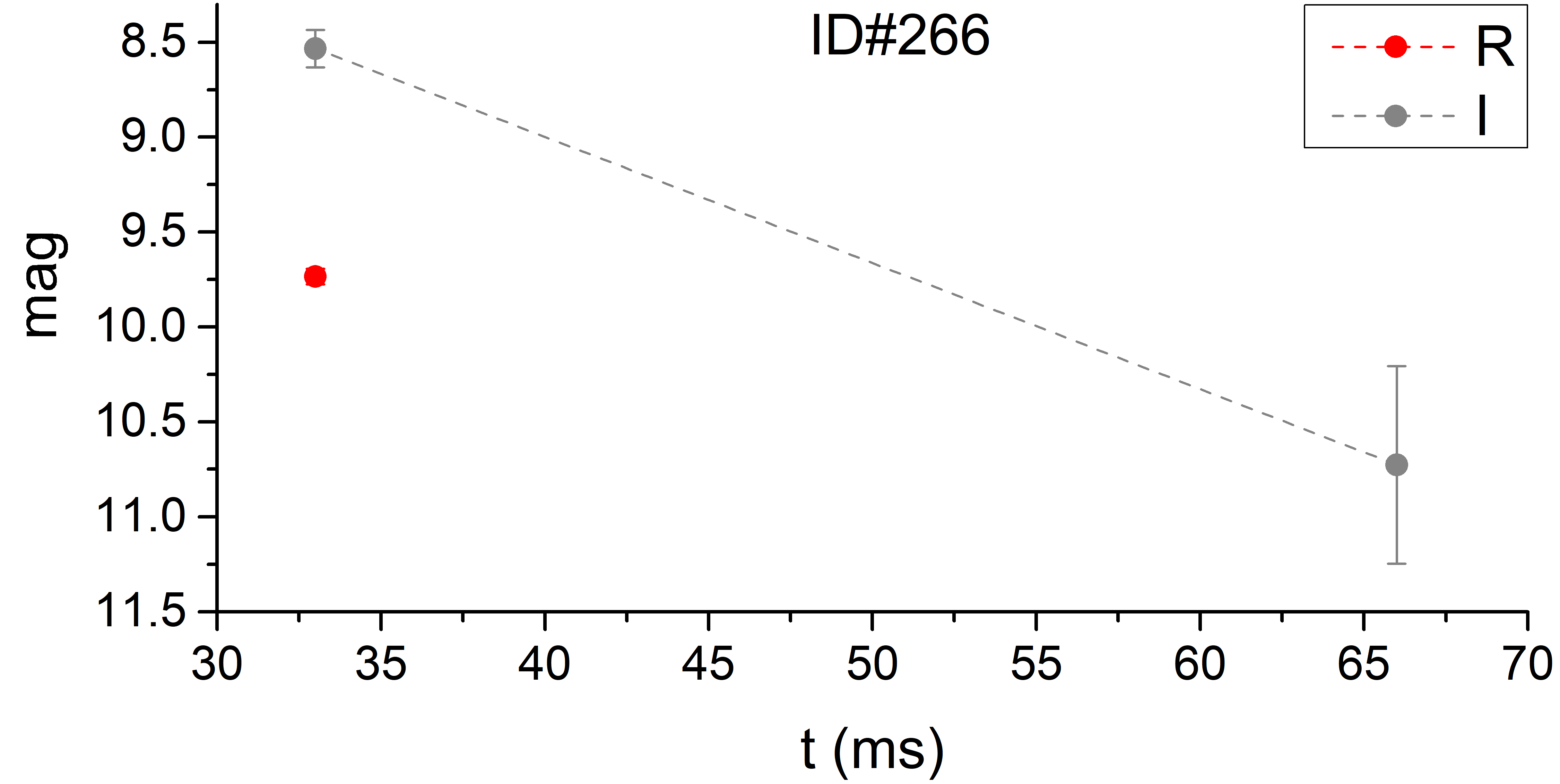}&\includegraphics[width=5.6cm]{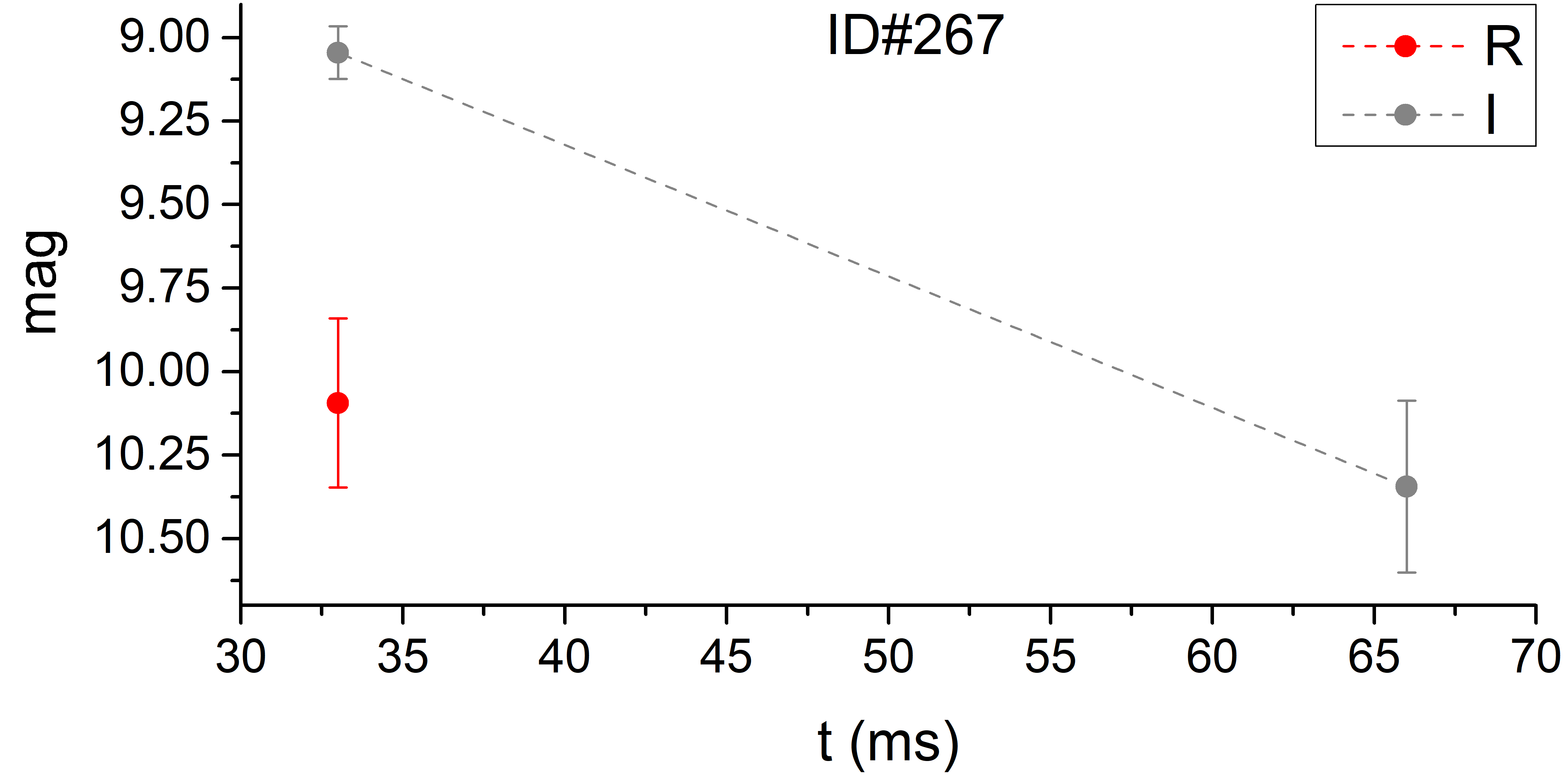}\\
\includegraphics[width=5.6cm]{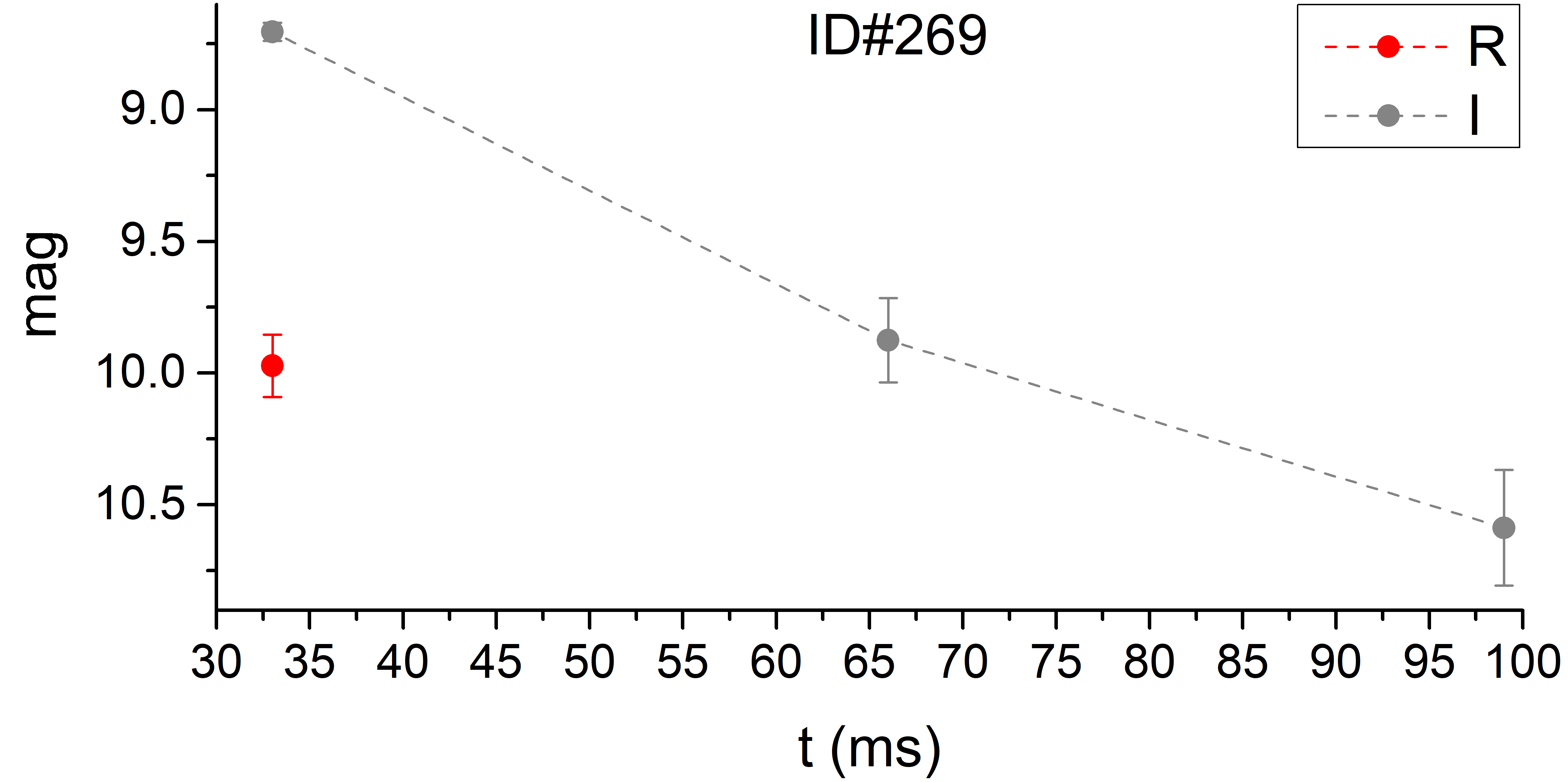}&\includegraphics[width=5.6cm]{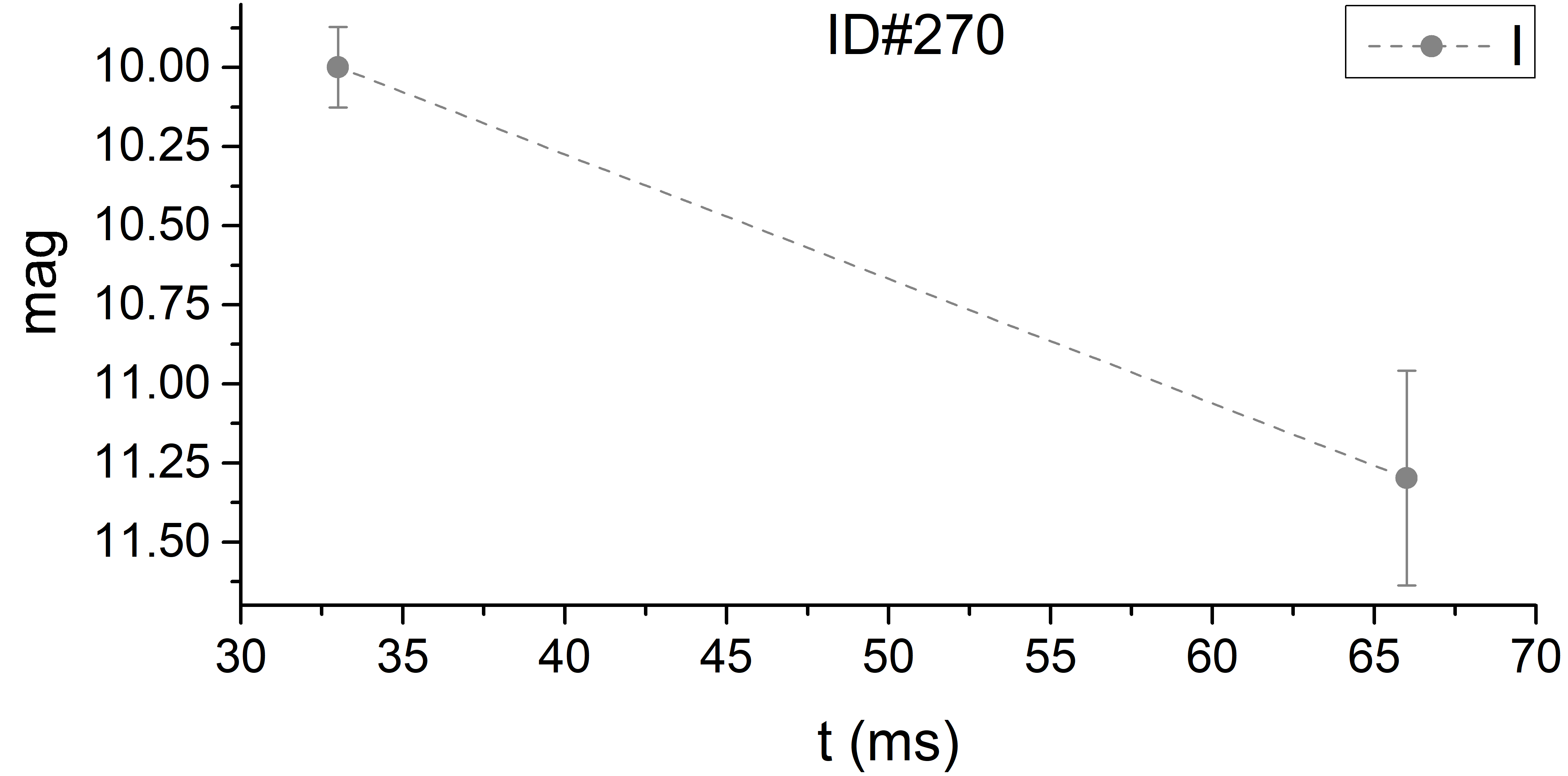}&\includegraphics[width=5.6cm]{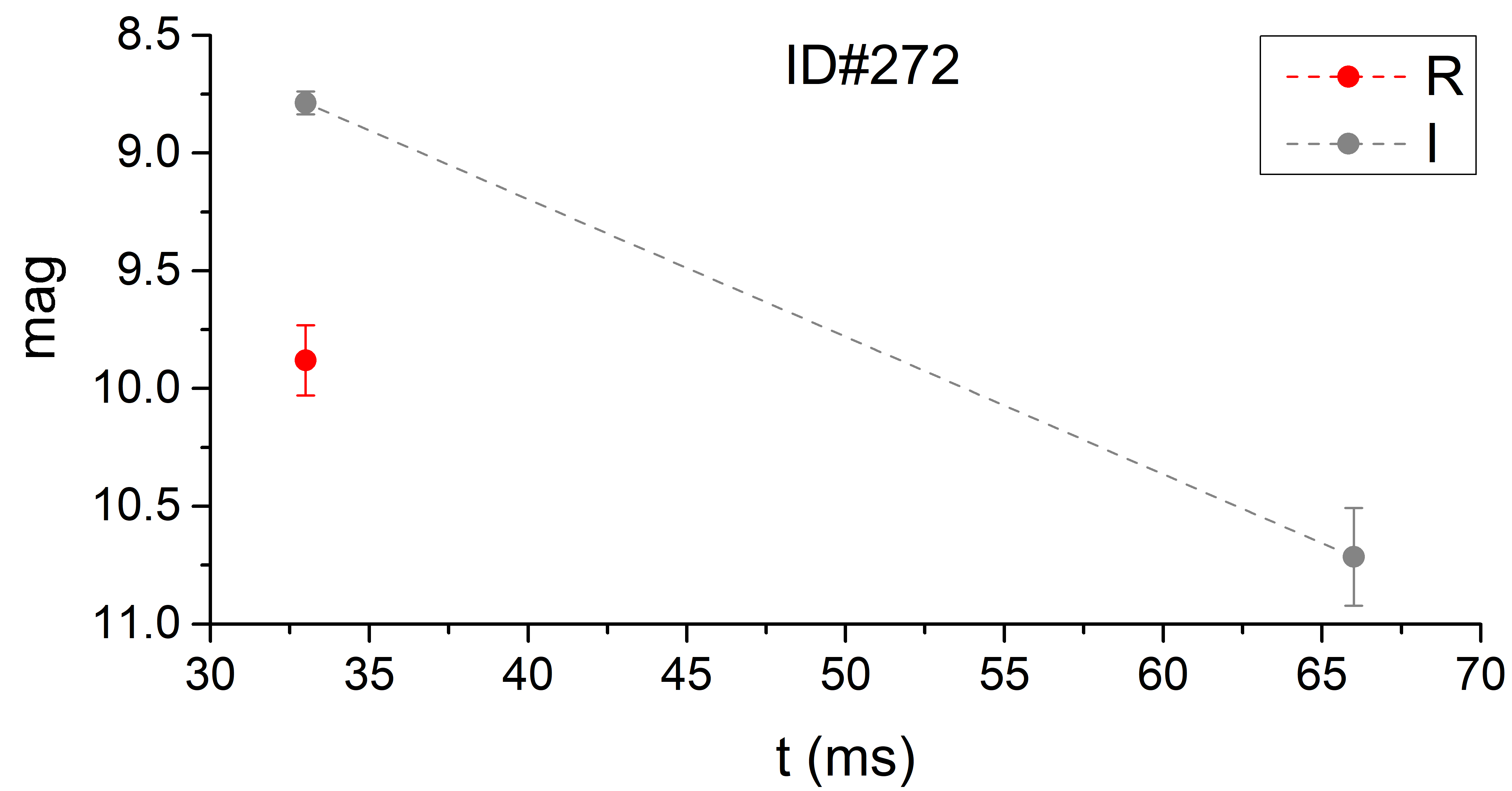}\\
\includegraphics[width=5.6cm]{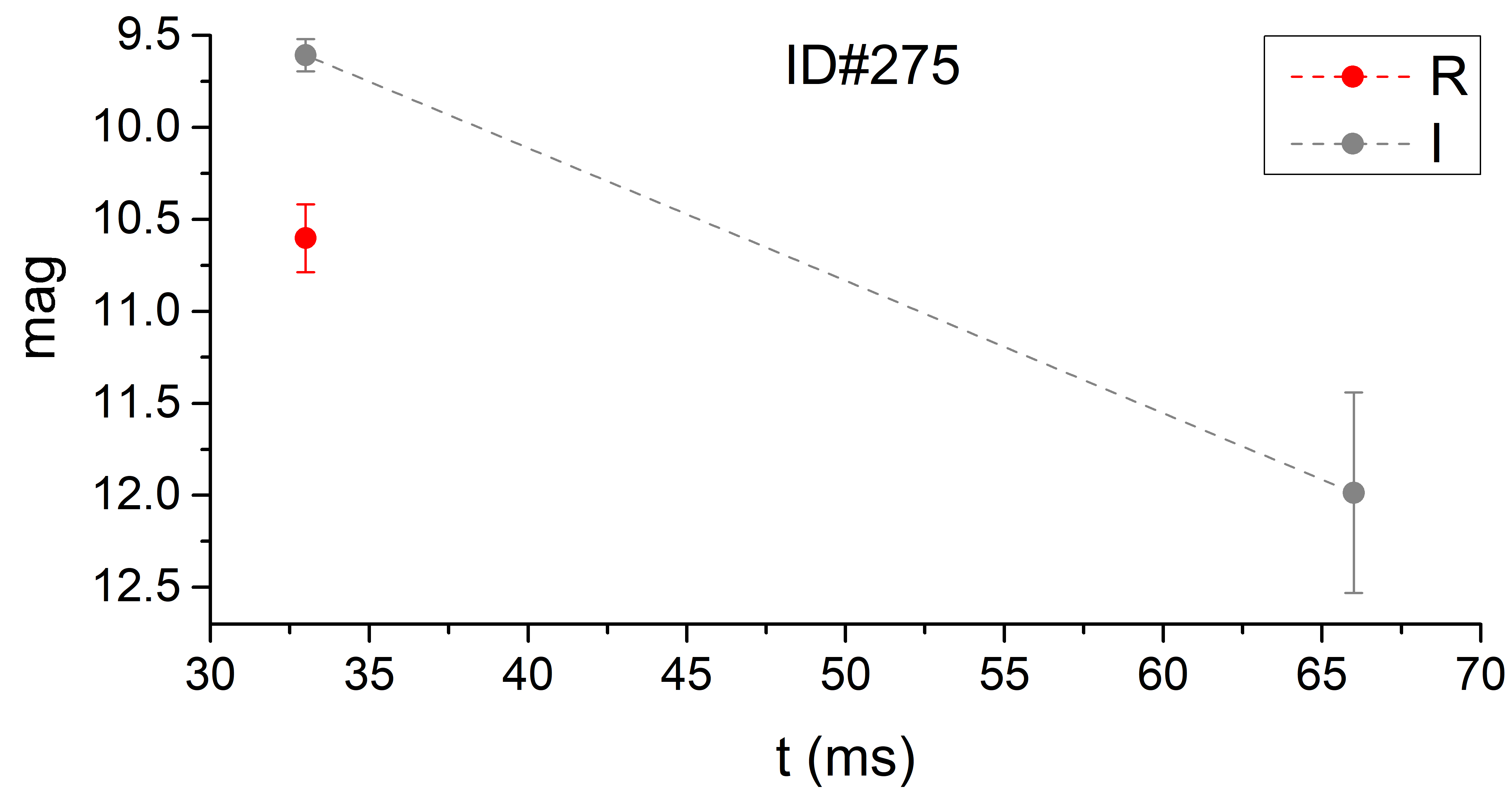}&\includegraphics[width=5.6cm]{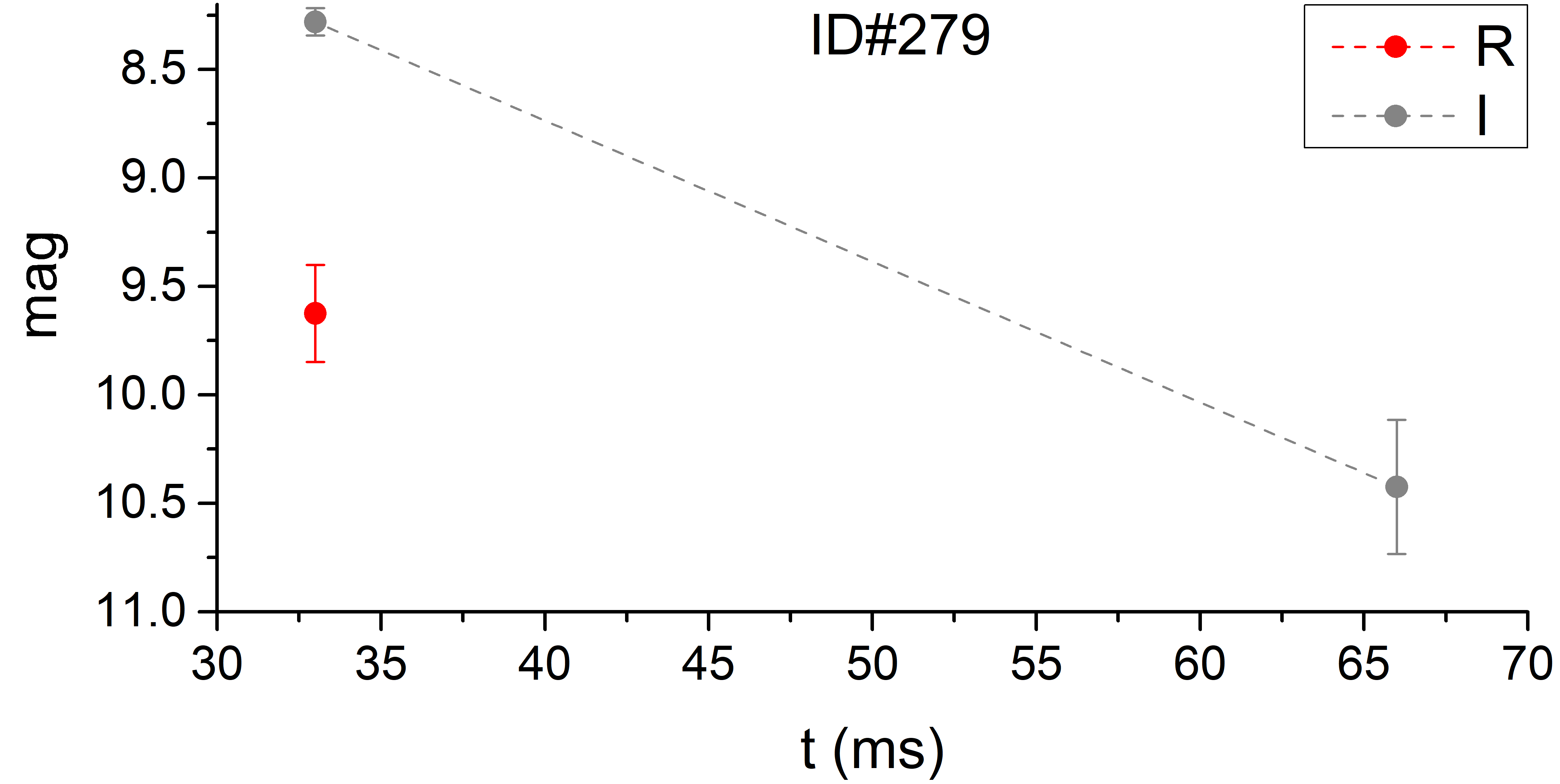}&\includegraphics[width=5.6cm]{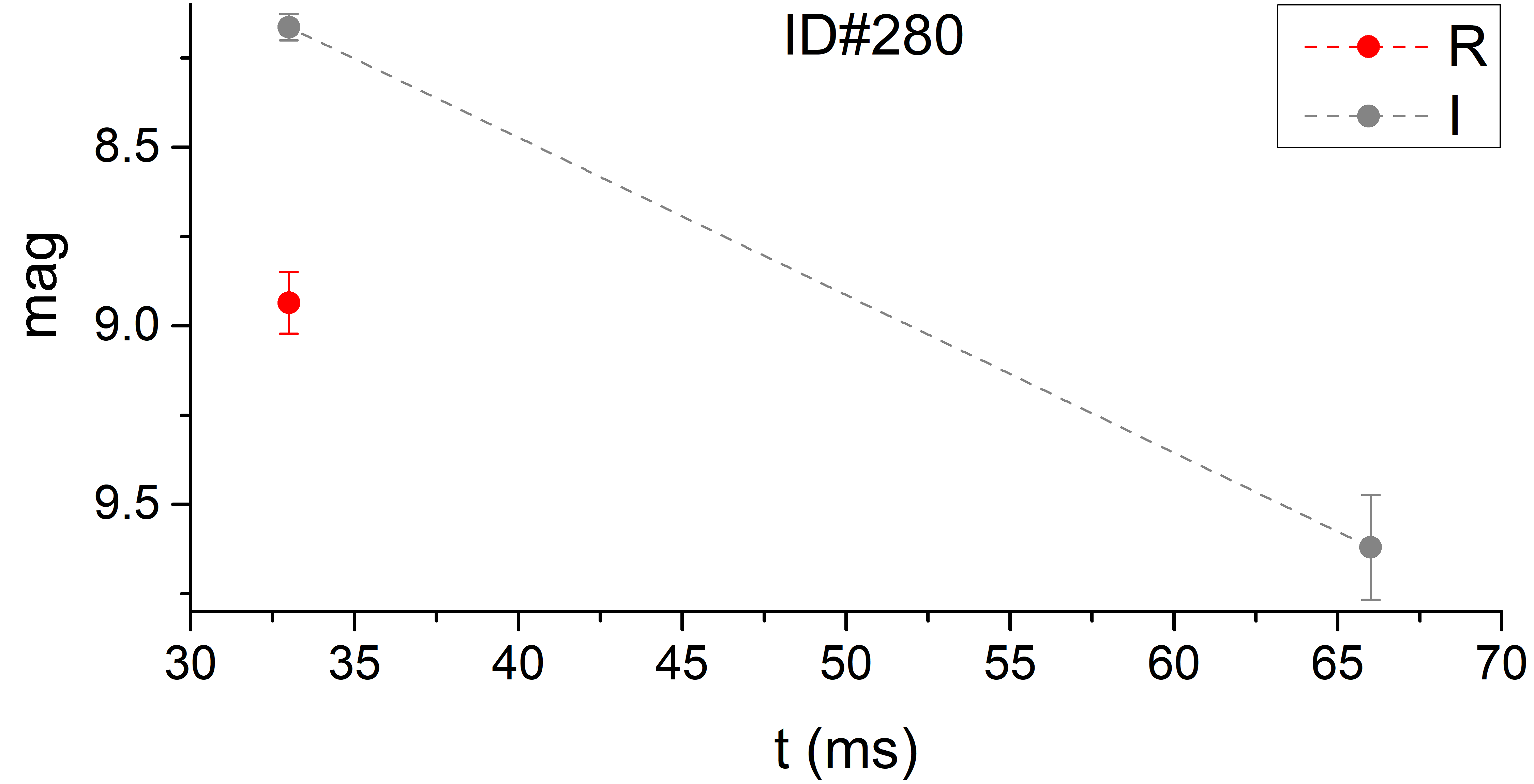}\\
\includegraphics[width=5.6cm]{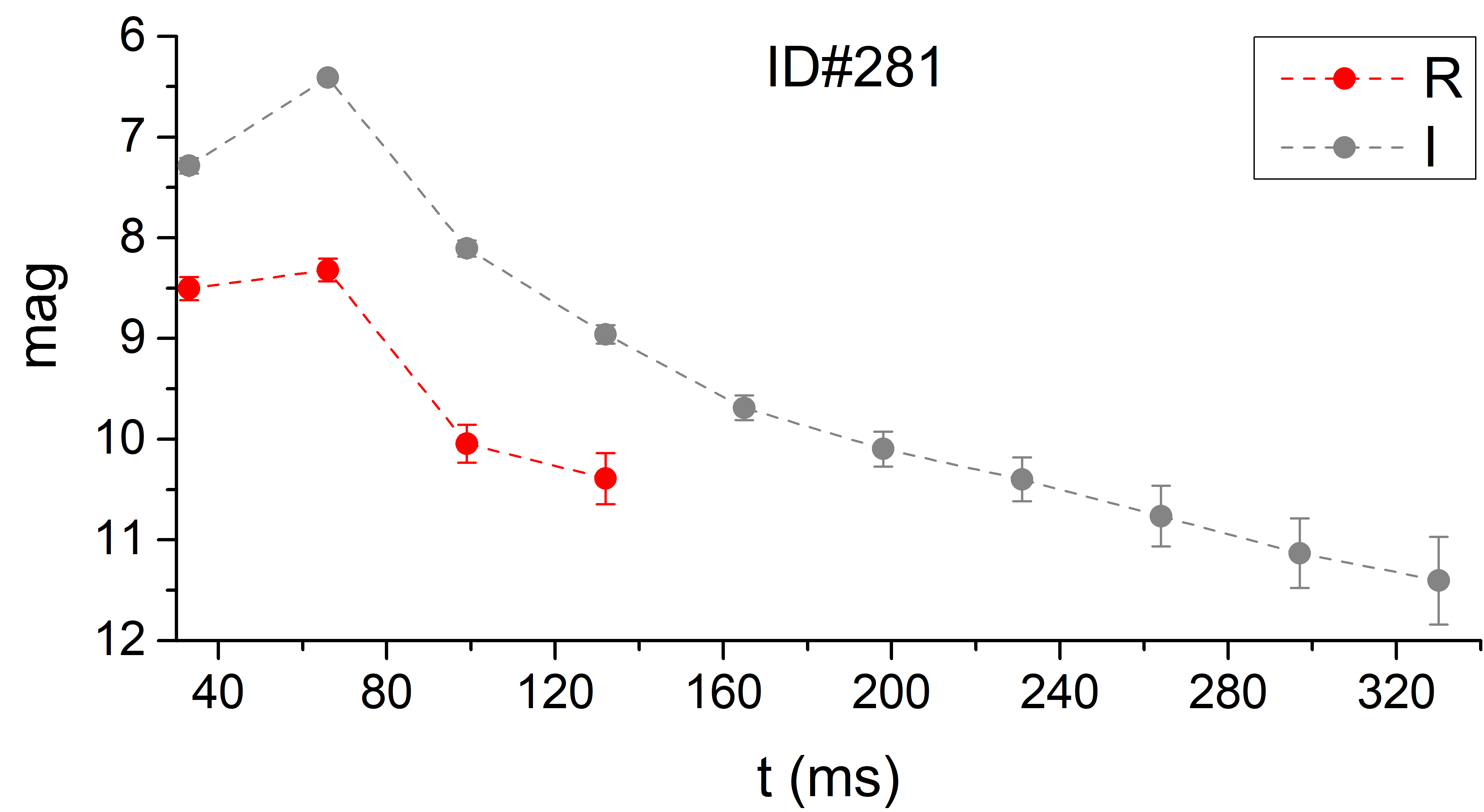}&\includegraphics[width=5.6cm]{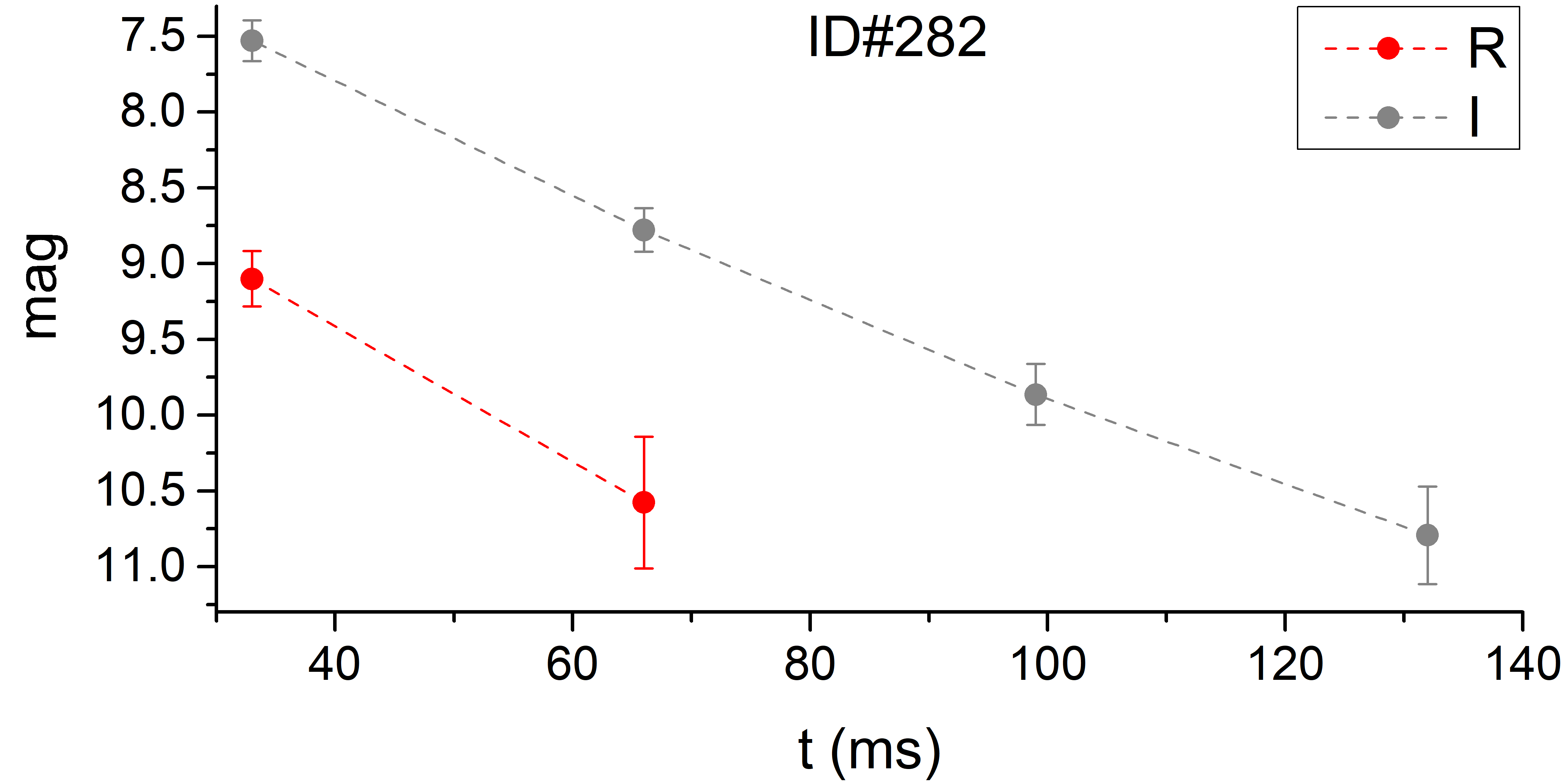}&\includegraphics[width=5.6cm]{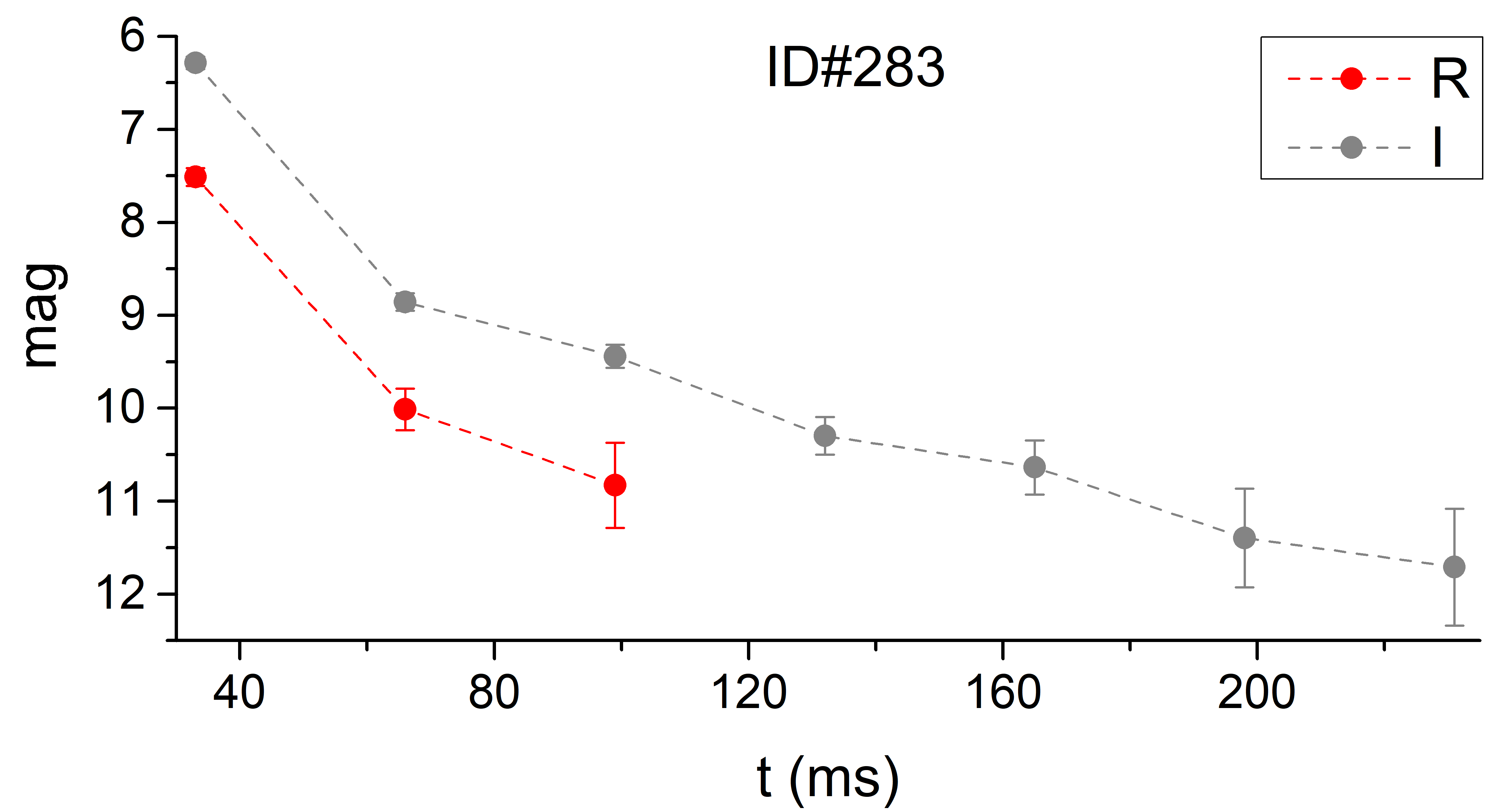}\\
\includegraphics[width=5.6cm]{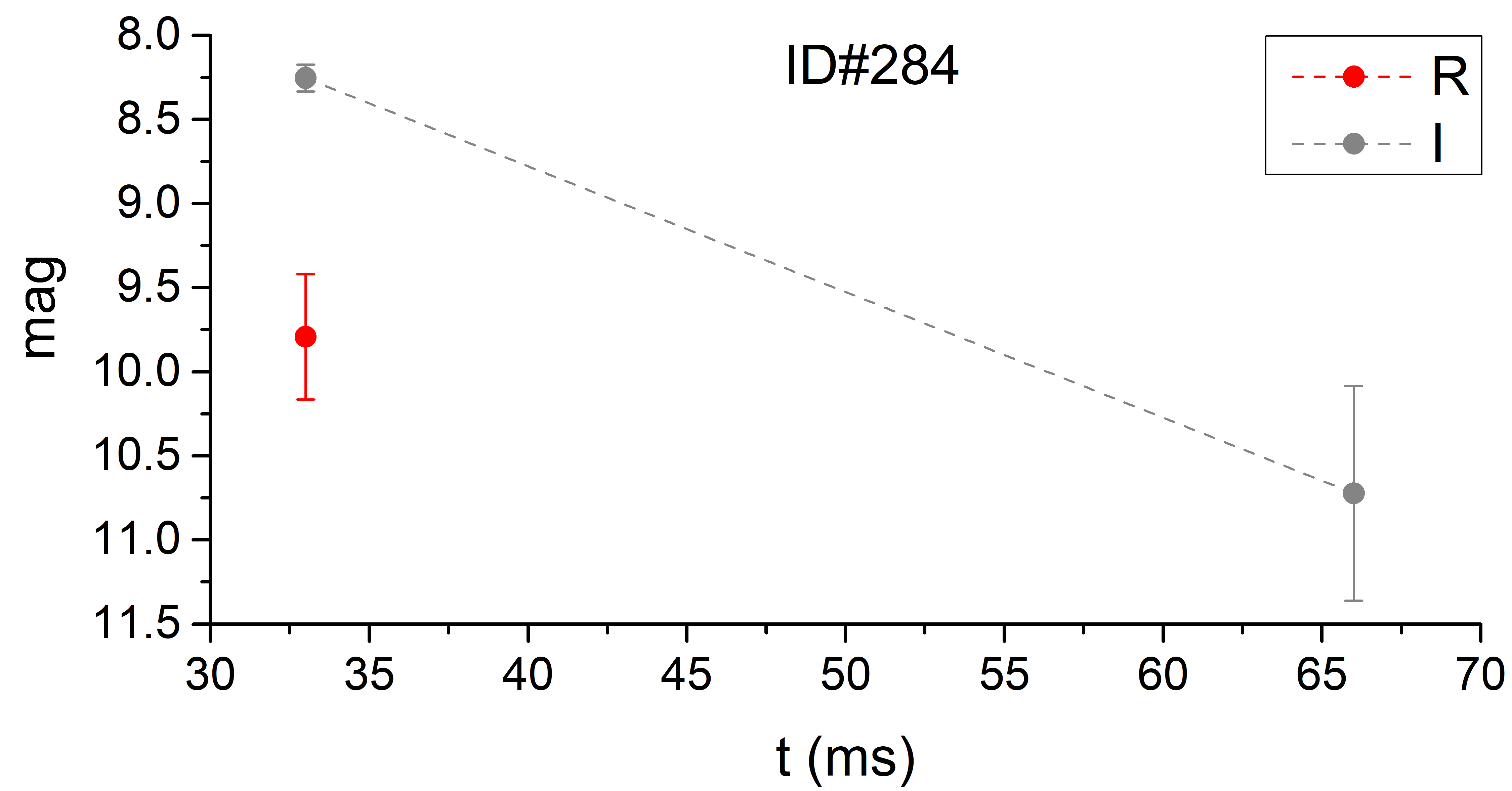}&\includegraphics[width=5.6cm]{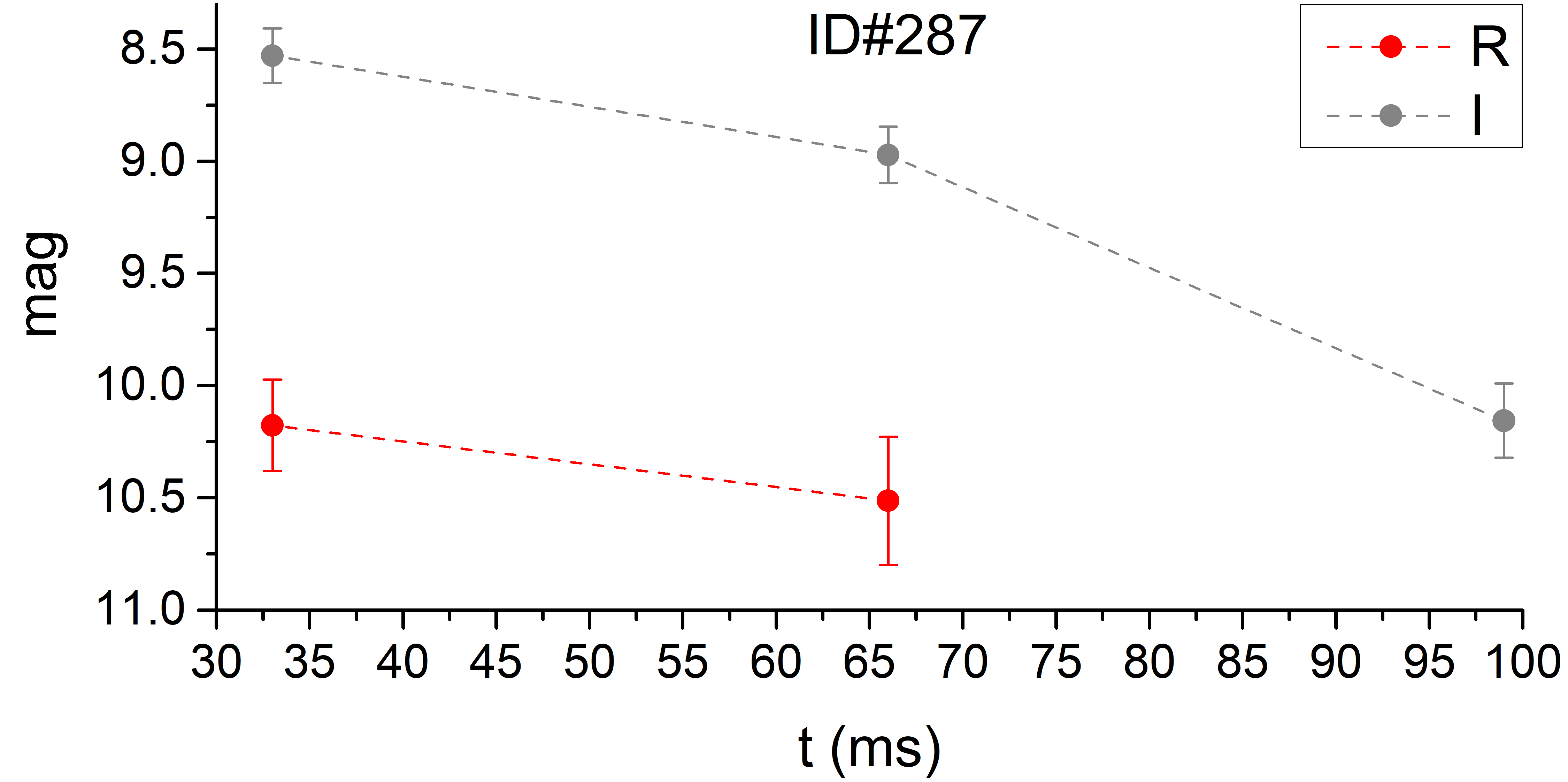}&\includegraphics[width=5.6cm]{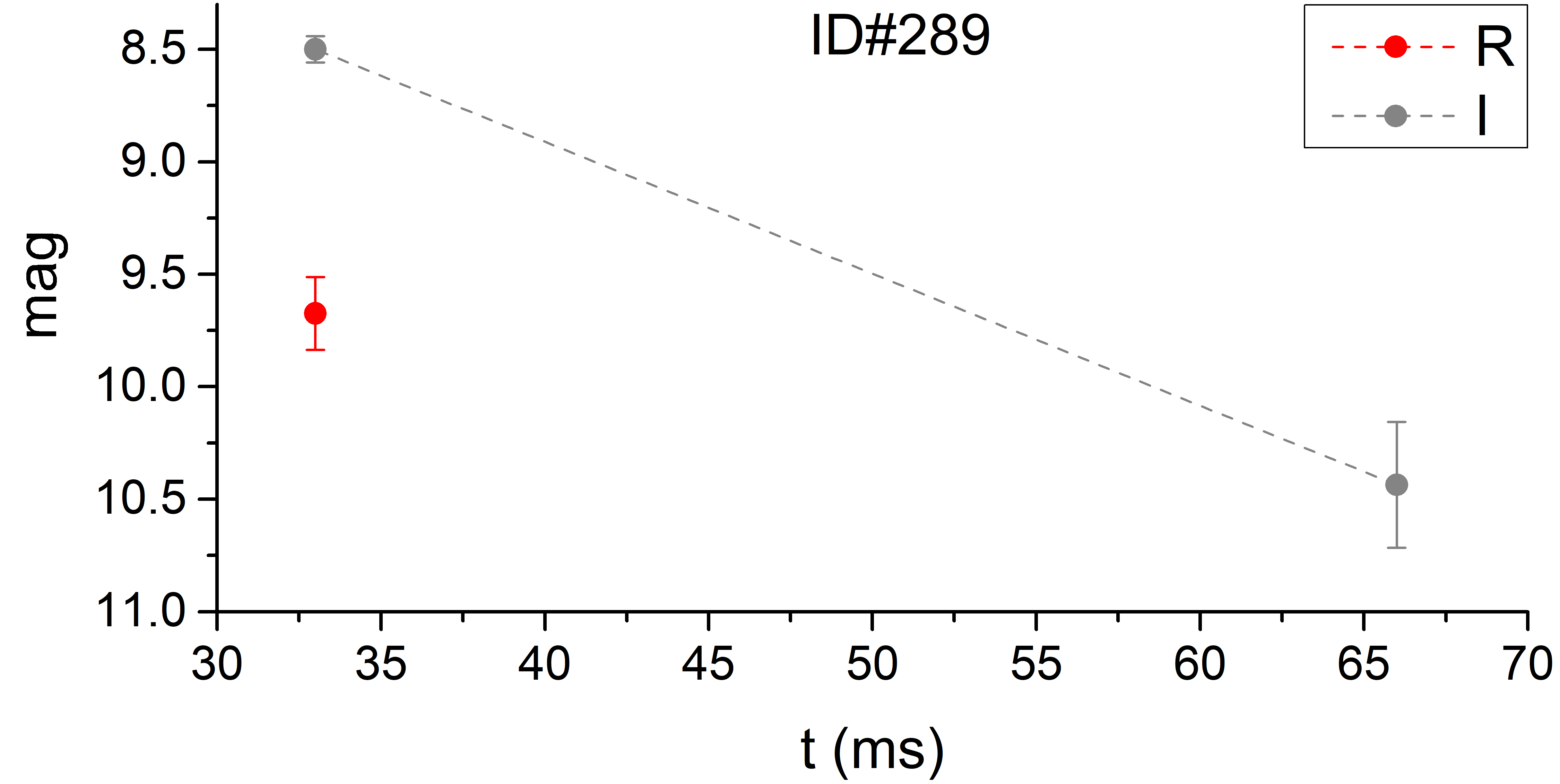}\\
\end{tabular}
\caption*{Fig.~\ref{fig:LCs1}~(cont'd)}
\end{figure*}

\begin{figure*}[h]
\begin{tabular}{ccc}
\includegraphics[width=5.6cm]{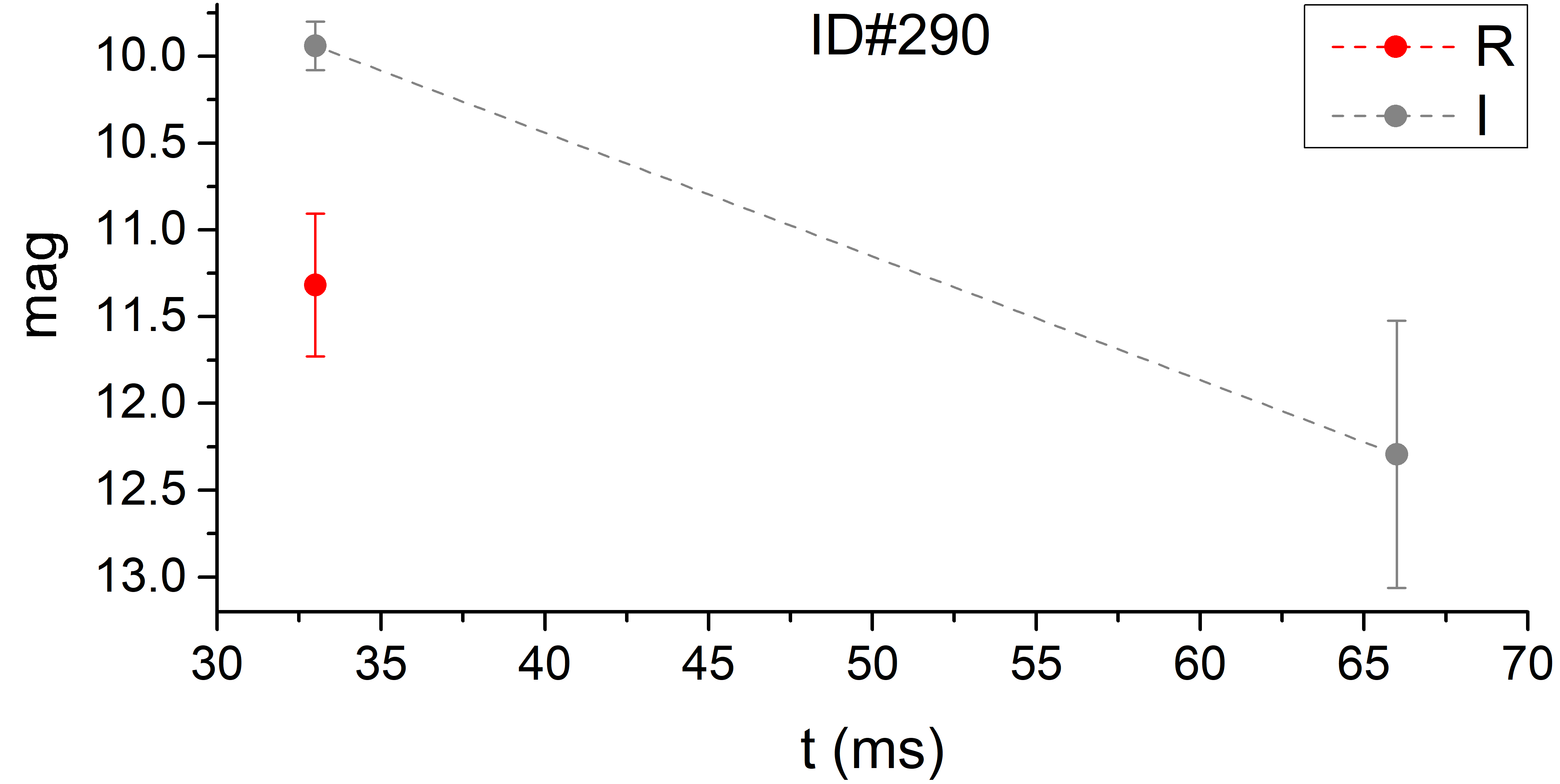}&\includegraphics[width=5.6cm]{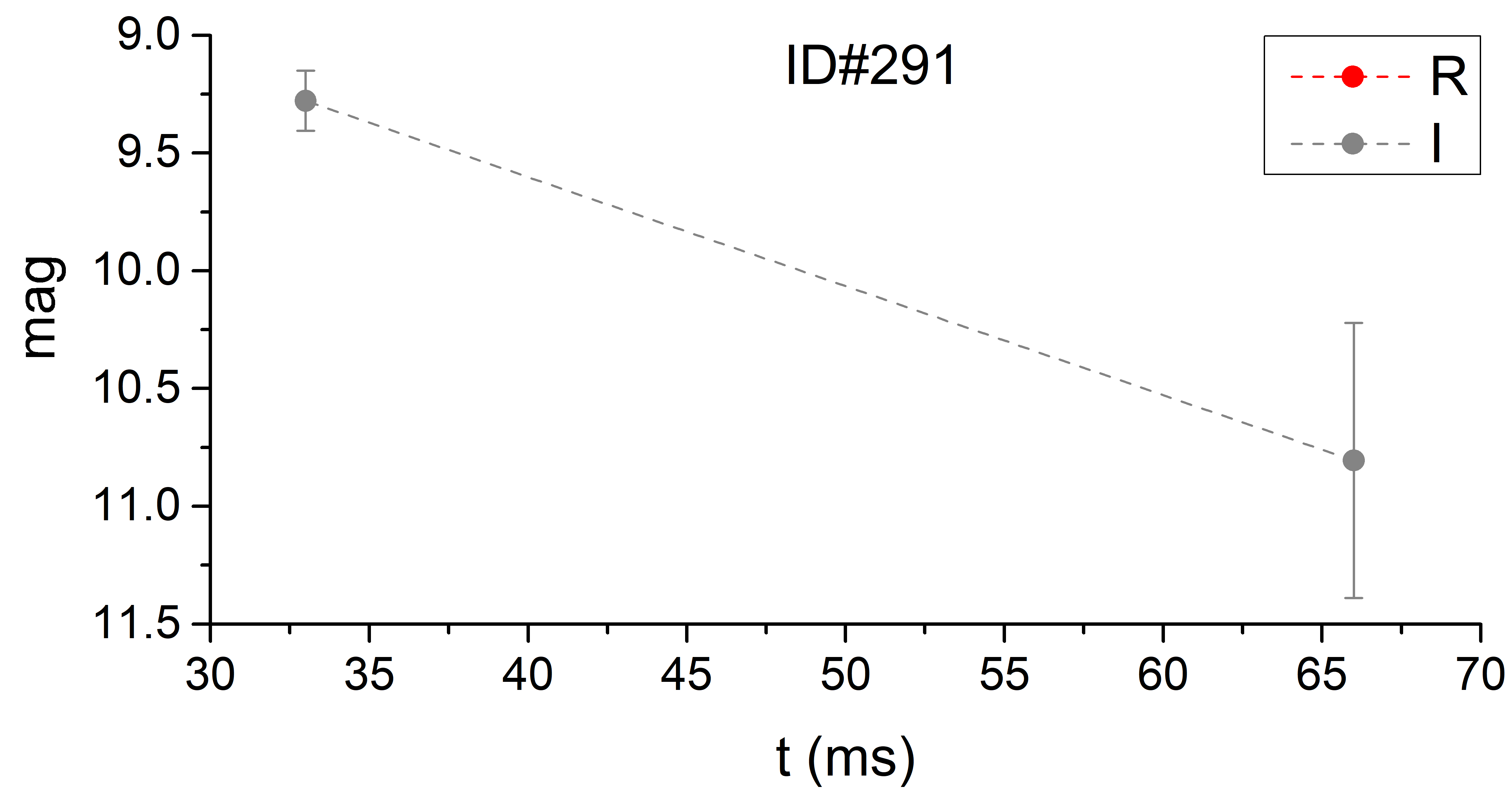}&\includegraphics[width=5.6cm]{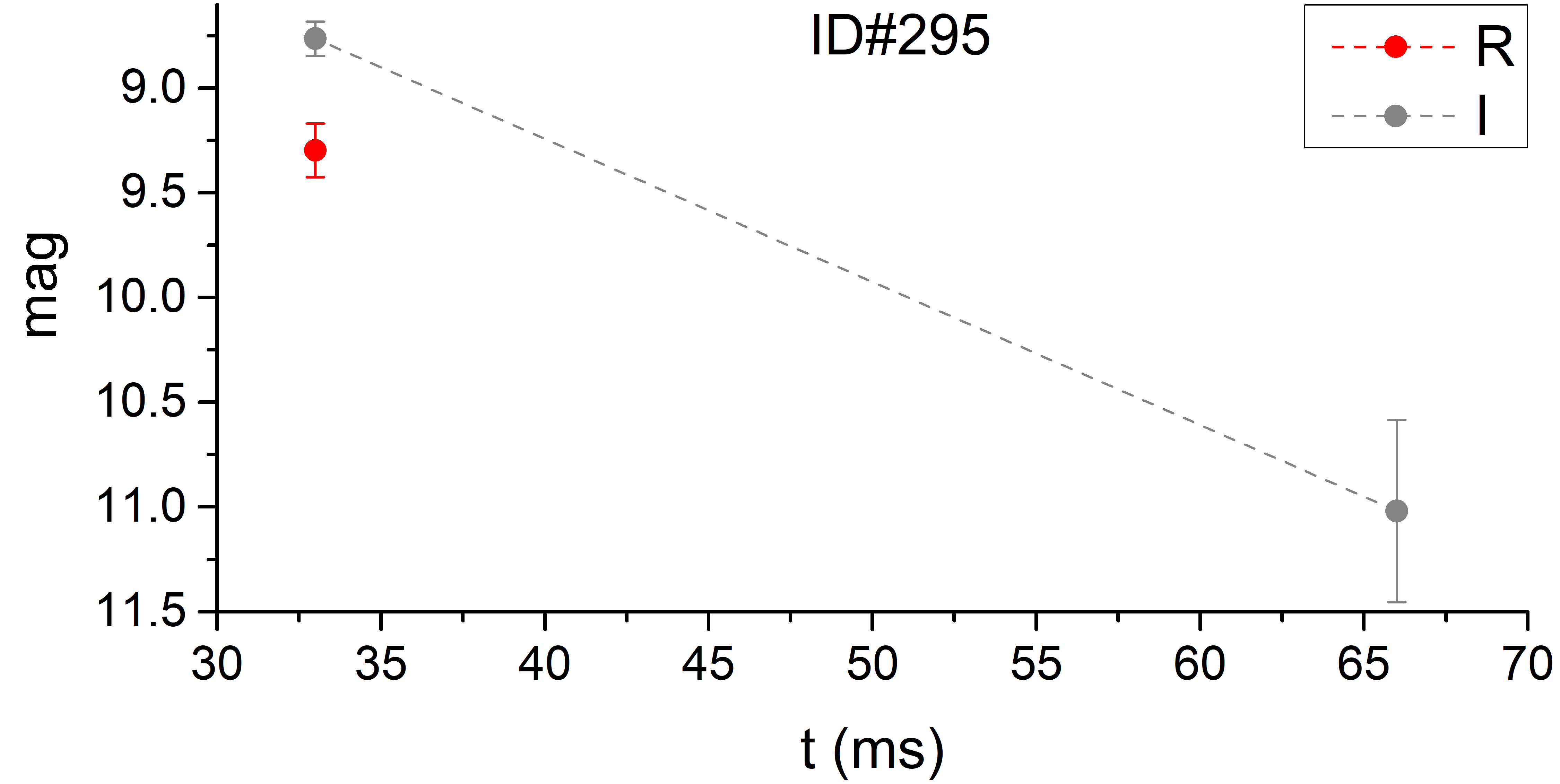}\\
\end{tabular}
\caption*{Fig.~\ref{fig:LCs1}~(cont'd)}
\end{figure*}

\begin{figure*}[h!]
\begin{tabular}{ccc}
\includegraphics[width=5.6cm]{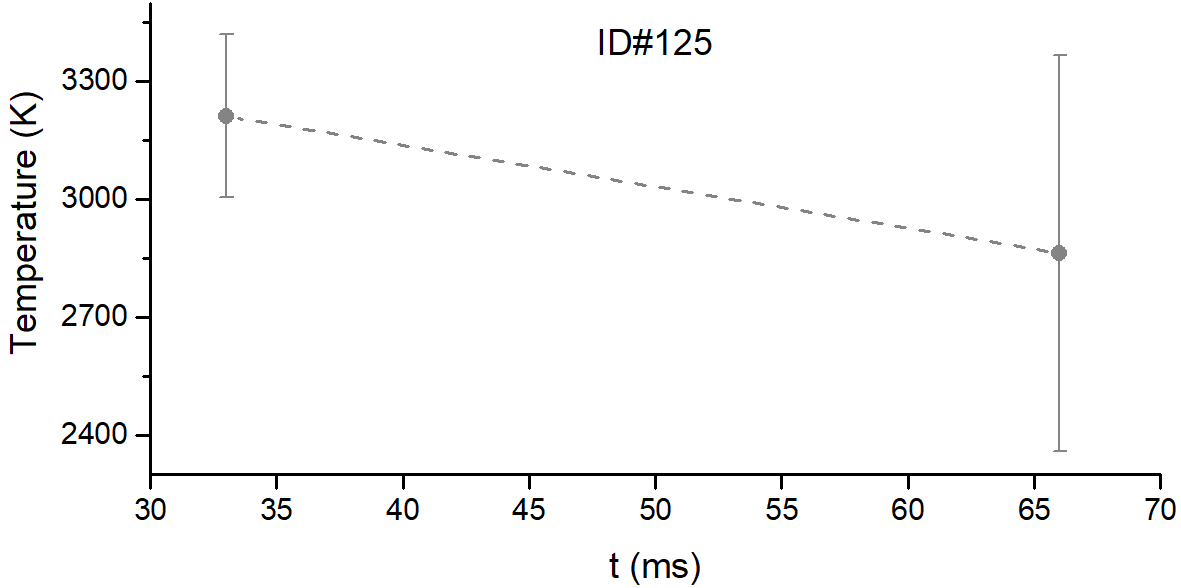}&\includegraphics[width=5.6cm]{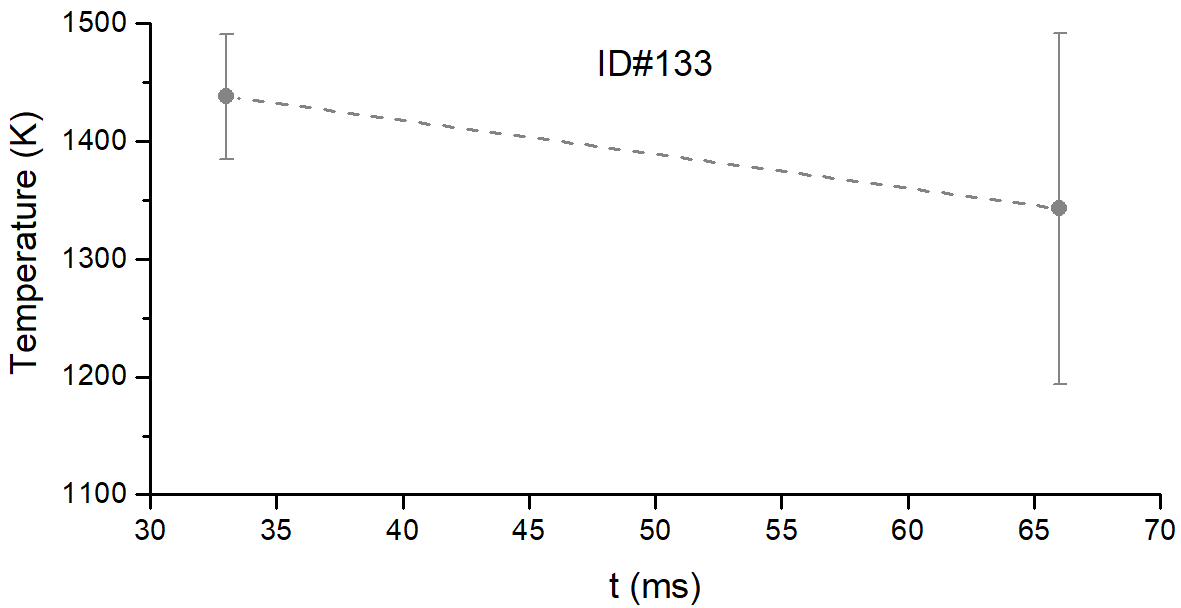}&\includegraphics[width=5.6cm]{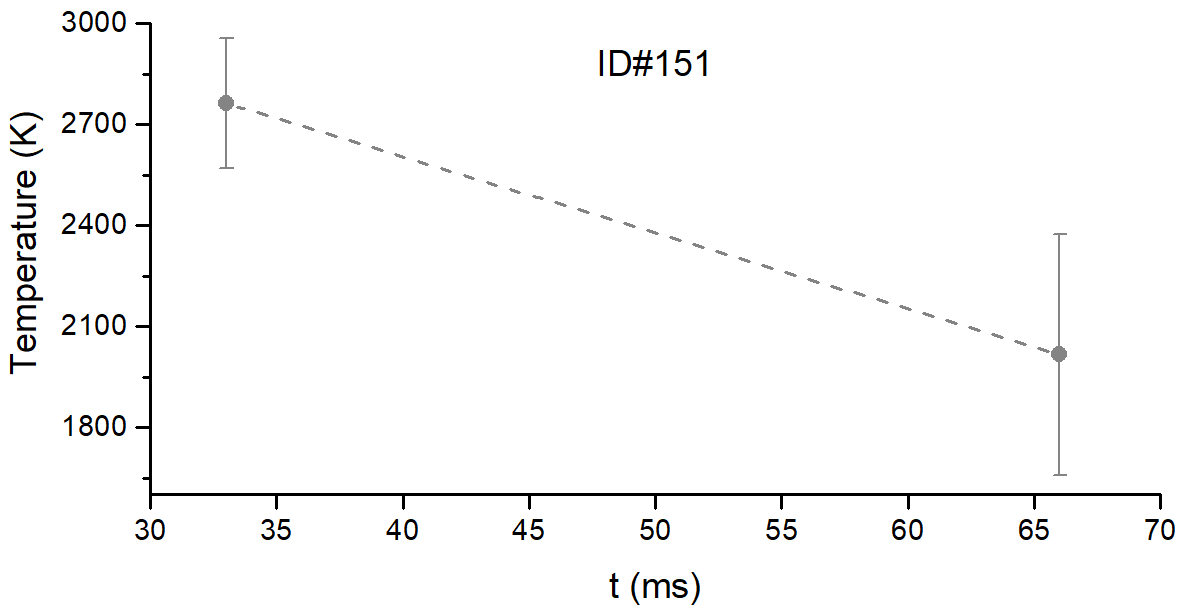}\\
\includegraphics[width=5.6cm]{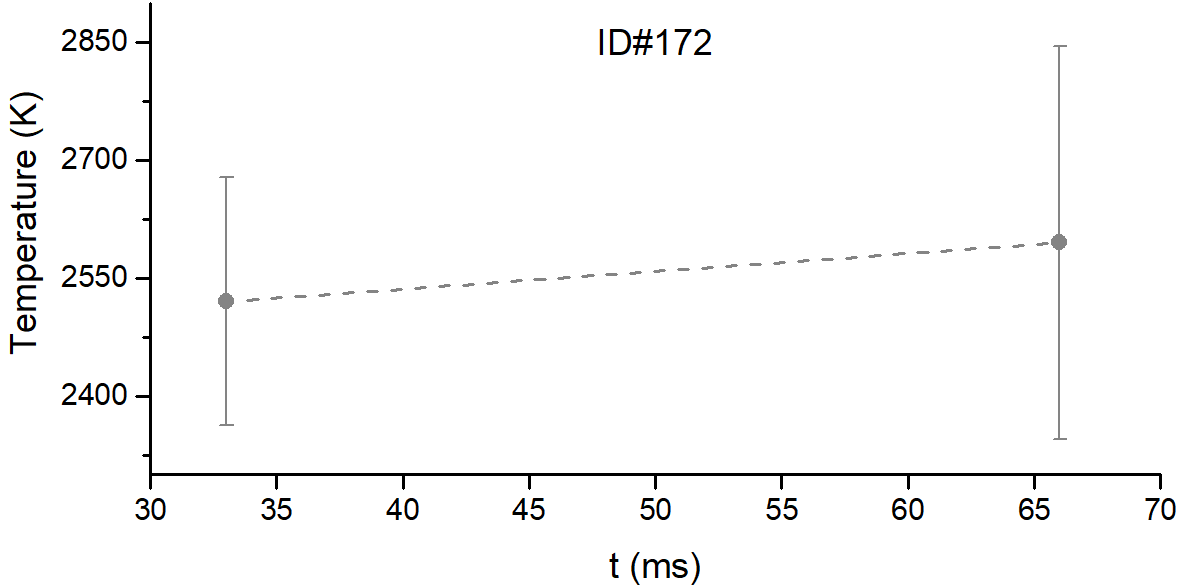}&\includegraphics[width=5.6cm]{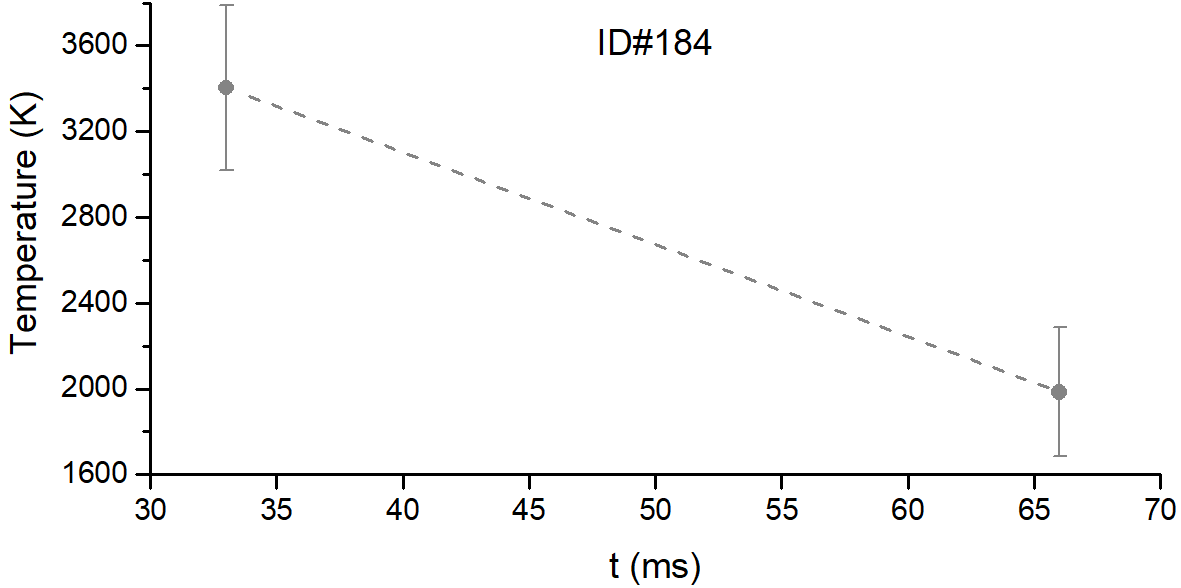}&\includegraphics[width=5.6cm]{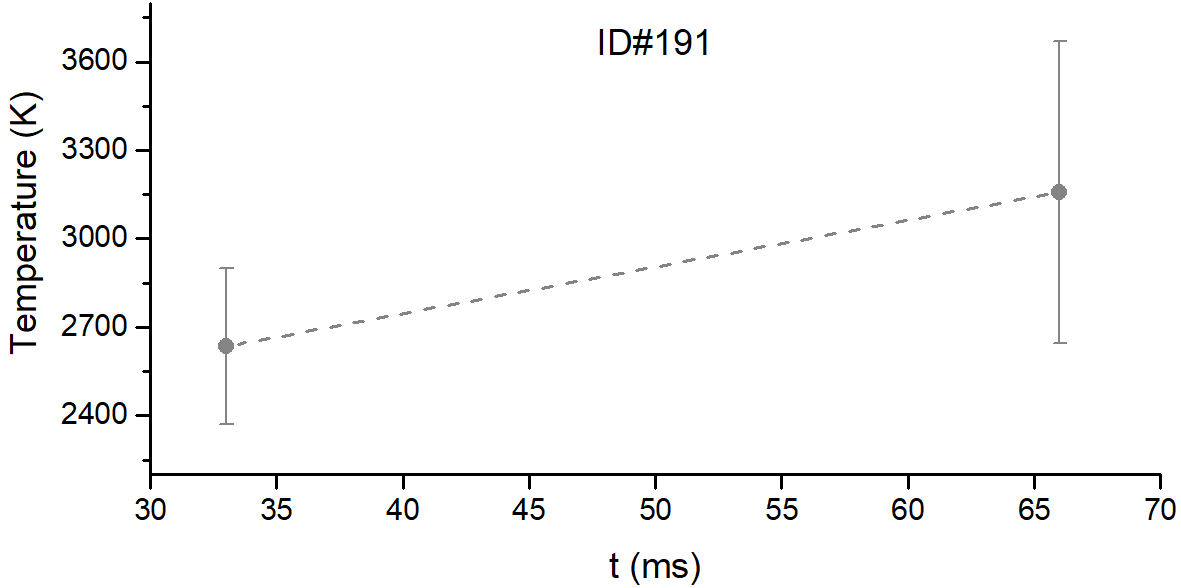}\\
\includegraphics[width=5.6cm]{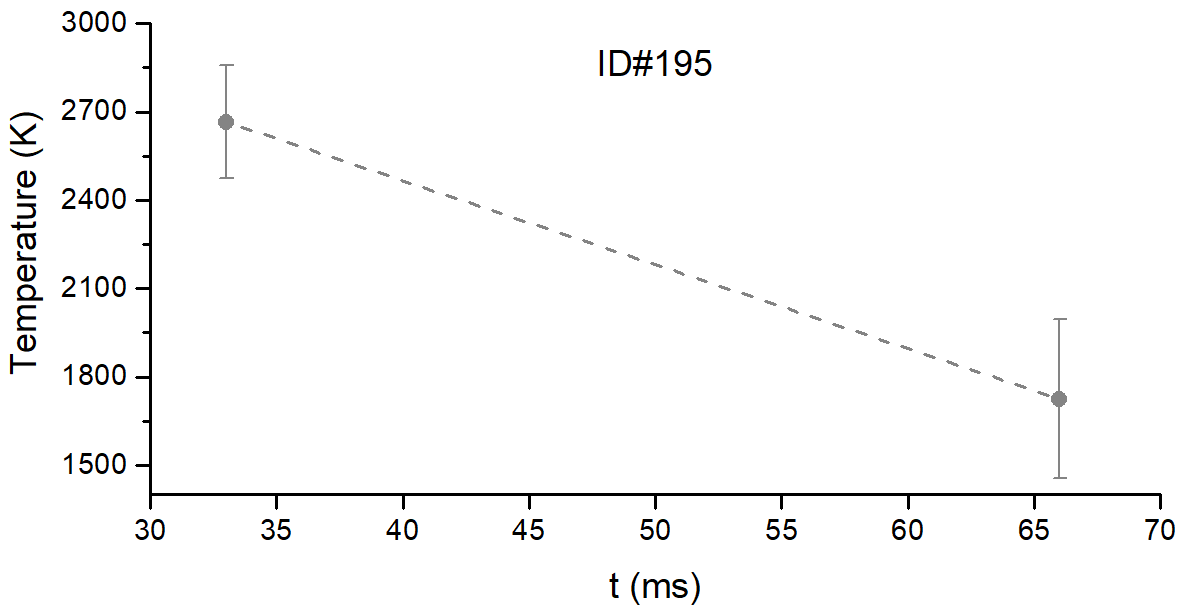}&\includegraphics[width=5.6cm]{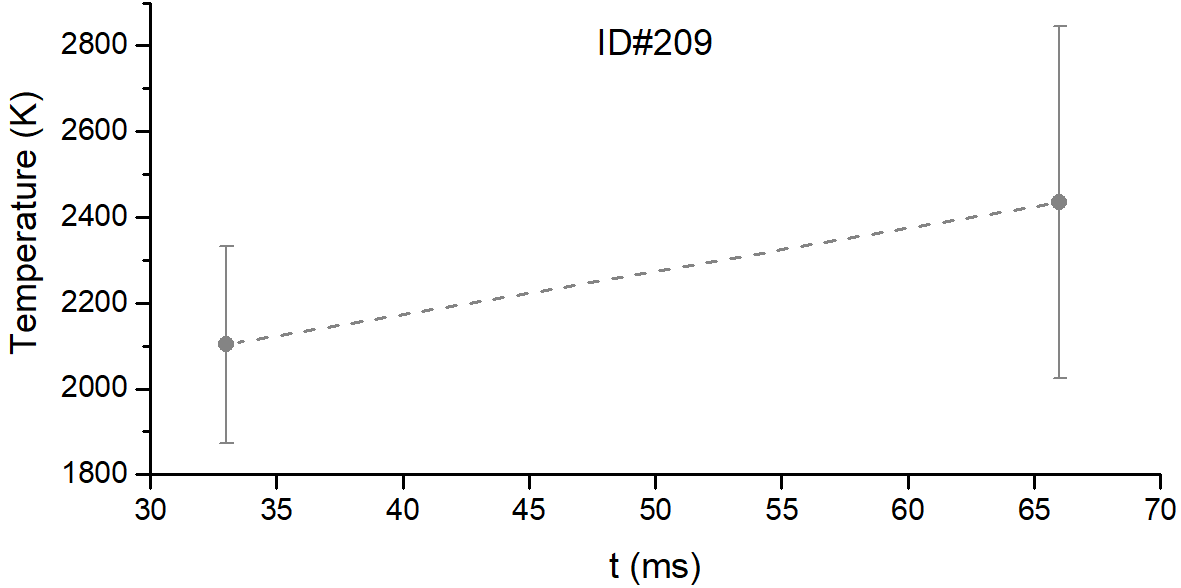}&\includegraphics[width=5.6cm]{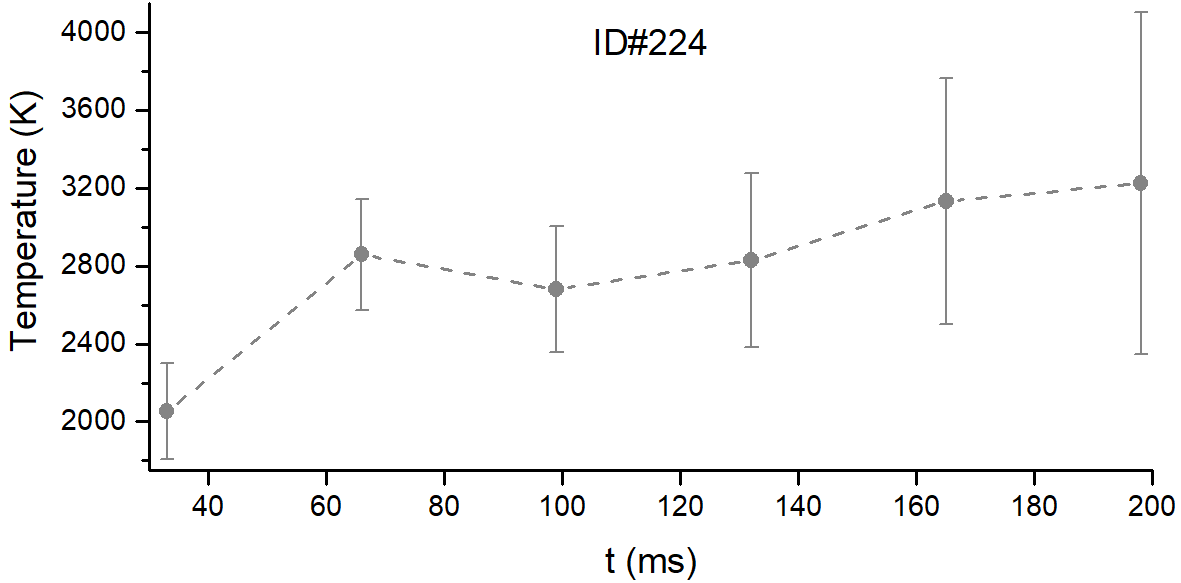}\\
\includegraphics[width=5.6cm]{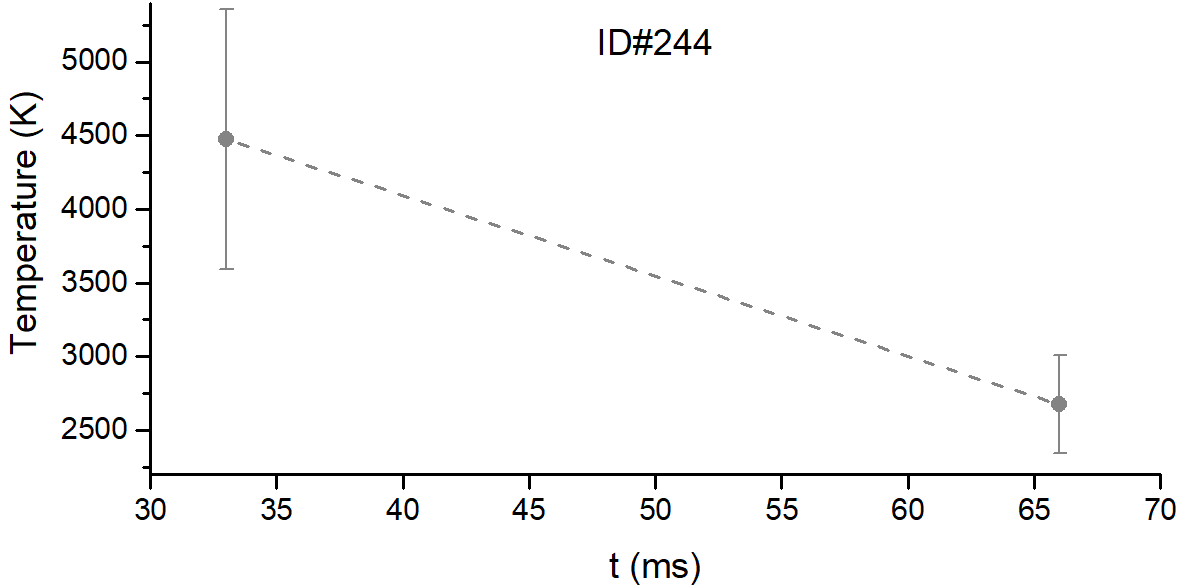}&\includegraphics[width=5.6cm]{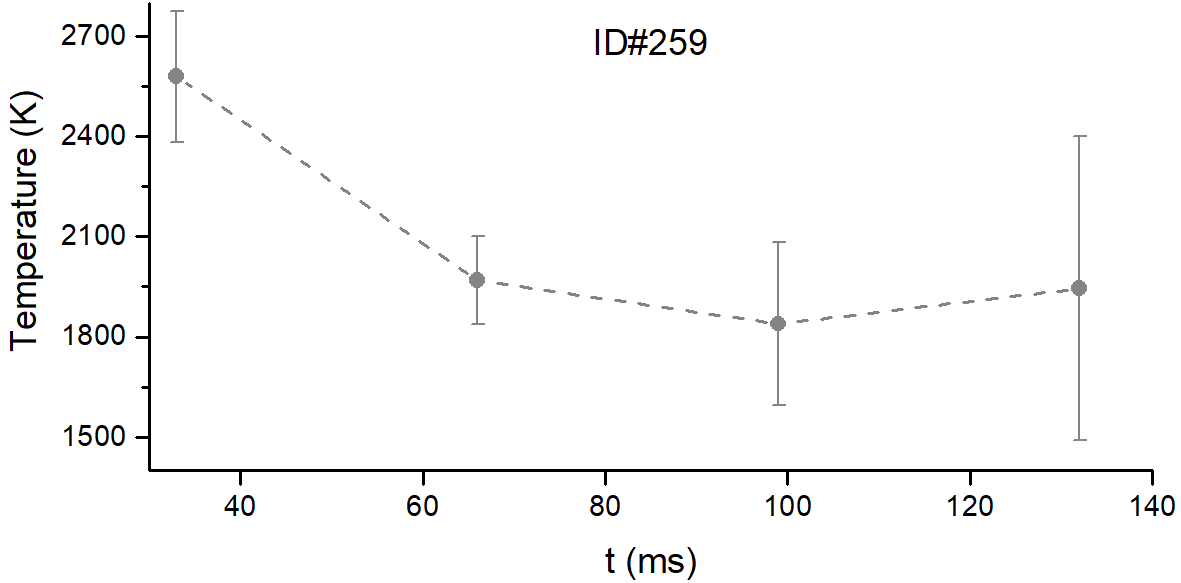}&\includegraphics[width=5.6cm]{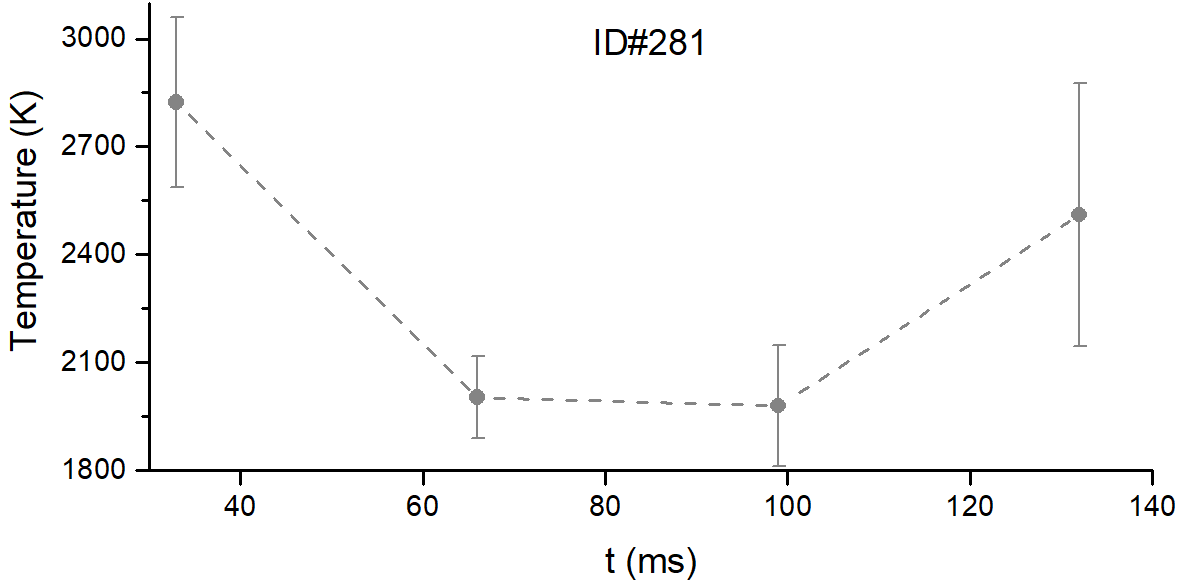}\\
\includegraphics[width=5.6cm]{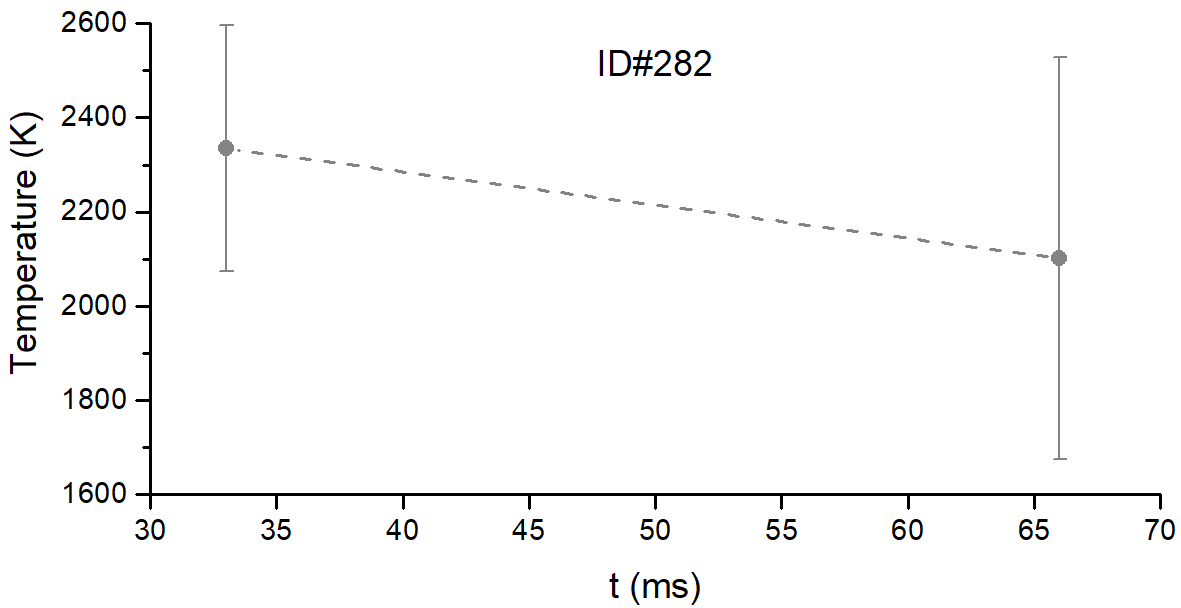}&\includegraphics[width=5.6cm]{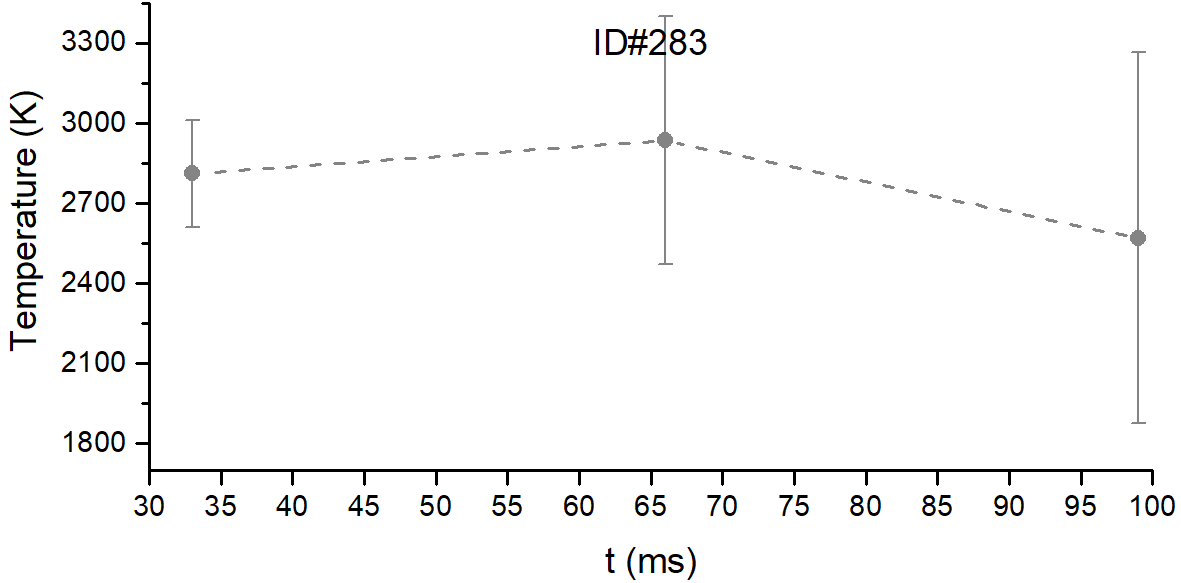}&\includegraphics[width=5.6cm]{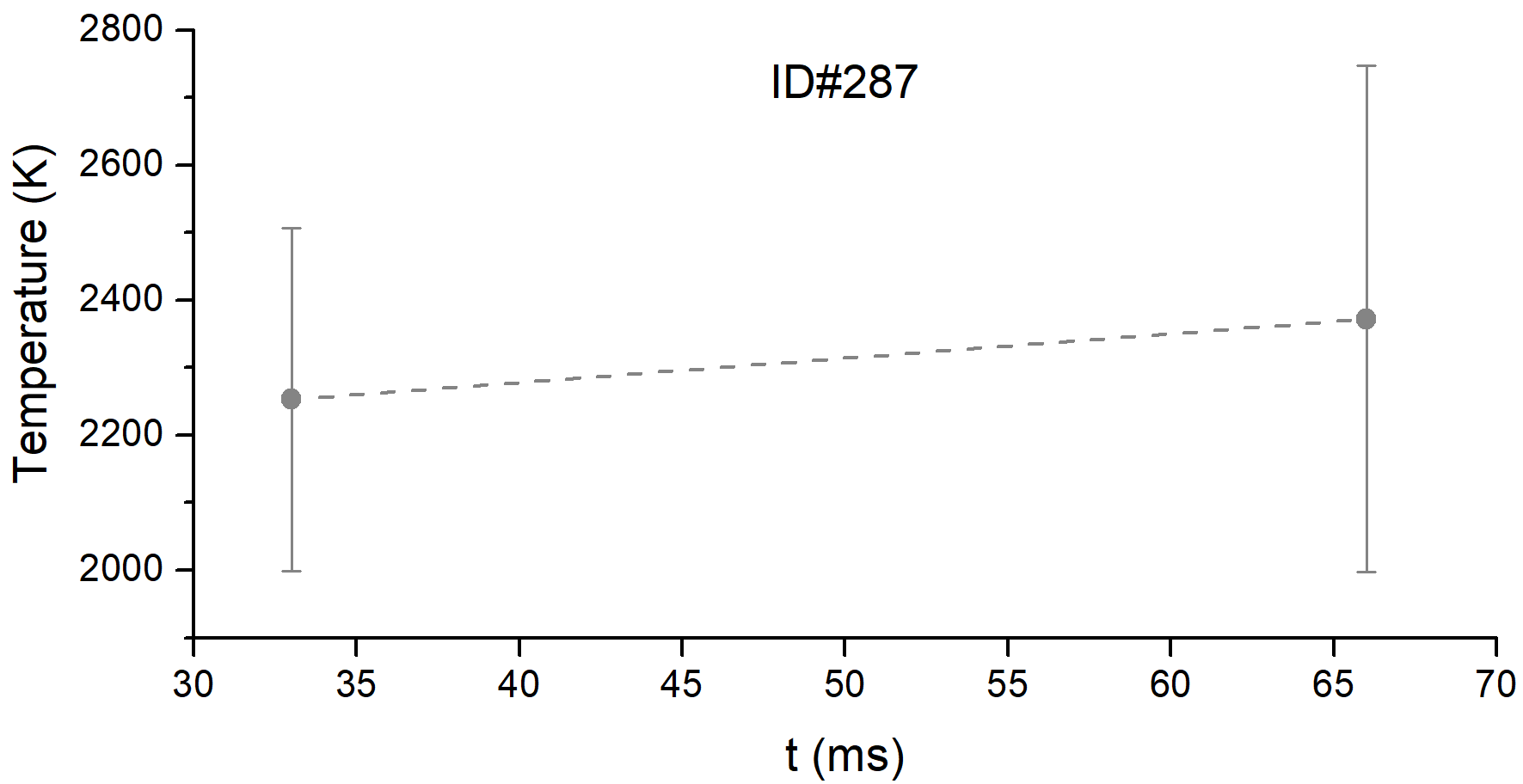}\\
\end{tabular}
\caption{Temperature evolution curves of the multiframe flashes.}
\label{fig:TCs1}
\end{figure*}

\section{Stream and sporadic meteoroid parameters}
\label{sec:metassoc}

\begin{table*}
\centering
\caption{Parameters of sporadic and stream meteoroids.}
\label{tab:streams}
\scalebox{0.85}{
\begin{tabular}{ccccc ccccc ccccc c}
\hline\hline																											
CODE	&	$\lambda_{\sun,\rm beg}$	&	$\lambda_{\sun,\rm max}$	&	$\lambda_{\sun,\rm end}$	&	$RA $	&	$Dec$	&	$l$	&	$b$	&	$V_{\rm p}$ 	&	$ZHR~(max)$	&	$B$	&	$m_0$	&	$s$	&	$r$	&	$\rho^a$	\\
	&	($\degr$)	&	($\degr$)	&	($\degr$)	&	($\degr$)	&	($\degr$)	&	($\degr$)	&	($\degr$)	&	(km~s$^{-1}$)	&	(hr$^{-1}$)	&	(deg$^{-1}$)	&	($\times10^{-7}$~kg)	&		&		&	(g~cm$^{-3}$)	\\
\hline																													
SPO	&		&		&		&		&		&		&		&	17	&		&		&	98.06	&	2.19	&	3	&	1.8(3)	\\
Lyr	&	29.2	&	31.7	&	34.2	&	272	&	33	&	273	&	56.4	&	49	&	18	&	0.22	&	1.09	&	1.81	&	2.1	&		\\
ETA	&	27.3	&	45.8	&	64.3	&	338	&	$-1$	&	339.3	&	7.64	&	66	&	50	&	0.08	&	0.31	&	1.95	&	2.4	&		\\
Ari	&	57	&	76	&	95	&	45	&	23	&	49.15	&	5.71	&	38	&	30	&	0.1	&	3.21	&	2.12	&	2.8	&		\\
SDA	&	99.4	&	124.9	&	150.4	&	339	&	$-17$	&	334.2	&	$-7.58$	&	41	&	25	&	0.091	&	2.33	&	1.99	&	2.5	&	2.4(6)	\\
Per	&	126.99	&	139.49	&	151.99	&	46	&	58	&	61.82	&	38.78	&	61	&	84	&	0.2	&	0.50	&	1.75	&	2	&	1.2(2)	\\
Ori	&	192.9	&	207.9	&	222.9	&	95	&	16	&	94.84	&	$-7.36$	&	67	&	20	&	0.12	&	0.31	&	1.99	&	2.5	&	0.9(5)	\\
Leo	&	230.4	&	234.4	&	238.4	&	153	&	22	&	147.06	&	10.15	&	71	&	23	&	0.39	&	0.24	&	1.92	&	2.33	&	0.4(1)	\\
Gem	&	234.4	&	261.4	&	288.4	&	112	&	32	&	108.8	&	9.99	&	36	&	88	&	0.39	&	4.56	&	1.95	&	2.4	&	2.9(6)	\\
QUA	&	253	&	283	&	313	&	231.5	&	48.5	&	203.3	&	63.3	&	41.7	&	120	&	1.8	&	2.33	&	2.30	&	1.9	&	1.9(2)	\\
Urs	&	268.3	&	270.3	&	272.3	&	223	&	78	&	120.42	&	72.5	&	35	&	11.8	&	0.61	&	4.56	&	2.19	&	3	&		\\
\hline																									
\end{tabular}}
\tablefoot{$^a$taken from \citet{BAD09}}
\end{table*}

\section{Extra figures}
\label{sec:extrafig}

\begin{figure}
\centering
\includegraphics[width=8.8cm]{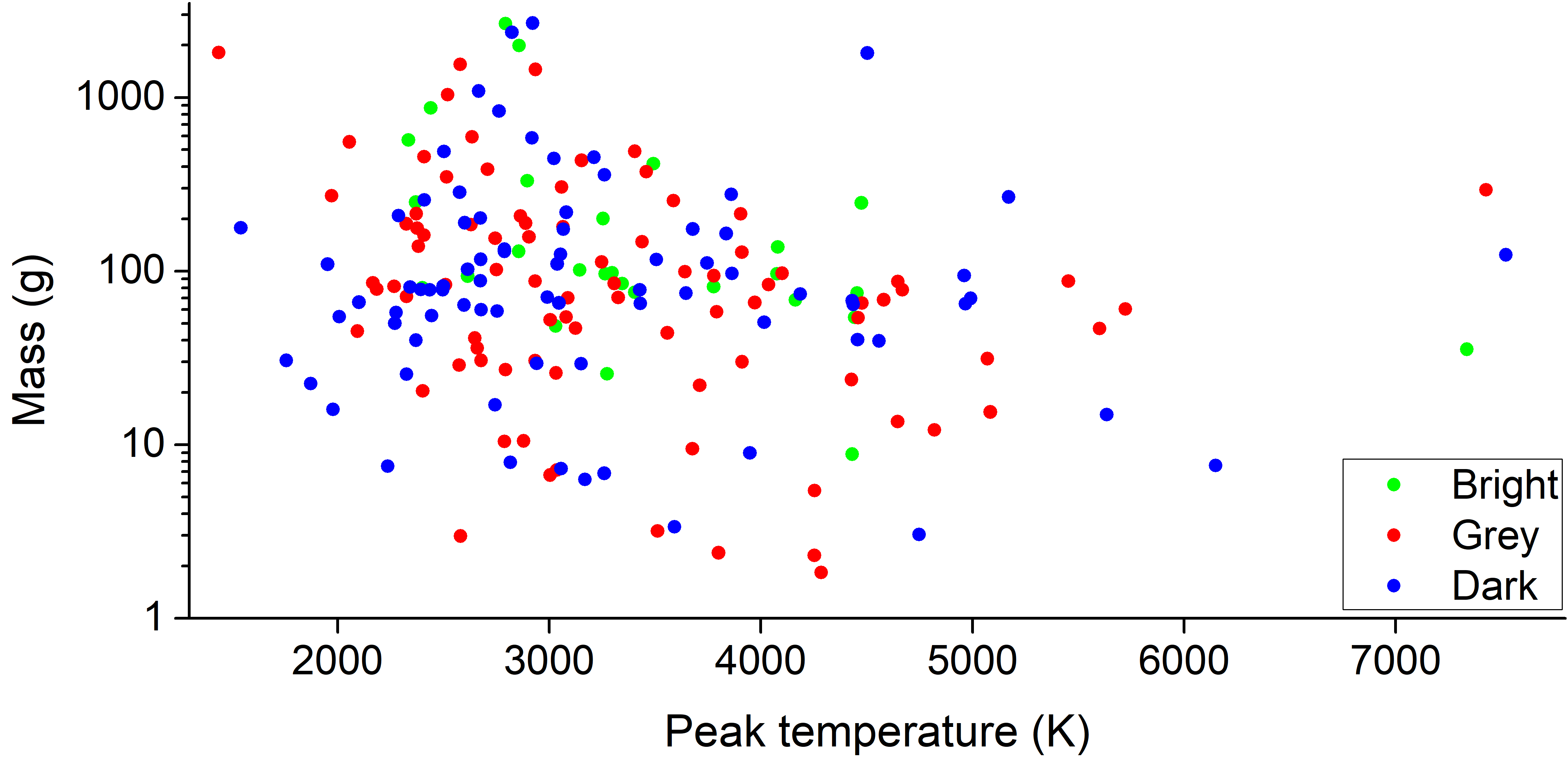}\\
\includegraphics[width=8.8cm]{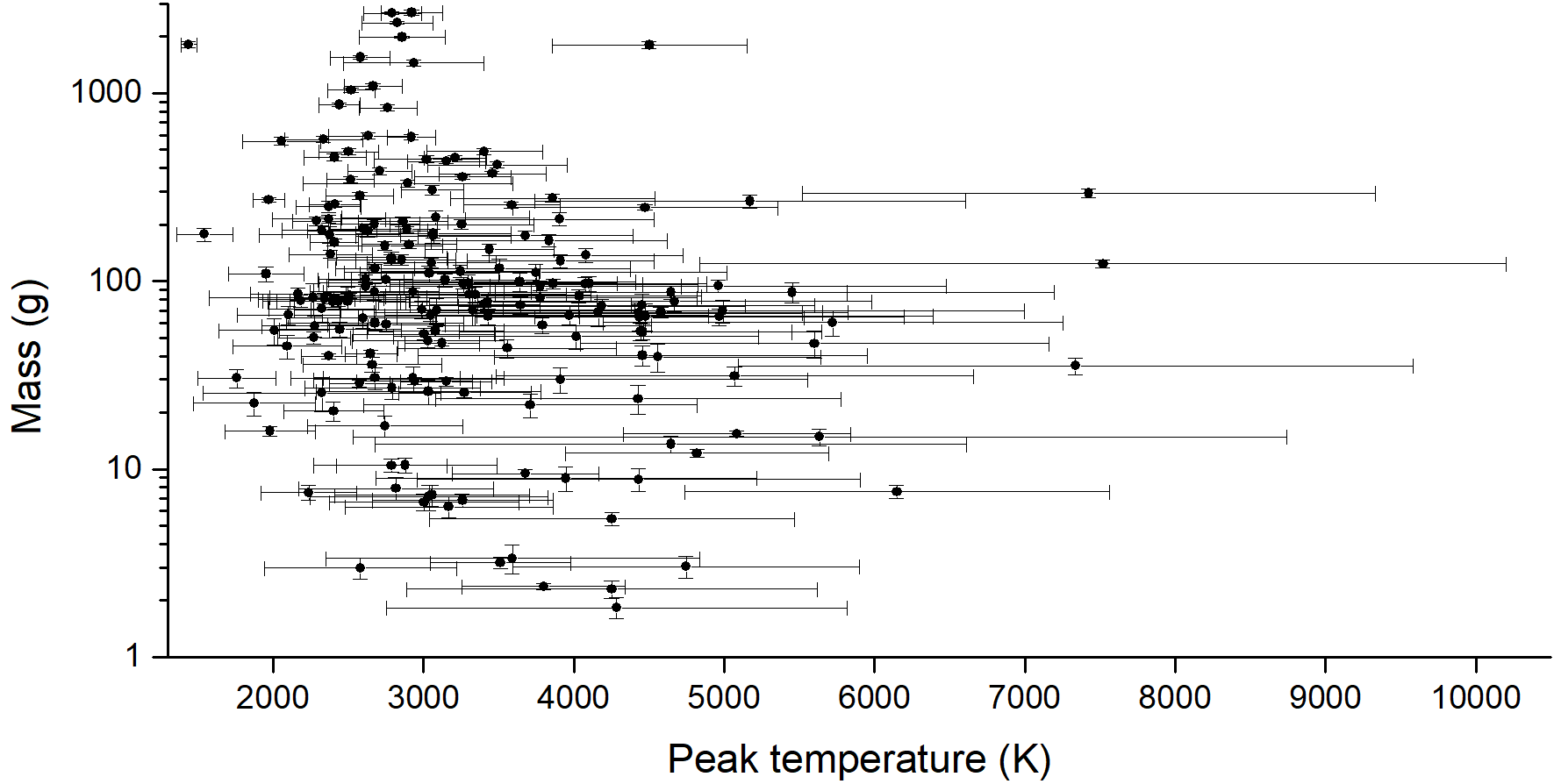}\\
\caption{Correlation attempt of masses of meteoroids with the peak temperatures evolved during the impacts. Colors in the upper panel denote the rough characterization of the albedo of the local lunar soil where the impacts occurred. The lower panel is the same with the upper and contains the error bars.}
\label{fig:MT}
\end{figure}

\begin{figure}
\centering
\includegraphics[width=8.8cm]{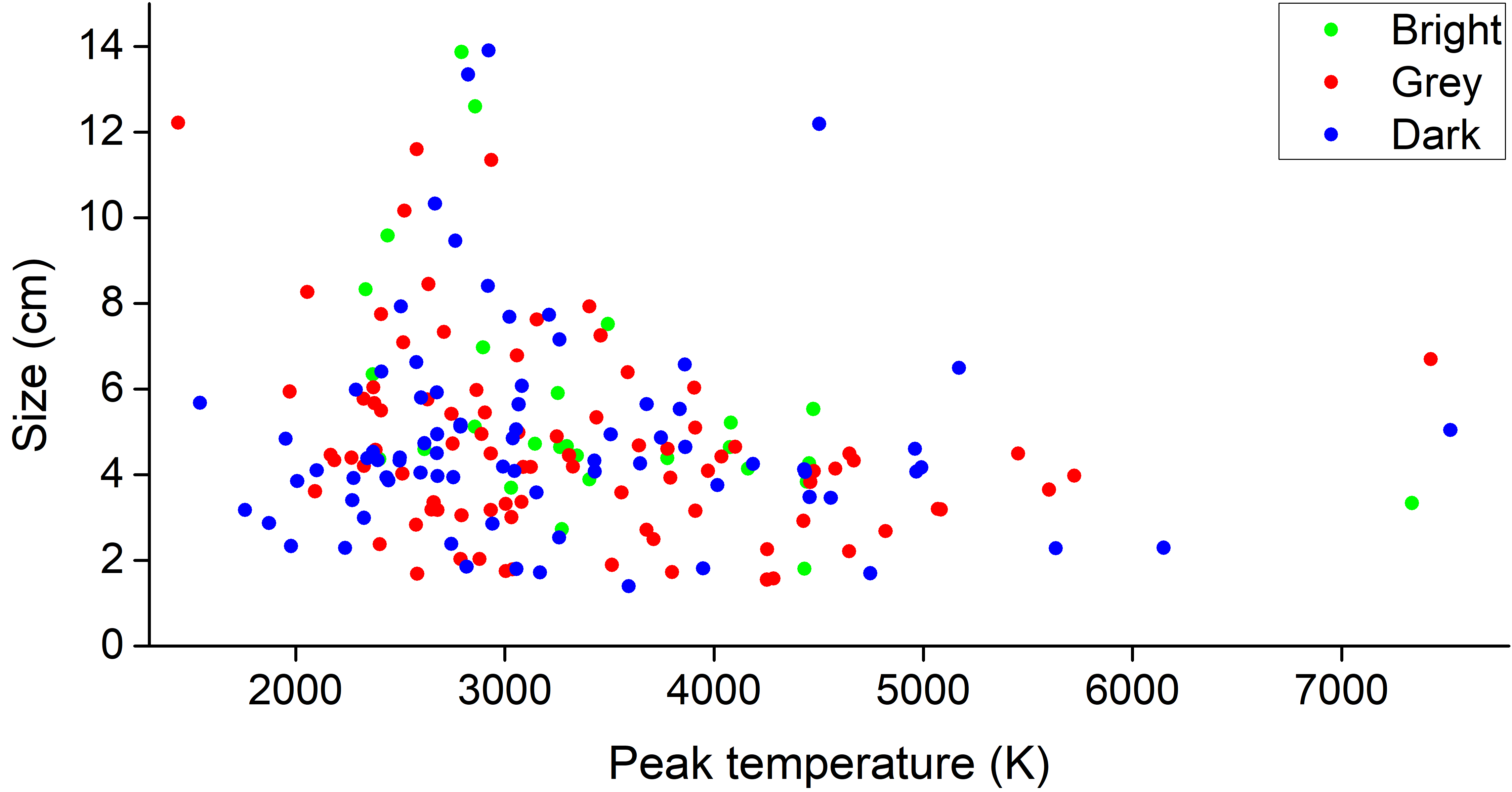}\\
\includegraphics[width=8.8cm]{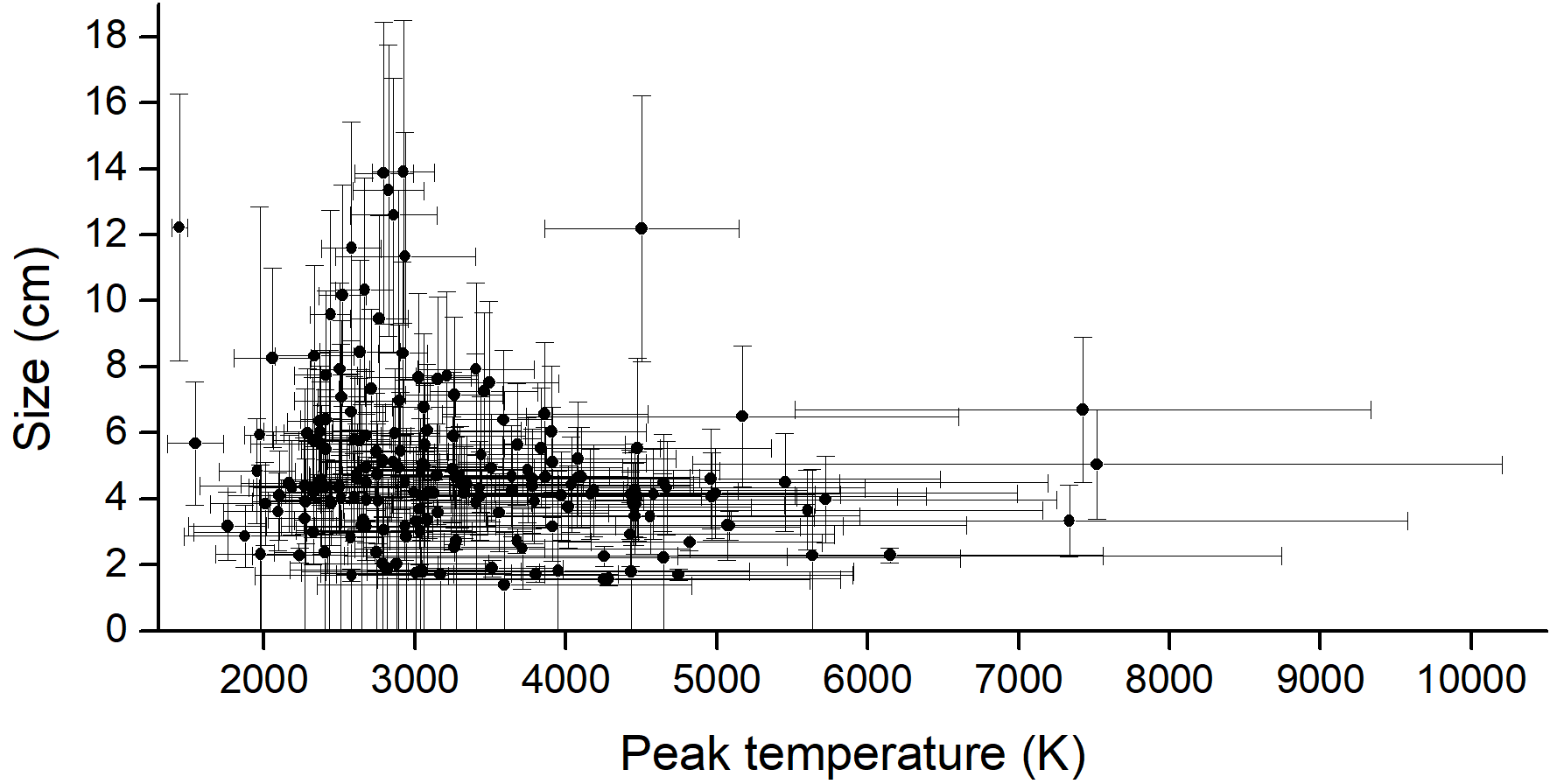}\\
\caption{Correlation attempt of radii of meteoroids with the peak temperatures evolved during the impacts. Panels and symbols contain the same as in Fig.~\ref{fig:MT}.}
\label{fig:RT}
\end{figure}

\end{appendix}
\end{document}